\documentclass{lmcs}
\pdfoutput=1
\usepackage[utf8]{inputenc}

\usepackage{lastpage}
\lmcsdoi{21}{2}{8}
\lmcsheading{}{\pageref{LastPage}}{}{}%
{Sep.~27,~2021}{Apr.~24,~2025}{}

\usepackage{graphicx}
\graphicspath{{diagrams/}}

\usepackage{amsmath}
\usepackage{algorithmic}
\usepackage{array}
\usepackage[caption=false,font=footnotesize]{subfig}
\usepackage{dblfloatfix}
\usepackage{url}
\usepackage{wrapfig}
\usepackage{lipsum}
\usepackage{doi}
\usepackage{etex}      
\usepackage{amssymb}		
\usepackage{cmll}
\usepackage{stmaryrd}
\usepackage{proof}
\usepackage[all]{xy}

\usepackage{wrapfig}
\usepackage{color}	
\usepackage{multirow}		
\usepackage{diagbox}
\usepackage{tikz}
\usetikzlibrary{automata, positioning, arrows, decorations.pathmorphing,decorations.pathreplacing,calc, shapes.geometric, positioning}
\usepackage{varwidth}
\usepackage{graphbox}

\usepackage{amsthm}
\theoremstyle{definition}
\newtheorem{definition}[thm]{Definition}

\newtheorem{notation}[definition]{Notation}
\newtheorem{terminology}[definition]{Terminology}
\theoremstyle{plain}

\newtheorem*{theorem*}{Theorem}

\newcommand{\N}{\mathbb{N}}

\newcommand{\Opr}{\mathbb{O}}
\newcommand{\MVar}{\mathbb{M}}

\newcommand{\FV}{\mathrm{FV}}
\newcommand{\FA}{\mathrm{FA}}

\newcommand{\keyw}[1]{\mathtt{#1}}

\newcommand{\lrapp}{\mathop{\overset{\rightarrow}{@}}}
\newcommand{\letin}[3]{\keyw{let}\ {#1} = {#2}\ \keyw{in}\ {#3}}

\newcommand{\contr}{
\mathrel{\raisebox{.2\height}{\scalebox{.6}{\boldmath $\otimes$}}}}

\newcommand{\focussed}[1]{{\dot{#1}}}

\newcommand{\dotrel}[1]{\mathrel{\dot{#1}}}

\newcommand{\Oprex}{\mathbb{O}^\mathrm{ex}}
\newcommand{\Oprlin}{\mathbb{O}^\mathrm{lin}}
\newcommand{\cbf}{{\text{-}\mathrm{bf}}}
\newcommand{\cdp}{{\text{-}\mathrm{dp}}}

\newcommand{\Mag}[1]{\textcolor{magenta}{#1}}

\newcommand{\To}{\Rightarrow}

\newcommand{\twoheadmapsto}{
\mathrel{\ooalign{$\twoheadrightarrow$\cr%
\kern-.15ex\raisebox{.2ex}{\scalebox{1}[0.8]{$\shortmid$}}\cr}}}
\newcommand{\longtwoheadmapsto}{
\mathrel{\ooalign{$\longtwoheadrightarrow$\cr%
\kern-.15ex\raisebox{.2ex}{\scalebox{1}[0.8]{$\shortmid$}}\cr}}}

\newcommand{\tobul}{\mathrel{\stackrel{
           \raisebox{.5ex}{$\scriptstyle\bullet\,$}}{
           \raisebox{0ex}[0ex][0ex]{$\rightarrow$}}}}

\definecolor{olive}{rgb}{0.5,0.5,0}

\newcommand{\defeq}{\stackrel{\mathit{def}}=}

\newcommand{\bqed}{\renewcommand\qedsymbol{$\blacksquare$} \qed}

\newcommand{\LV}{L}
\newcommand{\LE}{M}
\newcommand{\LVlin}{L_\mathrm{lin}}
\newcommand{\LElin}{M_\mathrm{lin}}
\newcommand{\LVgen}{L_\mathrm{gen}}
\newcommand{\LEgen}{M_\mathrm{gen}}
\newcommand{\HN}[2]{\mathcal{H}_\omega(#1,#2)}

\newcommand{\UAM}{\mathcal{U}}

\definecolor{linkcolor}{rgb}{0,0,0.5}

\usepackage{hyperref}
\hypersetup{
    colorlinks=true,
    linkcolor=linkcolor,
    filecolor=linkcolor,      
    citecolor=linkcolor,
}
\newcommand{\green}{\textcolor{green!50!black}}
\newcommand{\cyan}{\textcolor{cyan!70!black}}

\begin{document}

\title{A robust graph-based approach to observational equivalence}

\author[D. R.~Ghica]{Dan R. Ghica\lmcsorcid{0000-0002-4003-8893}}[a]
\author[K. Muroya]{Koko Muroya\lmcsorcid{0000-0003-0454-6900}}[b]
\author[T. Waugh Ambridge]{Todd Waugh Ambridge}[a]

\address{University of Birmingham, UK}
\email{d.r.ghica@cs.bham.ac.uk, t.waughambridge@bham.ac.uk}
\address{National Institute of Informatics, Japan}
\email{kmuroya@nii.ac.jp}

%



\begin{abstract}
 We propose a new step-wise approach to proving observational equivalence, and in particular reasoning about fragility of observational equivalence.
 Our approach is based on what we call local reasoning. The local reasoning exploits the graphical concept of neighbourhood, and it extracts a new, formal, concept of robustness as a key sufficient condition of observational equivalence.
 Moreover, our proof methodology is capable of proving a generalised notion of observational equivalence. The generalised notion can be quantified over syntactically restricted contexts instead of all contexts, and also quantitatively constrained in terms of the number of reduction steps.
 The operational machinery we use is given by a hypergraph-rewriting abstract machine inspired by Girard's Geometry of Interaction. The behaviour of language features, including function abstraction and application, is provided by hypergraph-rewriting rules.
 We demonstrate our proof methodology using the call-by-value lambda-calculus equipped with (higher-order) state. 
\end{abstract}


%

\maketitle

\section{Introduction} \label{sec:introduction}

\subsection{Context and motivation}

\emph{Observational equivalence}~\cite{morris1969lambda} is an old and central question in the study of programming languages. Two executable programs are observationally equivalent when they have the same behaviour. Observational equivalence between two program fragments (aka.\ terms) is the smallest congruence with respect to arbitrary program contexts. By formally establishing observational equivalence, one can justify compiler optimisation, and verify and validate programs.

There are two mathematical challenges in proving observational equivalence.
Firstly, universal quantification over contexts is unwieldy. This has led to various indirect approaches to observational semantics.
As an extremal case, \emph{denotational semantics} provides a model-theoretic route to observational equivalence. There are also hybrid approaches that employ both denotational and operational techniques, such as \emph{Kripke logical relations}~\cite{statman1985logical} and \emph{trace semantics}~\cite{jeffrey2005java}. Moreover, an operational and coinductive approach exists, under the name of \emph{applicative bisimilarity}~\cite{Abramsky90AppBisim}.

The second challenge is \emph{fragility} of observational equivalence.
The richer a programming language is, the more discriminating power program contexts have and hence the less observational equivalences hold in the language.
For example, the beta-law $(\lambda x. t)\ v \simeq t[v/x]$ is regarded as the fundamental observational equivalence in functional programming. However, it can be violated in the presence of a memory inspection feature like the one provided by the OCaml garbage-collection (Gc) module. A function that returns the size of a given program enables contexts to distinguish $(\lambda x. 0)\ 100$ from $0$, for example\footnote{See a concrete example written in OCaml, on the online platform \emph{Try It Online}: \url{https://bit.ly/3TqnGOW}}.

The fragility of observational equivalence extends to its proof methodologies.
There have been studies of the impact that language features have on semantics and hence on proof methodologies for observational equivalence.
The development of \emph{game semantics} made it possible to give combinatorial, syntax-independent and orthogonal characterisations for classes of features such as state and control, e.g.\ the so-called ``Abramsky cube''~\cite{abramsky1997game,Ghica23cube}, or to replace the syntactic notion of context by an abstracted \emph{adversary}~\cite{DBLP:journals/entcs/GhicaT12}.
A classification~\cite{DBLP:journals/jfp/DreyerNB12} and characterisation~\cite{Dreyer10} of language features and their impact on reasoning has also been undertaken using logical relations.
Applicative bisimilarity has been enriched to handle effects such as algebraic effects, local state, names and continuations~\cite{SangiorgiKS07,KoutavasLS11,DalLagoGL17,SimpsonV18}.

\subsection{Overview and contribution}

What is missing and desirable seems a general semantical framework with which one can directly analyse fragility, or robustness, of observational equivalences.
To this end, we introduce a graphical abstract machine that implements \emph{focussed hypernet rewriting}. We then propose a radically new approach to proving observational equivalence that is based on \emph{step-wise} and \emph{local} reasoning and centred around a concept of \emph{robustness}. All these concepts will be rigorously defined in the paper.

The main contribution of the paper is rather conceptual, showing how the graphical concept of neighbourhood can be exploited to reason about observational equivalence, in a new and advantageous way.
The technical development of the paper might seem quite elaborate, but this is because we construct a whole new methodology from scratch: namely, focussed hypernet rewriting and reasoning principles for it. These reasoning principles enable us to analyse fragility, or robustness, of observational equivalences in a formal way.

We introduce and use \emph{hypernets} to represent (functional) programs with effects.
Hypernets are an anonymised version of abstract syntax trees, where variables are simply represented as connections. Formally, hypernets are given by hierarchical hypergraphs. Hierarchy allows a hypergraph to be an edge label recursively.
An extensive introduction to hypernets and rewriting of them can be found in the literature~\cite{ZanasiG23tutorial}.

Given a hypernet that represents a term, its evaluation is modelled by step-by-step traversal and update of the hypernet. Traversal steps implement depth-first search on the hypernet for a redex, and each update step triggers application of a rewrite rule to the hypernet. Traversal and update are interwoven strategically using a \emph{focus} that is simply a dedicated edge passed around the hypernet. Importantly, updates are always triggered by a certain focus, and designed to only happen around the focus.
We call this model of evaluation \emph{focussed hypernet rewriting}.

There are mainly two differences compared with conventional reduction semantics. The first difference is the use of hypernets instead of terms. This makes renaming of variables irrelevant. The second difference is the use of the focus instead of evaluation contexts. In conventional reduction semantics, redexes are identified using evaluation contexts. Whenever the focus triggers an update of a hypernet, its position in the hypernet coincides with where the hole is in an evaluation context.

\subsubsection*{A new step-wise approach.}
This work takes a new coinductive, \emph{step-wise}, approach to proving observational equivalence.
We introduce a novel variant of the weak simulation dubbed \emph{counting simulation}. We demonstrate that, to prove observational equivalence, it suffices to construct a counting simulation that is closed under contexts by definition.
This approach is opposite to the known coinductive approach which uses applicative bisimilarity; one first constructs an applicative bisimulation and then proves that it is a congruence, typically using Howe's method~\cite{Howe96}.

\subsubsection*{Local reasoning.}
In combination with our new step-wise approach, focussed hypernet rewriting facilitates what we call \emph{local} reasoning. Our key observation is that, to obtain the counting simulation that is closed under contexts by definition, it suffices to simply trace sub-graphs and analyse their interaction with the focus. The interaction can namely be analysed by inspecting updates that happen around the focus and how these updates can interfere with the sub-graphs of interest. The reasoning principal here is the graphical concept of neighbourhood, or graph locality.

The local reasoning is a graph counterpart of analysing interaction between (sub-)terms and contexts using the conventional reduction semantics. In fact, it is not just a counterpart but an enhancement in two directions. Firstly, sub-graphs are more expressive than sub-terms; sub-graphs can represent parts of a program that are not necessarily well-formed. Secondly, the focus can indicate which part of a context is relevant in the interaction between the context and a term, which is not easy to make explicit in the conventional semantics that uses evaluation contexts.

\subsubsection*{Robustness.}
Finally, local reasoning extracts a formal concept of \emph{robustness} in proving observational equivalence. Robustness is identified as the key sufficient condition that ensures two sub-graphs that we wish to equate interact with updates of a hypernet, which is triggered by the focus, in the \emph{same} way; for example, if one sub-graph is duplicated (or discarded), the other is also duplicated (or discarded).

The concept of robustness helps us gain insights into fragility of observational equivalence. If robustness of two sub-graphs $G,H$ fails, we obtain a counterexample, which is given by a rewrite rule that interferes with the two sub-graphs in different ways. Let $G',H'$ be the two different results of interference (i.e.\ $G'$ is the result of updating $G$, and $H'$ is the result of updating $H$). There are two possibilities.
\begin{enumerate}
 \item The sub-graphs $G,H$ are actually observationally equivalent. In this case, the counterexample suggests that the two different results $G',H'$ should first be equated. The observational equivalence $G \simeq H$ we wish to establish is likely to depend on the ancillary observational equivalence $G' \simeq H'$.
 \item The observational equivalence $G \simeq H$ fails too. In this case, the counterexample provides the particular computation that violates the equivalence, in terms of a rewrite rule. We can conclude that the language feature that induces the computation violates the observational equivalence.
\end{enumerate}

\subsubsection*{Generalised contextual equivalence.}
Using focussed hypernet rewriting, we propose a \emph{generalised} notion of contextual equivalence.
The notion has two parameters: a class of contexts and a preorder on natural numbers.
The first parameter enables us to quantify over syntactically restricted
contexts, instead of all contexts as in the standard notion. This can
be used to identify a shape of contexts that respects or violates
certain observational equivalences, given that not necessarily all
arbitrarily generated contexts arise in program execution.
The second parameter, a preorder on natural numbers, deals with
numbers of steps it takes for program execution to terminate. Taking the
universal relation recovers the standard notion of contextual
equivalence.
Another instance is contextual equivalence with respect to the
greater-than-or-equal relation on natural numbers, which resembles the notion of
\emph{improvement}~\cite{Sands95,AccattoliLV20,AccattoliLV21} that is used to establish equivalence
and also to compare efficiency of abstract machines.
This instance of contextual equivalence is useful to establish that two
programs have the same observable execution result, and also that one program
terminates with fewer steps than the other.

\subsection{Organisation of the paper}

\autoref{sec:gentle-introduction} provides a gentle introduction to our graph-based approach to modelling program evaluation, and reasoning about observational equivalence with the key concepts of locality and robustness.
\autoref{sec:hypernets} formalises the graphs we use, namely hypernets.
The rest of the paper is in two halves.

In the first half, we develop our reasoning framework, targeting the linear lambda-calculus. Although linear lambda-terms have restricted expressive power, they are simple enough to demonstrate the development throughout.
\autoref{sec:linear-lambda-calc} presents the hypernet representation of linear lambda-terms.
\autoref{sec:focussed-hypernet-rewriting-UAM} then presents our operational semantics, i.e.\ focussed hypernet rewriting. \autoref{sec:UAM-details} formalises it as an abstract machine called \emph{universal abstract machine (UAM)}.

\autoref{sec:ctxt-refinement} sets the target of our proof methodology, introducing the generalised notion of contextual equivalence. \autoref{sec:partial-char-thm} presents our main technical contributions: it formalises the concept of robustness, and presents our main technical result which is the sufficiency-of-robustness theorem (\autoref{thm:MetaThm}).

In the second half, we extend our approach to the general (non-linear) lambda-calculus equipped with store.
\autoref{sec:variable-sharing-store} describes how the hypernet representation can be adapted.
\autoref{sec:copying-uam} shows how the UAM can be extended accordingly, and presents the \emph{copying} UAM.
\autoref{sec:obs-equiv-lambda} formalises observational equivalence between lambda-terms by means of contextual equivalence between hypernets.
\autoref{sec:ex-law} then demonstrates our approach by proving some example equivalences for the call-by-value lambda-calculus extended with state.
The choice of the language here is pedagogical; our methodology can accommodate other effects as long as they are deterministic.

Finally, \autoref{sec:related-future-work} discusses related and future work, concluding the paper.
Some details of proofs are presented in Appendix.

\section{A gentle introduction} \label{sec:gentle-introduction}

\subsection{Hypernets} \label{sec:gentle-introduction-hypernets}

Compilers and interpreters deal with programs mainly in the form of an abstract syntax tree (AST) rather than text. 
It is broadly accepted that such a data structure is easier to manipulate algorithmically. 
Somewhat curiously perhaps, reduction semantics (or small-step operational semantics), which is essentially a list of rules for program manipulation, is expressed using text rather than the tree form. In contrast, our graph-based semantics is expressed as algorithmic manipulations of the data structure that represents syntax.

\begin{figure}[p]
 \centering
 \subfloat[ASTs
 \label{fig:AST-beta}]{
 \includegraphics[width=.8\linewidth]{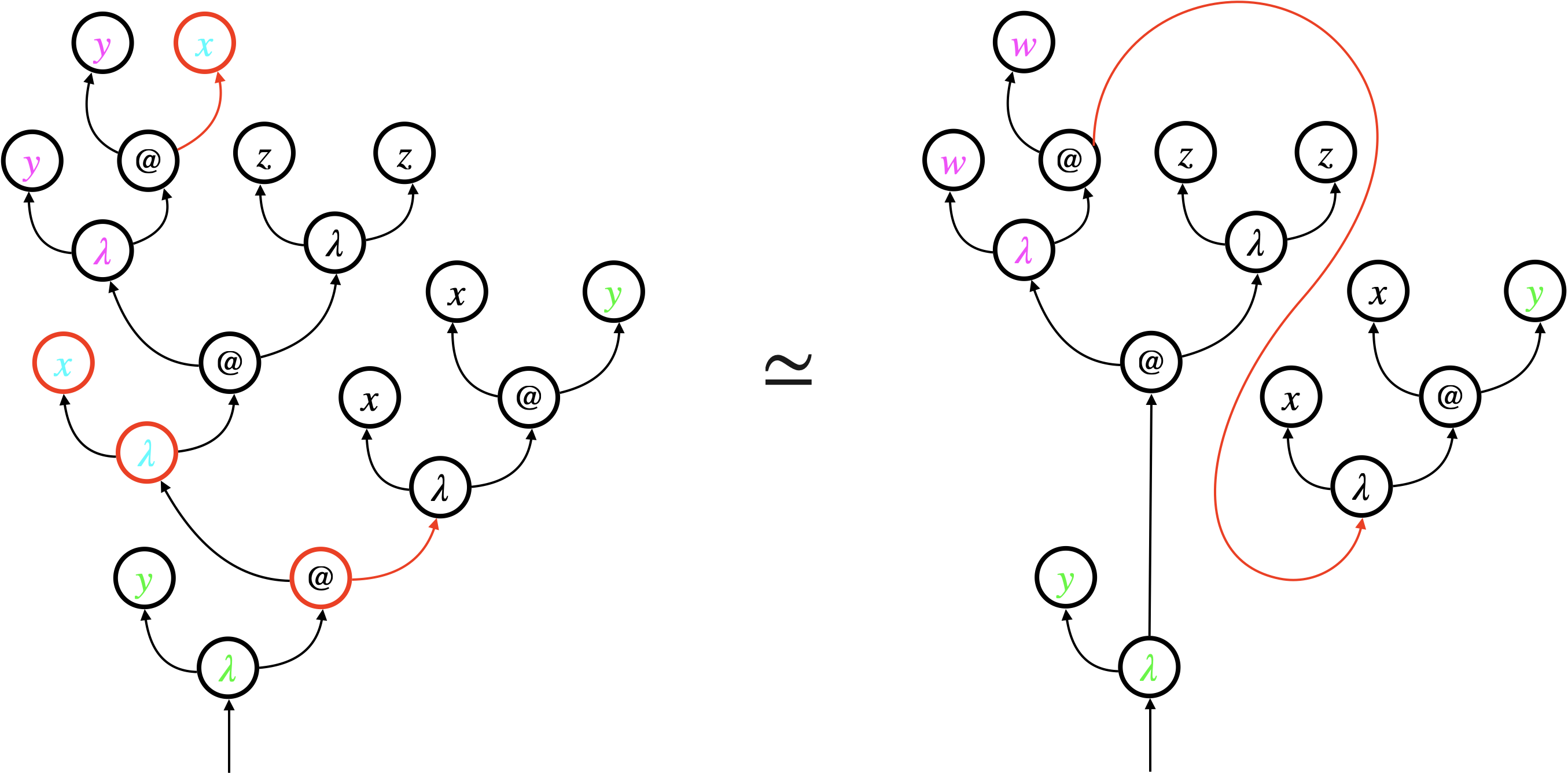}
 }
 
 \subfloat[Hypernets
 \label{fig:hypernet-beta}]{
 \includegraphics[width=.8\linewidth]{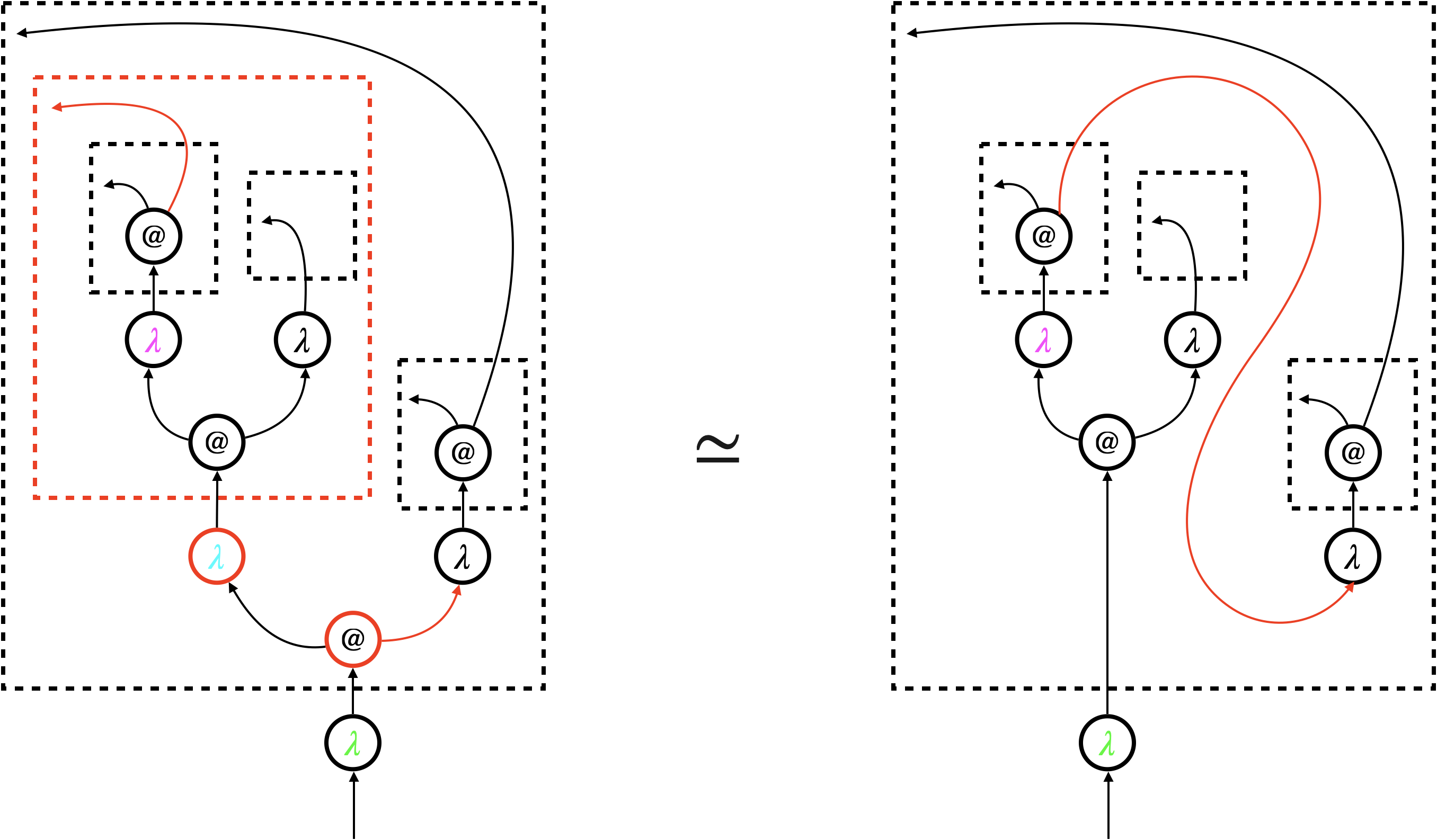}
 }
 
 \subfloat[Decorated hypernets
 \label{fig:decorated-hypernet-beta}]{
 \includegraphics[width=.8\linewidth]{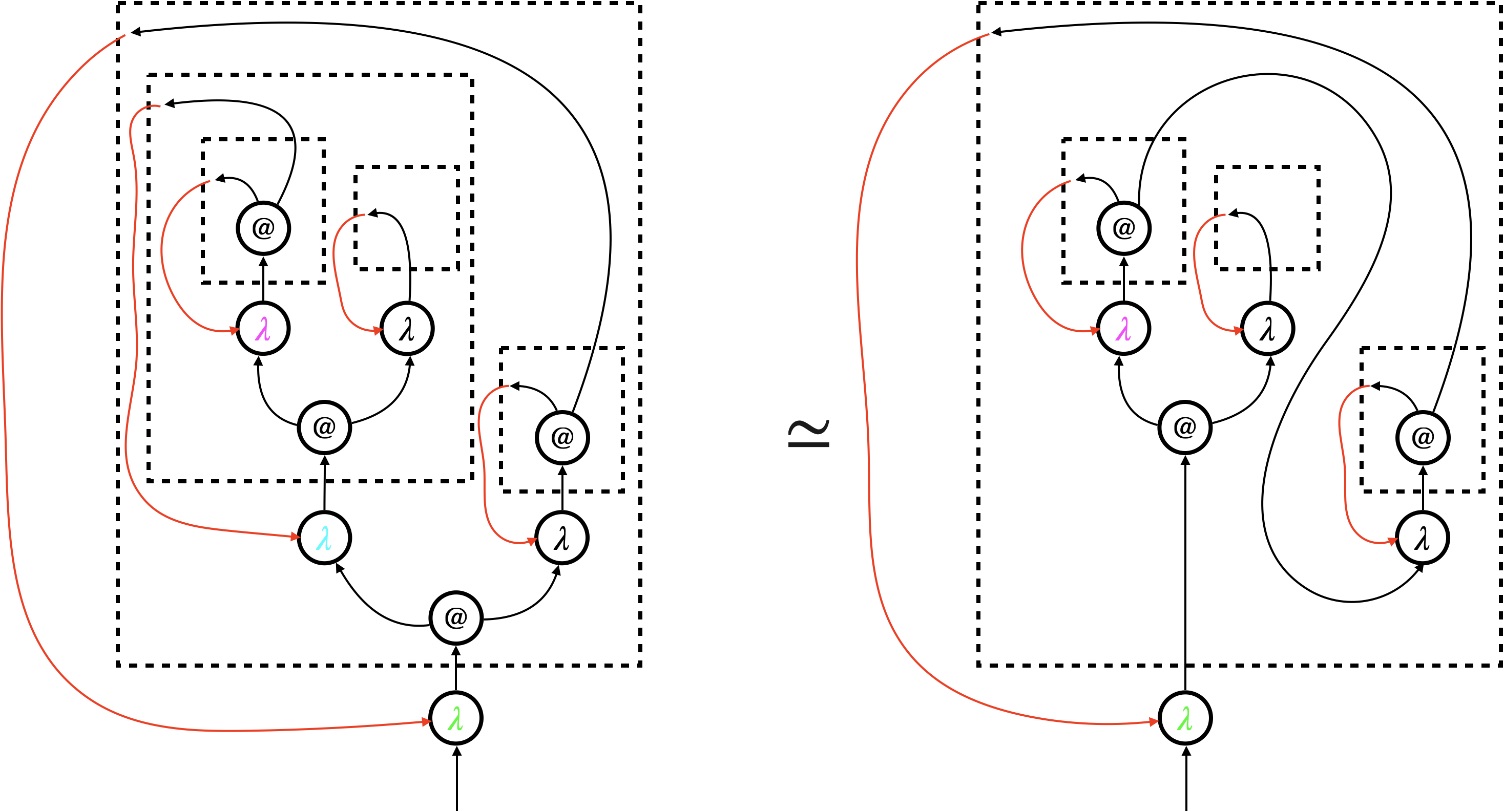}
 }
 \caption{The beta-law $
\green{\lambda y}. (\cyan{\lambda x}. (\textcolor{magenta}{\lambda y}. \textcolor{magenta}{y}\;\cyan{x})\;(\lambda z. z))\;(\lambda x. x\;\green{y})
 \simeq
 \green{\lambda y}. (\textcolor{magenta}{\lambda w}. \textcolor{magenta}{w}\;(\lambda x. x\;\green{y}))\;(\lambda z. z)
 $}
\end{figure}

Let us demonstrate our graphical representation, using the beta-law
\begin{equation}
 \green{\lambda y}. (\cyan{\lambda x}. (\textcolor{magenta}{\lambda y}. \textcolor{magenta}{y}\;\cyan{x})\;(\lambda z. z))\;(\lambda x. x\;\green{y})
  \simeq
  \green{\lambda y}. (\textcolor{magenta}{\lambda w}. \textcolor{magenta}{w}\;(\lambda x. x\;\green{y}))\;(\lambda z. z)
  \label{eq:linear-beta-ex}
\end{equation}
in the linear lambda-calculus.
We use colours to clarify some variable scopes.
This law substitutes $\lambda x. x\;\green{y}$ for the variable $\cyan{x}$, and in doing so, the bound variable $\textcolor{magenta}{y}$ has to be renamed to $\textcolor{magenta}{w}$ so it does not capture the variable $\green{y}$ in $\lambda x. x\;\green{y}$.

Our first observation is that ASTs are not satisfactory to represent syntax, when it comes to define operational semantics.
They contain more syntactic details than necessary, namely by representing variables using names. This makes an operation on terms like substitution a global affair. To avoid variable capturing, substitution needs to clarify the scope of each variable and appropriately rename some variables.

\autoref{fig:AST-beta} shows the beta-law (\ref{eq:linear-beta-ex}) using ASTs.
The scope of each variable is not obvious in the ASTs, which is why we keep using colours to distinguish variable scopes.
The law deletes the four red nodes of the left AST, and connects two red arrows to represent substitution for $\green{x}$. Additionally, all occurrences of the variable $\textcolor{magenta}{y}$ has to be renamed to $\textcolor{magenta}{w}$.

We propose hypernets as an alternative graph representation. Hypernets, inspired by \emph{proof nets}~\cite{Girard87LL}, replace variable names with virtual connection, and hence keep variables anonymous.
Binding structures and scopes are made explicit by (dashed) boxes around sub-graphs.

\autoref{fig:hypernet-beta} shows the same beta-law (\ref{eq:linear-beta-ex}) using hypernets instead of ASTs. Each bound variable is simply represented by an arrow that points at the left edge of the associated dashed box. For example, the upper one of the two red arrows in the left hypernet represents the bound variable $\green{x}$. It points at the left edge of the red dashed box that represents the scope of the variable. The dashed box is connected to the corresponding binder ($\green{\lambda}$).

The beta-law requires relatively local changes to hypernets. In \autoref{fig:hypernet-beta}, the two red nodes are deleted, the associated red dashed box is also deleted, and the two red arrows are connected to represent substitution for $\green{x}$. There is no need for renaming $\textcolor{magenta}{y}$, as it is simply represented by an anonymous arrow.

\begin{rem}[Arrows representing bound variables]
In hypernets, the arrow representing a bound variable points at the left edge of the associated dashed box.
In other graphical notations (e.g.\ proof nets~\cite{Girard87LL}), the bound variable would be connected to the corresponding binder ($\lambda$), as shown by red arrows in \autoref{fig:decorated-hypernet-beta}. We treat these red arrows as mere \emph{decorations}, and exclude them from the formalisation of hypernets. We find that boxes suffice to delimit the scope of variables and sub-terms. Exclusion of decorations also simplifies the formalisation by reducing the number of loops in each hypernet.
 \bqed
\end{rem}

\subsection{Focussed hypernet rewriting} \label{sec:gentle-introduction-focussed-hypernet-rewriting}

The main difference between a \emph{law} (an equation) and a \emph{reduction} is that the former can be applied in any context, at any time, whereas the latter must be applied \emph{strategically}, in a particular (evaluation) context and in a particular order.
Different reduction strategies, for instance, make different programming languages out of the same calculus.

The question to be addressed here is how to define strategies for determining redexes in hypernets.
Our operational semantics, i.e.\ \emph{focussed hypernet rewriting}, combines graph traversal with update, and exploits the traversal to search for a redex.

\begin{figure}[t]
 \includegraphics[width=.8\linewidth]{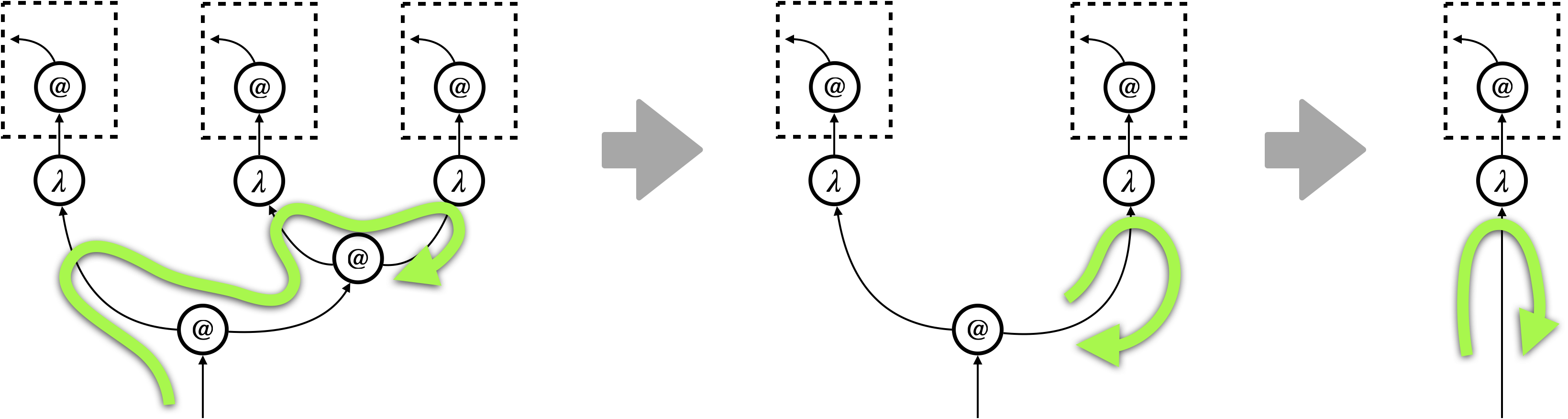}
 \caption{Reduction as graph traversal and update}
 \label{fig:traversal-and-update-ex}
\end{figure}

Let us illustrate focussed hypernet rewriting, using the call-by-value reduction of the linear lambda-term $(\lambda x. x)\;((\lambda y. y)\;(\lambda z. z))$ as shown in \autoref{fig:traversal-and-update-ex}.
The thicker green arrows are not part of the hypernets but they show the traversal.
The reduction proceeds as follows.
\begin{enumerate}
 \item The depth-first traversal witnesses that the abstraction $\lambda x. x$ is a value, and that the sub-term $(\lambda y. y)\;(\lambda z. z)$ contains two abstractions and it is ready for the beta-reduction.
       In the reduction, an application node ($@$) and its matching abstraction node ($\lambda$) are deleted, and the associated dashed box is removed. The argument $\lambda z. z$ is then connected to the bound variable $y$, yielding the second hypernet.
 \item The traversal continues on the resultant hypernet (representing $(\lambda x. x)\;(\lambda z. z)$), confirming that the result $\lambda z. z$ of the beta-reduction is a value. Note that the abstraction $\lambda x. x$ has already been inspected in the previous step, so the traversal does not repeat the inspection. It only witnesses the abstraction $\lambda z. z$ at this stage. The beta-reduction is then triggered, yielding the third and final hypernet representing $\lambda z. z$.
 \item The traversal confirms that the result $\lambda z. z$ of the beta-reduction is a value, and it finishes.
\end{enumerate}

\begin{figure}[t]
 \includegraphics[width=.8\linewidth]{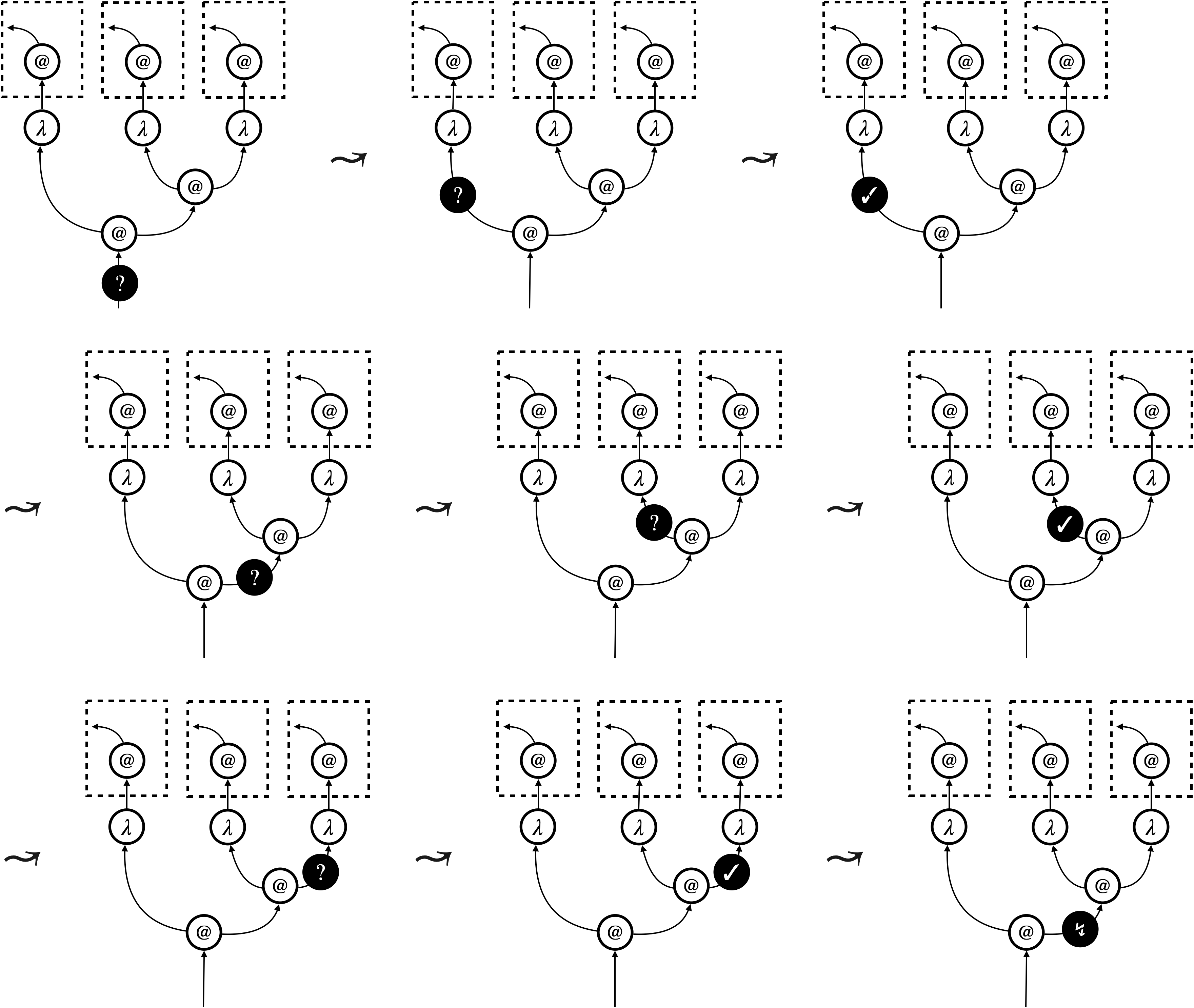}

 \includegraphics[width=.8\linewidth]{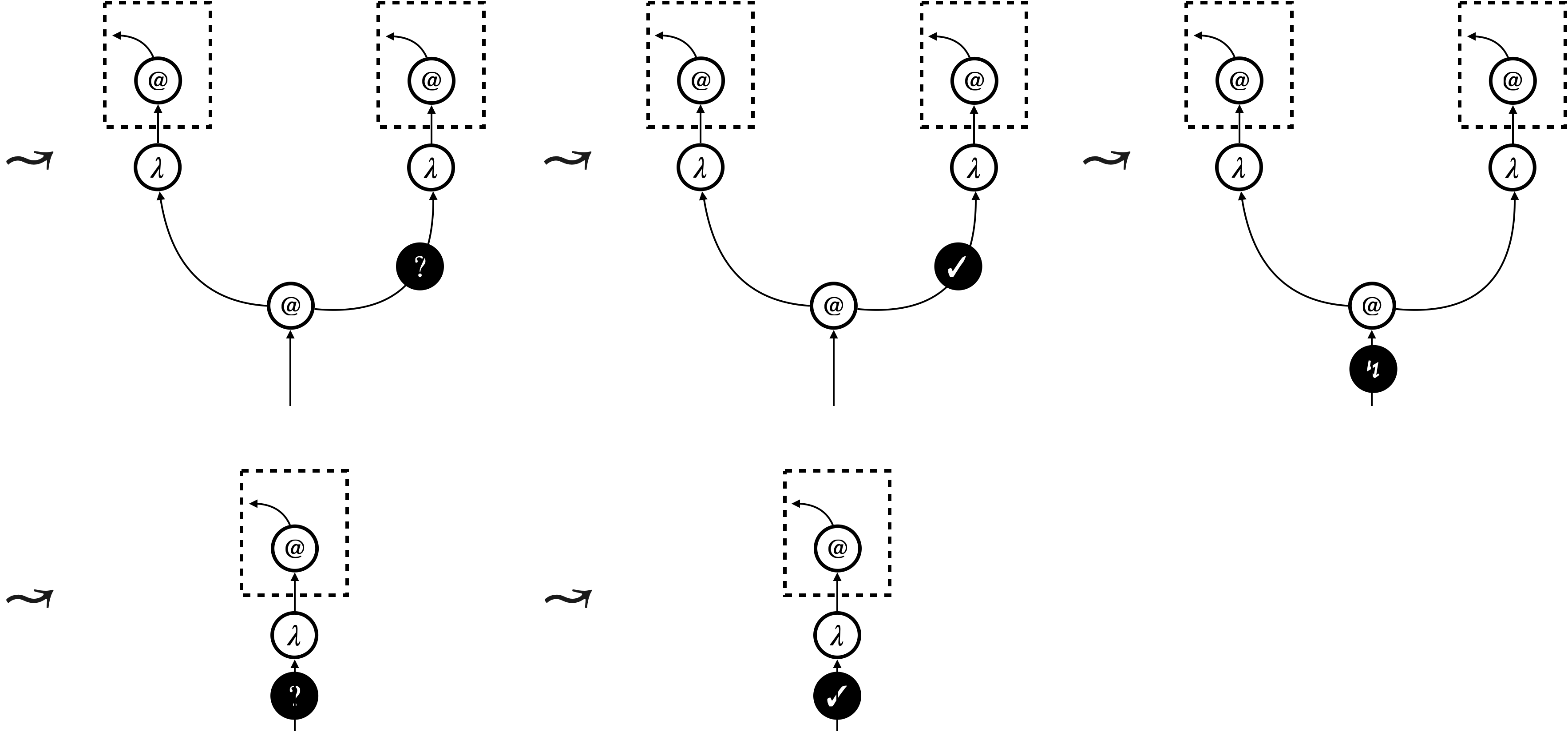}
 \caption{Graph traversal and update with a focus}
 \label{fig:focus-ex}
\end{figure}

We implement focussed hypernet rewriting, in particular the graph traversal (the thick green arrows in \autoref{fig:traversal-and-update-ex}), using a dedicated node dubbed \emph{focus}. A focus can be in three modes: searching ($?$), backtracking ($\checkmark$), and triggering ($\lightning$). The first two modes implement the depth-first traversal, and the last mode triggers update of the underlying hypernet.
\autoref{fig:focus-ex} shows how focussed hypernet rewriting actually proceeds\footnote{An interpreter and visualiser can be accessed online at \url{https://tnttodda.github.io/Spartan-Visualiser/}}, given the linear term $(\lambda x. x)\;((\lambda y. y)\;(\lambda z. z))$. The black nodes are the focus. The first eight steps $\rightsquigarrow$ altogether implement the thick green arrow in the first part of \autoref{fig:traversal-and-update-ex}. At the end of these steps, the focus changes to $\lightning$, signalling that the hypernet is ready for the beta-reduction. What follows is an update of the hypernet, which resets the focus to the searching mode ($?$), so the traversal continues and triggers further update.

Evaluation of a program $P$ starts, when the $?$-focus enters the hypernet representing $P$ from the bottom. Evaluation successfully finishes, when the $\checkmark$-focus exits a hypernet from the bottom.

We will formalise focussed hypernet rewriting as an abstract machine (see~\cite{DBLP:conf/ac/Pitts00} for a comprehensive introduction). The machine has two kinds of transitions: one for the traversal, and the other for the update. 
It is important that the focus governs transitions; a traversal transition or an update transition is selected according to the mode of the focus. It is the focus that implements the traversal, triggers the update, and hence realises the call-by-value reduction strategy.

\subsection{Step-wise local reasoning, and robustness} \label{sec:gentle-introduction-reasoning}

Finally we overview the reasoning principle that focussed hypernet rewriting enables, which leads to our main theorem, sufficiency-of-robustness theorem (\autoref{thm:MetaThm}).

Using focussed hypernet semantics, this work takes a new coinductive, step-wise, approach to proving observational equivalence. We will introduce a new variant of weak simulation dubbed \emph{counting simulation}. A counting simulation is a relation on \emph{focussed} hypernets that are hypernets with a focus. We write $\focussed{N}$ to indicate that a hypernet $N$ contains a focus.

Our proof of an observational refinement $N \preceq H$, which is the asymmetric version of observational equivalence $N \simeq H$, proceeds as follows.
\begin{enumerate}
 \item We start with the relation $\mathit{\vartriangleleft} := \{(N,H)\}$ on hypernets. We call it \emph{pre-template}.
 \item We take the \emph{contextual closure} $\overline{\vartriangleleft}$ of the pre-template $\vartriangleleft$. It is defined by $\focussed{C}[N,\ldots,N] \mathrel{\overline{\vartriangleleft}} \focussed{C}[H,\ldots,H]$ for an arbitrary focussed (multi-hole) context $\focussed{C}$.
 \item We show that $\overline{\vartriangleleft}$ is a counting simulation.
\end{enumerate}
Once we establish the counting simulation $\overline{\vartriangleleft}$,
soundness of counting simulation asserts that the pre-template $N \vartriangleleft H$ implies observational refinement $N \preceq H$.

The key part of the observational refinement proof is therefore showing that $\overline{\vartriangleleft}$ is a counting simulation. Put simply, this amounts to show the following:
for any $\focussed{C}[N,\ldots,N] \mathrel{\overline{\vartriangleleft}} \focussed{C}[H,\ldots,H]$ and a transition $\focussed{C}[N,\ldots,N] \to \focussed{P}$, there exists a focussed context $\focussed{C'}$ that satisfies the following.
\begin{equation}
 \xymatrix@R=2ex{
  \focussed{C}[N,\ldots,N] \ar[r] \ar@{.}[d]_{\overline{\vartriangleleft}}
  & \focussed{P} \ar@[magenta][r]^>*[magenta]{*}
  & \textcolor{magenta}{\focussed{C'}[N,\ldots,N]} \ar@{.}@[magenta][d]^*[magenta]{\overline{\vartriangleleft}} \\
 \focussed{C}[H,\ldots,H] \ar@[magenta][rr]^>*[magenta]{*}
  && \textcolor{magenta}{\focussed{C'}[H,\ldots,H]}
  }
  \label{eq:counting-sim-simple}
\end{equation}
Above, black parts are universally quantified, and magenta parts are existentially quantified. The arrow $\to^*$ represents an arbitrary number of transitions $\to$.
This situation (\ref{eq:counting-sim-simple}) asserts that, after a few transitions from $\focussed{P}$ and $\focussed{C}[H,\ldots,H]$, we can obtain two focussed hypernets that can be decomposed using the new context $\focussed{C'}$ and the sub-graphs $N,H$.

\begin{figure}[t]
 \newcommand{\hmargin}{0.3cm}
 \centering
 \subfloat[
 \label{fig:case-move}]{
 \hspace{\hmargin}
 \begin{tikzpicture}[scale=0.15]
  \draw [rounded corners, very thick] (-6.5,-4) rectangle (6.5,4)
  node [at start, anchor=south west] {$C$};
  \draw [rounded corners, very thick] (-0.5,-1) rectangle (5.5,3)
  node [midway] {$F$};
  \draw [rounded corners, very thick]
  (0,-6)--(0,-4);
  \fill (-1.5,-1.5) circle [radius = 0.5];
  \fill [magenta] (-3.5,1.5) circle [radius = 0.5];
  \draw [dotted, ->, magenta, very thick,%
  shorten <= 3pt, shorten >= 3pt]
  (-1.5,-1.5)--(-3.5,1.5);
 \end{tikzpicture}
 \hspace{\hmargin}
 }
 \subfloat[
 \label{fig:case-visit}]{
 \hspace{\hmargin}
 \begin{tikzpicture}[scale=0.15]
  \draw [rounded corners, very thick] (-6.5,-4) rectangle (6.5,4)
  node [at start, anchor=south west] {$C$};
  \draw [rounded corners, very thick] (-0.5,-1) rectangle (5.5,3)
  node [midway] {$F$};
  \draw [rounded corners, very thick]
  (0,-6)--(0,-4);
  \fill (2.5,-1) circle [radius = 0.5];
  \draw [dotted, ->, magenta, very thick,%
  shorten <= 3pt, shorten >= 3pt]
  (2.5,-1)--(5,2);
 \end{tikzpicture}
 \hspace{\hmargin}
 }
 \subfloat[
 \label{fig:case-visit-reduction}]{
 \hspace{\hmargin}
 \begin{tikzpicture}[scale=0.15]
  \draw [rounded corners, very thick] (-6.5,-4) rectangle (6.5,4)
  node [at start, anchor=south west] {$C$};
  \draw [rounded corners, very thick] (-0.5,-1) rectangle (5.5,3)
  node [midway] {$N$};
  \draw [rounded corners, very thick]
  (0,-6)--(0,-4);
  \fill (2.5,-1) circle [radius = 0.5];
  \draw [dotted, ->, magenta, very thick,%
  shorten <= 3pt, shorten >= 3pt]
  (2.5,-1)--(5,2);
 \end{tikzpicture}
 \begin{tikzpicture}[scale=0.15]
  \node at (0,0) {$\rightsquigarrow$};
  \node at (0,6) {};
  \node at (0,-4) {};
 \end{tikzpicture}
 \begin{tikzpicture}[scale=0.15]
  \draw [rounded corners, very thick] (-6.5,-4) rectangle (6.5,4)
  node [at start, anchor=south west] {$C$};
  \draw [rounded corners, very thick] (-0.5,-1) rectangle (5.5,3)
  node [midway] {$H$};
  \draw [rounded corners, very thick]
  (0,-6)--(0,-4);
  \fill (2.5,-1) circle [radius = 0.5];
  \draw [dotted, ->, magenta, very thick,%
  shorten <= 3pt, shorten >= 3pt]
  (2.5,-1)--(5,2);
 \end{tikzpicture}
 \hspace{\hmargin}
 }
 
 \subfloat[
 \label{fig:case-update}]{
 \hspace{\hmargin}
 \begin{tikzpicture}[scale=0.15]
  \draw [rounded corners, very thick] (-6.5,-4) rectangle (6.5,4)
  node [at start, anchor=south west] {$C$};
  \draw [rounded corners, very thick] (-0.5,-1) rectangle (5.5,3)
  node [midway] {$F$};
  \draw [rounded corners, very thick]
  (0,-6)--(0,-4);
  \fill (-2.5,1) circle [radius = 0.5];
  \draw [magenta, dashed, very thick,%
  rotate=-30] (-1,-0.5) ellipse (3cm and 2cm);
 \end{tikzpicture}
 \hspace{\hmargin}
 }
 \subfloat[
 \label{fig:case-update-irrelevant}]{
 \hspace{\hmargin}
 \begin{tikzpicture}[scale=0.15]
  \draw [rounded corners, very thick] (-6.5,-4) rectangle (6.5,4)
  node [at start, anchor=south west] {$C$};
  \draw [rounded corners, very thick] (-0.5,-1) rectangle (5.5,3)
  node [midway] {$F$};
  \draw [rounded corners, very thick]
  (0,-6)--(0,-4);
  \fill (-2.5,1) circle [radius = 0.5];
  \draw [magenta, dashed, very thick,%
  rotate=-45] (-2.5,-2) ellipse (3cm and 2cm);
 \end{tikzpicture}
 \begin{tikzpicture}[scale=0.15]
  \node at (0,0) {$\rightsquigarrow$};
  \node at (0,6) {};
  \node at (0,-4) {};
 \end{tikzpicture}
 \begin{tikzpicture}[scale=0.15]
  \draw [rounded corners, very thick] (-6.5,-4) rectangle (6.5,4)
  node [magenta, at start, anchor=south west] {$C'$};
  \draw [rounded corners, very thick] (-0.5,-1) rectangle (5.5,3)
  node [midway] {$F$};
  \draw [rounded corners, very thick]
  (0,-6)--(0,-4);
  \fill (-2.5,1) circle [radius = 0.5];
 \end{tikzpicture}
 \hspace{\hmargin}
 }
 \caption{Example scenarios of case analysis for (\ref{eq:counting-sim-simple}), where $F \in \{ N,H \}$}
 \label{fig:counting-sim-case-analysis}
\end{figure}

Our important observation is that (\ref{eq:counting-sim-simple}) can be established by elementary case analysis of interaction between the sub-graphs $N,H$ and what happens around the focus in $\focussed{C}[N,\ldots,N] \to \focussed{P}$. This is because updates of a hypernet always happen around the $\lightning$-focus, and the $?$-focus and the $\checkmark$-focus (representing the graph traversal) move according to its neighbourhood. The analysis is hence centred around the graphical concept of neighbourhood, or graph locality. There are three possible cases of the interaction.
\begin{description}
 \item[Case (i) Move inside the context]
	    The $?$-focus or the $\checkmark$-focus, which implements the depth-first traversal, simply moves inside the context $C$ (see \autoref{fig:case-move}). Because any move of the focus is only according to its neighbourhood, the move solely depends on the context $C$. In other words, the sub-graphs $N,H$ have no interaction with the focus. In this case, we can conclude that we are always in (\ref{eq:counting-sim-simple}).
 \item[Case (ii) Visit to the sub-graphs]
	    The $?$-focus visits the sub-graphs $N,H$ (see \autoref{fig:case-visit}). This is the case where the $?$-focus actually interacts with $N,H$; what happens after entering of the focus depends on $N,H$. We identify a sufficient condition of the pre-template $\vartriangleleft$, dubbed \emph{safety}, for (\ref{eq:counting-sim-simple}) to hold.
	    
	    A typical example of safe pre-templates is the pre-templates that are induced by rewrite rules of hypernets. \autoref{fig:case-visit-reduction} illustrates what happens to such a pre-template. The visit of the $?$-focus to $N$ triggers the rewrite rule, and actually turns $N$ into $H$.
	    
	    Note that the case where the $\checkmark$-focus visits $N,H$ boils down to the visit of the $?$-focus instead, because the $\checkmark$-focus implements backtracking of graph traversal.
 \item[Case (iii) Update of the hypernets]
	    The $\lightning$-focus triggers a rewrite rule and updates the hypernet (see \autoref{fig:case-update}). This is the case where the $\lightning$-focus interacts with $N,H$; the update may involve $N,H$ in a non-trivial manner. We identify a sufficient condition of the pre-template $\vartriangleleft$, relative to the triggered rewrite rule, dubbed \emph{robustness}, for (\ref{eq:counting-sim-simple}) to hold.

	    An example scenario of robustness is where the update only affects parts of the context $C$; see \autoref{fig:case-update-irrelevant}. In this scenario, the $\lightning$-focus does not really interact with $N,H$. The sub-graphs $N,H$ are preserved, and we can take the new context $C'$ with the same number of holes as $C$ that makes (\ref{eq:counting-sim-simple}) hold.

	    Another example scenario of robustness is where the update duplicates (or eliminates) $N,H$ without breaking them. We can take the context $C'$ that has more (or less) holes to make (\ref{eq:counting-sim-simple}) hold.
	    \bqed
\end{description}

The above case analysis reveals sufficient conditions, namely safety and robustness, to make (\ref{eq:counting-sim-simple}) hold and hence make the context closure $\overline{\vartriangleleft}$ a counting simulation. Combining this with soundness of counting simulation, we obtain our main theorem, sufficiency-of-robustness theorem (\autoref{thm:MetaThm}). It can be informally stated as follows.
\begin{theorem*}[Sufficiency-of-robustness theorem (\autoref{thm:MetaThm}), informally]
 A robust and safe pre-template $N \vartriangleleft H$ induces observational refinement $N \preceq H$.
\end{theorem*}

\section{Preliminaries: Hypernets} \label{sec:hypernets}

We start formalising the ideas described in the previous section, by first defining hypernets.
We opt for formalising hypernets as hypergraphs, following the literature \cite{BonchiGKSZ22a,BonchiGKSZ22b,BonchiGKSZ22c,Alvarez-PicalloGSZ22} on categorically formalising string diagram rewriting using hypergraphs.

Let $\N$ be the set of natural numbers.
Given a set $X$ we write by $X^*$ the set of elements of the free monoid over $X$.
Given a function $f:X\rightarrow Y$ we write $f^*:X^*\rightarrow Y^*$ for the pointwise application (map) of $f$ to the elements of $X^*$.

\subsection{Monoidal hypergraphs and hypernets} \label{sec:monoidal-hypergraphs}

Hypernets have a couple of distinctive features in comparison with ordinary graphs.
The first feature of hypernets is that they have ``dangling edges'' (see \autoref{fig:hypernet-beta}); a hypernet has one incoming arrow with no source, and it may have outgoing arrows with no targets. To model this, we use \emph{hypergraphs}---we formalise what we have been calling edges (i.e.\ arrows) as \emph{vertices} and what we have been calling nodes (i.e.\ circled objects) as \emph{hyperedges} (i.e.\ edges with arbitrary numbers of sources and targets).
More specifically, we use what we call \emph{interfaced labelled monoidal hypergraphs} that satisfies the following.
\begin{enumerate}
 \setcounter{enumi}{-1}
 \item Each arrow (modelled as a vertex) and each circled object (modelled as a hyperedge) are labelled.
 \item Each circled object (modelled as a hyperedge) is adjacent to distinct arrows (modelled as vertices).
 \item Each arrow (modelled as a vertex) is adjacent to at most two circled objects (modelled as hyperedges).
 \item The label of a circled object (modelled as a hyperedge) is always consistent with the number and labelling of its endpoints.
 \item Dangling arrows are ordered, and each arrow has at least a source or a target.
\end{enumerate}
\begin{defi}[Monoidal hypergraphs]
 \label{def:monoidal-hypergraphs}
 A \emph{monoidal hypergraph} is a pair $(V,E)$ of finite sets, \emph{vertices} and \emph{(hyper)edges} along with a pair of functions $S:E\rightarrow V^*$, $T:E\rightarrow V^*$ defining the \emph{source list} and \emph{target list}, respectively, of an edge.
\end{defi}
\begin{defi}[Interfaced labelled monoidal hypergraphs]
 \label{def:labelled-monoidal-hypergraphs}
An \emph{interfaced labelled monoidal hypergraph} consists of a monoidal hypergraph, a set of vertex labels $L_V$, a set of edge labels $L_E$, and labelling functions $f_V \colon V\rightarrow L_V, f_E \colon E\rightarrow L_E$ such that:
\begin{enumerate}
 \item For any edge $e \in E$, its source list $S(e)$ consists of
       distinct vertices, and its target list $T(e)$ also consists of
       distinct vertices.
 \item For any vertex $v\in V$ there exists at most one edge $e\in E$ such that $v\in S(e)$ and at most one edge $e'\in E$ such that $v\in T(e')$.
 \item \label{item:hypergraphs-type}
       For any edges $e_1, e_2\in E$ if $f_E(e_1)=f_E(e_2)$ then $f_V^*\bigl(S(e_1)\bigr)=f_V^*\bigl(S(e_2)\bigr)$, and $f_V^*\bigl(T(e_1)\bigr)=f_V^*\bigl(T(e_2)\bigr)$.
 \item If a vertex belongs to the target (resp.\ source) list of no edge we
       call it an \emph{input} (resp.\ \emph{output}).
       Inputs and
       outputs are respectively ordered, and no vertex is both an input and
       an output.
\end{enumerate}
\end{defi}
\begin{notation}[Types of circled objects (i.e.\ hyperedges)]
 \autoref{def:labelled-monoidal-hypergraphs}~(\ref{item:hypergraphs-type}) makes it possible to use labels of arrows (i.e.\ vertices) as \emph{types} for labels of circled objects (i.e.\ hyperedges).
 For each $m \in L_E$, we can associate it a type and write $m \colon x \To x'$, where $x,x' \in L_V^*$ satisfy
$x = f_V^*\bigl(S(e)\bigr)$ and
$x' = f_V^*\bigl( T(e)\bigr)$ for any
$e\in E$ such that $f_E(e)=m$.
\end{notation}
\begin{notation}[Types of interfaced labelled monoidal hypergraphs]
 The concept of type can be extended to a whole interfaced labelled monoidal hypergraph $G$. Let $I,O$ be the lists of inputs and outputs, respectively, of $G$. We associate $G$ a type and write $G \colon f_V^*(I)\To f_V^*(O)$.
 In the syntax for lists of inputs and outputs we use $\otimes$ to denote concatenation and define $\epsilon$ to be the empty list, and $A^{\otimes 0} := \epsilon$, $A^{\otimes(n+1)} := A\otimes A^{\otimes n}$ for any label $A$ and $n \in \N$.
\end{notation}

In the sequel, when we say hypergraphs we always mean interfaced labelled monoidal hypergraphs.

We sometimes permute inputs and outputs of a hypergraph. Such permutation yields another hypergraph.
\begin{defi}[Interface permutation] \label{def:permut-HG}
Let $G$ be a hypergraph with an input list $i_1,\ldots,i_n$ and an
output list $o_1,\ldots,o_m$.
Given two bijections $\rho$ and $\rho'$ on sets $\{ 1,\ldots,n \}$ and
$\{ 1,\ldots,m \}$, respectively, we write $\Pi^{\rho'}_\rho(G)$ to
denote the hypergraph that is defined by the same data as $G$ except
for the input list $i_{\rho(1)},\ldots,i_{\rho(n)}$ and the output
list $o_{\rho'(1)},\ldots,o_{\rho'(m)}$.
\end{defi}

The second distinctive feature of hypernets is that they have dashed boxes that indicate the scope of variable bindings (see \autoref{fig:hypernet-beta}). We formalise these dashed boxes by introducing \emph{hierarchy} to hypergraphs.
The hierarchy is implemented by allowing a hypergraph to be the label of a hyperedge.
As a result, informally, hypernets are nested hypergraphs, up to some finite depth,
using the same sets of labels.
We here present a relatively intuitive definition of hypernets;
\autoref{sec:hypernets-alt} discusses an alternative definition of hypernets\footnote{Another, slightly different, definition of hypernets is given in~\cite[Section~4]{DBLP:conf/csl/Alvarez-Picallo23}. The difference is motivated by desired support of categorical graph rewriting, which requires certain properties to hold. These properties are sensitive to the definition.}.
\begin{defi}[Hypernets]
 \label{def:Hypernets}
  Given a set of vertex labels $\LV$ and edge labels $\LE$ we write $\mathcal H(\LV,\LE)$ for the set of hypergraphs with these labels; we also call these \emph{level-0 hypernets} $\mathcal H_0(\LV,\LE)$. 
  We call \emph{level-$(k{+}1)$ hypernets} the set of hypergraphs
  \[ \mathcal H_{k+1}(L, M)=\mathcal H\Bigl(\LV,\LE\cup \bigcup_{i\leq k}\mathcal H_i(L, M)\Bigr).\]
  We call \emph{hypernets} the set $\mathcal H_\omega(\LV,\LE)=\bigcup_{i\in \mathbb N}\mathcal H_i(\LV,\LE)$.
\end{defi}
\begin{terminology}[Boxes and depth]
An edge labelled with a hypergraph is called \emph{box} edge, and a
hypergraph labelling a box edge is called \emph{content}.
Edges of a hypernet $G$ are said to be \emph{shallow}. Edges of
nesting hypernets of $G$, i.e.\ edges of hypernets that recursively
appear as edge labels, are said to be \emph{deep} edges of $G$. Shallow
edges and deep edges of a hypernet are altogether referred to as edges
\emph{at any depth}.
\end{terminology}

\subsection{Graphical conventions} \label{sec:graphical-conventions}

\begin{figure}[t]
 \centering
 \subfloat[\label{fig:hypergraph-normal}]{
 \includegraphics[width=.15\linewidth]{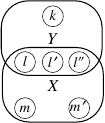}
 }
 \hfil
 \subfloat[\label{fig:hypergraph-graph}]{
 \includegraphics[scale=.25]{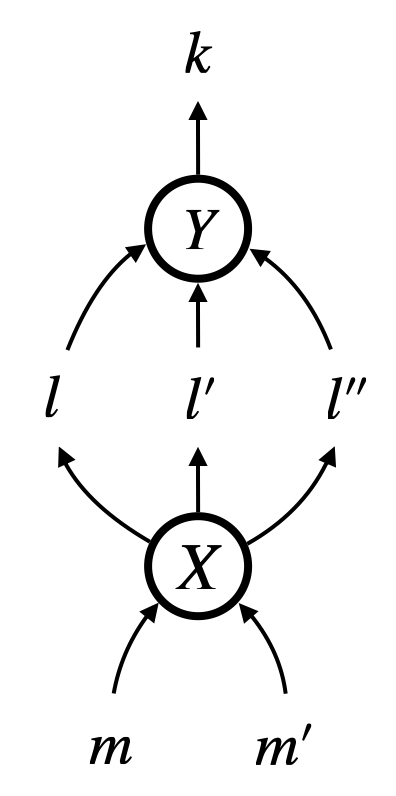}
 }
 \hfil
 \subfloat[\label{fig:hypergraph-simple}]{
 \hspace{.01\linewidth}
 \includegraphics[scale=.25]{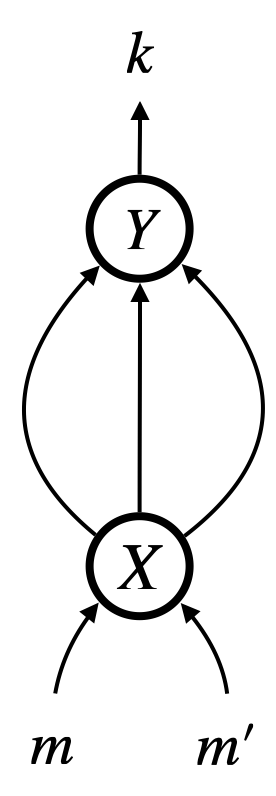}
 }
 \hfil
 \subfloat[\label{fig:hypergraph-sub-all}]{
 \includegraphics[scale=.25]{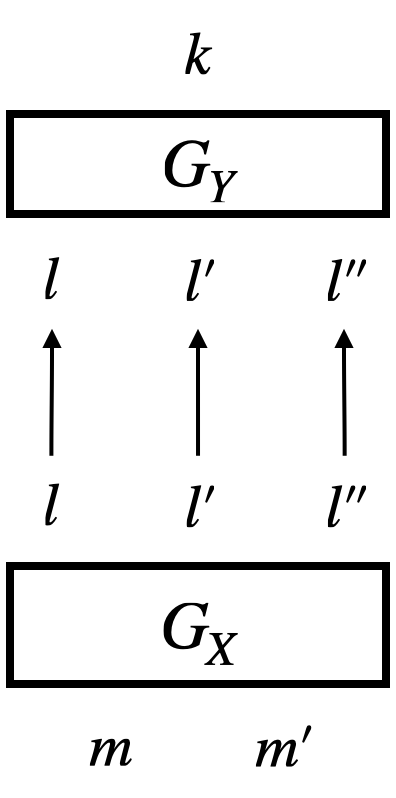}
 }
 \hfil
 \subfloat[\label{fig:hypergraph-sub-arrows}]{
 \includegraphics[scale=.25]{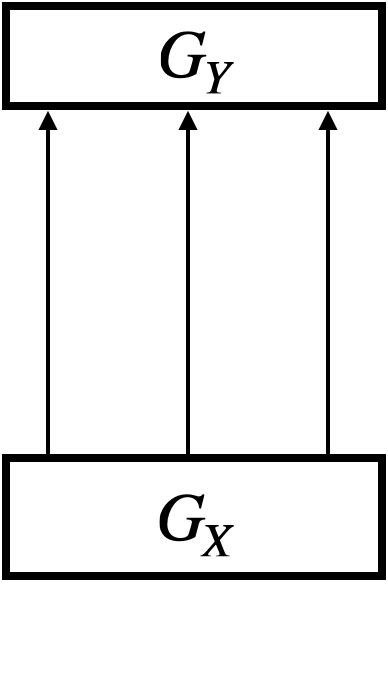}
 }
 \hfil
 \subfloat[\label{fig:hypergraph-sub-arr}]{
 \includegraphics[scale=.25]{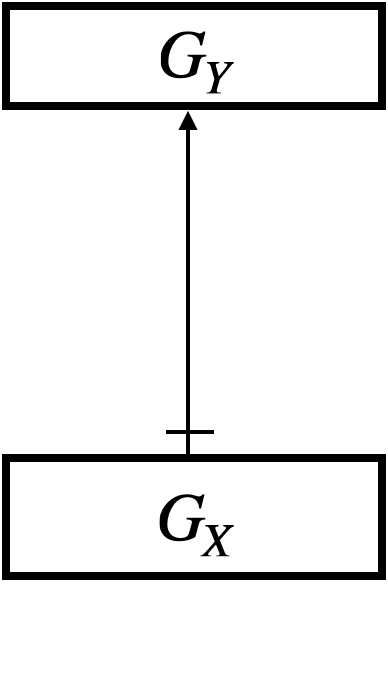}
 }
 \caption{The hypergraph $G_\mathrm{ex}$}
 \label{fig:hypergraph}
\end{figure}

A hypergraph $G_\mathrm{ex}$ with vertices
$V=\{v_0,v_1,v_2,v_3,v_4,v_5\}$ and edges $E=\{e_1,e_2\}$ such that
\begin{align*}
  S(e_0) &= \{v_0, v_1\} \\
  T(e_0) &= S(e_1)= \{v_2, v_3, v_4\}\\
  T(e_1) &= \{v_5\}\\
  f_V &= \{v_0\mapsto m, v_1\mapsto m', v_2\mapsto l, v_3\mapsto l', v_4\mapsto l'', v_5\mapsto k\}\\
  f_E &= \{e_0\mapsto X, e_1\mapsto Y\}
\end{align*}
is normally represented as \autoref{fig:hypergraph-normal}.
However, we find this style of representing hypergraphs awkward for understanding their structure. We will often graphically represent hypergraphs as graphs, as in \autoref{fig:hypergraph-graph}, by (i) marking vertices with their labels and mark hyperedges with their labels circled, and (ii) connecting input vertices and output vertices with a hyperedge using arrows.

Recall that node labels are often determined in hypergraphs, thanks to typing, e.g.\ $X \colon m \otimes m' \To l \otimes l' \otimes l''$. We accordingly omit node labels to avoid clutter, as in \autoref{fig:hypergraph-simple}, letting arrows connect circles directly.

Sometimes we draw a hypergraph by connecting its sub-graphs using extra arrows. Sub-graphs are depicted as boxes. For the hypergraph $G_\mathrm{ex}$, we can think of the sub-graph $G_X = (\{v_0,v_1,v_2,v_3,v_4\},\{e_0\})$ and the sub-graph $G_Y = (\{v_2,v_3,v_4,v_5\},\{e_1\})$. We may draw $G_\mathrm{ex}$ in three ways, as in \autoref{fig:hypergraph-sub-all}--\ref{fig:hypergraph-sub-arr}:
\begin{itemize}
 \item In \autoref{fig:hypergraph-sub-all}, we use extra arrows connecting node labels $l,l',l''$ directly, with intention that two occurrences of $l$ (or, $l',l''$) are graphical representations of the same node $v_2$ (or $v_3,v_4$).
 \item In \autoref{fig:hypergraph-sub-arrows}, we omit node labels entirely, assuming that they are obvious from context.
 \item In \autoref{fig:hypergraph-sub-arr}, we further replace the three extra arrows with a single arrow, given that the entire output type of $G_X \colon m \otimes m' \To l \otimes l' \otimes l''$ matches the input type of $G_Y \colon l \otimes l' \otimes l'' \To k$. The single arrow comes with a dash across, which indicates that the arrow represents a bunch of parallel arrows.
\end{itemize}

The final convention is about box edges; a box edge (i.e.\ an edge labelled by a hypernet) is depicted by a dashed box decorated with its content (i.e.\ the labelling hypernet).

\section{Representation of the call-by-value untyped linear lambda-calculus} \label{sec:linear-lambda-calc}

In this section we introduce specific label sets that we use to represent lambda-terms as hypernets. We begin with the untyped linear lambda-calculus extended with arithmetic. This is a fairly limited language in terms of expressive power. Although simple, it is interesting enough to demonstrate our reasoning framework. We present a translation $(-)^\dag$ of linear lambda-terms in this section, and will adapt it to the general lambda-terms in \autoref{sec:variable-sharing-store}.

Lambda-terms are defined by the BNF $t ::= x \mid \lambda x. t \mid t\;t \mid n \mid t \mathop{\mathit{op}} u$, where $n \in \N$ and $\mathit{op} \in \{+,-\}$. We assume alpha-equivalence on terms, and assume that bound variables are distinct in a term. A term $t$ is \emph{linear} when each variable appears exactly once in $t$.


\begin{figure}[t]
 \begin{gather*}
  x^\dag \ =\ \includegraphics[align=c,scale=.25]{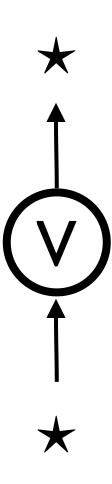}
  \hspace{4em}
  (\lambda x. t)^\dag \ =\ \includegraphics[align=c,scale=.25]{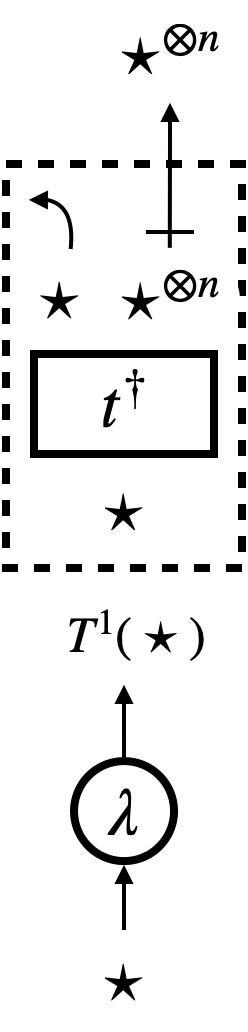}
  \hspace{4em}
  (t\;u)^\dag \ =\ \includegraphics[align=c,scale=.25]{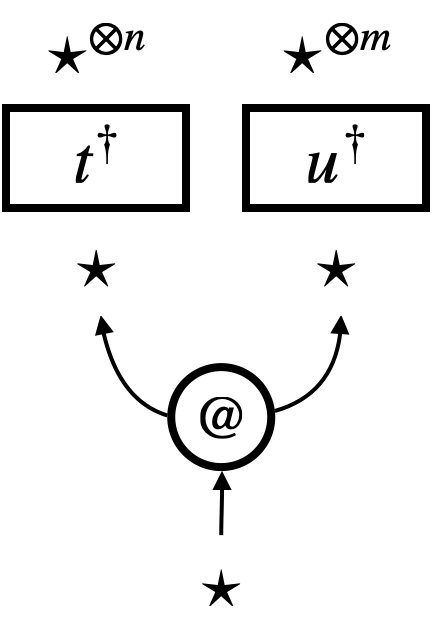} \\
  n^\dag \ =\ \includegraphics[align=c,scale=.25]{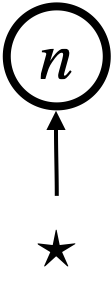}
  \hspace{4em}
  (t \mathop{\mathit{op}} u)^\dag \ =\ \includegraphics[align=c,scale=.25]{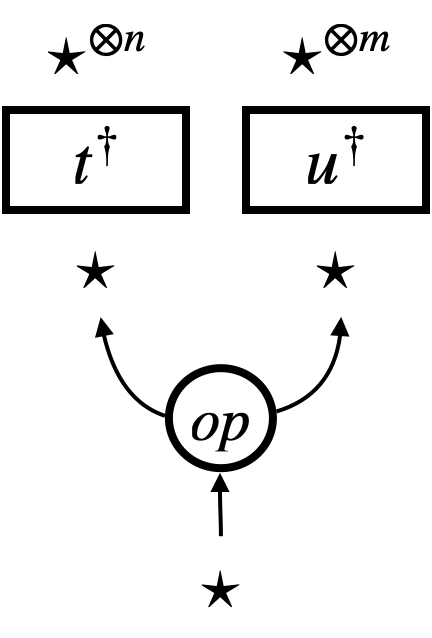}
 \end{gather*}
 \caption{Inductive translation $(-)^\dag$ of linear lambda-terms}
 \label{fig:transl-linear}
\end{figure}

First, recall that each edge label $m \in \LE$ of a hypernet $G \in \HN{\LV}{\LE}$ comes with a \emph{type} $m \colon X \To X'$ where $X,X' \in \LV^*$. Even though lambda-terms are untyped here, we use edges to represent term constructors, and use edge types (i.e.\ node labels) to distinguish thunks from terms. Namely, the term type is denoted by $\star$, and a thunk type with $n$ bound variables is denoted by $T^n(\star)$.

\autoref{fig:transl-linear} shows inductive translation of linear lambda-terms to hypernets $\HN{\LVlin}{\LElin}$ where
\begin{align}
 \LVlin &= \{\star\} \cup \{T^n(\star) \mid n \in \N\}, \label{eq:transl-linear-node-labels} \\
 \LElin &= \{
 \mathsf{V} \colon \star \To \star,\;
 \lambda \colon \star \To T^1(\star),\;
 \lrapp \colon \star \To \star^{\otimes 2},\;
 \mathit{+} \colon \star \To \star^{\otimes 2},\;
 \mathit{-} \colon \star \To \star^{\otimes 2}
 \} \notag \\
 &\quad \cup \{ n \colon \star \To \epsilon \mid n \in \N \}. \label{eq:transl-linear-edge-labels}
\end{align}
In general, a term $t$ with $n$ free variables is translated into $t^\dag \colon \star \To \star^{\otimes n}$. Each term constructor is turned into an edge as follows.
\begin{itemize}
 \item Any variable becomes an anonymous edge $\mathsf{V} \colon \star \To \star$. We represented a variable as a single arrow in \autoref{sec:gentle-introduction} (see e.g.\ \autoref{fig:hypernet-beta} and \autoref{fig:focus-ex}), but this means a variable would become an empty graph (i.e.\ a hypernet with no edge). We rather use the anonymous edge $\mathsf{V} \colon \star \To \star$ to prevent an empty graph from labelling a box edge and hence being a box content. This is for technical reasons as they simplify our development.
 \item Abstraction becomes $\lambda \colon \star \To T^1(\star)$; it constructs a term, taking one thunk that has one bound variable.
       A thunk that has one bound variable and $n$ free variables is represented by a box edge of type $T^1(\star) \To \star^{\otimes n}$ whose content is $t^\dag \colon \star \To \star \otimes \star^{\otimes n}$. Note that the box has one less output than its content. We emphasise this graphically, by bending the arrow that is connected to the leftmost $\star$ of the type $\star \otimes \star^{\otimes n}$.
 \item Application becomes $\lrapp \colon \star \To \star^{\otimes 2}$; it constructs a term, taking two terms as arguments. This translation is for the call-by-value evaluation strategy; both of the arguments are not thunks, and hence they will be evaluated before the application is computed.
 \item Each natural number $n \in \N$ becomes $n \colon \star \To \epsilon$; it is a term, taking no arguments.
 \item Arithmetic operations becomes $\mathit{+} \colon \star \To \star^{\otimes 2}$ and $\mathit{-} \colon \star \To \star^{\otimes 2}$; they construct a term, taking two terms as arguments. The translation for these operations has the same shape as that for application.
\end{itemize}

\begin{figure}[t]
 \includegraphics[scale=.25]{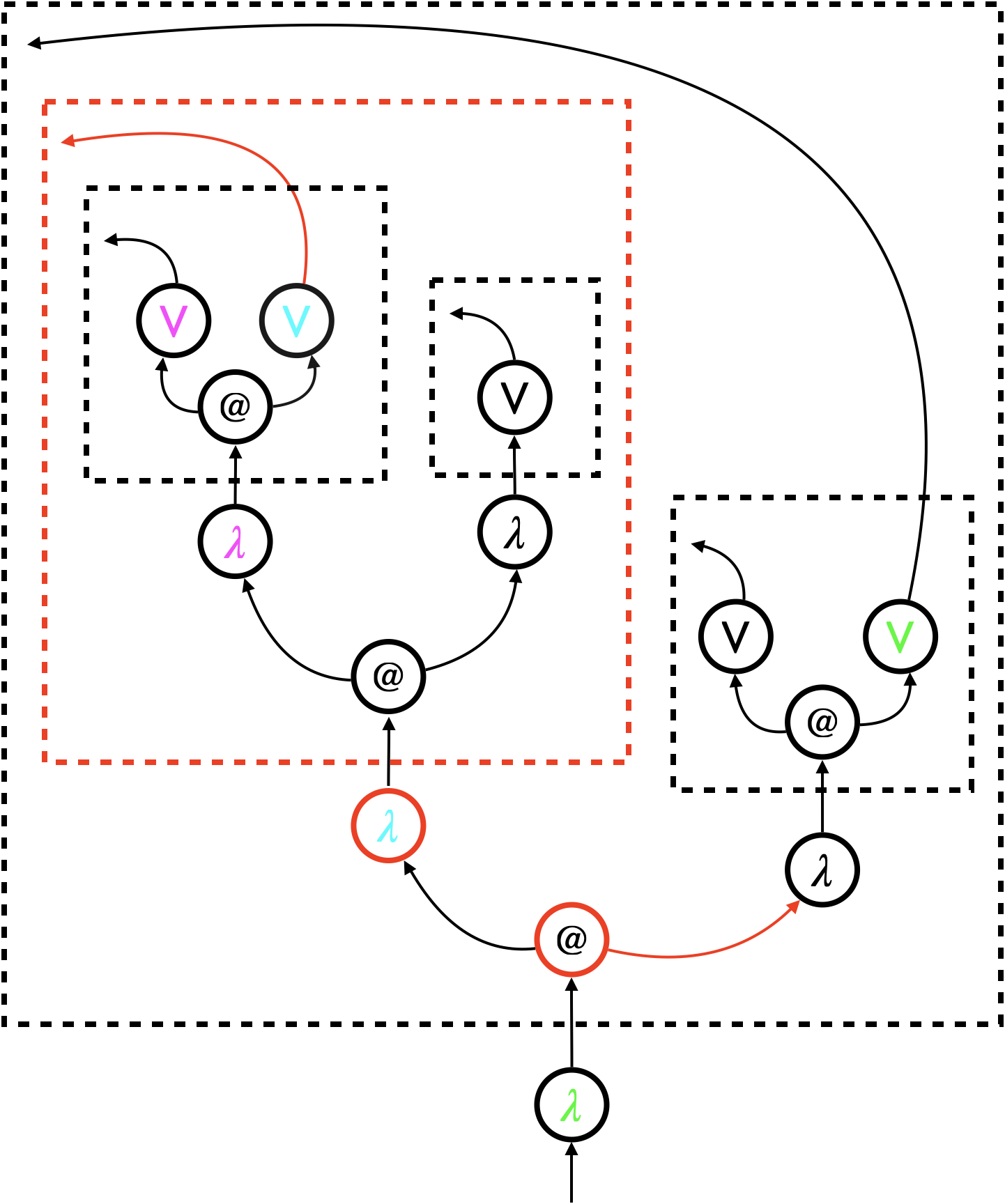}
 \caption{The hypernet $\bigl(\green{\lambda y}. (\cyan{\lambda x}. (\textcolor{magenta}{\lambda y}. \textcolor{magenta}{y}\;\cyan{x})\;(\lambda z. z))\;(\lambda x. x\;\green{y})\bigr)^\dag \colon \star \To \epsilon$}
 \label{fig:hypernet-beta-V-left}
\end{figure}
\autoref{fig:hypernet-beta-V-left} shows an example of the translation. Note that this is different from the left hand side of \autoref{fig:hypernet-beta}; each variable is now represented by the anonymous $\mathsf{V}$-edge.

\section{Focussed hypernet rewriting---the Universal Abstract Machine} \label{sec:focussed-hypernet-rewriting-UAM}

In this section we present focussed hypernet rewriting. It will be formalised as an abstract machine dubbed \emph{universal abstract machine (UAM)}.
This abstract machine is ``universal'' in a sense of the word similar to the way it is used in ``universal algebra'' rather than in ``universal Turing machine''. It is a general abstract framework in which a very wide range of concrete abstract machines can be instantiated by providing \emph{operations} and their \emph{behaviour}.

\subsection{Operations and focus} \label{sec:operations-focus}

The first parameter of the UAM is given by a set of \emph{operations} $\Opr = \Opr_\checkmark \uplus \Opr_\lightning$. Operations are classified into two: \emph{passive} operations $\Opr_\checkmark$ that construct evaluation results (i.e.\ values) and \emph{active} operations $\Opr_\lightning$ that realise computation.
We let $\phi,\phi_\checkmark,\phi_\lightning$ range over $\Opr,\Opr_\checkmark,\Opr_\lightning$ respectively.
The edge labels in $L_\mathrm{lin}$ (\ref{eq:transl-linear-edge-labels}), except for $\mathsf{V}$, are examples of operations:
\begin{center}
 \begin{tabular}{|l|l|} \hline
  passive operations & active operations \\ \hline
  $\lambda \colon \star \To T^1(\star)$ & $\lrapp \colon \star \To \star^{\otimes 2}$ \\
  $n \colon \star \To \epsilon$ for each $n \in \N$ & $\mathit{op} \colon \star \To \star^{\otimes 2}$ where $\mathit{op} \in \{+,-\}$ \\ \hline
 \end{tabular}
\end{center}

In general, each operation $\phi \in \Opr$ has a type $\star \To \star^{\otimes m} \otimes \bigotimes_{i=1}^k T^{n_i}(\star)$ where $m,k,n_1,\ldots,$ $n_k \in \N$. This means that each operation takes $m$ arguments and $k$ thunks, and the $i$-th thunk has $n_i$ bound variables.

Given the operation set $\Opr$, the UAM acts on hypernets $\HN{\LVlin}{\LElin(\Opr)}$ where
\begin{equation}
 \LElin(\Opr) = \Opr \cup \{ \mathsf{V} \colon \star \To \star \}
  \cup \{ ? \colon \star \To \star,\;\checkmark \colon \star \To \star,\;\lightning \colon \star \To \star \}.
 \label{eq:UAM-linear-labels}
\end{equation}
We let $\ell$ range over $\LVlin$.
For box edges, we impose the following type discipline: each box edge must have a type $T^n(\star) \To \star^{\otimes m}$ with its content having a type $\star \To \star^{\otimes (n+m)}$.

The last three elements of~(\ref{eq:UAM-linear-labels}) are \emph{focuses}.
In the UAM, focuses are edges with the dedicated labels $?,\checkmark,\lightning$ of type $\star \To \star$. Each label represents one of the three modes of the focus:
\begin{center}
 \begin{tabular}{|c|c|} \hline
  label & mode \\ \hline \hline
  $?$ & searching \\ \hline
  $\checkmark$ & backtracking \\ \hline
  $\lightning$ & triggering an update \\ \hline
 \end{tabular}
\end{center}
As illustrated in \autoref{sec:gentle-introduction-focussed-hypernet-rewriting}, focussed hypernet rewriting implements program evaluation by combining (i) depth-first graph traversal and (ii) update (rewrite) of a hypernet. Focuses are the key element of this combination; they determine which action (i.e.\ traversal or rewrite) to be taken next, and they indicate where a redex of the rewrite is.

We refer to a hypernet that contains one focus as \emph{focussed} hypernet. Given a focussed hypernet, we refer to the hypernet without the focus as \emph{underlying} hypernet.

\subsection{Transitions---overview} \label{sec:transitions-overview}

\begin{table}[t]
 \centering
 \begin{tabular}{|c|c||c|c|} \hline
  \multicolumn{2}{|c||}{transitions} & focus & provenance \\ \hline \hline
  \multicolumn{2}{|c||}{search transitions} & $?,\checkmark$ & \multirow{2}{*}{intrinsic} \\ \cline{1-3}
  \multirow{2}{*}{rewrite transitions} & substitution transitions & \multirow{2}{*}{$\lightning$} & \\ \cline{2-2}\cline{4-4}
  & behaviour $B_\Opr$ && extrinsic \\
  & (compute transitions) && \\ \hline
 \end{tabular}
 \caption{Transitions of the UAM}
 \label{tab:UAM-transitions}
\end{table}
\autoref{tab:UAM-transitions} summarises classifications of transitions of the UAM.

The first classification is according to the focuses: \emph{search} transitions for the $?$-focus and the $\checkmark$-focus, implementing the depth-first search of redexes, and \emph{rewrite} transitions for the $\lightning$-focus, implementing rewrite of the underlying hypernet.
Rewrite transitions are further classified into two: \emph{substitution} transitions for edges labelled by $\mathsf{V}$, and \emph{behaviour} $B_\Opr$ for (active) operations.

The next classification is according to provenance: \emph{intrinsic} transitions that are inherent to the UAM, and \emph{extrinsic} transitions that are not. While search transitions and substitution transitions constitute intrinsic transitions, the behaviour $B_\Opr$ solely provides extrinsic transitions. In fact, the behaviour $B_\Opr$ is the second parameter of the UAM.

\begin{figure}[t]
 \includegraphics[scale=.2]{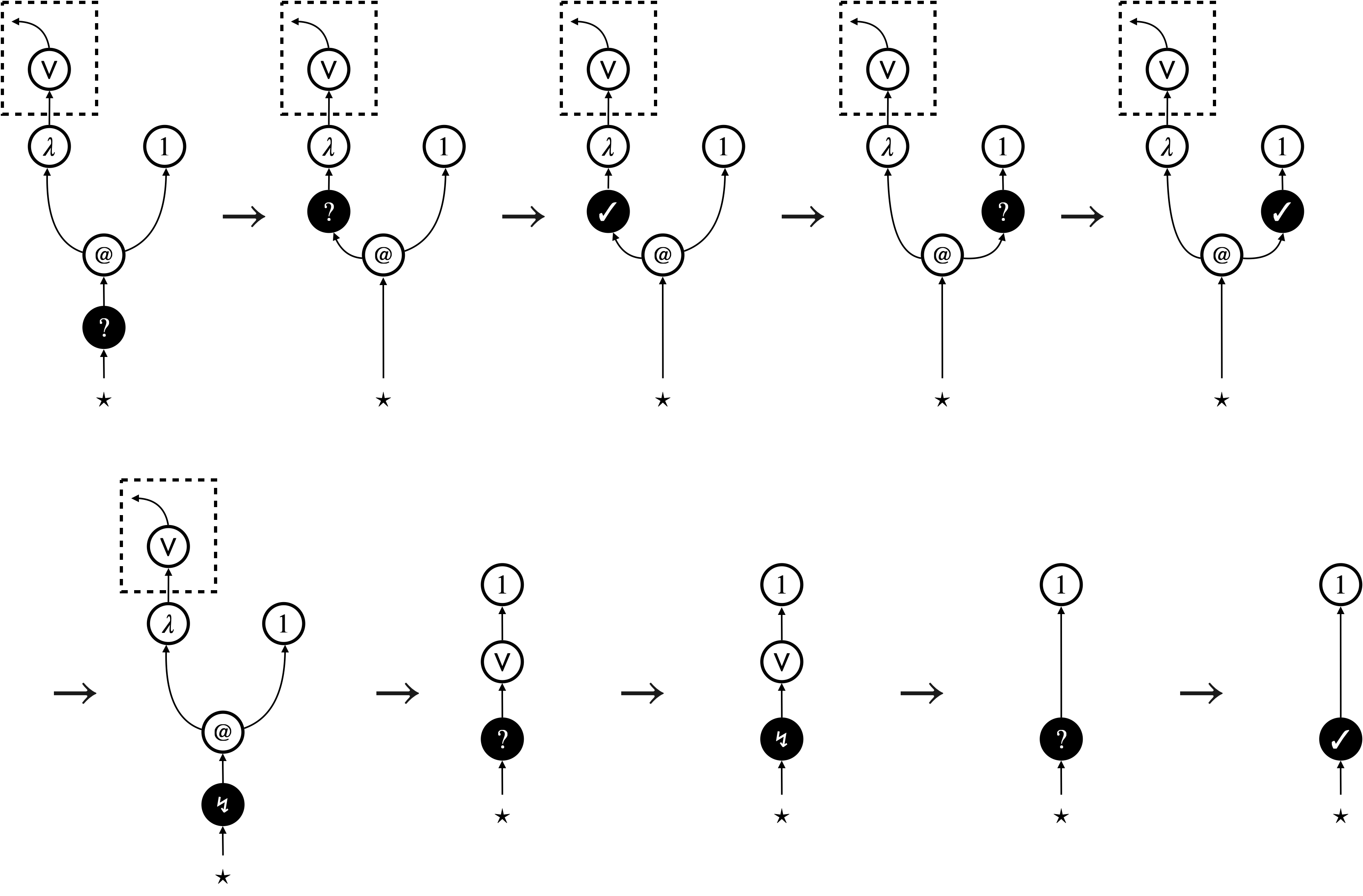}
 \caption{An example of the UAM execution on $((\lambda x. x)\;1)^\dag$}
 \label{fig:UAM-ex}
\end{figure}
\autoref{fig:UAM-ex} shows an example of the UAM execution. This evaluates the linear-term $(\lambda x. x)\;1$, and the execution starts with the underlying hypernet $((\lambda x. x)\;1)^\dag$. It demonstrates all the three kinds of transitions: search transitions, substitution transitions, and the behaviour of application ($\lrapp$).

\subsection{Intrinsic transitions} \label{sec:intrinsic-trans}

\begin{figure}[t]
 \centering
 \subfloat[\label{fig:interaction-V}]{
 \includegraphics[scale=.2]{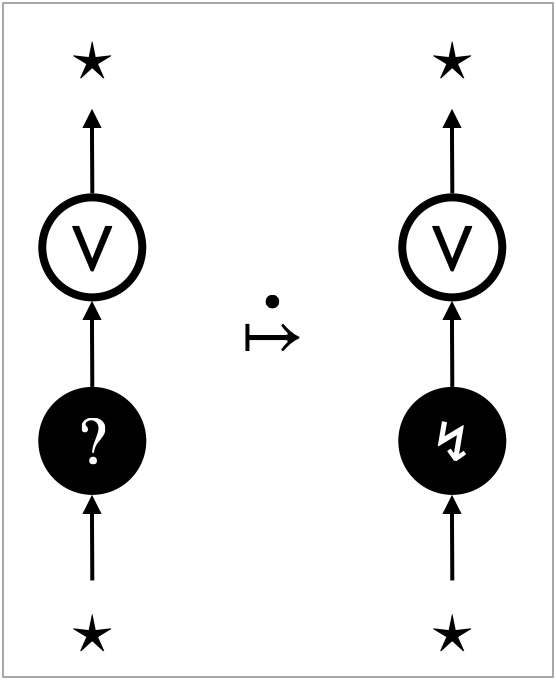}
 }
 \hfil
 \subfloat[\label{fig:interaction-opr-first}]{
 \includegraphics[scale=.2]{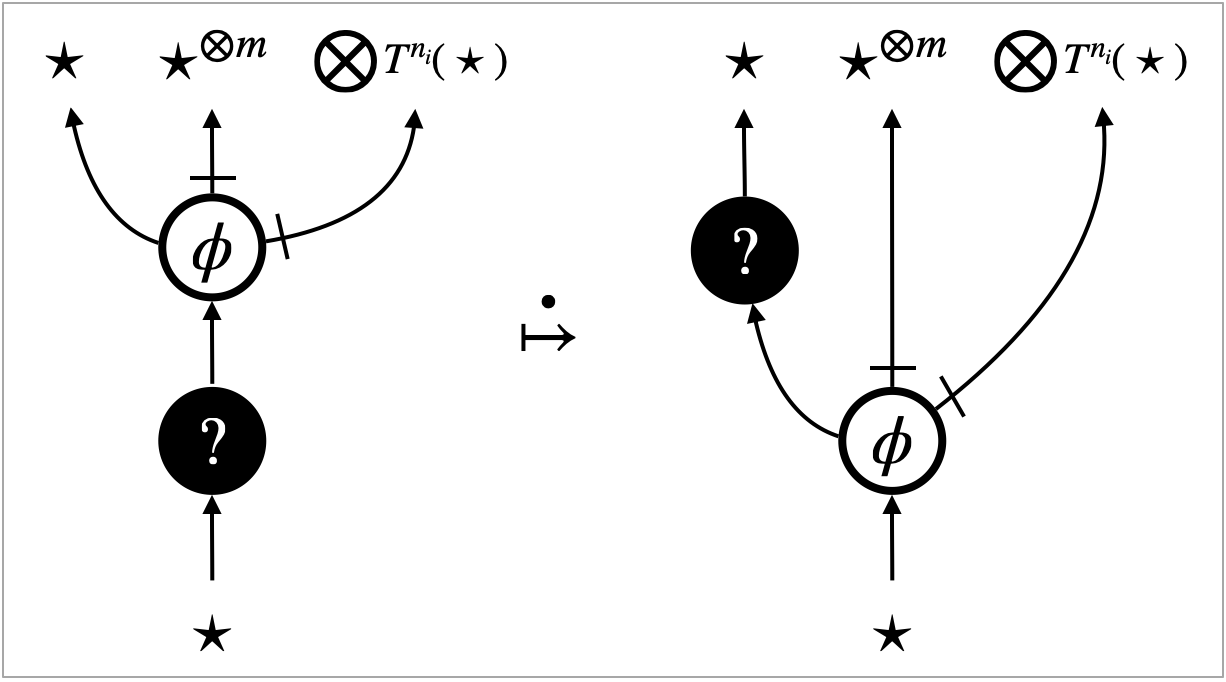}
 }
 \hfil
 \subfloat[\label{fig:interaction-opr-next}]{
 \includegraphics[scale=.2]{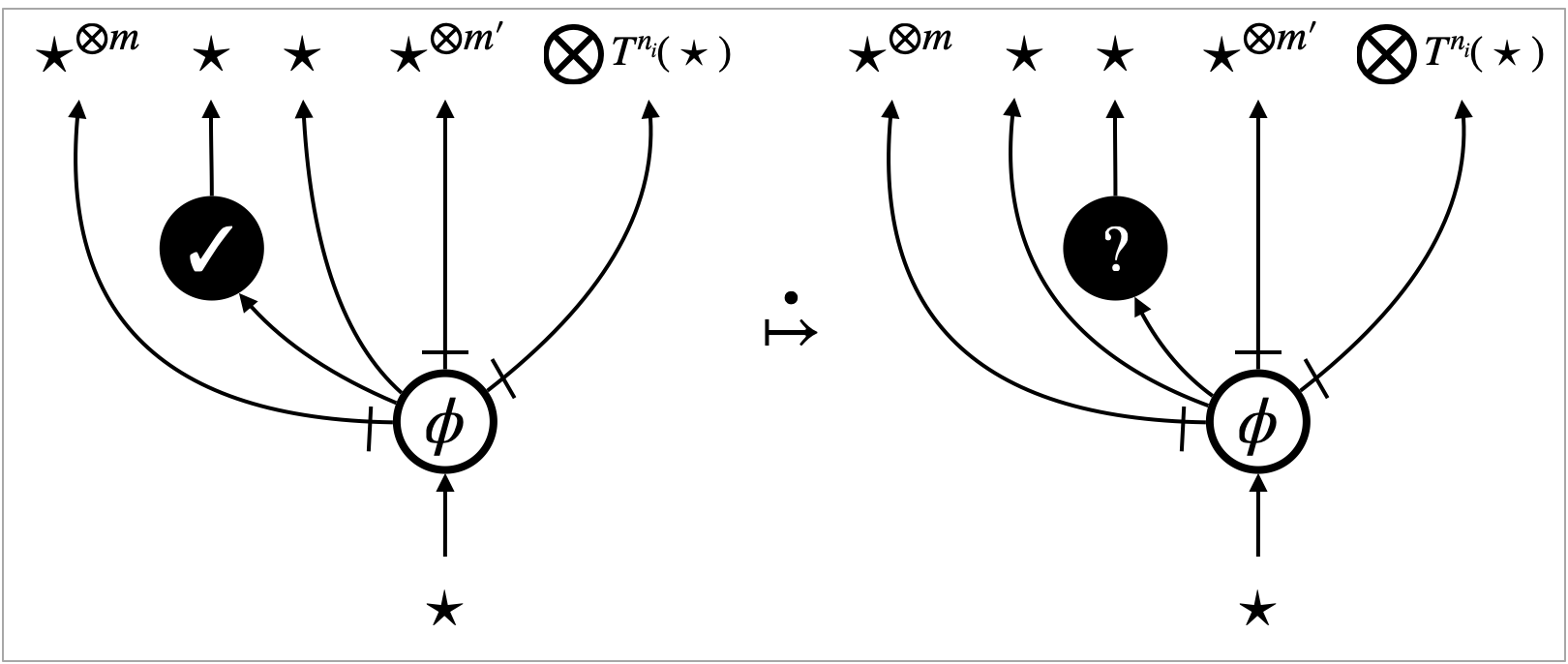}
 }
 \\
 \hfil
 \subfloat[\label{fig:interaction-opr-last}]{
 \includegraphics[scale=.2]{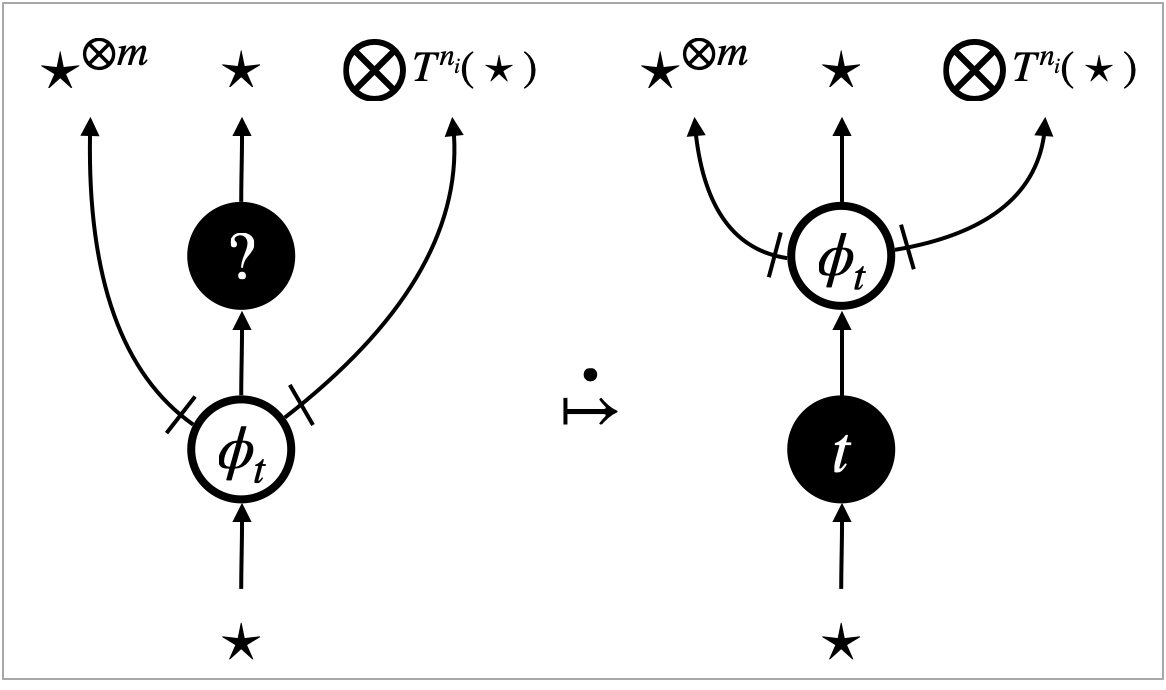}
 }
 \hfil
 \subfloat[\label{fig:interaction-opr-none}]{
 \includegraphics[scale=.2]{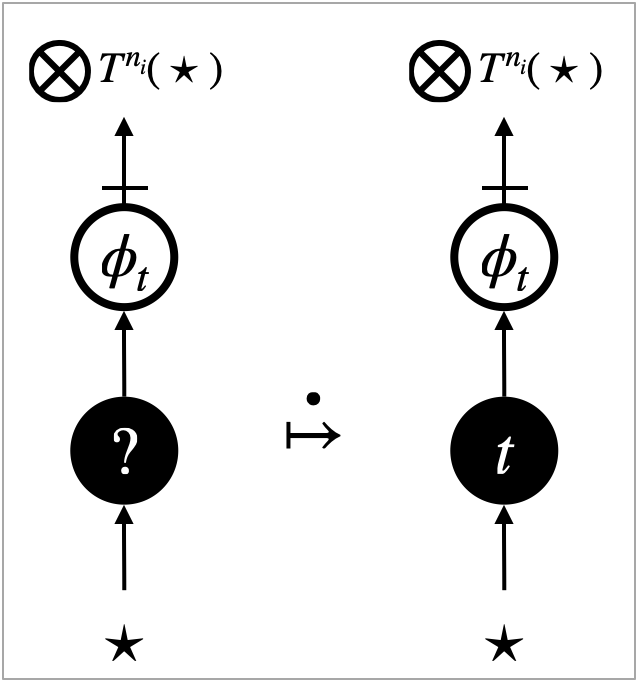}
 }
 \hfil
 \subfloat[\label{fig:subst-rule}]{
 \includegraphics[scale=.2]{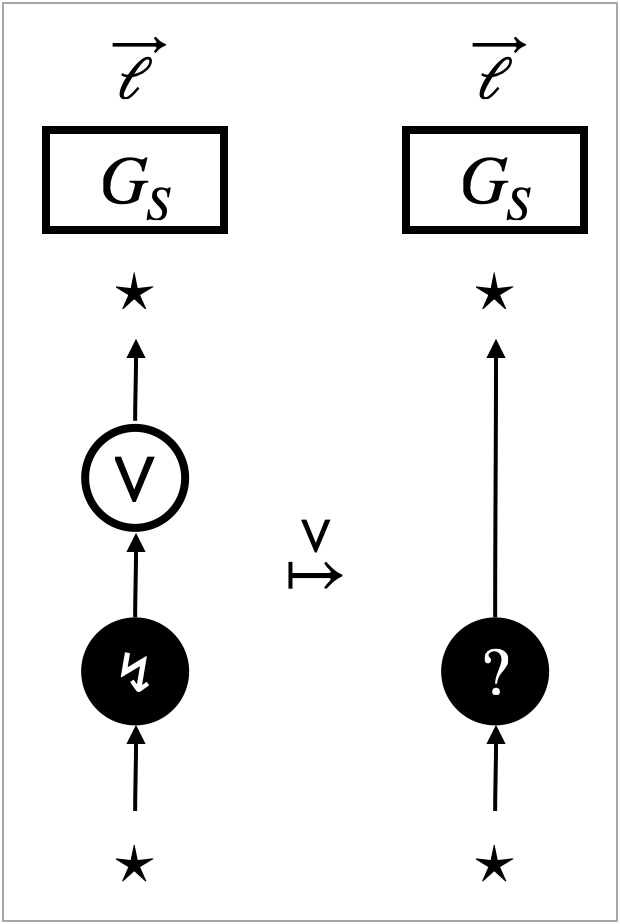}
 }
 \caption{Interaction rules \protect\subref{fig:interaction-V}--\protect\subref{fig:interaction-opr-none} and the substitution rule \protect\subref{fig:subst-rule}, where $t \in \{\checkmark,\lightning\}$ and $G_S$ is a hypernet}
 \label{fig:interaction-subst-rules}
\end{figure}
Search transitions are possible for the $?$-focus and the $\checkmark$-focus, and they implement the depth-first search of redexes. They are specified by \emph{interaction rules} depicted in \autoref{fig:interaction-V}--\ref{fig:interaction-opr-none}. The $?$-focus interacts with what is connected above, and the $\checkmark$-focus interacts with what is connected below. From the perspective of program evaluation, the interaction rules specify the left-to-right call-by-value evaluation of arguments.
\begin{description}
 \item[\autoref{fig:interaction-V}] When the $?$-focus encounters a variable ($\mathsf{V}$), it changes to the $\lightning$-focus. What is connected above the variable will be substituted for the variable, in a subsequent substitute transition.
 \item[\autoref{fig:interaction-opr-first}] When the $?$-focus encounters an operation $\phi$ with at least one argument, it proceeds to the first argument.
 \item[\autoref{fig:interaction-opr-next}] After inspecting the $(m+1)$-th argument, the $\checkmark$-focus changes to the $?$-focus and proceeds to the next argument.
 \item[\autoref{fig:interaction-opr-last}] After inspecting all the arguments, the $\checkmark$-focus finishes redex search and changes to a focus depending on the operation $\phi_t$: to the $\checkmark$-focus for a passive operation $\phi_\checkmark$, and to the $\lightning$-focus for an active operation $\phi_\lightning$.
 \item[\autoref{fig:interaction-opr-none}] When the $?$-focus encounters an operation that takes no arguments but only thunks, it immediately finishes redex search and changes to a focus depending on the operation $\phi_t$, like in \autoref{fig:interaction-opr-last}.
\end{description}

The first kind of rewrite transitions, namely substitution transitions, implements substitution by simply removing a variable edge ($\mathsf{V}$). These transitions are specified by a \emph{substitution rule} depicted in \autoref{fig:subst-rule}.
What is connected above the variable edge ($\mathsf{V}$) is computation bound to the variable. By removing the variable edge, the bound computation gets directly connected to the $?$-focus, ready for redex search.

\subsection{Extrinsic transitions: behaviour of operations} \label{sec:extrinsic-trans}

The second kind of rewrite transitions are for operations $\Opr$, in particular active operations $\Opr_\lightning$. These transitions are \emph{extrinsic}; they are given as the second parameter $B_\Opr$, called \emph{behaviour} of $\Opr$, of the UAM.

\begin{figure}[t]
 \centering
 \subfloat[Arithmetic\label{fig:rewrite-arith}]{
 \includegraphics[scale=.2]{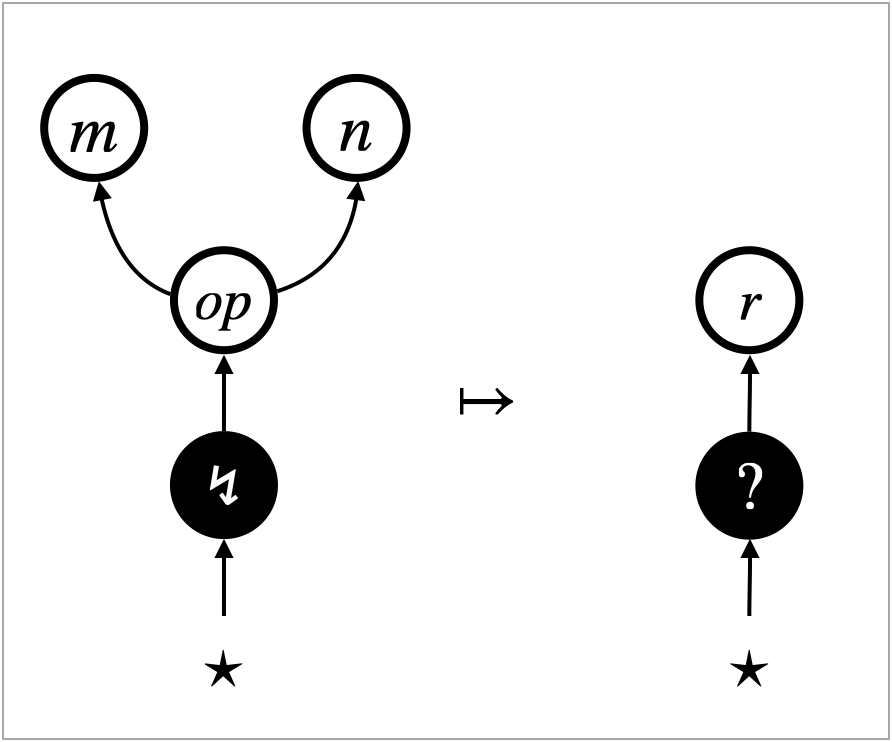}
 }
 \hfil
 \subfloat[Micro-beta\label{fig:rewrite-micro-beta}]{
 \includegraphics[scale=.2]{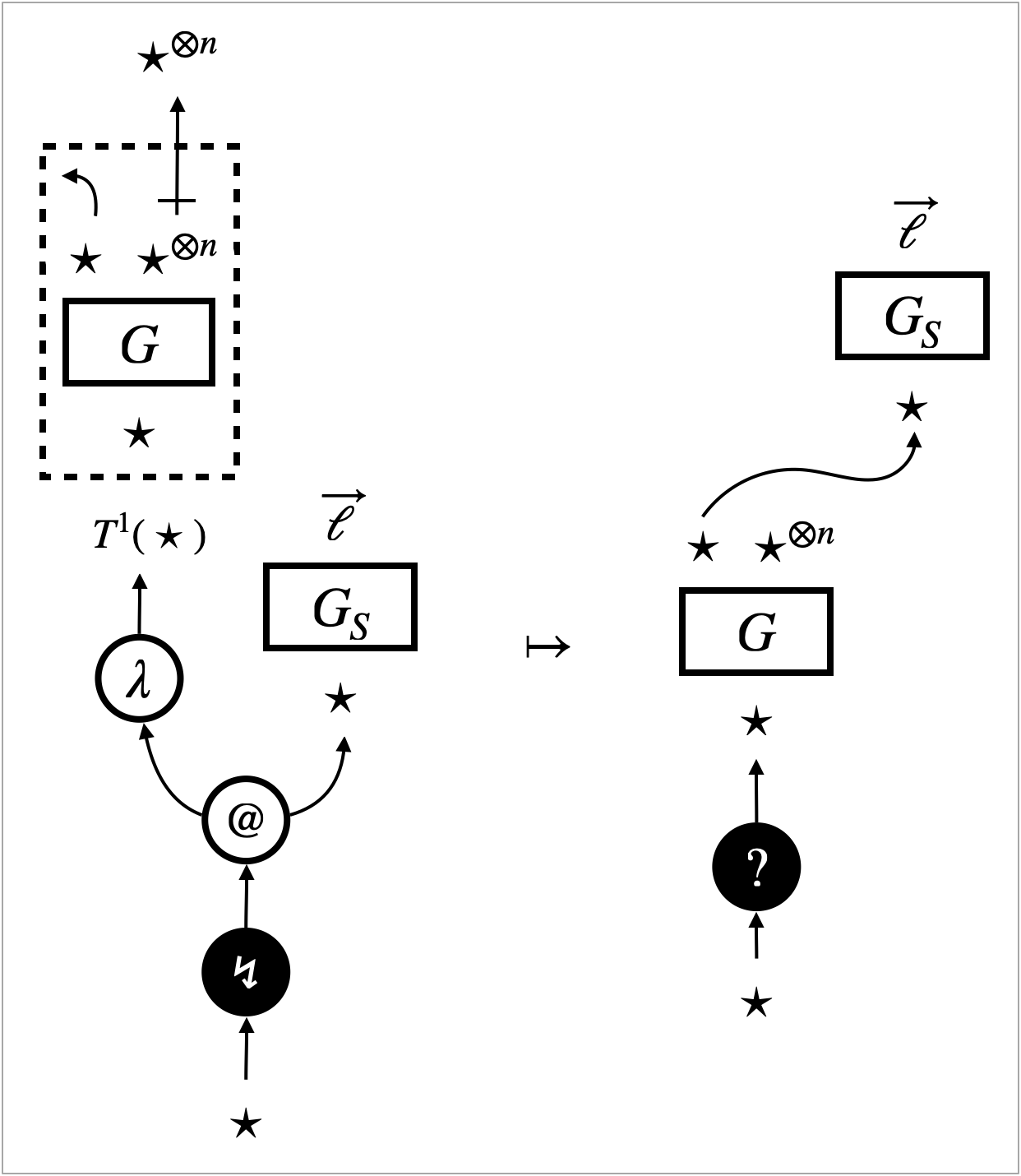}
 }
 \hfil
 \subfloat[$\keyw{stat}$\label{fig:rewrite-stat}]{
 \includegraphics[scale=.2]{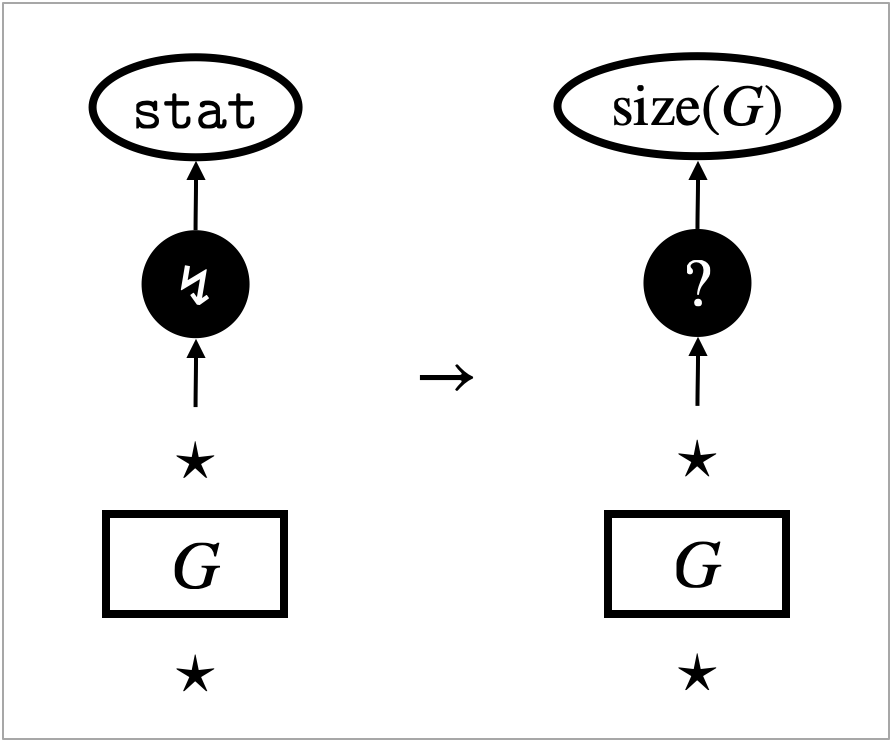}
 }
 \caption{The example behaviour $B_{\{+,-,\lrapp,\keyw{stat}\}}$ where $r = t \mathop{\mathit{op}} u, \mathit{op} \in \{+,-\}$ and $G,G_S$ are hypernets}
 \label{fig:behaviour-ex}
\end{figure}
\autoref{fig:behaviour-ex} shows an example of the behaviour, namely that for the following active operations:
\begin{equation*}
 \mathit{+} \colon \star \To \star^{\otimes 2}, \quad
 \mathit{-} \colon \star \To \star^{\otimes 2}, \quad
  \lrapp \colon \star \To \star^{\otimes 2}, \quad
 \keyw{stat} \colon \star \To \epsilon.
\end{equation*}
For some of these operations, their behaviour is specified locally by \emph{rewrite rules}.
The rewrite transitions for the four active operations are enabled when the $\lightning$-focus encounters one of the operations. The $\lightning$-focus then changes to the $?$-focus and resumes redex search.
\begin{description}
 \item[\autoref{fig:rewrite-arith}] This rewrite rule specifies the behaviour of arithmetic ($+,-$). The rule eliminates all three edges ($m,n,\mathit{op}$) and replace them with a new edge ($r$) such that $r = t \mathop{\mathit{op}} u$.
 \item[\autoref{fig:rewrite-micro-beta}] This rewrite rule specifies the behaviour of application ($\lrapp$), namely the (micro-)\linebreak[5] beta-reduction. It is \emph{micro} in the sense that it delays substitution. It only eliminates the constructors ($\lambda,@$), opens the box whose content is $G$, and connects $G_S$ which represents a function argument to the body $G$ of the function.
 \item[\autoref{fig:rewrite-stat}] This is a rewrite transition, not a rewrite rule. It is for the operation $\keyw{stat}$ that inspects memory usage. Namely, $\keyw{stat}$ counts the number $\mathrm{size}(G)$ of edges in the hypernet $G$. The transition replaces the operation edge ($\keyw{stat}$) with the result ($\mathrm{size}(G)$) of counting.
\end{description}

\section{A formal definition of the UAM} \label{sec:UAM-details}

The UAM, and hence the definitions below, are all globally parameterised by the operation set $\Opr$ and its behaviour $B_\Opr$.

\subsection{Auxiliary definitions} \label{sec:auxiliary-def}

We use the terms \emph{incoming} and \emph{outgoing} to characterise the incidence relation between neighbouring edges. Conventionally incidence is defined relative to nodes, but we find it helpful to extend this notion to edges. 
\begin{defi}[Incoming and outgoing edges]
An \emph{incoming} edge of an edge $e$ has a target that is a source of
the edge $e$. An \emph{outgoing} edge of the edge $e$ has a source
that is a target of the edge $e$.
\end{defi}

The notions of \emph{path} and \emph{reachability} are standard.
Our technical development will heavily rely on these graph-theoretic notions. Note that these are the notions that are difficult to translate back into the language of terms.
\begin{defi}[Paths and reachability] \label{def:path-reachability}
 \noindent
 \begin{enumerate}
  \item \label{item:def:path}
A \emph{path} in a hypergraph is given by a non-empty sequence of
edges, where an edge $e$ is followed by an edge $e'$ if the edge $e$
is an incoming edge of the edge $e'$.
  \item \label{item:def:reachability}
A vertex $v'$ is \emph{reachable} from a vertex $v$ if $v = v'$
holds, or there exists a path from the vertex $v$ to the vertex $v'$.
 \end{enumerate}
\end{defi}
Note that, in general, the first edge (resp.\ the last edge) of a path
may have no source (resp.\ target).
A path is said to be \emph{from} a vertex $v$,
if $v$ is a source of the first edge of the path.
Similarly, a path is said to be \emph{to} a vertex $v'$,
if $v'$ is a target of the last edge of the path.
A hypergraph $G$ is itself said to be a path, if all edges of $G$
comprise a path from an input (if any) and an output (if any) and
every vertex is an endpoint of an edge.

During focussed hypernet rewriting, operations are the only edges that the $?$-focus can ``leave behind''. The $?$-focus is always at the end of an \emph{operation path}.
\begin{defi}[Operation paths]
A path whose edges are all labelled with operations is called
\emph{operation path}.
\end{defi}

We shall introduce a few classes of hypernets below.
The first is \emph{box} hypernets that are simply single box edges.
\begin{defi}[Box hypernets]
If a hypernet is a path of only one box edge, it is called \emph{box}
hypernet.
\end{defi}
The second is \emph{stable} hypernets, in which a focus can never trigger a rewrite (i.e.\ a focus never changes to the $\lightning$-focus). Stable hypernets can be seen as a graph-based notion of values/normal form. For example, the hypernet that consists of an abstraction edge ($\lambda$) only is a stable hypernet.
\begin{defi}[Stable hypernets] \label{def:stable-HN}
A \emph{stable} hypernet is a hypernet
$(G : \star \To \otimes_{i=1}^m \ell_i) \in
\mathcal{H}(\LVlin, \Opr_\checkmark)$, such that
$\otimes_{i=1}^m \ell_i\in
(\{ T^n(\star) \mid n \in \N \})^m$
and each vertex is reachable from the unique input.
\end{defi}
The last is \emph{one-way} hypernets, which will play an important role in local reasoning. These specify sub-graphs to which a focus enters only from the bottom (i.e.\ the $?$-focus through an input), never from the top (i.e.\ the $\checkmark$-focus through an output). Should the $\checkmark$-focus enter from the top, it must have traversed upwards the sub-graph and left an operation path behind. One-way hypernets are defined by ruling out such operation paths.
\begin{defi}[One-way hypernets] \label{def:one-way-HN}
A hypernet $H$ is \emph{one-way} if, for any pair $(v_i,v_o)$ of an
input and an output of $H$ such that $v_i$ and $v_o$ both have type
$\star$, any path from $v_i$ to $v_o$ is not an operation path.
\end{defi}
For example, the underlying hypernet $H$ of the left hand side of the micro-beta rewrite rule (\autoref{fig:rewrite-micro-beta}) is a one-way hypernet, if $G_S$ is stable.
Should the $\checkmark$-focus enters to $H$ from the top, it must be backtracking the depth-first search, and hence the $?$-focus must have been visited $H$ from the bottom. In the presence of the micro-beta rewrite rule, such visit must result in a rewrite transition, and therefore, the backtracking of the $\checkmark$-focus cannot be possible.

\subsection{Focussed hypernets} \label{sec:focussed-hypernets}

Focussed hypernets are those that contain a focus. We impose some extra conditions as below, to ensure that the focus is outside a box and not isolated.
\begin{defi}[Focussed hypernets]
 \noindent
 \begin{enumerate}
  \item 
A focus in a hypergraph is said to be \emph{exposed} if its
source is an input and its target is an output, and
\emph{self-acyclic} if its source and its target are different
vertices.
  \item 
\emph{Focussed} hypernets (typically ranged over by
$\focussed{G},\focussed{H},\focussed{N}$) are those that contain only
one focus and the focus is shallow, self-acyclic and not exposed.
 \end{enumerate}
\end{defi}

\emph{Focus-free} hypernets are given by
$\HN{\LVlin}{\LElin(\Opr) \backslash \{?,\checkmark,\lightning\}}$,
i.e.\ hypernets without a focus.

\begin{notation}[Removing, replacing and attaching a focus]
 \noindent
 \begin{enumerate}
  \item 
A focussed hypernet $\focussed{G}$ can be turned into an
\emph{underlying} focus-free hypernet $|\focussed{G}|$ with the same
type, by removing its
unique focus and identifying the source and the target of the
focus.
  \item 
When a focussed hypernet $\focussed{G}$ has a
$\mathsf{t}$-focus, then changing
the focus label $\mathsf{t}$ to another one $\mathsf{t}'$ yields a
focussed hypernet denoted by
$\langle\focussed{G}\rangle_{\mathsf{t}'/\mathsf{t}}$.
  \item 
Given a focus-free hypernet $G$,
a focussed hypernet $\mathsf{t} ;_i G$ with the same type
can be yielded by connecting a
$\mathsf{t}$-focus to the $i$-th input of $G$ if the input has type
$\star$. Similarly,
a focussed hypernet $G ;_i \mathsf{t}$ with the same type
can be yielded by connecting a
$\mathsf{t}$-focus to the $i$-th output of $G$ if the output has type
$\star$. 
If it is not ambiguous, we omit the index $i$ in the notation $ ;_i $.
 \end{enumerate}
\end{notation}
The source (resp.\ target) of a focus is called ``focus source''
(resp.\ ``focus target'') in short.

\subsection{Contexts} \label{sec:contexts}

We next formalise a notion of \emph{context}, which is hypernets with \emph{holes}.
We use a set $\MVar$ of hole labels, and contexts are allowed to contain an arbitrary number of holes.
Hole labels are typed, and typically ranged over by
$\chi : \vec{\ell} \To \vec{\ell'}$.
\begin{defi}[(Simple) contexts]
 \noindent
 \begin{enumerate}
  \item 
\emph{Holed} hypernets (typically ranged over by
$\mathcal{C}$) are given by
$\HN{\LVlin}{\LElin(\Opr) \cup \MVar}$, where the edge label set $\LElin(\Opr)$ is
extended by the set $\MVar$.
  \item 
 A holed hypernet $\mathcal{C}$ is said to be \emph{context}
 if each hole label appears at most once (at any depth) in
 $\mathcal{C}$.
  \item 
 A \emph{simple} context is a context that contains a single hole,
 which is shallow.
 \end{enumerate}
\end{defi}

By what we call \emph{plugging}, we can replace a hole of a context with a hypernet, and obtain a new context. We here provide a description of plugging and fix a notation. A formal definition of plugging can be found in \autoref{app:plugging}.
\begin{notation}[Plugging of contexts]
 \noindent
 \begin{enumerate}
  \item
When $\vec{\chi}$ gives a list of all and only hole labels that appear
in a context $\mathcal{C}$, the context can also be written as
$\mathcal{C}[\vec{\chi}]$.
A hypernet in $\HN{\LVlin}{\LElin(\Opr)}$ can be seen as a context without a hole and written as $\mathcal{C}[\, ]$.
  \item 
Let $\mathcal{C}[\vec{\chi^1},\chi,\vec{\chi^2}]$ and
$\mathcal{C}'[\vec{\chi^3}]$ be contexts, such that
the hole $\chi$ and the latter context $\mathcal{C}'$ have the same
type and
$\vec{\chi^1} \cap \vec{\chi^2} \cap \vec{\chi^3} = \emptyset$.
A new context
$\mathcal{C}[\vec{\chi^1},\mathcal{C}',\vec{\chi^2}]
\in \HN{\LVlin}{\LElin \cup \vec{\chi^1} \cup \vec{\chi^3} \cup \vec{\chi^2}})$ can be obtained by \emph{plugging} $\mathcal{C}'$
into $\mathcal{C}$:
namely, by replacing the (possibly deep) hole edge of
$\mathcal{C}$ that has
label $\chi$ with the context $\mathcal{C}'$, and by identifying
each input (resp.\ output) of $\mathcal{C}'$ with its corresponding
source (resp.\ target) of the hole edge.
 \end{enumerate}
\end{notation}

Each edge of
the new context $\mathcal{C}[\vec{\chi^1},\mathcal{C}',\vec{\chi^3}]$
is inherited from either $\mathcal{C}$ or $\mathcal{C}'$, keeping the
type; this implies that the new context is indeed a context with hole
labels $\vec{\chi^1},\vec{\chi^3},\vec{\chi^2}$.
Inputs and outputs of
the new context 
coincide with those of the original context $\mathcal{C}$,
and hence these two
contexts have the same type.

The plugging is associative in two senses:
plugging two contexts into two holes of a context yields the same
result regardless of the order, i.e.\
$\mathcal{C}[\vec{\chi^1},\mathcal{C}',\vec{\chi^2},\mathcal{C}'',\vec{\chi^3}]$
is well-defined;
and nested plugging yields the same result regardless of the order,
i.e.\
$\mathcal{C}[\vec{\chi^1},\mathcal{C}'[\vec{\chi^3},\mathcal{C}'',\vec{\chi^4}],\vec{\chi^2}]
=
(\mathcal{C}[\vec{\chi^1},\mathcal{C}',\vec{\chi^2}])[\vec{\chi^1},\vec{\chi^3},\mathcal{C}'',\vec{\chi^4},\vec{\chi^2}]$.

The notions of focussed and focus-free hypernets can be naturally
extended to contexts.
We use the terms \emph{entering} and \emph{exiting} to refer to a focus that is adjacent to a hole. A focus may be both entering and exiting.
\begin{defi}[Entering/exiting focuses]
In a focussed context $\focussed{\mathcal{C}}[\vec{\chi}]$, the focus is said to
be \emph{entering} if it is an incoming edge of a hole, and
\emph{exiting} if it is an outgoing edge of a hole.
\end{defi}

\subsection{States and transitions} \label{sec:states-transitions}

We now define the UAM as a state transition system.
States are hypernets that represent closed terms and hence have type $\star \To \epsilon$.
\begin{defi}[States]
 \noindent
 \begin{enumerate}
  \item A \emph{state} is given by a focussed hypernet $\focussed{G} \in \HN{\LVlin}{\LElin(\Opr)}$ of type $\star \To \epsilon$.
  \item A state $\focussed{G}$ is called \emph{initial} if $\focussed{G} = ? ; |\focussed{G}|$, and \emph{final} if $\focussed{G} = \checkmark ; |\focussed{G}|$.
 \end{enumerate}
\end{defi}
The following will be apparent once transitions are defined:
initial states are indeed initial in the sense that no search
transition results in an initial state; and final states are indeed final
in the sense that no transition is possible from a final state.

Intrinsic transitions, which consists of search transitions and substitution transitions, are specified by the interaction rules and the substitution rule in \autoref{fig:interaction-subst-rules}. Each intrinsic transition applies a rule outside a box and at one place.
\begin{defi}[Intrinsic transitions]
 \noindent
 \begin{enumerate}
  \item For each interaction rule $\focussed{G} \overset{\bullet}{\mapsto} \focussed{G'}$, if there exists a focus-free simple context $\mathcal{C}[\chi] : \star \To \epsilon$ such that $\mathcal{C}[\focussed{G}]$ and $\mathcal{C}[\focussed{G'}]$ are states,
$\mathcal{C}[\focussed{G}] \to \mathcal{C}[\focussed{G'}]$ is a \emph{search transition}.
  \item For each substitution rule $\focussed{G} \overset{\mathsf{V}}{\mapsto} \focussed{G'}$, if there exists a focus-free simple context $\mathcal{C}[\chi] : \star \To \epsilon$ such that $\mathcal{C}[\focussed{G}]$ and $\mathcal{C}[\focussed{G'}]$ are states, $\mathcal{C}[\focussed{G}] \to \mathcal{C}[\focussed{G'}]$ is a \emph{substitution transition}.
 \end{enumerate}
\end{defi}
When a sequence $\focussed{G} \to^* \focussed{G'}$ of transitions
consists of search transitions only, it is annotated by the symbol
$\bullet$ as $\focussed{G} \tobul^* \focussed{G'}$.

Extrinsic transitions $B_\Opr$ must have a specific form; namely they must be a \emph{compute transition}.
\begin{defi}[Compute transitions]
 A transition
 $\focussed{G} \to \focussed{G'}$ is a \emph{compute transition} if: (i) the first
 state $\focussed{G}$ has the $\lightning$-focus that is an incoming
 edge of an active operation edge; and (ii) the second
 state $\focussed{G'}$ has the $?$-focus.
\end{defi}
We can observe that substitution transitions or compute transitions are possible if and only if a state has the $\lightning$-focus, and they always change the focus to the $?$-focus. We refer to substitution transitions and compute transitions altogether as \emph{rewrite} transitions (cf.\ \autoref{tab:UAM-transitions}).

Compute transitions may be specified
locally, by \emph{rewrite rules}, in the same manner as the intrinsic
transitions. \autoref{fig:rewrite-arith}~\&~\ref{fig:rewrite-micro-beta} shows examples of rewrite rules. We leave it entirely open what the actual rewrite
associated to some operation is, by having the behaviour $B_\Opr$ as
parameter of the UAM as well as the operation set $\Opr$.
This is part of the semantic
flexibility of our framework. We do not specify a meta-language for
encoding effects as particular transitions. Any \emph{algorithmic}
state transformation (e.g.\ the compute transition for $\keyw{stat}$ in \autoref{fig:rewrite-stat}) is acceptable.

We can now define the UAM as follows.
\begin{defi}[the UAM]
 Given two parameters $\Opr$ and $B_\Opr$, the \emph{universal abstract machine (UAM)} $\UAM(\Opr,B_\Opr)$ is given by data $(S_\Opr, T \uplus B_\Opr)$ such that:
 \begin{itemize}
  \item $S_\Opr \subseteq \HN{\LVlin}{\LElin(\Opr)}$ is a set of states,
  \item $T \subseteq S_\Opr \times S_\Opr$ is a set of intrinsic transitions, and
  \item $B_\Opr \subseteq S_\Opr \times S_\Opr$ is a set of compute transitions.
 \end{itemize}
\end{defi}
We refer to elements of $B_\Opr$ as \emph{extrinsic} transitions, as well as \emph{compute} transitions; we use these two terms interchangeably.

An \emph{execution} of the UAM starts with a focus-free hypernet that represents a closed term, e.g.\ a result of the translation $(-)^\dag$ (cf.\ \autoref{fig:transl-linear}). It is successful if it terminates with a final state, and not if it gets stuck with a non-final state.
\begin{defi}[Execution and stuck states]
 \noindent
 \begin{enumerate}
  \item An \emph{execution} on a focus-free hypernet $G : \star \To \epsilon$ is a sequence of transitions starting from the initial state $? ; {G}$.
  \item A state is said to be \emph{stuck} if it is not final and cannot be followed by any transition.
 \end{enumerate}
\end{defi}

Recall that we have a notion of \emph{stable hypernet} (\autoref{def:stable-HN}) that is a graphical counterpart of values.
An execution on any stable hypernet
terminates successfully at a final state, with only search transitions
(cf.\ \autoref{lem:AnswerAccessiblePaths}(\ref{item:ValueToken}) which is proved for the ``non-linear'' UAM).

\section{Contextual equivalence on hypernets} \label{sec:ctxt-refinement}

In this section, we set the target of our proof methodology. First, we define contextual equivalence in a general manner. Next, we clarify what kind of operations $\Opr$ our proof methodology applies to.

\subsection{Generalised contextual equivalence} \label{sec:general-ctxt-refinement}

We propose notions of contextual refinement and equivalence that
check for successful termination of execution\footnote{%
We opt for the very basic notion of contextual refinement that concerns termination only. Richer observation (e.g.\ evaluation results, output, probability, nondeterminism) would require different definitions of contextual refinement, and these are out of the scope of this paper.
}.
Our notion of contextual refinement (and hence contextual equivalence) generalise the standard notions in two ways.
\begin{itemize}
 \item The notion of contextual refinement can be flexible in terms of a class of contexts in which it holds. Namely, contextual refinement is parameterised by a set $\mathbb{C} \subseteq \HN{\LVlin}{\LElin(\Opr) \cup \MVar}$ of focus-free contexts. The standard contextual refinement can be recovered by setting $\mathbb{C}$ to be the set of all focus-free contexts.
 \item The notion of contextual refinement can count and compare the number of transitions. Namely, contextual refinement is parameterised by a preorder $Q$ on natural numbers. The standard contextual refinement can be obtained by setting $Q$ to be the total relation $\N \times \N$. Other typical examples of the preorder $Q$ are the greater-than-or-equal relation $\geq_\N$ and the equality $=_\N$. With these preorders, one can prove that two terms are contextually equivalent, and moreover, one takes a less number of transitions to terminate than the other (with $\geq_\N$) or the two terms take exactly the same number of transitions to terminate (with $=_\N$).
\end{itemize}
We require the parameter $\mathbb{C}$ to be closed under plugging,
i.e.\ for any contexts
$\mathcal{C}[\vec{\chi^1},\chi,\vec{\chi^2}],$ $\mathcal{C}' \in
\mathbb{C}$
such that
$\mathcal{C}[\vec{\chi^1},\mathcal{C}',\vec{\chi^2}]$ is defined,
$\mathcal{C}[\vec{\chi^1},\mathcal{C}',\vec{\chi^2}] \in \mathbb{C}$.
\begin{defi}[State refinement and equivalence]
 \label{def:StateRefEquiv}
 Let $Q$ be a preorder on $\N$, and
 $\focussed{G_1}$ and $\focussed{G_2}$ be two states.
 \begin{itemize}
  \item $\focussed{G_1}$ is said to \emph{refine} $\focussed{G_2}$
	up to $Q$, written as
	$B_\Opr \models
	(\focussed{G_1} \dotrel{\preceq}_Q \focussed{G_2})$,
	if for any number $k_1 \in \N$ and any final state
	$\focussed{N_1}$ such that
	$\focussed{G_1} \to^{k_1} \focussed{N_1}$,
	there exist a number $k_2 \in \N$ and a final state
	$\focussed{N_2}$ such that
	$k_1 \mathrel{Q} k_2$ and
	$\focussed{G_2} \to^{k_2} \focussed{N_2}$.
  \item $\focussed{G_1}$ and $\focussed{G_2}$ are said to be
	\emph{equivalent} up to $Q$, written as
	$B_\Opr \models
	(\focussed{G_1} \dotrel{\simeq}_Q \focussed{G_2})$, if
	$B_\Opr \models
	(\focussed{G_1} \dotrel{\preceq}_Q \focussed{G_2})$
	and $B_\Opr \models
	(\focussed{G_2} \dotrel{\preceq}_Q \focussed{G_1})$.
 \end{itemize}
\end{defi}
\begin{defi}[Contextual refinement and equivalence]
 \label{def:CtxtRefEquiv}
 Let $\mathbb{C}$ be a set of contexts that is closed under plugging,
 $Q$ be a preorder on $\N$, and
 $H_1$ and $H_2$ be focus-free hypernets of the same type.
 \begin{itemize}
  \item $H_1$ is said to \emph{contextually refine} $H_2$
	in $\mathbb{C}$ up to $Q$,
	written as $B_\Opr \models (H_1 \preceq^\mathbb{C}_Q H_2)$,
	if any focus-free context $\mathcal{C}[\chi] \in \mathbb{C}$,
	such that
	$? ; \mathcal{C}[H_1]$ and $? ; \mathcal{C}[H_2]$ are states,
	yields refinement
	$B_\Opr \models
	({? ; \mathcal{C}[H_1]} \dotrel{\preceq}_Q
	{? ; \mathcal{C}[H_2]})$.
  \item $H_1$ and $H_2$ are said to be \emph{contextually equivalent}
	in $\mathbb{C}$ up to $Q$,
	written as $B_\Opr \models (H_1 \simeq^\mathbb{C}_Q H_2)$,
	if $B_\Opr \models (H_1 \preceq^\mathbb{C}_Q H_2)$ and
	$B_\Opr \models (H_2 \preceq^\mathbb{C}_Q H_1)$.
 \end{itemize}
\end{defi}
In the sequel, we simply write
$\focussed{G_1} \dotrel{\preceq}_Q \focussed{G_2}$ etc., making the
parameter $B_\Opr$ implicit.

Because $Q$ is a preorder,
$\dotrel{\preceq}_Q$ and $\preceq^\mathbb{C}_Q$ are indeed preorders,
and accordingly, equivalences
$\dotrel{\simeq}_Q$ and $\simeq^\mathbb{C}_Q$ are indeed equivalences
(\autoref{lem:QReflTransSym}).

When the relation $Q$ is the universal relation
$\N \times \N$, the notions concern successful termination, and the
number of transitions is irrelevant.
If all compute transitions are deterministic, contextual equivalences
$\simeq^\mathbb{C}_{\geq_\N}$ and $\simeq^\mathbb{C}_{=_\N}$ coincide
for any $\mathbb{C}$
(as a consequence of \autoref{lem:BoundedIdentityCtxtEquiv}).

Because $\mathbb{C}$ is closed under plugging,
the contextual notions $\preceq^\mathbb{C}_Q$ and
$\simeq^\mathbb{C}_Q$ indeed become congruences.
Namely, for
any $H_1 \mathrel{\square}^\mathbb{C} H_2$ and
$\mathcal{C} \in \mathbb{C}$ such
that $\mathcal{C}[H_1]$ and $\mathcal{C}[H_2]$ are defined,
$\mathcal{C}[H_1] \mathrel{\square}^\mathbb{C} \mathcal{C}[H_2]$,
where $\square \in \{ \preceq_Q, \simeq_Q \}$.

As the parameter $\mathbb{C}$, we will use the set $\mathbb{C}_{\Opr}$ of all focus-free contexts, for the time being. We will use another set in \autoref{sec:obs-equiv-lambda}.
The standard notions of contextual refinement and equivalence can be
recovered as $\preceq^{\mathbb{C}_\Opr}_{\N \times \N}$ and
$\simeq^{\mathbb{C}_\Opr}_{\N \times \N}$.

\subsection{Determinism and refocusing} \label{sec:determ-refocusing}

We will focus on operations $\Opr$ whose behaviour $B_\Opr$ makes the UAM both \emph{deterministic} and \emph{refocusing} in the following sense.
\begin{defi}[Determinism and refocusing] \label{def:determ-refocusing}
 \noindent
 \begin{enumerate}
  \item A UAM $\UAM(\Opr,B_\Opr)$ is \emph{deterministic} if the following holds: if two transitions $\focussed{G} \to \focussed{G'}$ and $\focussed{G} \to \focussed{G''}$ are possible, it holds that $\focussed{G'} = \focussed{G''}$ up to graph isomorphism.
  \item A state $\focussed{G}$ is \emph{rooted} if
	$?;|\focussed{G}| \tobul^* \focussed{G}$.
  \item A UAM $\UAM(\Opr,B_\Opr)$ is
	\emph{refocusing} if every transition preserves the rooted
	property.
 \end{enumerate}
\end{defi}

In a refocusing UAM, the rooted property becomes an invariant, because any initial state is trivially rooted.
The invariant ensures the following: 
 starting the search process (i.e.\ a search transition) from a
 state $\focussed{N'}$ with the $?$-focus
 can be seen as \emph{resuming} the search process
 $? ; |\focussed{N'}| \tobul^* \focussed{N'}$, from an initial state,
 on the underlying hypernet $|\focussed{N'}|$.
 Resuming redex search after a rewrite, rather than starting from
 scratch, is an important aspect of abstract machines.
 In the case of the lambda-calculus,
 enabling the resumption is identified as one of the key steps (called
 ``refocusing'') to synthesise abstract machines from reduction
 semantics by Danvy et al.~\cite{DanvyMMZ12}.
 In our setting, it is preservation of the rooted property that
 justifies the resumption.

For many rewrite transitions that are specified by a local rewrite rule, it suffices to check the shape of the rewrite rule, in order to conclude that the corresponding rewrite transition preserves the rooted property.
Intuitively, a rewrite rule $\focussed{H} \mapsto \focussed{H'}$ (or the substitution rule) should satisfy the following.
\begin{enumerate}
 \item No $\checkmark$-focus can encounter $|\focussed{H}|$ prior to application of the rewrite rule $\focussed{H} \mapsto \focussed{H'}$.
 \item The rewrite rule changes only edges above the $\lightning$-focus, and turns the focus into the $?$-focus without moving it around.
 \item If the $?$-focus encounters $|\focussed{H}|$, subsequent search transitions yield the $\lightning$-focus.
\end{enumerate}
The above idea is formalised as a notion of \emph{stationary} rewrite transition. If a rewrite transition is \emph{stationary} in the following sense, it preserves the rooted property.
\begin{defi}[Stationary rewrite transitions]
 \label{def:StationaryRW}
 A rewrite transition $\focussed{G} \to \focussed{G'}$ is
 \emph{stationary} if there exist a focus-free simple context
 $\mathcal{C}$, focus-free hypernets $H$ and $H'$, and a number
 $i \in \N$, such that the following holds.
 \begin{enumerate}
  \item $H$ is one-way,
  \item $\focussed{G} = \mathcal{C}[\lightning ;_i H]$ and
 $\focussed{G'} = \mathcal{C}[? ;_i H']$, and
  \item 
 for any $j \in \N \backslash \{ i \}$, such that
 $\mathcal{C}[? ;_j H]$ is a state, there exists a state
 $\focussed{N}$ with the $\lightning$-focus, such that
 $\mathcal{C}[? ;_j H] \tobul \focussed{N}$.
 \end{enumerate}
\end{defi}
\begin{lem}[\autoref{lem:StationaryRWPreserveRooted}] \label{lem:stationary}
 If a rewrite transition $\focussed{G} \to \focussed{G'}$ is
 stationary, it preserves the rooted property, i.e.\
 $\focussed{G}$ being rooted implies $\focussed{G'}$ is also rooted.
\end{lem}

Finally, determinism and refocusing of a UAM boil down to those of extrinsic transitions $B_\Opr$ under a mild condition.
\begin{lem}[Determinism and refocusing] \label{lem:determ-refocusing}
 \noindent
 \begin{itemize}
  \item A universal abstract machine $\UAM(\Opr,B_\Opr)$ is deterministic if
	extrinsic transitions $B_\Opr$ are deterministic.
  \item Suppose that $G_S$ in the substitution rule (\autoref{fig:subst-rule}) is a stable hypernet.
	A universal abstract machine $\UAM(\Opr,B_\Opr)$ is refocusing if
	extrinsic transitions $B_\Opr$ preserve the rooted property.
 \end{itemize}
\end{lem}
\begin{proof}
 Intrinsic transitions and extrinsic transitions are
 mutually exclusive, and intrinsic transitions are all deterministic for the following reasons.
 \begin{itemize}
  \item Search transitions are deterministic, because at most one interaction
 rule can be applied at any state.
  \item 
 Although two different substitution rules may be possible at a state,
 substitution transitions are still deterministic.
 Namely, if two different substitution rules
 $\focussed{G} \overset{\mathsf{V}}{\mapsto} \focussed{G'}$ and
 $\focussed{H} \overset{\mathsf{V}}{\mapsto} \focussed{H'}$ can be applied to the same state,
 i.e.\ there exist focus-free simple contexts $\mathcal{C}_G$ and
 $\mathcal{C}_H$ such that
 $\mathcal{C}_G[\focussed{G}] = \mathcal{C}_H[\focussed{H}]$,
 then these two rules yield the same transition, by satisfying
 $\mathcal{C}_G[\focussed{G'}] = \mathcal{C}_H[\focussed{H'}]$.
 \end{itemize}
 Therefore, if extrinsic transitions are deterministic, all transitions become
 deterministic.

 Search transitions trivially preserve the rooted property. Substitution transitions
 also preserve the rooted property, because they are stationary, under the assumption that $G_S$ in \autoref{fig:subst-rule} is stable.
 Therefore, all transitions but extrinsic transitions already preserve the rooted
 property.
\end{proof}

\section{A sufficiency-of-robustness theorem} \label{sec:partial-char-thm}

In this section, we will state a \emph{sufficiency-of-robustness theorem} (\autoref{thm:MetaThm}), by identifying sufficient conditions (namely \emph{safety} and \emph{robustness}) for contextual refinement $N \preceq^\mathbb{C}_Q H$.
Throughout this section, we will use the micro-beta law, which looks like \autoref{fig:template-micro-beta}, as a leading example. It is derived from the micro-beta rewrite rule (\autoref{fig:rewrite-micro-beta}) by removing focuses, and it is the core of a graphical counterpart of the beta-law $(\lambda x. t)\ v \preceq t[v/x]$. We will introduce relevant notions and concepts (such as safety and robustness), by illustrating what it takes for the micro-beta law to hold.

\subsection{Pre-templates and specimens}

\begin{figure}[t]
 \includegraphics[scale=.2]{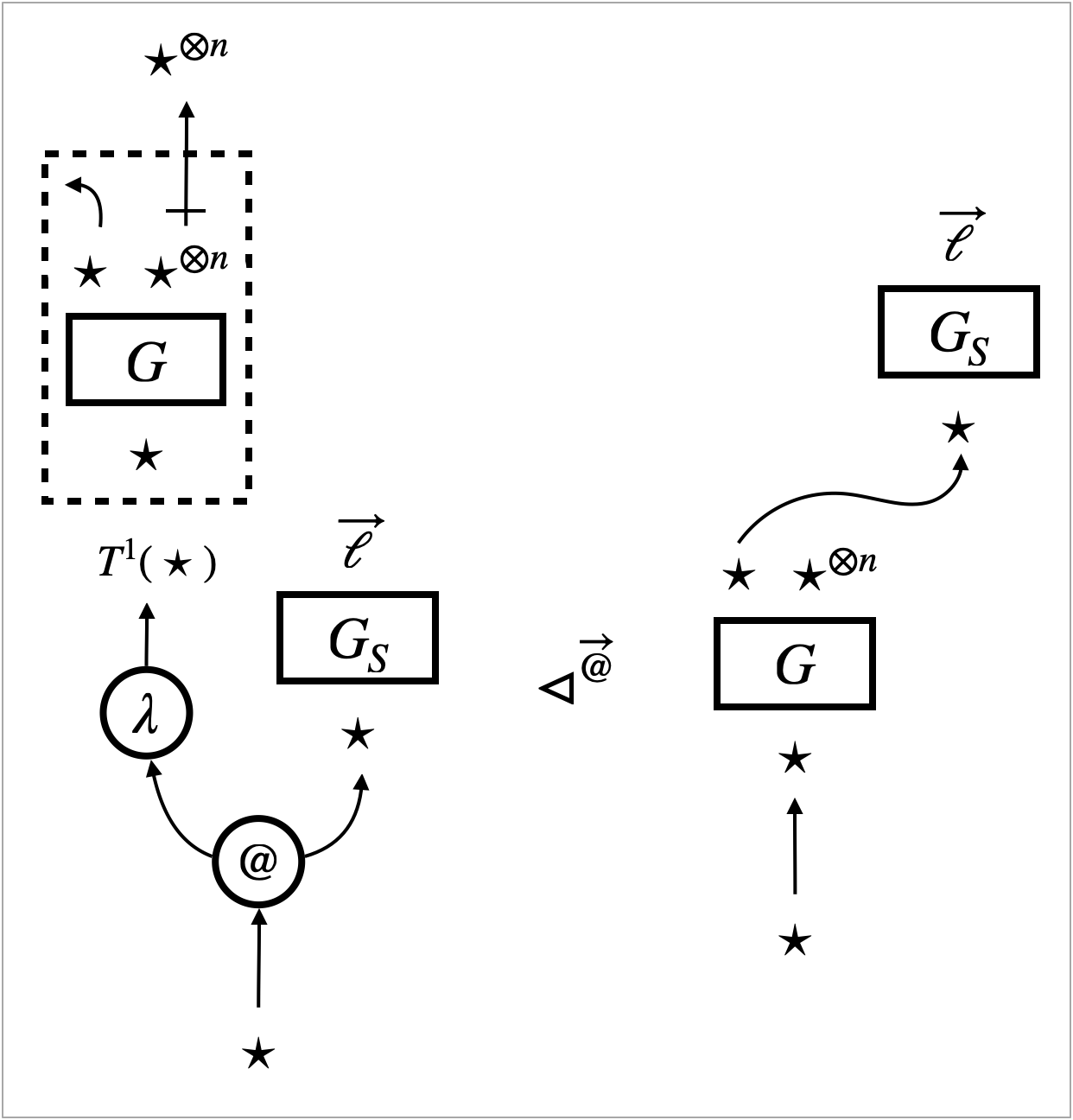}
 \caption{The micro-beta pre-template, where $G$ is a hypernet and $G_S$ is a stable hypernet}
 \label{fig:template-micro-beta}
\end{figure}

We begin with an observation that we are often interested in a \emph{family} of contextual refinements.
Syntactically, indeed, the beta-law $(\lambda x. t)\ v \preceq t[v/x]$ represents a family of (concrete) contextual refinements such as $(\lambda x. x)\ (\lambda y. y) \preceq \lambda y. y$. Graphically, it is the same; the micro-beta law represents a family of (concrete) laws. In the micro-beta law (\autoref{fig:template-micro-beta}), $G$ can be arbitrary and $G_S$ can be any stable hypernet.
The sufficiency-of-robustness theorem will therefore take a family $\vartriangleleft$ of pairs of focus-free hypernets (i.e.\ a relation $\vartriangleleft$ on focus-free hypernets), and identifies sufficient conditions for the family to imply a family of (concrete) contextual refinements.

Moreover, the relation $\vartriangleleft$ must be well-typed, i.e.\ each pair $(N,H) \in \mathit{\vartriangleleft}$ must share the same type. This in particular means that, if $N$ represents a term (or a thunk), $H$ must represents a term (or resp.\ a thunk) with the same number of free variables. We therefore formalise $\vartriangleleft$ as a type-indexed family of relations on focus-free hypernets, and call it \emph{pre-template}. A pre-template is our candidate of (a family of) contextual refinement.
\begin{defi}[Pre-templates]
A \emph{pre-template} is given by a union
$\mathord{\vartriangleleft} :=
{\cup_{I \in \mathcal{I}} \vartriangleleft_I}$
of a type-indexed family $\{ \vartriangleleft_I \}_{I \in \mathcal{I}}$, where $\mathcal{I}$ is a set of types.
 Each $\vartriangleleft_I$ is a binary relation on focus-free hypernets
 such that,
for any $G_1 \vartriangleleft_I G_2$ where $I \in \mathcal{I}$,
$G_1$ and $G_2$ are focus-free hypernets with type
$G_1 : I$ and $G_2 : I$.
\end{defi}

\begin{exa}[Micro-beta pre-template $\vartriangleleft^{\lrapp}$]
 \label{ex:BetaPreTemplate}
 As a leading example, we consider the \emph{micro-beta} pre-template
 $\vartriangleleft^{\lrapp}$, depicted in \autoref{fig:template-micro-beta}, derived from the micro-beta rewrite rule
 (\autoref{fig:rewrite-micro-beta}).
 The pre-template additionally requires $G_S$ to be stable, compared to the micro-beta rewrite rule; this amounts to require $v$ to be a value in the beta-law $(\lambda x. t)\ v \preceq t[v/x]$.
 For each $N \vartriangleleft^{\lrapp} H$,
 the hypernets $N$ and $H$ have the same type
 $\star \To
 \star^{\otimes n} \otimes \vec{\ell}$,
 where $n \in \N$ and the sequence $\vec{\ell}$ of types can be arbitrary as long as $G_S$ is stable.
 \bqed
\end{exa}

To directly prove that a pre-template $N \vartriangleleft H$ implies contextual refinement $N \preceq^{\mathbb{C}}_Q H$, one would need to compare states $\focussed{C}[N], \focussed{C}[H]$ for any focussed context $\focussed{C}$. We use data dubbed \emph{specimen} to provide such a pair.
Sometimes we prefer relaxed comparison, between two states $\focussed{P_1},\focussed{P_2}$ such that $\focussed{P_1} \mathrel{R} \focussed{C}[N]$ and $\focussed{C}[H] \mathrel{R'} \focussed{P_2}$ for some binary relations $R,R'$ on states. Such comparison can be specified by what we call \emph{quasi-specimen up to $(R,R')$}.
The notion of (quasi-)specimen is relative to the set $\mathbb{C}$ of focus-free contexts, which is one of the parameters of contextual refinement.
\begin{defi}[(Quasi-)specimens]
 \label{def:SpecimenQuasiSpecimen}
 Let $\vartriangleleft$ be a pre-template,
 and $R$ and $R'$ be binary relations on states.
 \begin{enumerate}
  \item \label{item:Specimen}
	A triple
	$(\focussed{\mathcal{C}}[\vec{\chi}];\vec{H^1};\vec{H^2})$ is
	a \emph{$\mathbb{C}$-specimen} of $\vartriangleleft$ if the
	following hold:
	\\ (A) $|\focussed{\mathcal{C}}[\vec{\chi}]| \in \mathbb{C}$,
	and the three sequences $\vec{\chi}, \vec{H^1}, \vec{H^2}$ have the same
	length $n$.
	\\ (B) $H^1_i \vartriangleleft H^2_i$ for each
	$i \in \{ 1,\ldots,n \}$.
	\\ (C) $\focussed{\mathcal{C}}[\vec{H^p}]$ is a state
	for each $p \in \{ 1,2 \}$.
  \item \label{item:QuasiSpecimen}
	A pair $(\focussed{N_1},\focussed{N_2})$ of states is
	a \emph{quasi-$\mathbb{C}$-specimen} of $\vartriangleleft$
	\emph{up to $(R,R')$}, if there exists a $\mathbb{C}$-specimen
	$(\focussed{\mathcal{C}};\vec{H^1};\vec{H^2})$ of
	$\vartriangleleft$ such that the following hold:
        \\ (A) The focuses of
	$\focussed{\mathcal{C}}$, $\focussed{N_1}$ and
	$\focussed{N_2}$ all have the same label.
        \\ (B) If $\focussed{N_1}$ and $\focussed{N_2}$ are rooted,
	then $\focussed{\mathcal{C}}[\vec{H^1}]$ and
	$\focussed{\mathcal{C}}[\vec{H^2}]$ are also rooted,
	$\focussed{N_1} \mathrel{R}
	\focussed{\mathcal{C}}[\vec{H^1}]$, and
	$\focussed{\mathcal{C}}[\vec{H^2}] \mathrel{R'}
	\focussed{N_2}$.
  \item A $\mathbb{C}$-specimen
$(\focussed{\mathcal{C}}[\vec{\chi}];\vec{H^1};\vec{H^2})$ is said to be
\emph{single} if the sequence $\vec{\chi}$ only has one element, i.e.\ the context
$\focussed{\mathcal{C}}$ has exactly one hole edge (at any depth).
 \end{enumerate}
\end{defi}
We can refer to \emph{the} focus label of a $\mathbb{C}$-specimen and
a quasi-$\mathbb{C}$-specimen.
Any $\mathbb{C}$-specimen $(\focussed{\mathcal{C}};\vec{H^1};\vec{H^2})$
gives a quasi-$\mathbb{C}$-specimen
$(\focussed{\mathcal{C}}[\vec{H^1}], \focussed{\mathcal{C}}[\vec{H^2}])$
up to $(=,=)$.

As described in \autoref{sec:gentle-introduction-reasoning}, the key part of proving contextual refinement $N \preceq^{\mathbb{C}}_Q H$ for each $N \vartriangleleft H$ is to show (\ref{eq:counting-sim-simple}). With the notion of specimen in hand, (\ref{eq:counting-sim-simple}) can be rephrased as follows:
\begin{itemize}
 \item for any $\mathbb{C}$-specimen $(\focussed{\mathcal{C}}[\vec{\chi}];\vec{N};\vec{H})$ of $\vartriangleleft$ and a transition $\focussed{\mathcal{C}}[\vec{N}] \to \focussed{P}$,
 \item there exist a $\mathbb{C}$-specimen $(\focussed{\mathcal{C}'}[\vec{\chi'}];\vec{N'};\vec{H'})$ and $k,l \in \N$ such that $(1+k) \mathrel{Q} l$ and the following holds.
\end{itemize}
\begin{equation}
 \xymatrix@R=2ex{
  \focussed{\mathcal{C}}[\vec{N}] \ar[r] \ar@{.}[d]_{\overline{\vartriangleleft}}
  & \focussed{P} \ar@[magenta][r]^>*[magenta]{k}
  & \textcolor{magenta}{\focussed{\mathcal{C}'}[\vec{N'}]} \ar@{.}@[magenta][d]^*[magenta]{\overline{\vartriangleleft}} \\
 \focussed{\mathcal{C}}[\vec{H}] \ar@[magenta][rr]^>*[magenta]{l}
  && \textcolor{magenta}{\focussed{\mathcal{C}'}[\vec{H'}]}
  }
  \label{eq:counting-sim}
\end{equation}

To establish (\ref{eq:counting-sim}), it suffices to perform case analysis on the transition $\focussed{\mathcal{C}}[\vec{N}] \to \focussed{P}$.
As explained in \autoref{sec:gentle-introduction-reasoning}, there are three possible cases.
The first case is a trivial case, where the $?$-focus or $\checkmark$-focus moves just inside the context $|\focussed{\mathcal{C}}|$. In this case, we can take the updated context $\focussed{\mathcal{C}'}$ such that $|\focussed{\mathcal{C}}| = |\focussed{\mathcal{C}'}|$.
For the other two cases, we identify sufficient conditions for (\ref{eq:counting-sim}), namely safety and robustness.

\subsection{Safety}

The second case of case analysis for (\ref{eq:counting-sim}) is when the $?$-focus or the $\checkmark$-focus encounters one of $\vec{N}$ (and $\vec{H}$); see \autoref{fig:case-visit}.
Let us look at how the micro-beta pre-template $\vartriangleleft^{\lrapp}$ (cf.\ \autoref{fig:rewrite-micro-beta}; mind that the focuses are dropped in the pre-template) satisfies (\ref{eq:counting-sim}) in this case.
Let $N_i \vartriangleleft^{\lrapp} H_i$ be the ones the focus is encountering. There are three sub-cases.
\begin{description}
 \item[Case (ii-1) Searching] The $?$-focus enters $N_i, H_i$, i.e.\ we have $?;N_i$ and $?;H_i$ inside states $\focussed{\mathcal{C}}[\vec{N}], \focussed{\mathcal{C}}[\vec{H}]$.
	    On $?;N_i$, a few search transitions will be followed by a compute transition that applies the micro-beta rewrite rule, because what is connected to the right target of the application edge ($@$) (i.e.\ $G_S$ in \autoref{fig:template-micro-beta}) is required to be stable. The result is exactly $?;H_i$.

	    This means that we can take a $\mathbb{C}$-specimen $(\focussed{\mathcal{C}'}[\vec{\chi'}];\vec{N'};\vec{H'})$ where: $\focussed{\mathcal{C}'}$ is obtained by replacing the $i$-th hole of $\focussed{\mathcal{C}}$ with $H_i$; and $\vec{N'},\vec{H'}$ are obtained by removing $N_i,H_i$ from $\vec{N},\vec{H}$.
	    One observation is that the focus in $\focussed{\mathcal{C}'}$ is not entering (i.e.\ pointing at a hole), because the focus is the $?$-focus pointing at $H_i$.
 \item[Case (ii-2) Backtracking on the box]  The $\checkmark$-focus is on top of the box (i.e.\ $G$ in \autoref{fig:template-micro-beta}). This case is in fact impossible in a refocusing UAM. To make states $\focussed{\mathcal{C}}[\vec{N}], \focussed{\mathcal{C}}[\vec{H}]$ rooted, the $\checkmark$-focus must be at the end of an operation path. However, the $\checkmark$-focus in the state $\focussed{\mathcal{C}}[\vec{N}]$ is adjacent to a box edge, and cannot be at the end of an operation path.
 \item[Case (ii-3) Backtracking on the stable argument] The $\checkmark$-focus is on top of what is connected to the right target of the application edge ($@$) (i.e.\ $G_S$ in \autoref{fig:template-micro-beta}). This case is also impossible, because of types. The stable hypernet $G_S$ has type $\star \To \otimes_{i=1}^m T^{n_i}(\star)$ (cf.\ \autoref{def:stable-HN}), and the $\checkmark$-focus has type $\star \To \star$. The $\checkmark$-focus cannot be on top of $G_S$.
\end{description}

From Case (ii-1), we extract a sufficient condition for (\ref{eq:counting-sim}), dubbed \emph{input-safety}, as follows. The micro-beta pre-template $\vartriangleleft^{\lrapp}$ falls into (II) below.
\begin{defi}[Input-safety]
 \label{def:InputSafe}
 A pre-template $\vartriangleleft$ is
 \emph{$(\mathbb{C},Q,Q')$-input-safe} if, for any
 $\mathbb{C}$-specimen $(\focussed{\mathcal{C}};\vec{H^1};\vec{H^2})$
 of $\vartriangleleft$ such that
 $\focussed{\mathcal{C}}$ has the entering $?$-focus, one of the
 following holds.
\\
 (I)
 There exist two stuck states $\focussed{N_1}$ and $\focussed{N_2}$
 such that
 $\focussed{\mathcal{C}}[\vec{H^p}] \to^* \focussed{N_p}$
 for each $p \in \{ 1,2 \}$.
\\
 (II)
 There exist a $\mathbb{C}$-specimen
 $(\focussed{\mathcal{C}'};\vec{H'^1};\vec{H'^2})$
 of $\vartriangleleft$
 and two numbers $k_1,k_2 \in \N$, such that
 the focus of
 $\focussed{\mathcal{C}'}$ is the $\checkmark$-focus or the non-entering $?$-focus, 
 $(1 + k_1) \mathrel{Q} k_2$,
 $\focussed{\mathcal{C}}[\vec{H^1}] \to^{1 + k_1}
 \focussed{\mathcal{C}'}[\vec{H'^1}]$, and
 $\focussed{\mathcal{C}}[\vec{H^2}] \to^{k_2}
 \focussed{\mathcal{C}'}[\vec{H'^2}]$.
\\
 (III)
 There exist a quasi-$\mathbb{C}$-specimen
 $(\focussed{N_1},\focussed{N_2})$ of
 $\vartriangleleft$ up to $(\dot{\simeq}_{Q'},\dot{\simeq}_{Q'})$,
 whose focus is not the $\lightning$-focus,
 and two numbers $k_1,k_2 \in \N$,
 such that $(1 + k_1) \mathrel{Q} (1 + k_2)$,
 $\focussed{\mathcal{C}}[\vec{H^1}] \to^{1 + k_1} \focussed{N_1}$, and
 $\focussed{\mathcal{C}}[\vec{H^2}] \to^{1 + k_2} \focussed{N_2}$.
\end{defi}

Case (ii-2) and (ii-3) tells us that types and the rooted property prevents the $\checkmark$-focus from visiting $N_i,H_i$ such that $N_i \vartriangleleft^{\lrapp} H_i$. This situation can be captured by one-way hypernets (\autoref{def:one-way-HN}), resulting in another sufficient condition dubbed \emph{output-closure}.
\begin{defi}[Output-closure]
 \label{def:OutputClose}
 A pre-template $\vartriangleleft$ is \emph{output-closed} if, for any
 hypernets $H_1 \vartriangleleft H_2$, either $H_1$ or $H_2$ is
 one-way.
\end{defi}

Input-safety and output-closure are the precise safety conditions. When a pre-template is safe, we simply call it \emph{template}.
\begin{defi}[Templates]
 A pre-template $\vartriangleleft$ is a
 \emph{$(\mathbb{C},Q,Q')$-template}, if it
 is $(\mathbb{C},Q,Q')$-input-safe and also output-closed.
\end{defi}

The micro-beta pre-template $\vartriangleleft^{\lrapp}$ is indeed safe, and hence a template. The third parameter of input-safety is irrelevant here, and we can simply set it as the equality $=$.
Recall that $\mathbb{C}_\Opr$ is the set of all focus-free contexts using operations $\Opr$.
\begin{lem}[Safety of $\vartriangleleft^{\lrapp}$]
 Let $\Oprlin_\checkmark$ be the set $\N \cup \{\lambda\}$, $\Oprlin_\lightning$ be the set $\{+,-,\lrapp,\keyw{stat}\}$, and $\Oprlin = \Oprlin_\checkmark \uplus \Oprlin_\lightning$.
 The micro-beta pre-template $\vartriangleleft^{\lrapp}$ is a $(\mathbb{C}_{\Oprlin},\geq,=)$-template.
 \qed
\end{lem}

\subsection{Robustness} \label{sec:robustness}

The last case of case analysis for (\ref{eq:counting-sim}) is when the $\lightning$-focus triggers a rewrite transition in $\focussed{\mathcal{C}}$ (and hence in $\focussed{\mathcal{C}'}$).
Does the micro-beta template satisfy (\ref{eq:counting-sim}) in this case? It depends on the rewrite transition $\focussed{\mathcal{C}}[\vec{N}] \to \focussed{P}$.
Let us consider an operation set $\Oprlin = \Oprlin_\checkmark \uplus \Oprlin_\lightning$ where $\Oprlin_\checkmark = \N \cup \{\lambda\}$ and $\Oprlin_\lightning = \{+,-,\lrapp,\keyw{stat}\}$. We can perform case analysis in terms of which rewrite is triggered by the $\lightning$-focus.
\begin{description}
 \item[Case (iii-1) Substitution] The $\lightning$-focus triggers application of the substitution rule (\autoref{fig:subst-rule}). The rule simply removes a variable edge ($\mathsf{V}$), and the same rule can be applied to both states $\focussed{\mathcal{C}}[\vec{N}], \focussed{\mathcal{C}}[\vec{H}]$. Sub-graphs related by $\vartriangleleft^{\lrapp}$ may be involved, as part of $G_S$ in \autoref{fig:subst-rule}, but they are kept unchanged.
	    Therefore we can take a $\mathbb{C}_{\Oprlin}$-specimen $(\focussed{\mathcal{C}'}[\vec{\chi'}];\vec{N'};\vec{H'})$ where $\focussed{\mathcal{C}'}$ has the same number of holes as $\focussed{\mathcal{C}}$.
 \item[Case (iii-2) Arithmetic] The $\lightning$-focus triggers application of the arithmetic rewrite rule (\autoref{fig:rewrite-arith}). The rewrite rule only involves three edges, and it can never involves sub-graphs related by $\vartriangleleft^{\lrapp}$.
	    We can also take a $\mathbb{C}_{\Oprlin}$-specimen $(\focussed{\mathcal{C}'}[\vec{\chi'}];\vec{N'};\vec{H'})$ where $\focussed{\mathcal{C}'}$ has the same number of holes as $\focussed{\mathcal{C}}$.
 \item[Case (iii-3) Micro-beta] The $\lightning$-focus triggers application of the micro-beta rewrite rule (\autoref{fig:rewrite-micro-beta}).
	    Sub-graphs related by $\vartriangleleft^{\lrapp}$ may be involved in the rewrite rule, as part of the box content $G$ in \autoref{fig:rewrite-micro-beta}, but they are kept unchanged.
	    We can also take a $\mathbb{C}_{\Oprlin}$-specimen $(\focussed{\mathcal{C}'}[\vec{\chi'}];\vec{N'};\vec{H'})$ where $\focussed{\mathcal{C}'}$ has the same number of holes as $\focussed{\mathcal{C}}$.
 \item[Case (iii-4) $\keyw{stat}$] The $\lightning$-focus triggers application of the $\keyw{stat}$ rewrite rule (\autoref{fig:rewrite-stat}).
	    The rewrite rule inevitably involves sub-graphs related by $\vartriangleleft^{\lrapp}$, and as a result, can yield different results, i.e.\ $\mathrm{size}(|\focussed{\mathcal{C}}[\vec{N}]|)$ vs. $\mathrm{size}(|\focussed{\mathcal{C}}[\vec{H}]|)$. The same rewrite rule can be applied to states $\focussed{\mathcal{C}}[\vec{N}], \focussed{\mathcal{C}}[\vec{H}]$ and yield $\focussed{P}, \focussed{P'}$, respectively, and these states $\focussed{P}, \focussed{P'}$ can contain different natural-number edges (i.e.\ $n$ vs. $m$ such that $n \geq m$). We cannot take a specimen of $\vartriangleleft^{\lrapp}$ to make (\ref{eq:counting-sim}) happen.
\bqed
\end{description}

We can therefore observe the following: the micro-beta template $\vartriangleleft^{\lrapp}$ satisfies (\ref{eq:counting-sim}) relative to active operations $\{+,-,\lrapp\} = \Oprlin_\lightning \backslash \{\keyw{stat}\}$ (formalised in \autoref{lem:micro-beta-equiv} below), and does not satisfy (\ref{eq:counting-sim}) relative to the active operation $\keyw{stat}$. We say the micro-beta template $\vartriangleleft^{\lrapp}$ is \emph{robust} only relative to $\Oprlin_\lightning \backslash \{\keyw{stat}\}$.
Robustness, as a sufficient condition for (\ref{eq:counting-sim}), can be formalised as follows.
\begin{defi}[Robustness]
 \label{def:Robustness}
 A pre-template $\vartriangleleft$ is
 \emph{$(\mathbb{C},Q,Q',Q'')$-robust relative to}
 a rewrite transition $\focussed{N} \to \focussed{N'}$ if, for
 any $\mathbb{C}$-specimen
 $(\focussed{\mathcal{C}};\vec{H^1};\vec{H^2})$
 of $\vartriangleleft$, such that
 $\focussed{\mathcal{C}}[\vec{H^1}] = \focussed{N}$ and
 the focus of
 $\focussed{\mathcal{C}}$ is the $\lightning$-focus and not entering,
 one of the following holds.
\\[1.5ex]
 (I) $\focussed{\mathcal{C}}[\vec{H^1}]$ or
 $\focussed{\mathcal{C}}[\vec{H^2}]$ is not rooted.
\\[1.5ex]
 (II)
 There exists a stuck state $\focussed{N''}$ such that
 $\focussed{N'} \to^* \focussed{N''}$.
\\[1.5ex]
 (III)
 There exist a quasi-$\mathbb{C}$-specimen
 $(\focussed{N''_1},\focussed{N''_2})$ of
 $\vartriangleleft$ up to $(\dot{\preceq}_{Q'},\dot{\preceq}_{Q''})$,
 whose focus is not the $\lightning$-focus,
 and two numbers $k_1,k_2 \in \N$,
 such that $(1 + k_1) \mathrel{Q} k_2$,
 $\focussed{N'} \to^{k_1} \focussed{N''_1}$, and
 $\focussed{\mathcal{C}}[\vec{H^2}] \to^{k_2} \focussed{N''_2}$.
\end{defi}

We can now formally say that the micro-beta pre-template $\vartriangleleft^{\lrapp}$ is robust relative to the substitution transitions and the extrinsic transitions for $\Oprlin \backslash \{\keyw{stat}\}$.
The third and fourth parameters of robustness are irrelevant here\footnote{These parameters will be used in \autoref{sec:ex-law} (See \autoref{tab:TemplateAnalysis}).}, and we can simply set them as the equality $=$.
\begin{lem}[Robustness of $\vartriangleleft^{\lrapp}$] \label{lem:micro-beta-equiv}
 The micro-beta pre-template $\vartriangleleft^{\lrapp}$ is $(\mathbb{C}_{\Oprlin \backslash \{\keyw{stat}\}},=,=,=)$-robust relative to the substitution transitions and the extrinsic transitions for $\Oprlin_\lightning \backslash \{\keyw{stat}\}$.
 \qed
\end{lem}

\subsection{A sufficiency-of-robustness theorem}

We have collected definitions of safety and robustness. We can finally state the sufficiency-of-robustness theorem.
The theorem incorporates the so-called \emph{up-to} technique: it enables us to prove contextual refinement $\preceq^\mathbb{C}_Q$ that depends on (or, that is \emph{up to}) state refinements $\dot{\preceq}_{Q'}$ and $\dot{\preceq}_{Q''}$.
Safety (\autoref{def:InputSafe}) and robustness (\autoref{def:Robustness}) accordingly incorporates the up-to technique by means of quasi-specimens up-to.
We cannot use arbitrary preorders $Q',Q''$ in combination with the preorder $Q$, though; they must be \emph{reasonable} in the following sense.

\begin{defi}[Reasonable triples]
 \label{def:reasonable-triple}
 A triple $(Q,Q',Q'')$ of preorders on $\N$ is
 \emph{reasonable} if the following hold:
\\[1.5ex]
 (A) $Q$ is closed under addition, i.e.\
 any $k_1 \mathrel{Q} k_2$ and $k'_1 \mathrel{Q} k'_2$ satisfy
 $(k_1 + k'_1) \mathrel{Q} (k_2 + k'_2)$.
\\[1.5ex]
 (B) $Q' \subseteq \mathord{\geq_\N}$, and $Q' \subseteq Q''$.
\\[1.5ex]
 (C) $Q' \circ Q \circ Q'' \subseteq Q$, where $\circ$ denotes
 composition of binary relations.
\end{defi}
Examples of a reasonable triple $(Q,Q',Q'')$ include:
$(\mathord{\N \times \N}, \mathord{\geq_\N}, \mathord{\N \times \N})$,
$(\mathord{\N \times \N}, \mathord{\geq_\N},$ $\mathord{\geq_\N})$,
$(\mathord{\N \times \N}, \mathord{=_\N}, \mathord{=_\N})$,
$(\mathord{\geq_\N}, \mathord{\geq_\N}, \mathord{\geq_\N})$,
$(\mathord{\geq_\N}, \mathord{=_\N}, \mathord{\geq_\N})$,
$(\mathord{\leq_\N}, \mathord{=_\N}, \mathord{\leq_\N})$,
$(\mathord{\geq_\N}, \mathord{=_\N}, \mathord{=_\N})$,
$(\mathord{\leq_\N}, \mathord{=_\N},$ $\mathord{=_\N})$,
$(\mathord{=_\N}, \mathord{=_\N}, \mathord{=_\N})$.

The sufficiency-of-robustness theorem has two clauses (\ref{item:forward}) and (\ref{item:backward}).
To prove that a pre-template $\vartriangleleft$ implies contextual equivalence (not refinement), one can use (\ref{item:forward}) twice with respect to $\vartriangleleft$ and $\vartriangleleft^{-1}$, or alternatively use (\ref{item:forward}) and (\ref{item:backward}) with respect to $\vartriangleleft$. The alternative approach is often more economical. This is because it involves proving input-safety of $\vartriangleleft$ for both parameters $Q$ and $Q^{-1}$, which typically boils down to a single proof for the smaller one of $Q,Q^{-1}$, thanks to monotonicity of input-safety with respect to $Q$.
\begin{thm}[Sufficiency-of-robustness theorem]
 \label{thm:MetaThm}
 If an universal abstract machine $\UAM(\Opr,B_\Opr)$ is
 deterministic and refocusing, it satisfies the following property. 
 For any set
 $\mathbb{C} \subseteq \HN{\LVlin}{\LElin \cup \MVar}$
 of contexts that is closed under plugging,
 any reasonable triple $(Q,Q',Q'')$,
 and any pre-template $\vartriangleleft$ on focus-free hypernets
 $\HN{\LVlin}{\LElin(\Opr) \backslash \{?,\checkmark,\lightning\}}$:
 \begin{enumerate}
  \item \label{item:forward}
	If $\vartriangleleft$ is a $(\mathbb{C},Q,Q')$-template and
	$(\mathbb{C},Q,Q',Q'')$-robust relative to all rewrite
	transitions, then $\vartriangleleft$ implies contextual
	refinement in $\mathbb{C}$ up to $Q$,
	i.e.\ any $G_1 \vartriangleleft G_2$ implies
	$G_1 \preceq^\mathbb{C}_Q G_2$.
  \item \label{item:backward}
	If $\vartriangleleft$ is a $(\mathbb{C},Q^{-1},Q')$-template
	and the converse $\vartriangleleft^{-1}$ is
	$(\mathbb{C},Q,Q',Q'')$-robust relative to all rewrite
	transitions, then $\vartriangleleft^{-1}$ implies contextual
	refinement in $\mathbb{C}$ up to $Q$,
	i.e.\ any $G_1 \vartriangleleft G_2$ implies
	$G_2 \preceq^\mathbb{C}_Q G_1$.
 \end{enumerate}
\end{thm}
\begin{proof}
 This is a consequence of
 \autoref{prop:TemplateToSimulation},
 \autoref{prop:SimulationToRefinement} and
 \autoref{prop:StateToCtxt} in \autoref{sec:ProofMetaThm}.

 The proof is centred around a notion of \emph{counting simulation} (\autoref{def:QSimulation}). We first prove that, for each robust template $\vartriangleleft$, its contextual closure $\overline{\vartriangleleft}$ is a counting simulation (\autoref{prop:TemplateToSimulation}). We then prove soundness of counting simulation with respect to state refinement (\autoref{prop:SimulationToRefinement}). We can further prove that, if $\overline{\vartriangleleft}$ implies state refinement, $\vartriangleleft$ implies contextual refinement (\autoref{prop:StateToCtxt}).
\end{proof}

We can use \autoref{thm:MetaThm}~(\ref{item:forward}) to conclude that the micro-beta pre-template implies contextual refinement $\preceq^{\mathbb{C}_{\Oprlin \backslash \{\keyw{stat}\}}}_{\geq_\N}$.
Note that The micro-beta pre-template $\vartriangleleft^{\lrapp}$ is $(\mathbb{C}_\Opr,=,=,\linebreak[5]=)$-robust, and therefore $(\mathbb{C}_\Opr,\geq,=,=)$-robust as well, because $\mathit{=} \subseteq \mathit{\geq}$.
\begin{prop} \label{prop:micro-beta-equiv}
 For the operation set $\Oprlin \backslash \{\keyw{stat}\}$, the micro-beta pre-template $\vartriangleleft^{\lrapp}$ implies contextual refinement $\preceq^{\mathbb{C}_{\Oprlin \backslash \{\keyw{stat}\}}}_{\geq_\N}$.
 \qed
\end{prop}

The above proposition establishes contextual refinement in the absence of the operation $\keyw{stat}$. This is due to the observation that the micro-beta template $\vartriangleleft^{\lrapp}$ is not robust relative to the rewrite rule of $\keyw{stat}$ (\autoref{fig:rewrite-stat}).
Can we say more than just failure of robustness, in the presence of $\keyw{stat}$?

\begin{figure}[t]
 \includegraphics[scale=.3]{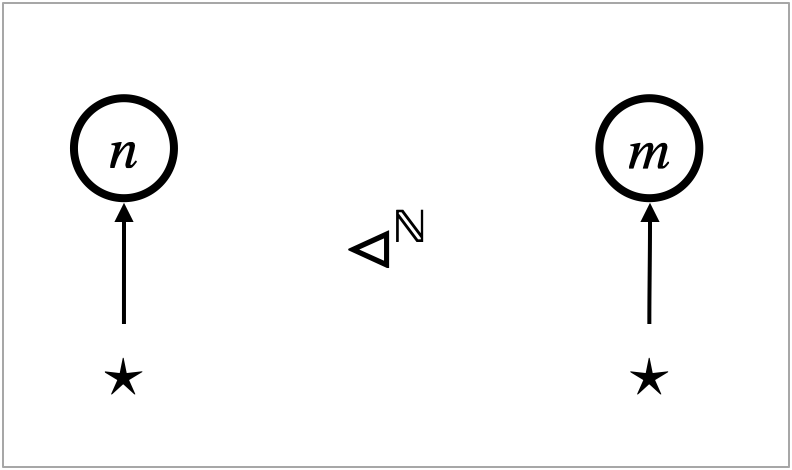}
 \caption{An auxiliary pre-template, where $n,m \in \N$}
 \label{fig:template-N}
\end{figure}

The answer is yes.
For the specific operation set $\Oprlin$, we can actually show that the micro-beta template $\vartriangleleft^{\lrapp}$ implies contextual refinement $\preceq^{\mathbb{C}_{\Oprlin}}_{\geq_\N}$.
\begin{lem}[Robustness of $\vartriangleleft^{\lrapp}$]
 The micro-beta pre-template $\vartriangleleft^{\lrapp}$ is $(\mathbb{C}_{\Oprlin},=,=,=)$-robust relative to the rewrite transitions for $\keyw{stat}$.
\end{lem}
\begin{proof}
 To prove robustness, we use the auxiliary pre-template $\vartriangleleft^{\N}$, shown in \autoref{fig:template-N}, that identifies all natural number edges.
 The pre-template is a $(\mathbb{C}_{\Oprlin},=,=)$-template, and it is $(\mathbb{C}_{\Oprlin},=,=,=)$-robust relative to the substitution transitions and the extrinsic transitions for $\Oprlin_\lightning$ including $\keyw{stat}$. Therefore, by \autoref{thm:MetaThm}~(\ref{item:forward}), it implies contextual refinement $\preceq^{\mathbb{C}_{\Oprlin}}_{=_\N}$. The pre-template is symmetric, and consequently, it implies contextual equivalence $\simeq^{\mathbb{C}_{\Oprlin}}_{=_\N}$.

 In Case (iii-4) in \autoref{sec:robustness}, we could not take a specimen of $\vartriangleleft^{\lrapp}$ that induces the states $\focussed{P},\focussed{P'}$. The difference between these states is given by $\vartriangleleft^{\lrapp}$ as well as different natural number edges; in other words, the difference is given by $\vartriangleleft^{\lrapp}$ and $\vartriangleleft^{\N}$.

 In fact, we \emph{can} take a \emph{quasi-}specimen of $\vartriangleleft^{\lrapp}$ up to $(\dot{\simeq}^{\mathbb{C}_{\Oprlin}}_{=_\N},\dot{\simeq}^{\mathbb{C}_{\Oprlin}}_{=_\N})$, which is namely the pair $(\focussed{P},\focussed{P'})$. There exists a state $\focussed{P''}$ such that: (a) $\focussed{P} \mathrel{\dot{\simeq}}^{\mathbb{C}_{\Oprlin}}_{=_\N} \focussed{P''}$ thanks to the symmetric robust template $\vartriangleleft^{\N}$, and (b) the pair $(\focussed{P''},\focussed{P'})$ is induced by a specimen of $\vartriangleleft^{\lrapp}$.

 As a result, with the help of the symmetric robust template $\vartriangleleft^{\N}$, the micro-beta template $\vartriangleleft^{\lrapp}$ is robust also relative to $\keyw{stat}$.
\end{proof}

In the presence of conditional branching and divergence, however, the micro-beta template $\vartriangleleft^{\lrapp}$ would not be robust relative to $\keyw{stat}$.
The above lemma crucially relies on robustness of the auxiliary template $\vartriangleleft^{\N}$, which would be violated by conditional branching; in other words, the template $\vartriangleleft^{\N}$ would not be robust relative to conditional branching.
Conditional branching and divergence, combined with $\keyw{stat}$, would enable us to construct a specific context that distinguishes the left- and the right hand sides of $\vartriangleleft^{\lrapp}$.

\section{Representation of variable sharing and store} \label{sec:variable-sharing-store}

\begin{figure}[t]
 \begin{align*}
  &\text{Constants:} &&
  c \mathrel{::=}
  n \tag{natural numbers} \\
  &&& \hspace{1.5em}
  \mid \keyw{tt} \mid \keyw{ff} \tag{booleans} \\
  &&& \hspace{1.5em}
  \mid () \tag{unit value} \\
  &\text{Unary operations:} &&
  \$_1 \mathrel{::=}
  -_1 \tag{negation of natural numbers} \\
  &&& \hspace{2em}
  \mid \keyw{ref} \tag{reference creation} \\
  &&& \hspace{2em}
  \mid \mathit{!} \tag{dereferencing} \\
  &\text{Binary operations:} &&
  \$_2 \mathrel{::=}
  + \mid - \tag{summation/subtraction of natural numbers} \\
  &&& \hspace{2em}
  \mid \mathit{:=} \tag{assignment} \\
  &&& \hspace{2em}
  \mid \mathit{=} \tag{equality testing of atoms} \\
  &\text{Terms:} &&
  t,u \mathrel{::=}
  x \mid \lambda x.\,t \mid t\ u \tag{the lambda-calculus terms} \\
  &&& \hspace{2.5em}
  \mid a \tag{atoms} \\
  &&& \hspace{2.5em}
  \mid c \mid \mathop{\$_1} t \mid t \mathbin{\$_2} u
  \tag{constants, unary/binary operations} \\[1ex]
  &\text{Formation rules:} &&
  \infer{\Gamma_1,x,\Gamma_2 \mid \Delta \vdash x}{} \qquad
  \infer{\Gamma \mid \Delta_1,a,\Delta_2 \vdash a}{} \\[1ex]
  &&&\infer{\Gamma \mid \Delta \vdash \lambda x.\,t}%
  {x,\Gamma \mid \Delta \vdash t} \qquad
  \infer{\Gamma \mid \Delta \vdash t\ u}%
  {\Gamma \mid \Delta \vdash t
  & \Gamma \mid \Delta \vdash u} \qquad
  \infer{\Gamma \mid \Delta \vdash c}%
  {} \\[1ex]
  &&&\infer{\Gamma \mid \Delta \vdash \mathop{\$_1} t}%
  {\Gamma \mid \Delta \vdash t} \quad
  \infer{\Gamma \mid \Delta \vdash t \mathbin{\$_2} u}%
  {\Gamma \mid \Delta \vdash t
  & \Gamma \mid \Delta \vdash u} \\
  &\text{Syntactic sugar:} &&
  \letin{x}{u}{t} \enspace\defeq\enspace (\lambda x.\, t) \ u \\
  &&& u;t \enspace\defeq\enspace (\lambda z.\, t) \ u
  \qquad \text{(where $z$ is a fresh variable)}
 \end{align*}
 \caption{An extended call-by-value lambda-calculus with variable sharing and store}
 \label{fig:Lamex}
\end{figure}

This section adapts the translation $(-)^\dag$ of the linear lambda-calculus, presented in \autoref{sec:linear-lambda-calc}, to accommodate variable sharing and (general, untyped) store.
\autoref{fig:Lamex} shows an extension of the untyped call-by-value lambda-calculus that accommodates arithmetic as well as:
\begin{itemize}
 \item atoms (or, locations of store) $a$,
 \item reference-related operations: reference creation $\keyw{ref}$, dereferencing $!$, assignment $:=$, equality testing of atoms $=$,
 \item and their return values: booleans $\keyw{tt},\keyw{ff}$ and the unit value $()$.
\end{itemize}
In \autoref{fig:Lamex}, $\Gamma$ is a finite sequence of (free) variables and $\Delta$ is a finite sequence of (free) atoms.

\subsection{Variable sharing} \label{sec:variable-sharing}

In an arbitrary term, a variable may occur many times, or it may not occur at all. To let sub-terms share and discard a variable, we introduce \emph{contraction} edges $\contr^\star_\mathsf{C} \colon \star^{\otimes 2} \To \star$ and \emph{weakening} edges $\contr^\star_\mathsf{W} \colon \epsilon \To \star$. We can construct a binary tree with an arbitrary number of leaves using these contraction and weakening edges. We call such a tree \emph{contraction tree}.

There are many ways to construct a contraction tree of $m$ leaves, but we fix a \emph{canonical}\footnote{Each $D_{1,m}^\star$ is canonical, in the sense that any contraction tree that contains at least one weakening and has $m$ leaves can be simplified to $D_{1,m}^\star$ using the laws in \autoref{fig:structural} (namely $\vartriangleleft^{\lrapp\mathrm{Assoc}},\vartriangleleft^{\lrapp\mathrm{Comm}},\vartriangleleft^{\lrapp\mathrm{Idem}}$).} tree $D_{1,m}^\star \colon \star^{\otimes m} \To \star$ as below.
\begin{equation*}
 D_{1,0}^\star = \includegraphics[align=c,scale=.25]{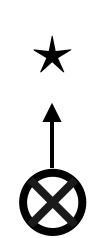} \hspace{4em}
 D_{1,1}^\star = \includegraphics[align=c,scale=.25]{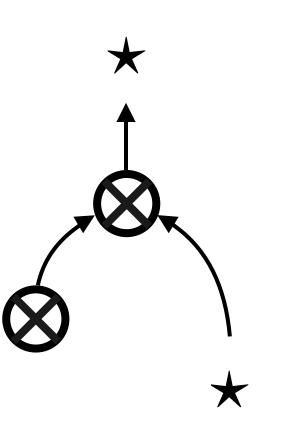} \hspace{4em}
 D_{1,m+2}^\star = \includegraphics[align=c,scale=.25]{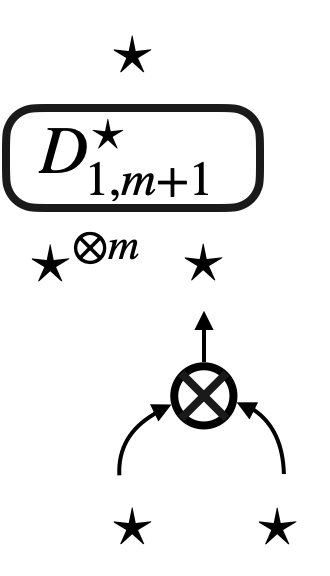}
\end{equation*}
Every canonical tree $D_{1,m}^\star$ contains exactly one weakening edge and $m$ contraction edges.

\begin{figure}[t]
 \subfloat[Original
 \label{fig:transl-lambda-ex-original}]{
 \includegraphics[scale=.25]{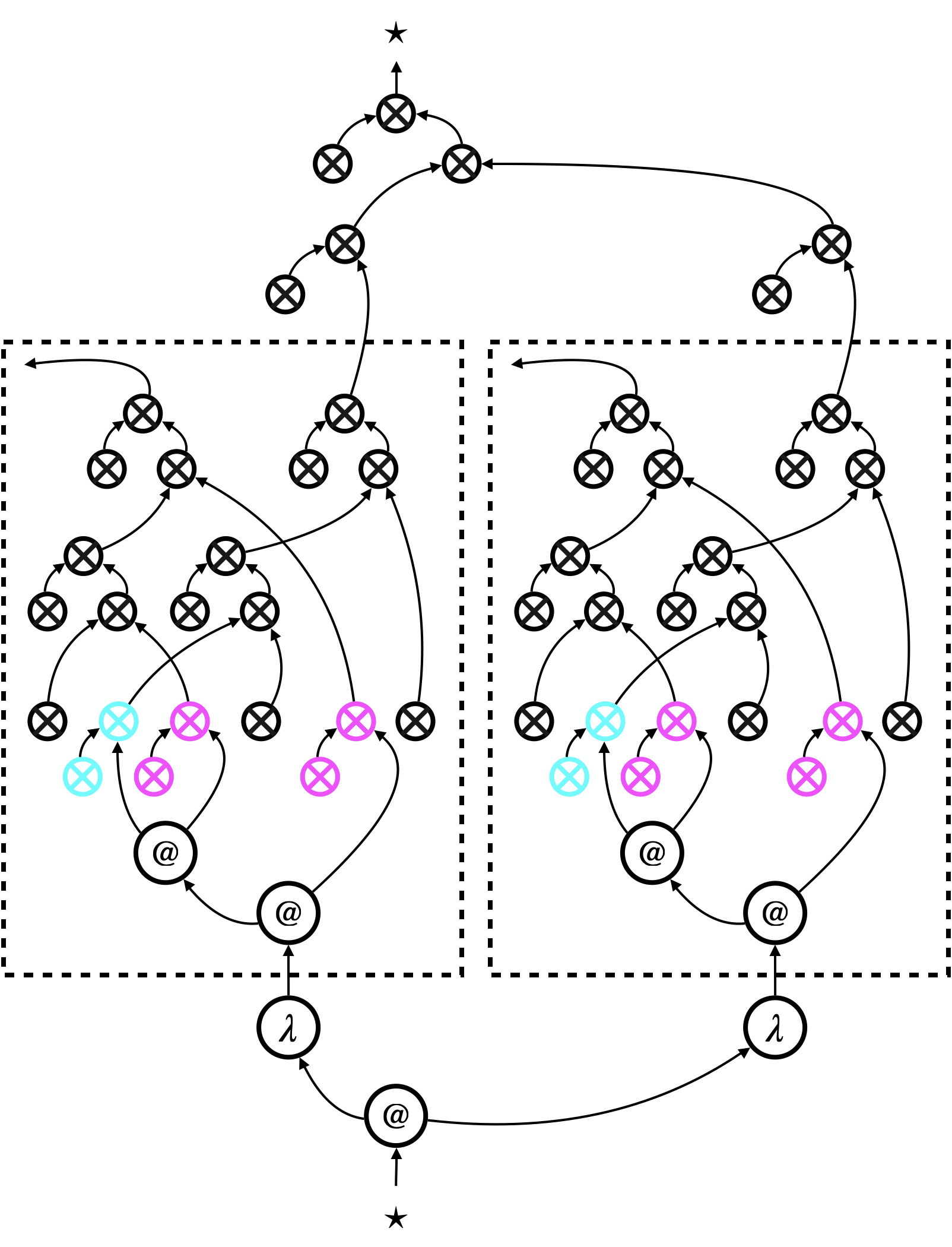}
 }
 \hfil
 \subfloat[Simplified
 \label{fig:transl-lambda-ex-simple}]{
 \includegraphics[scale=.25]{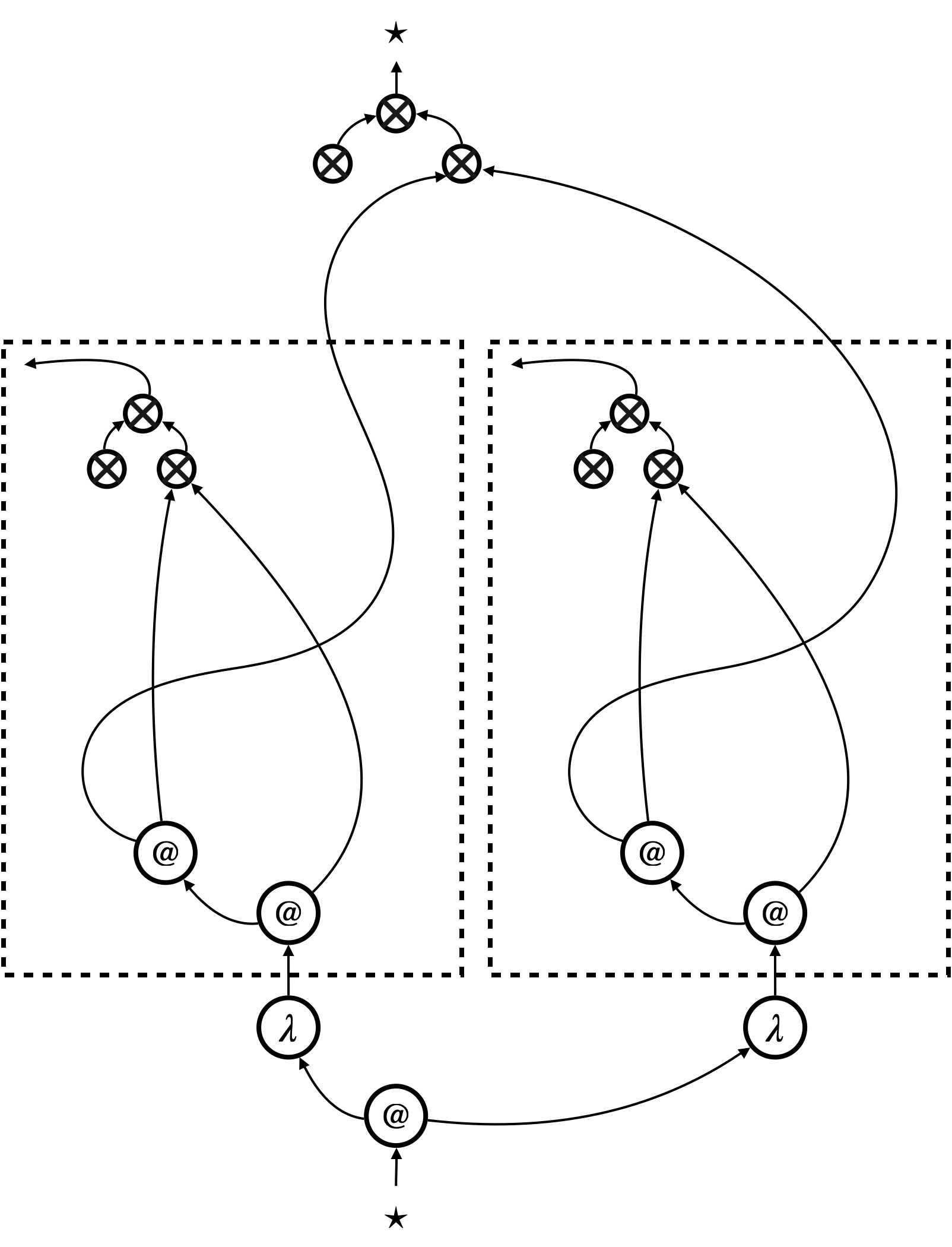}
 }
 \caption{The hypernet $\bigl(f \vdash (\lambda x. \cyan{f}\;\textcolor{magenta}{x}\;\textcolor{magenta}{x})\;(\lambda x. \cyan{f}\;\textcolor{magenta}{x}\;\textcolor{magenta}{x})\bigr)^\ddag \colon \star \To \star$ and its simplification}
 \label{fig:transl-lambda-ex}
\end{figure}
The translation $(-)^\dag$ of linear lambda-terms needs to be adapted to yield a translation $(-)^\ddag$ of general lambda-terms. The adaptation amounts to inserting canonical trees appropriately. Additionally, we replace the edges $\mathsf{V} \colon \star \To \star$ with canonical trees $D_{1,1}^\star$, to obtain uniform translation.

\begin{figure}[t]
 \begin{gather*}
  (\Gamma_1,x,\Gamma_2 \mid \Delta \vdash x)^\ddag \ =\ \includegraphics[align=c,scale=.2]{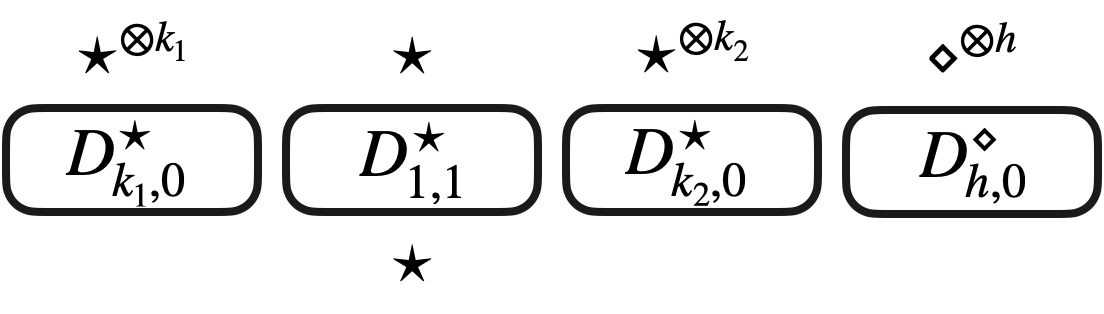}
  \hspace{4em}
  (\Gamma \mid \Delta \vdash c)^\ddag \ ~=\ \includegraphics[align=c,scale=.2]{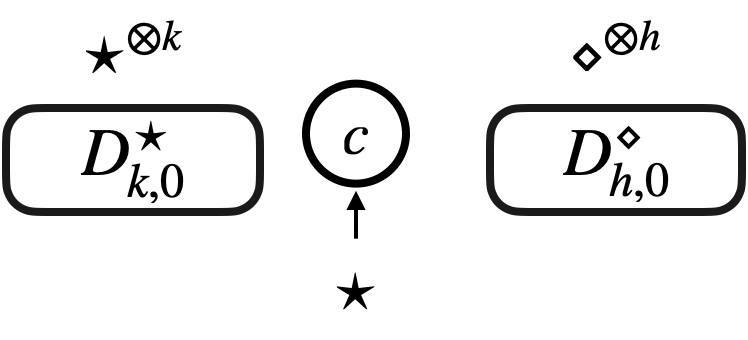}
  \\ \hline
  (\Gamma \mid \Delta_1,a,\Delta_2 \vdash a)^\ddag \ =\ \includegraphics[align=c,scale=.2]{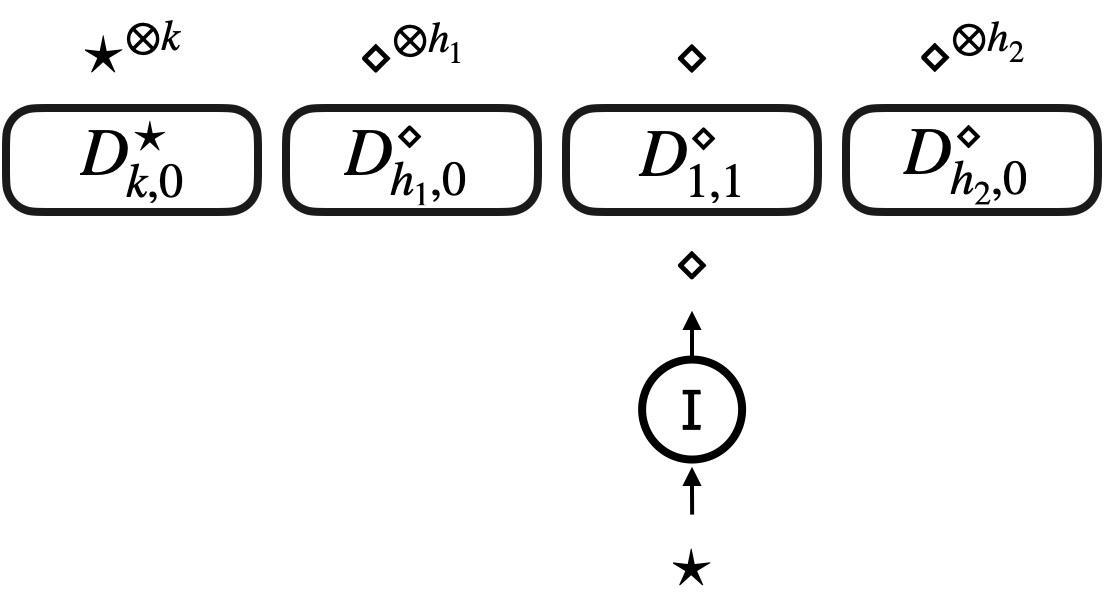}
  \\ \hline
  (\Gamma \mid \Delta \vdash \lambda x.\,t)^\ddag \ ~=\ \includegraphics[align=c,scale=.2]{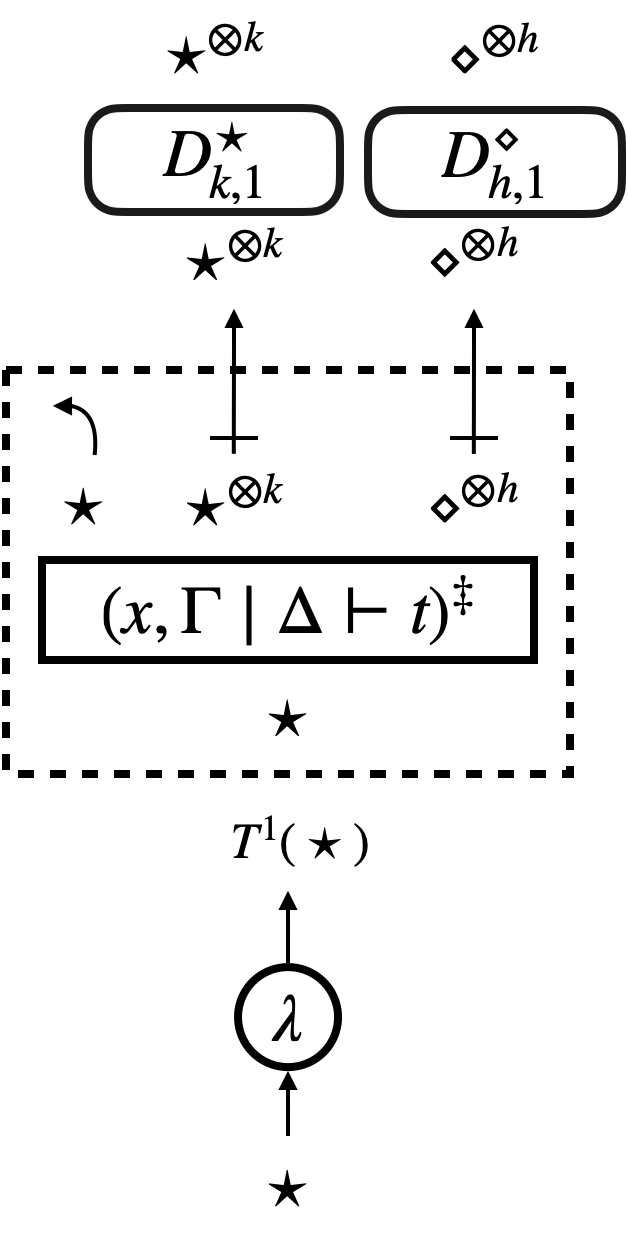}
  \hspace{4em}
  (\Gamma \mid \Delta \vdash t\ u)^\ddag \ ~=\ \includegraphics[align=c,scale=.2]{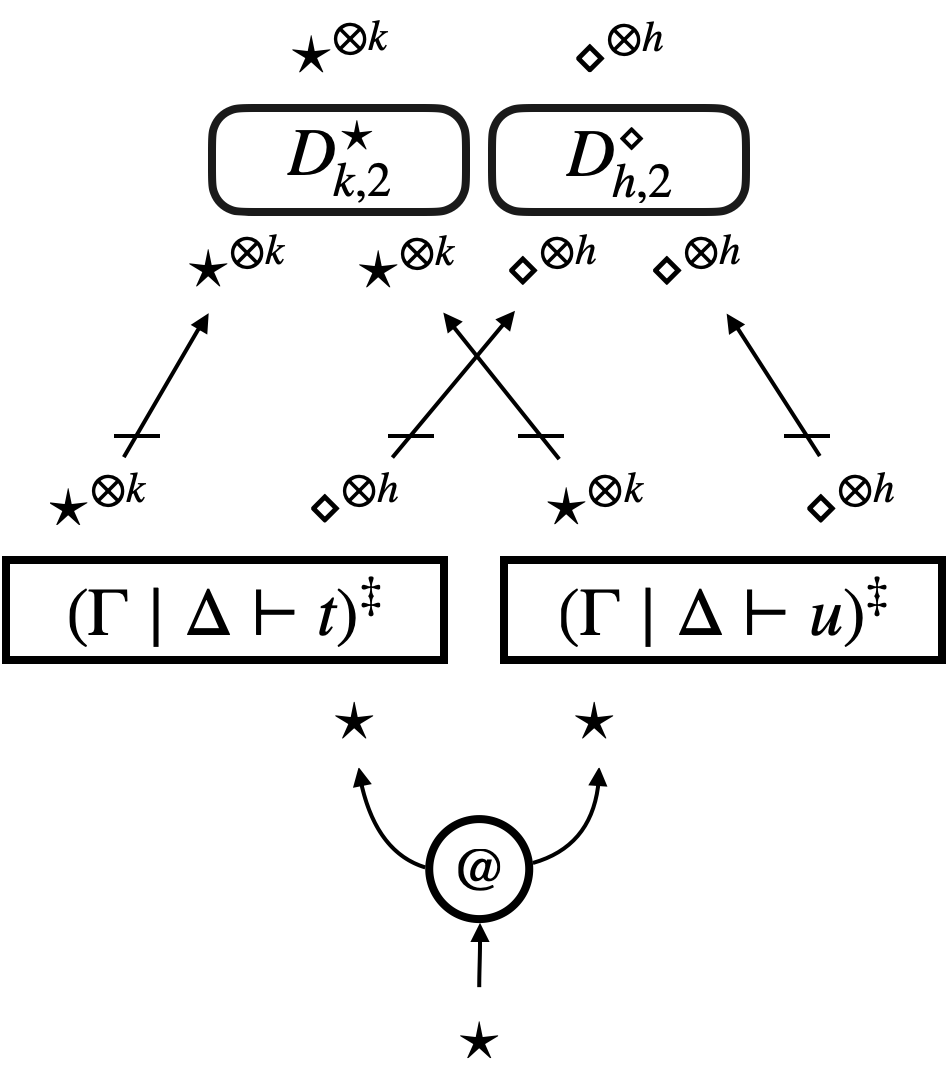}
  \\ \hline
  (\Gamma \mid \Delta \vdash \mathop{\$_1} t)^\ddag \ ~=\ \includegraphics[align=c,scale=.2]{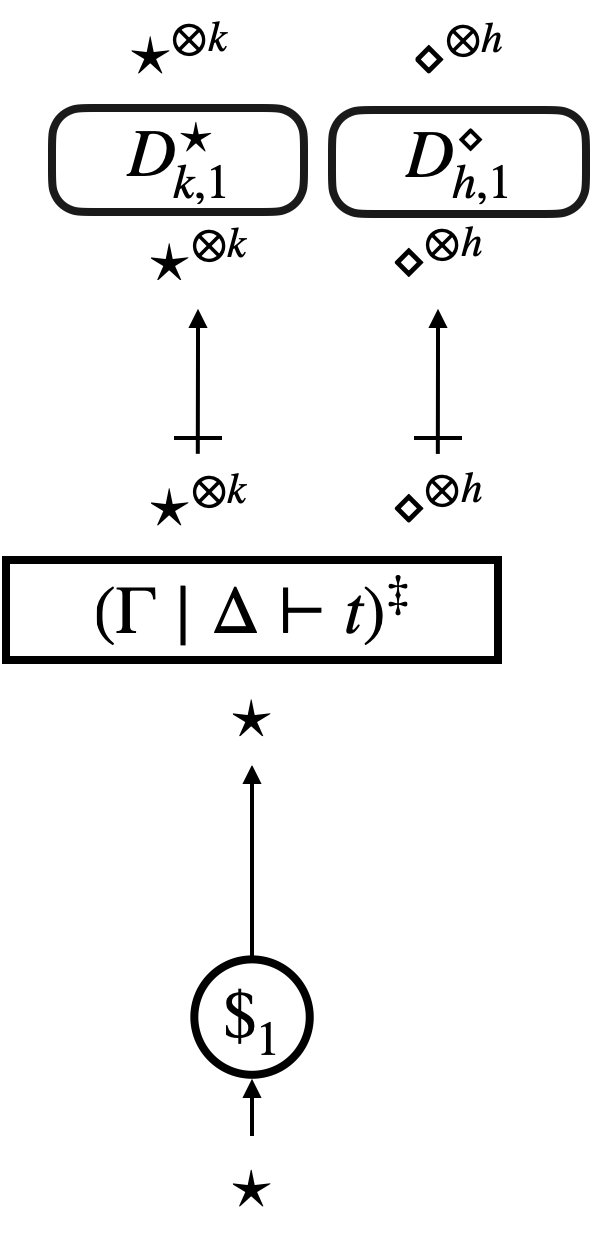}
  \hspace{4em}
  (\Gamma \mid \Delta \vdash t \mathbin{\$_2} u)^\ddag \ ~=\ \includegraphics[align=c,scale=.2]{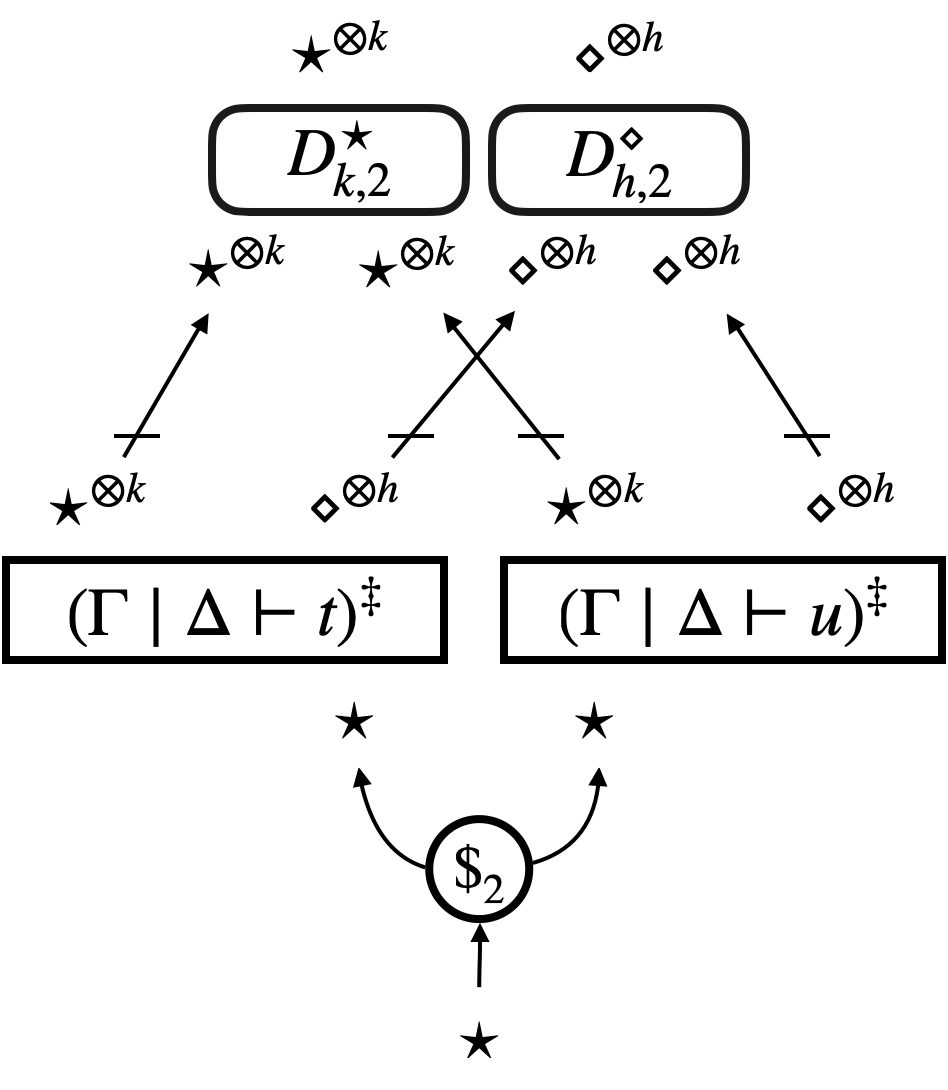}
 \end{gather*}
 where $k,k_1,k_2,h,h_1,h_2$ are the lengths of
 $\Gamma,\Gamma_1,\Gamma_2,\Delta,\Delta_1,\Delta_2$ respectively.
 \caption{Inductive translation $(-)^\ddag$ of well-formed lambda-terms}
 \label{fig:transl-general}
\end{figure}

We will present the exact translation $(-)^\ddag$ in \autoref{sec:inductive-transl} (see \autoref{fig:transl-general}), and here we just show an example in \autoref{fig:transl-lambda-ex}.
\autoref{fig:transl-lambda-ex-original} shows the hypernet $\bigl(f \vdash (\lambda x. \cyan{f}\;\textcolor{magenta}{x}\;\textcolor{magenta}{x})\;(\lambda x. \cyan{f}\;\textcolor{magenta}{x}\;\textcolor{magenta}{x})\bigr)^\ddag \colon$ $\star \To \star$. Canonical trees $D_{1,0}^\star, D_{1,1}^\star, D_{1,2}^\star$ are inserted for each term constructor, but among those, coloured canonical trees $D_{1,1}^\star$ replace the edges $\mathsf{V} \colon \star \To \star$ that would represent variable occurrences in the linear translation $(-)^\dag$.
\autoref{fig:transl-lambda-ex-simple} shows a possible simplification of the hypernet; such simplification is validated only by proving certain observational equivalences on contraction trees (namely, those in \autoref{fig:structural}).

\subsection{Store and atom sharing} \label{sec:store-atom-sharing}

\begin{figure}[t]
 \subfloat[The term alone
 \label{fig:transl-ref-ex-alone}]{
 \hspace{5em}\includegraphics[scale=.25]{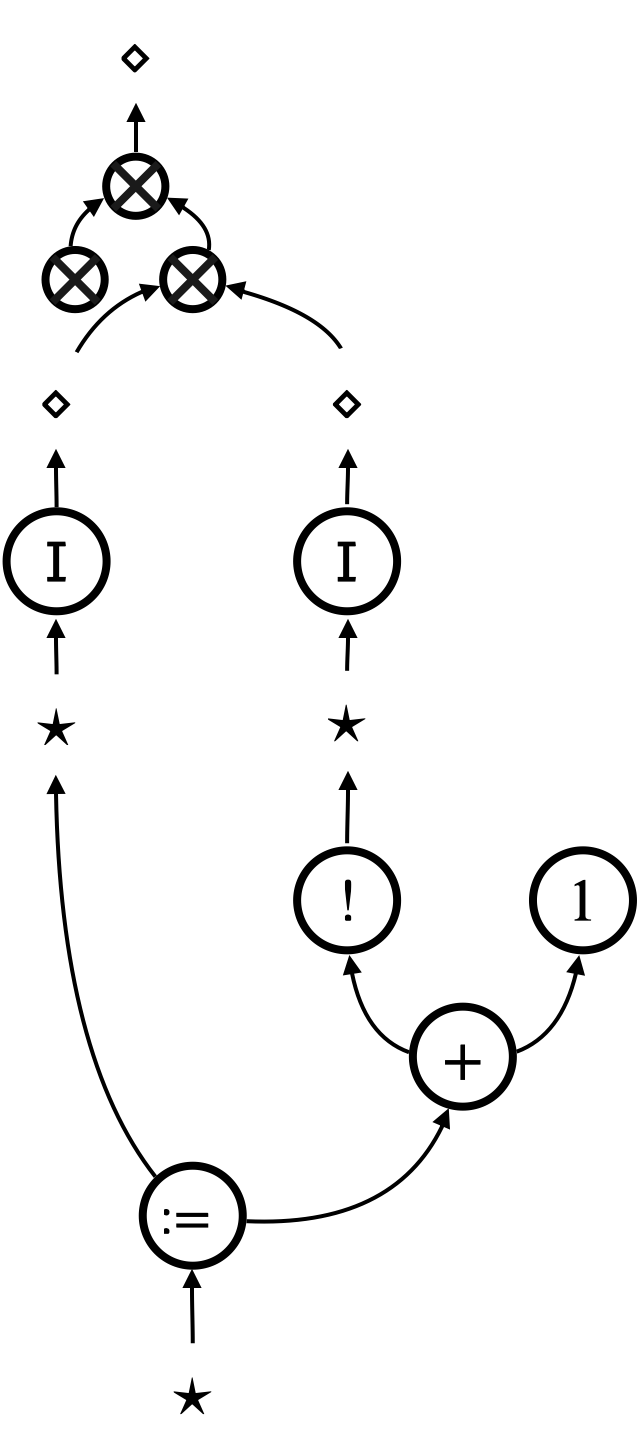}\hspace{5em}
 }
 \hfil
 \subfloat[With store
 \label{fig:transl-ref-ex-store}]{
 \hspace{5em}\includegraphics[scale=.25]{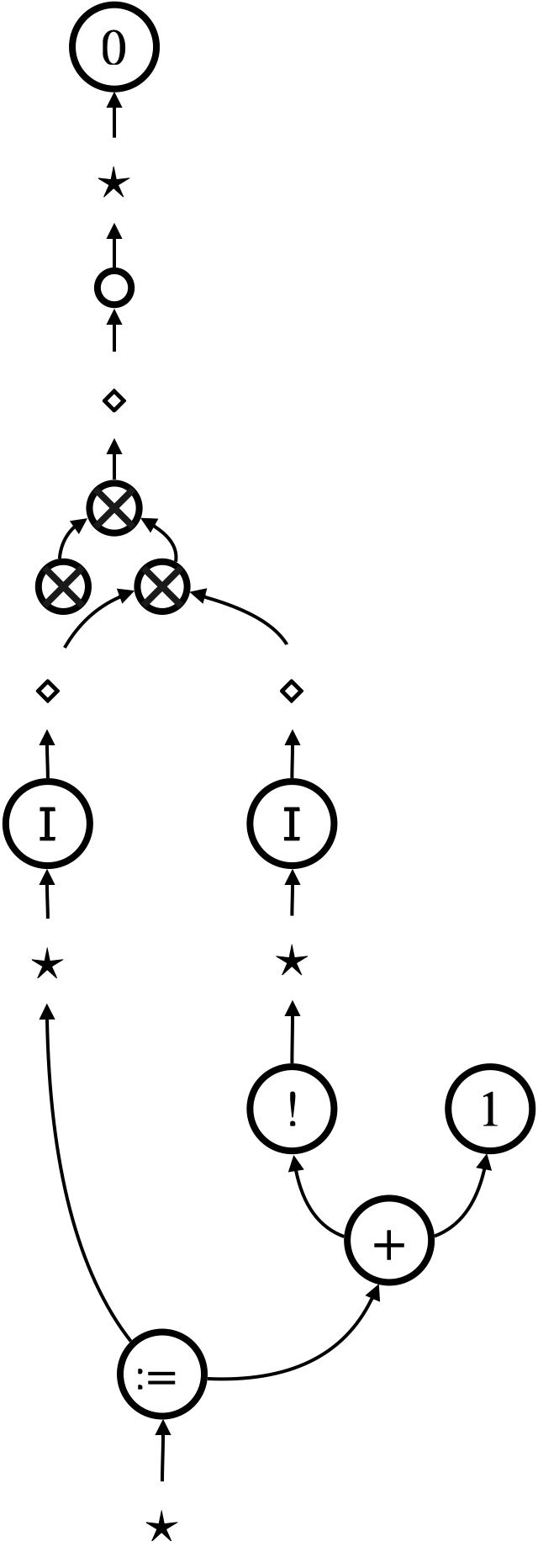}\hspace{5em}
 }
 \caption{The hypernet $(a \vdash a := (!a + 1))^\ddag \colon \star \To \diamond$ with store $\{a \mapsto 0\}$, simplified}
 \label{fig:transl-ref-ex}
\end{figure}

We next look at terms that contain references to store. An example term is $a := (!a + 1)$ that increments the value stored at atom (location) $a$. We represent such a term, together with store (e.g.\ $\{a \mapsto 0\}$), altogether as a single hypernet; see \autoref{fig:transl-ref-ex}. The hypernets in the figure are simplified in the same manner as \autoref{fig:transl-lambda-ex-simple}.

To represent store and atom sharing, we begin with an observation about the difference between variable sharing and atom sharing.
Atoms, like variables, may occur arbitrarily many times in a single term. However, whereas terms bound to a variable can be duplicated, each store associated to the atom should not be duplicated. For example, the atom $a$ appears twice in the term $a := (!a + 1)$, but this does not mean the store $\{a \mapsto 0\}$ is accordingly duplicated. Both occurrences of the atom $a$ points at the single store $\{a \mapsto 0\}$.

This observation leads us to introduce one new vertex label, and some edge labels, of our hypernets.
Firstly, we introduce a new vertex label $\diamond$ that represents store, in addition to the labels we have used (i.e.\ $\star$ for terms and $T^n(\star)$ for thunks with $n$ bound variables). While terms and thunks can both be duplicated, store cannot be duplicated; this is the reason why we need a new vertex label.

Secondly, we introduce another version of contraction and weakening, namely $\contr^\diamond_\mathsf{C} \colon \diamond^{\otimes 2} \To \diamond$ and $\contr^\diamond_\mathsf{W} \colon \epsilon \To \diamond$, for the type $\diamond$. We call these contraction and weakening as well. Contractions and weakenings for $\star$ and $\diamond$ appear the same, but they will have different behaviours (see \autoref{sec:new-behaviour}).

Finally, we introduce edge labels for type conversion between $\star$ and $\diamond$: namely, anonymous atom edges called \emph{instance} $\mathsf{I} \colon \star \To \diamond$, and anonymous store edges $\circ \colon \diamond \To \star$. Their roles in the hypernet representation are best understood looking at the example in \autoref{fig:transl-ref-ex}. Each occurrence of an atom is represented by the anonymous instance edge `$\mathsf{I}$', and instances of the same atom is connected to a contraction tree of type $\diamond$. The contraction tree is then connected to the part of the hypernet in \autoref{fig:transl-ref-ex-store} that represents the store $\{a \mapsto 0\}$, which consists of the store edge `$\circ$' and the stored value $0$. In \autoref{fig:transl-ref-ex-store}, the part of the hypernet that is typed by $\diamond$ connects instances of an atom to its stored value.

To represent sharing of (instances of) atoms, we will use \emph{canonical} contraction trees $D_{1,m}^\diamond$ that are constructed in the same manner as the canonical trees $D_{1,m}^\star$.
Every canonical tree $D_{1,m}^\diamond$ again contains exactly one weakening edge and $m$ contraction edges.

\subsection{Inductive translation $(-)^\ddag$} \label{sec:inductive-transl}

We now present the translation $(-)^\ddag$ of the extended lambda-terms into hypernets $\HN{\LVgen}{\LEgen}$ where
\begin{align}
 \LVgen &= \{\star, \diamond\} \cup \{T^n(\star) \mid n \in \N\}, \label{eq:transl-general-node-labels} \\
 \LEgen &= \{
 \lambda \colon \star \To T^1(\star),\;
 \lrapp \colon \star \To \star^{\otimes 2},\;
 \mathit{+} \colon \star \To \star^{\otimes 2},\;
 \mathit{-}_2 \colon \star \To \star^{\otimes 2},\;
 \mathit{-}_1 \colon \star \To \star,
 \} \notag \\
 &\quad \cup \{
 \keyw{ref} \colon \star \To \star,\;
 \mathit{!} \colon \star \To \star,\;
 \mathit{=} \colon \star \To \star^{\otimes 2},\;
 \mathit{:=} \colon \star \To \star^{\otimes 2}
 \} \notag \\
 &\quad \cup \{
 \keyw{tt} \colon \star \To \epsilon,\;
 \keyw{ff} \colon \star \To \epsilon,\;
 () \colon \star \To \epsilon
 \} \notag \\
 &\quad \cup \{ n \colon \star \To \epsilon \mid n \in \N \} \notag \\
 &\quad \cup \{
 \contr^\star_\mathsf{C} \colon \star^{\otimes 2} \To \star,\;
 \contr^\star_\mathsf{W} \colon \epsilon \To \star,\;
 \contr^\diamond_\mathsf{C} \colon \diamond^{\otimes 2} \To \diamond,\;
 \contr^\diamond_\mathsf{W} \colon \epsilon \To \diamond
 \}. \label{eq:transl-general-edge-labels}
\end{align}
In general, a judgement $\Gamma \mid \Delta \vdash t$ is translated into $t^\ddag \colon \star \To \star^{\otimes k} \otimes \diamond^{\otimes h}$ where the lengths of $\Gamma,\Delta$ are $k,h$ respectively.

The translation $(-)^\ddag$ uses a generalisation of canonical trees $D^\ell_{1,m} \colon \ell^{\otimes m} \To \ell$ (where $\ell \in \{\star,\diamond\}$), namely a forest $D^\ell_{k,m}$ of canonical trees dubbed \emph{distributor}.
A distributor $D^\ell_{k,m} \colon \ell^{\otimes km} \To \ell^{\otimes k}$ is a forest of $k$ canonical trees $D^\ell_{1,m}$ with some permutation of inputs. It is inserted in the translation for sharing $k$ variables/atoms (depending on $\ell \in \{\star,\diamond\}$) among $m$ sub-terms.
For instance, \[ D^\diamond_{3,2}=\vcenter{\hbox{\includegraphics[scale=.25]{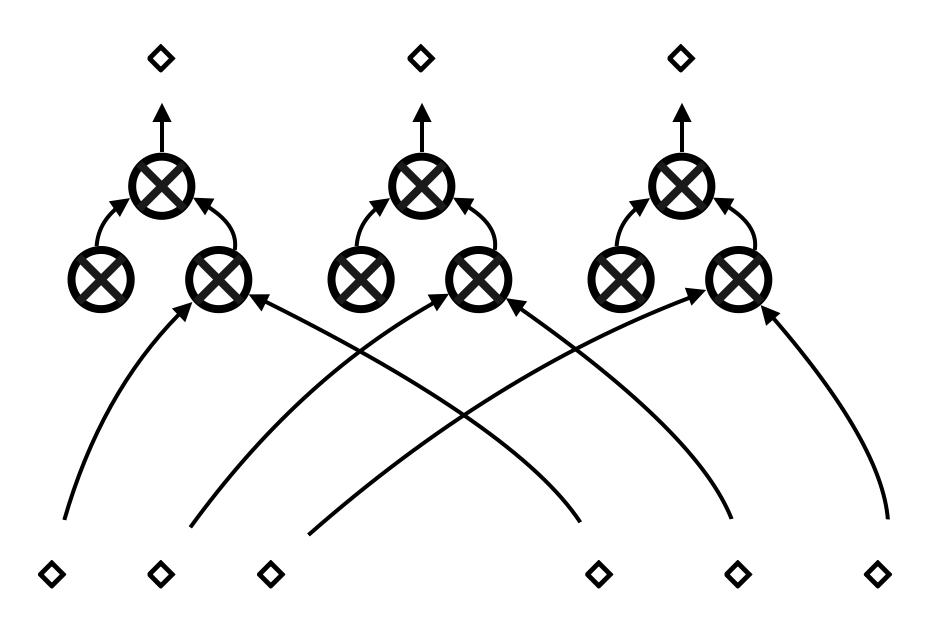}}} \text{ and }
D^\star_{4,0}=\vcenter{\hbox{\includegraphics[scale=.25]{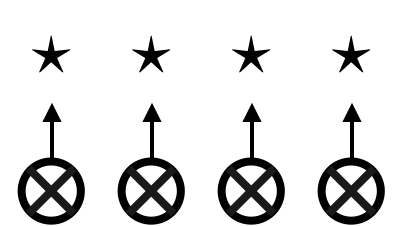}}}.\]
The precise definition of distributors will be given in \autoref{sec:aux-def}.

\section{The copying UAM} \label{sec:copying-uam}

\subsection{New behaviour} \label{sec:new-behaviour}

We equip the UAM with \emph{copying}, which is the behaviour of contraction edges $\contr^\star_\mathsf{C} \colon \star^{\otimes 2} \To \star$.
The copying UAM $\UAM(\Opr,B_\Opr)$ acts on hypernets $\HN{\LVgen}{\LEgen(\Opr)}$ where
\begin{align}
 \LEgen(\Opr) &= \Opr
  \cup \{ ? \colon \star \To \star,\;\checkmark \colon \star \To \star,\;\lightning \colon \star \To \star \} \\ \notag
  &\quad \cup \{
  \contr^\star_\mathsf{C} \colon \star^{\otimes 2} \To \star,\;
  \contr^\star_\mathsf{W} \colon \epsilon \To \star,\;
  \contr^\diamond_\mathsf{C} \colon \diamond^{\otimes 2} \To \diamond,\;
  \contr^\diamond_\mathsf{W} \colon \epsilon \To \diamond
  \}.
 \label{eq:UAM-general-labels}
\end{align}
For box edges, we impose the following type discipline: each box edge must have a type $T^n(\star) \To \star^{\otimes m} \otimes \diamond^{\otimes h}$ with its content having a type $\star \To \star^{\otimes (n+m)} \otimes \diamond^{\otimes h}$.

\begin{table}[t]
 \centering
 \begin{tabular}{|c|c||c|c|} \hline
  \multicolumn{2}{|c||}{transitions} & focus & provenance \\ \hline \hline
  \multicolumn{2}{|c||}{search transitions} & $?,\checkmark$ & \multirow{2}{*}{intrinsic} \\ \cline{1-3}
  \multirow{2}{*}{rewrite transitions} & \emph{copy} transitions & \multirow{2}{*}{$\lightning$} & \\ \cline{2-2}\cline{4-4}
  & behaviour $B_\Opr$ && extrinsic \\
  & (compute transitions) && \\ \hline
 \end{tabular}
 \caption{Transitions of the copying UAM}
 \label{tab:CUAM-transitions}
\end{table}

\begin{figure}[t]
 \centering
 \includegraphics[scale=.2]{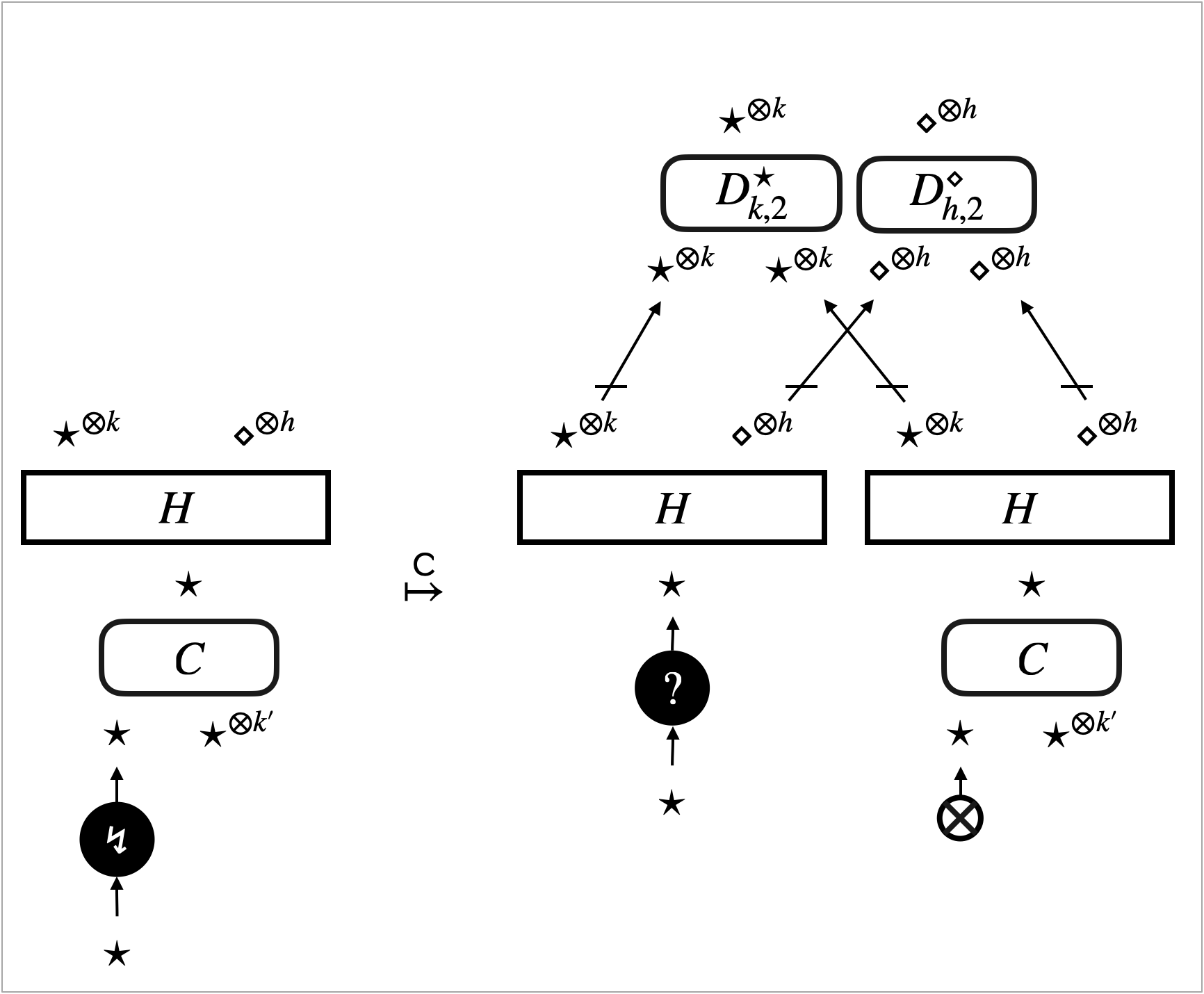}
 \caption{The copy rule where $H$ is a copyable hypernet}
 \label{fig:copy-rule}
\end{figure}
\begin{figure}[t]
 \centering
 \subfloat[\label{fig:interaction-contr-L}]{
 \includegraphics[scale=.2]{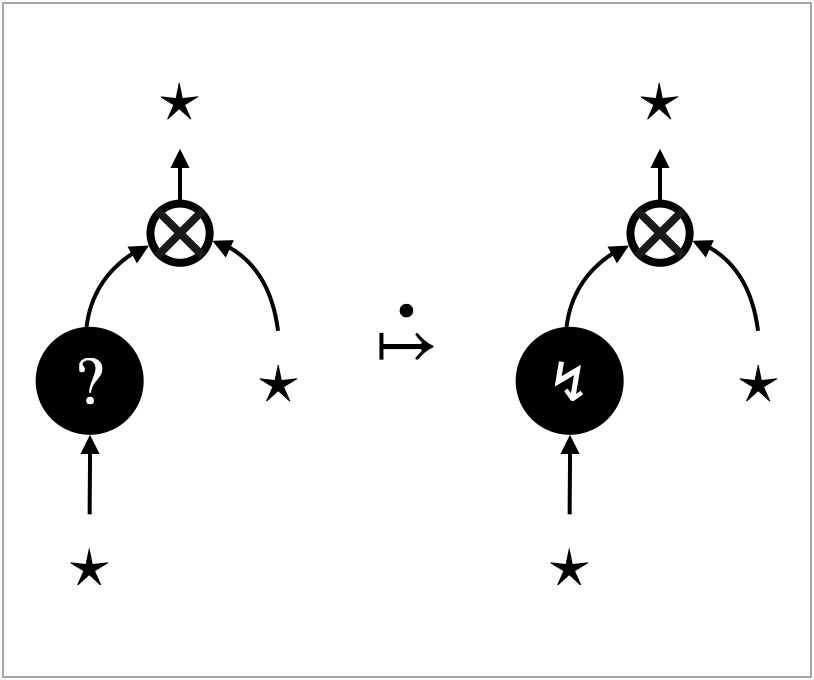}
 }
 \hfil
 \subfloat[\label{fig:interaction-contr-R}]{
 \includegraphics[scale=.2]{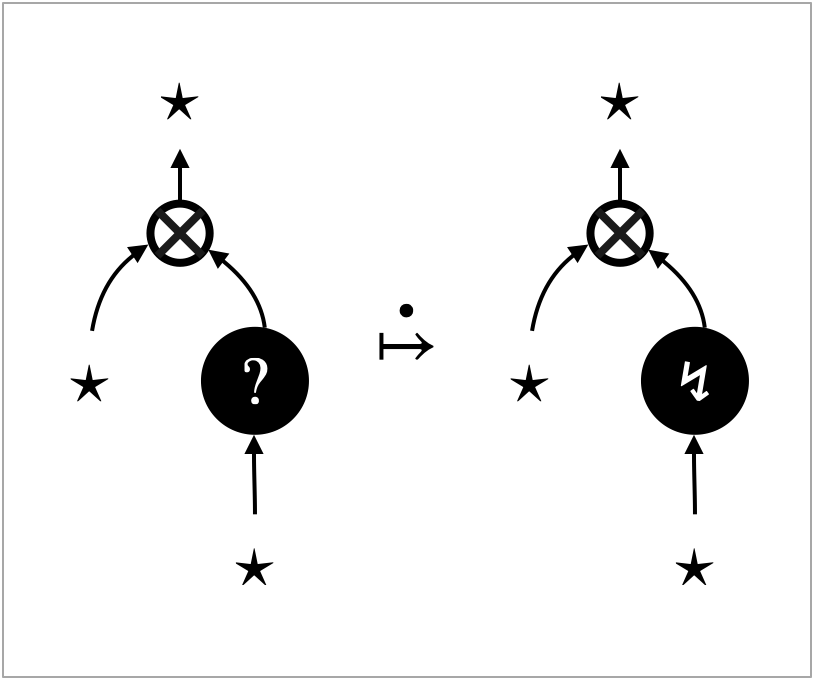}
 }
 \hfil
 \subfloat[\label{fig:interaction-I}]{
 \includegraphics[scale=.2]{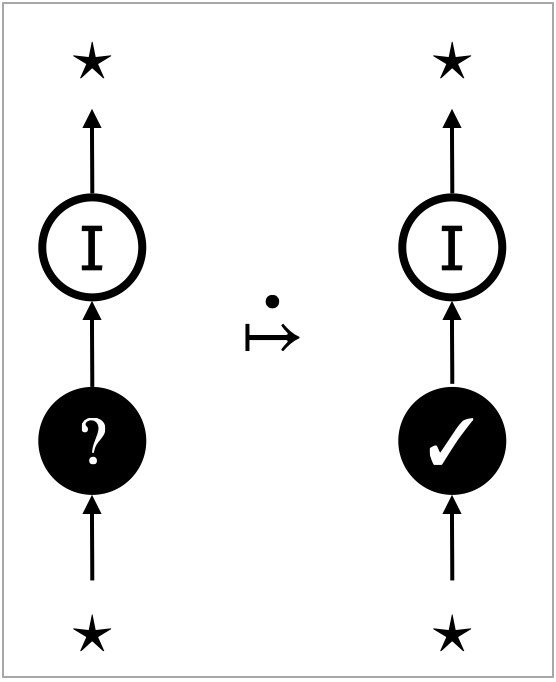}
 }
 \caption{Extra interaction rules}
 \label{fig:interaction-rules-extra}
\end{figure}

\autoref{tab:CUAM-transitions} summarises transitions of the copying UAM.
The copying UAM has \emph{copy} transitions instead of substitution transitions (cf.\ \autoref{tab:UAM-transitions}). Copy transitions are specified by the copy rule shown in \autoref{fig:copy-rule}. The copy rule duplicates a \emph{copyable} hypernet, and inserts distributors and a weakening edge. A copyable hypernet consists of instance edges, operation edges, and box edges. The precise definition of copyable hypernets will be given in \autoref{sec:aux-def}.
Note that the copy rule only applies to contraction of type $\star$; contraction of type $\diamond$, which are for atoms, do not have a corresponding copy rule. This reflects the fact that atoms are never duplicated unless it is inside a thunk.

Accordingly, the copying UAM has extra search transitions concerning contractions (`$\contr$') and instances (`$\mathsf{I}$'). These are specified by the interaction rules shown in \autoref{fig:interaction-rules-extra}.

\begin{figure}[t]
 \centering
 \subfloat[Reference creation\label{fig:rewrite-ref}]{
 \includegraphics[scale=.2]{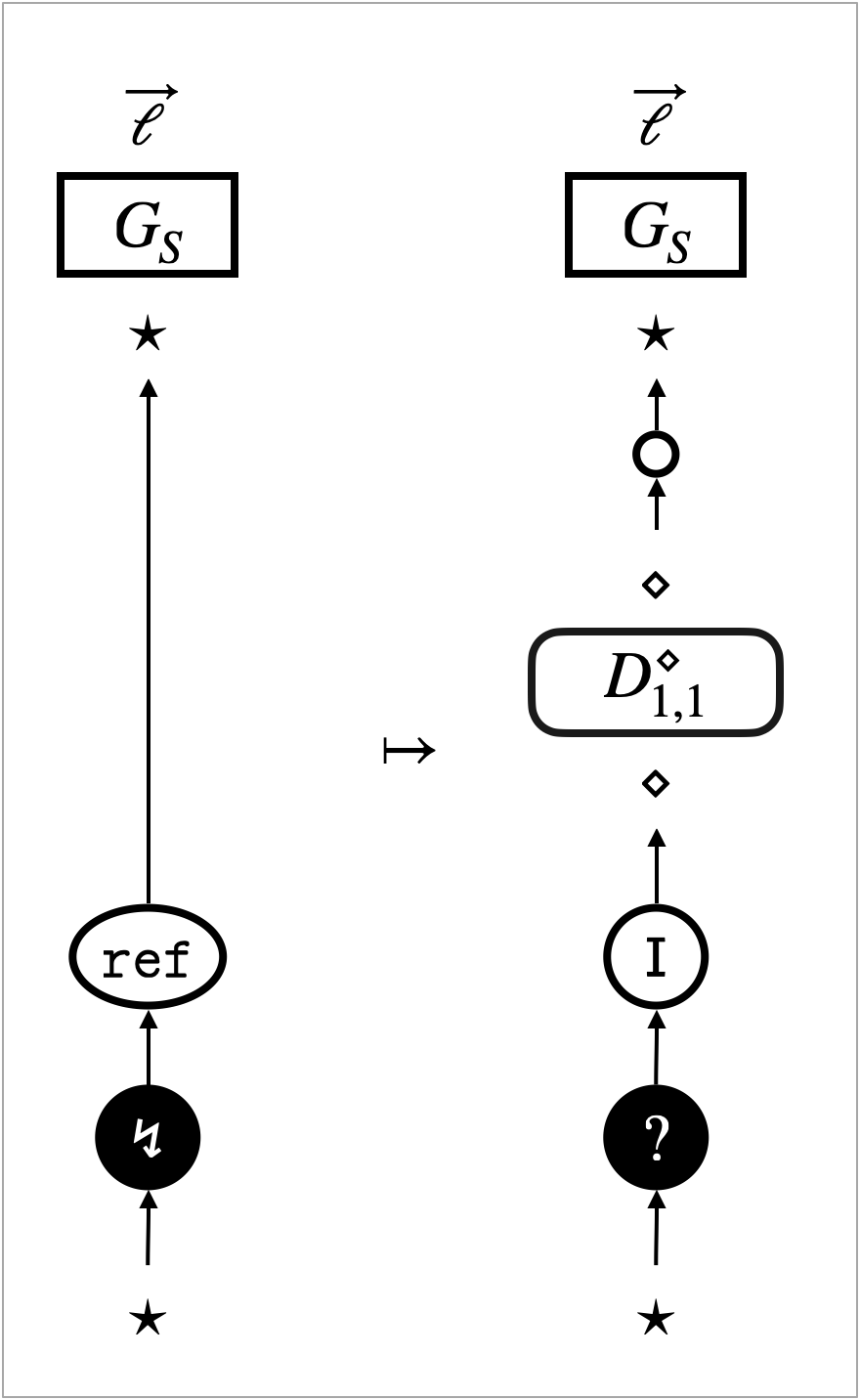}
 }
 \hfil
 \subfloat[Dereferencing\label{fig:rewrite-deref}]{
 \includegraphics[scale=.2]{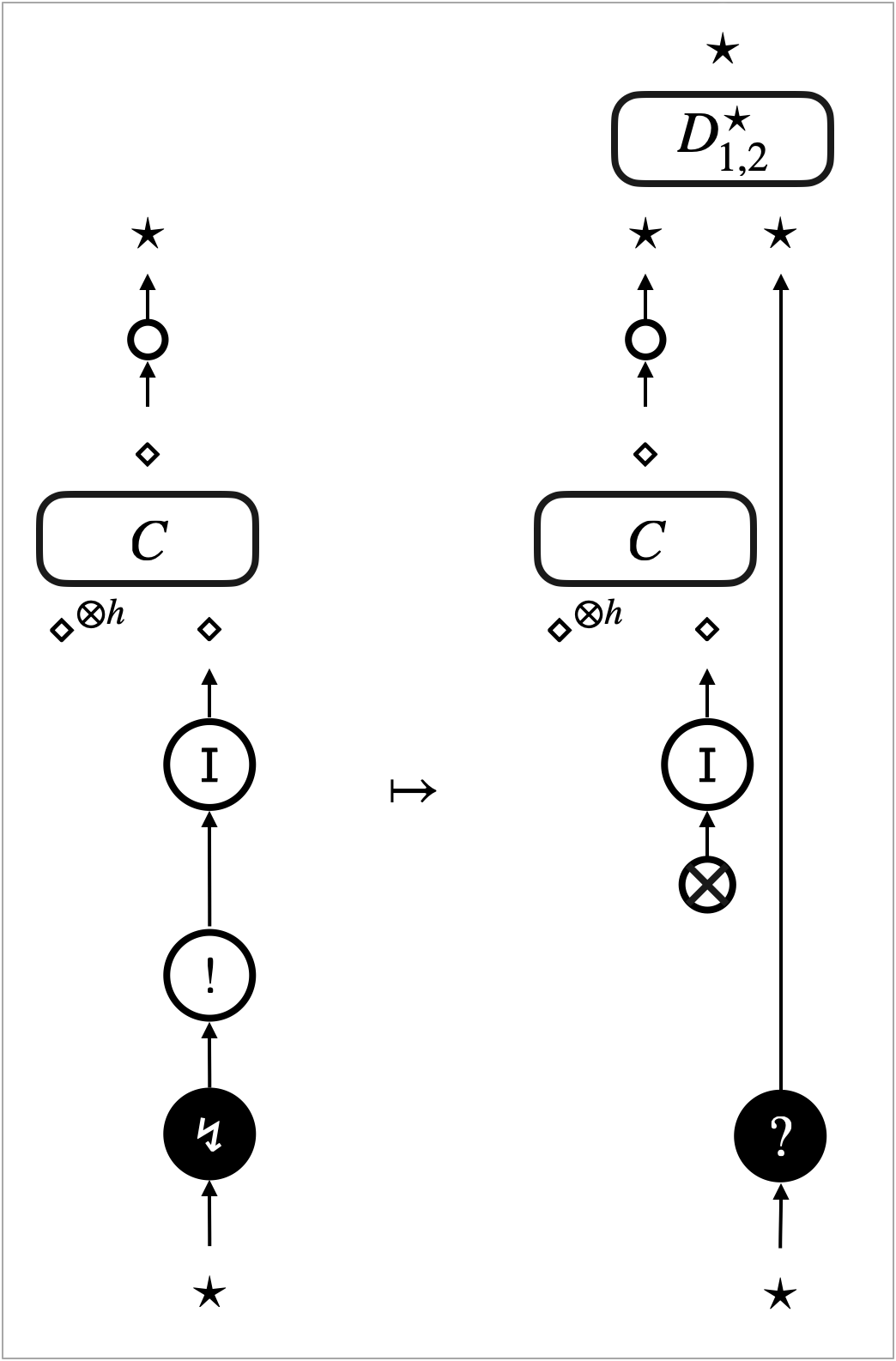}
 }
 \hfil
 \subfloat[Assignment\label{fig:rewrite-assign}]{
 \includegraphics[scale=.2]{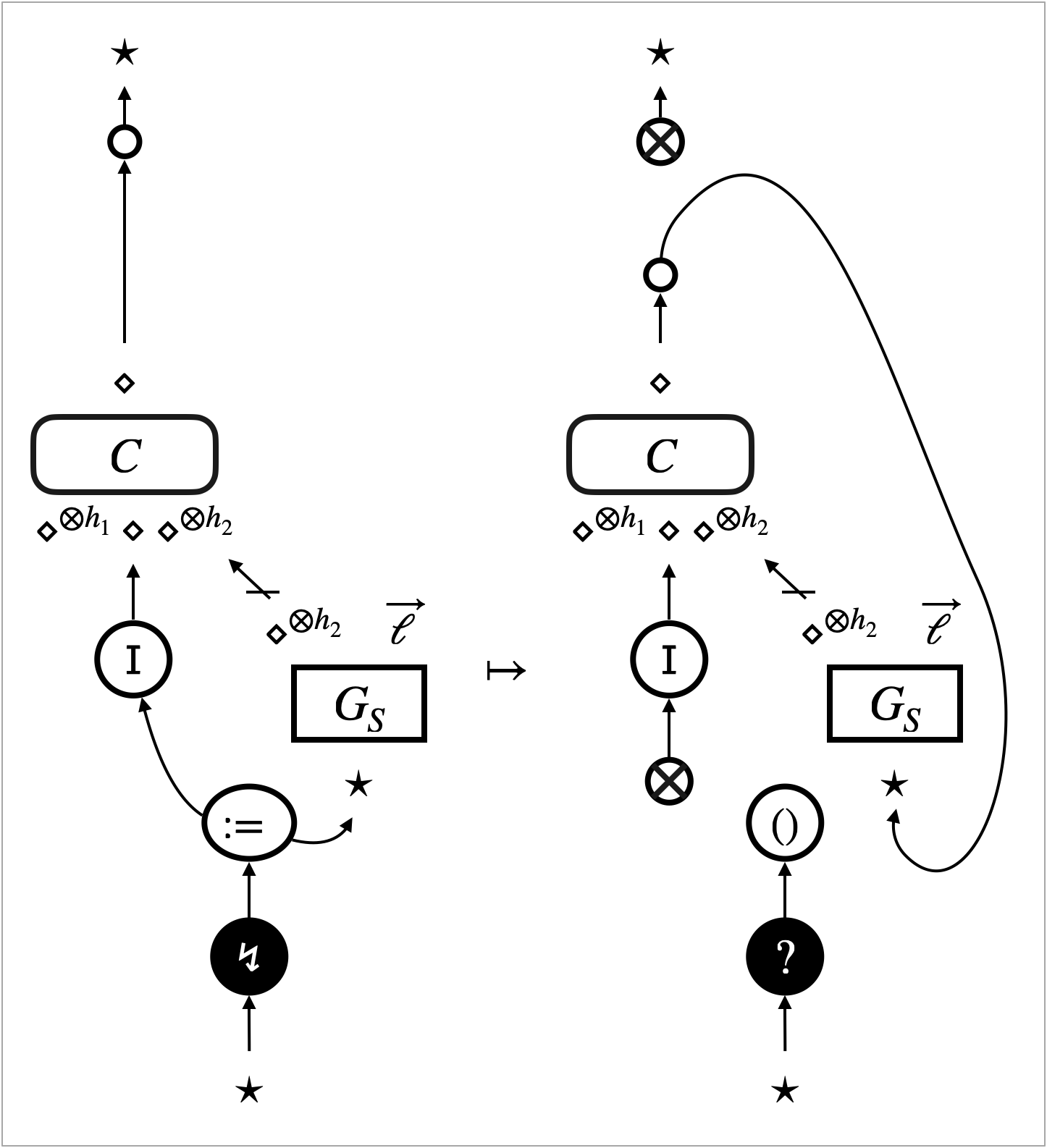}
 }
 \\
 \subfloat[Equality testing\label{fig:rewrite-eq}]{
 \includegraphics[scale=.2]{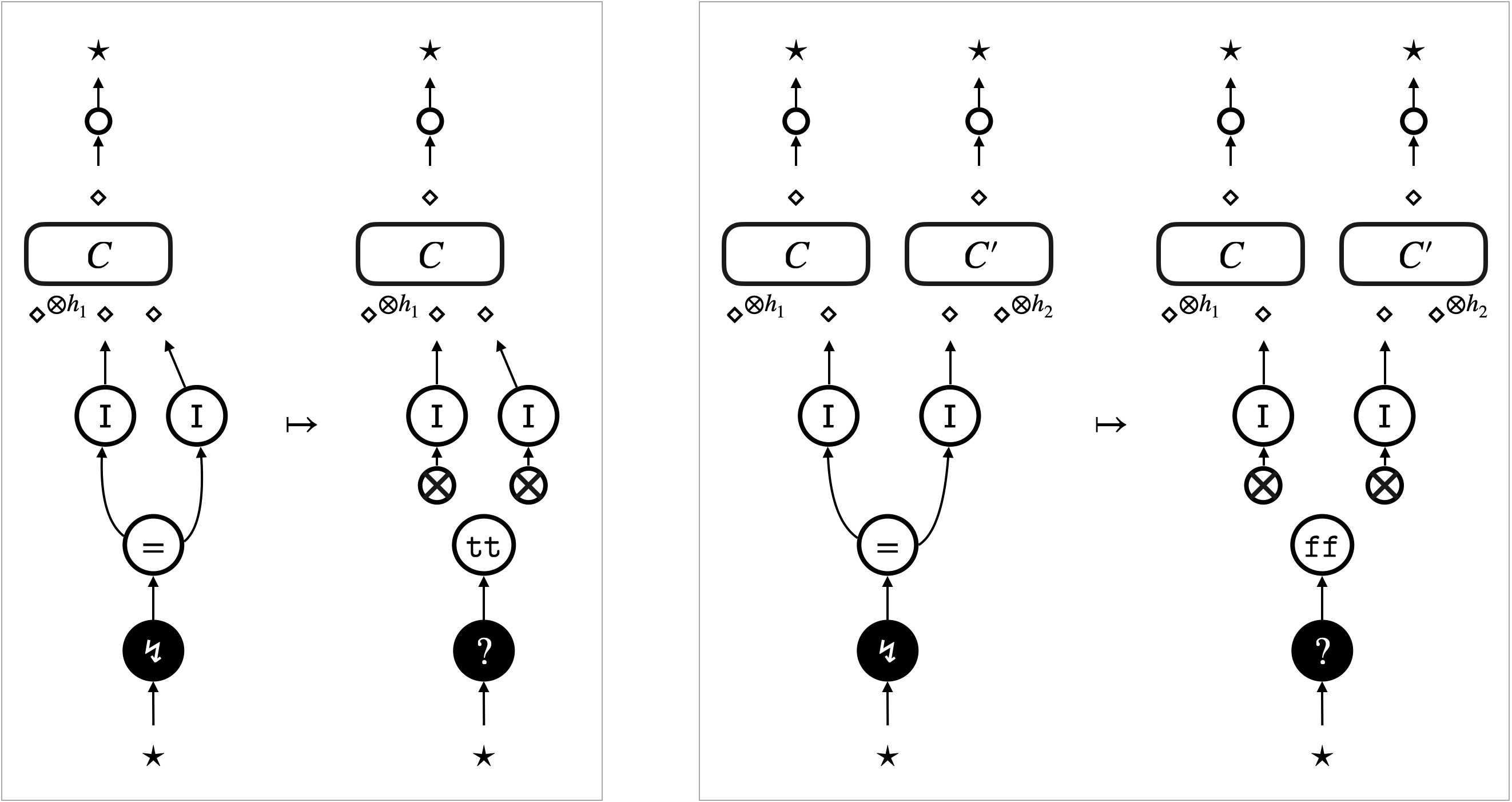}
 }
 \caption{The example behaviour $B_{\{\keyw{ref},\mathit{!},\mathit{:=},\mathit{=}\}}$ where $C,C'$ are contraction trees and $G_S$ is a hypernet}
 \label{fig:behaviour-ex-store}
\end{figure}
\autoref{fig:behaviour-ex-store} shows an example of the behaviour $B_\Opr$, namely that for the following active operations for store:
\begin{equation*}
 \keyw{ref} \colon \star \To \star, \quad
 \mathit{!} \colon \star \To \star, \quad
 \mathit{:=} \colon \star \To \star^{\otimes 2}, \quad
 \mathit{=} \colon \star \To \star^{\otimes 2}.
\end{equation*}
Their behaviour is specified locally by rewrite rules.
\begin{description}
 \item[\autoref{fig:rewrite-ref}] This rewrite rule specifies the behaviour of reference creation. The rule introduces store (`$\circ$') and its instance (`$\mathsf{I}$'). In between these two edges, a canonical tree $D^\diamond_{1,1}$ is inserted so that there is always a contraction tree between any store and instances.
 \item[\autoref{fig:rewrite-deref}] This rewrite rule specifies the behaviour of dereferencing. When dereferencing is triggered, the store edge (`$\circ$') gets attached to a canonical tree $D^\star_{1,2}$, so that its stored value will be copied by subsequent transitions.
 \item[\autoref{fig:rewrite-assign}] This rewrite rule specifies the behaviour of assignment. It identifies the store edge (`$\circ$') by tracing the contraction tree $C$ connected to the instance edge (`$\mathsf{I}$'). Once the store edge is identified, the current stored value gets disconnected and the new value $G_S$ gets connected to the store edge. Note that the value $G_S$ may contain other instances referring to the same store; this means the rewrite rule may introduce a cycle (via $\diamond^{\otimes h_2}$).
 \item[\autoref{fig:rewrite-eq}] This rewrite rule specifies the behaviour of equality testing for atoms. It determines whether two instance edges (`$\mathsf{I}$') are connected to the same store edge (`$\circ$') or not, by tracing contraction trees $C,C'$. If they are connected to the same store, the result $\keyw{tt}$ is introduced and attached to the $?$-focus. If not, the result $\keyw{ff}$ is introduced instead.
\end{description}

\subsection{Auxiliary definitions} \label{sec:aux-def}

We here present a few definitions of the concepts introduced so far.
The first is \emph{distributors}. A distributor $D^\ell_{k,m}$ ($\ell \in \{\star,\diamond\}$) is given by a forest of canonical trees $D^\ell_{1,m}$ whose inputs are permuted. The permutation is realised by \emph{interface permutation} (\autoref{def:permut-HG}).
\begin{defi}[Distributors and canonical trees]\label{def:distributors}
 For $\ell \in \{\star,\diamond\}$ and $k,m \in \N$, a \emph{distributor} is defined inductively as follows.
 \begin{gather*}
  D_{0,m}^\ell = \emptyset \hspace{4em}
  D_{1,0}^\ell = \includegraphics[align=c,scale=.25]{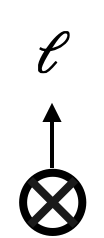} \hspace{4em}
  D_{1,1}^\ell = \includegraphics[align=c,scale=.25]{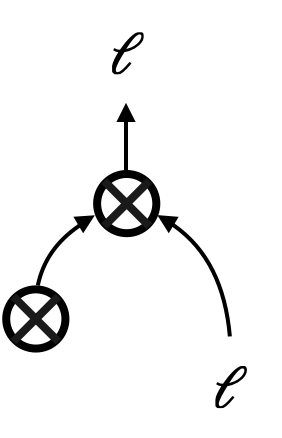} \hspace{4em}
  D_{1,m+2}^\ell = \includegraphics[align=c,scale=.25]{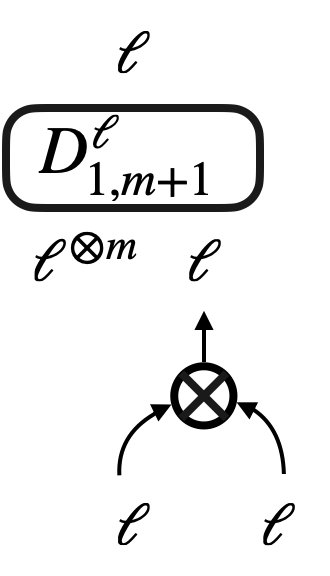} \\
  D_{k+1,m}^\ell = \Pi_\rho^{\mathrm{id}}\left(\vcenter{\hbox{\includegraphics[align=c,scale=.25]{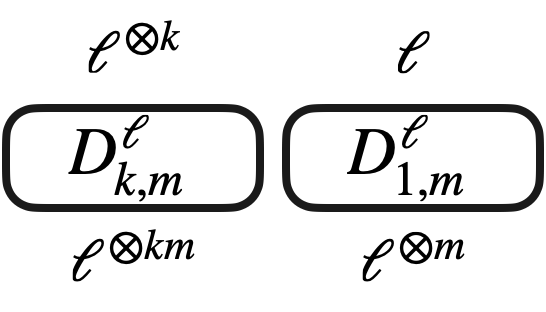}}}\right)
 \end{gather*}
 where $\emptyset$ denotes the empty hypernet,
 $\mathrm{id}$ is the identity map, and $\rho$ is a bijection
 such that, for each
 $j \in \{ 1,\ldots,k \}$ and $i \in \{ 1,\ldots,m \}$,
 $\rho(j + (k+1)(i-1)) = j + k(i-1)$ and
 $\rho((k+1)i) = km + i$.

 When $k = 1$, the distributor $D^\ell_{1,m}$ is called \emph{canonical tree}.
\end{defi}

The second is \emph{copyable} hypernets. The copy rule (\autoref{fig:copy-rule}) duplicates a single copyable hypernet at a time.
\begin{defi}[Copyable hypernets]\label{def:copyable-HN}
A hypernet
$H : \star \To \star^{\otimes k} \otimes \diamond^{\otimes h}$
is called \emph{copyable} if it is given by
\[\includegraphics[align=c,scale=.25]{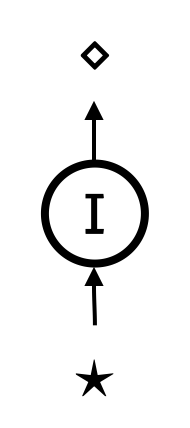} \text{ or }
\includegraphics[align=c,scale=.25]{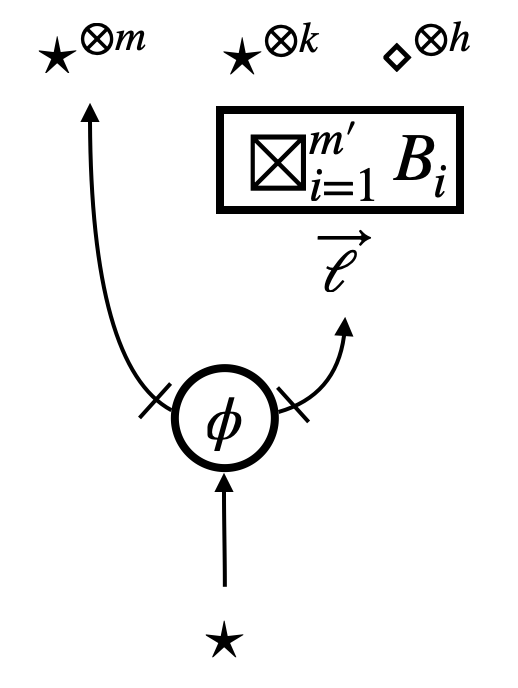}\]
where $\phi \in \Opr$ and $\boxtimes_{i=1}^{m'}B_i$ is a (possibly empty) parallel juxtaposition of box hypernets.
\end{defi}

Finally, the notion of \emph{stable hypernet} (\autoref{def:stable-HN}), which corresponds to values, needs to change to incorporate store.
\begin{defi}[Stable hypernets] \label{def:stable-HN-store}
A \emph{stable} hypernet is a hypernet
$(G : \star \To \otimes_{i=1}^m \ell_i) \in
\mathcal{H}(\LVgen, \{\mathsf{I}\} \cup \Opr_\checkmark)$, such that
$\otimes_{i=1}^m \ell_i\in
(\{\diamond\} \cup \{ T^n(\star) \mid n \in \N \})^m$
and each vertex is reachable from the unique input.
\end{defi}

\subsection{Determinism and refocusing}

The properties of determinism and refocusing can be defined for the copying UAM in the same way as the original UAM (cf.\ \autoref{def:determ-refocusing}). We can again use the stationary property as a sufficient condition for refocusing (\autoref{lem:stationary}).
Determinism and refocusing of a copying UAM also boil down to those of extrinsic transitions $B_\Opr$ under a mild condition.

\begin{lem}[Determinism and refocusing] \label{lem:determ-refocusing-copying}
 \noindent
 \begin{itemize}
  \item A copying universal abstract machine $\UAM(\Opr,B_\Opr)$ is deterministic if
	extrinsic transitions $B_\Opr$ are deterministic.
  \item Suppose that $G_S$ in the substitution rule (\autoref{fig:subst-rule}) is a stable hypernet.
	A copying universal abstract machine $\UAM(\Opr,B_\Opr)$ is refocusing if
	extrinsic transitions $B_\Opr$ preserve the rooted property.
 \end{itemize}
\end{lem}
\begin{proof}
 The proof is that of \autoref{lem:determ-refocusing} equipped with analysis of copying transitions.
 Copy transitions are all
 deterministic, because different contraction rules applied to a single
 state result in the same state. The choice of the contraction tree ($C$ in \autoref{fig:copy-rule}) is irrelevant.
 Copy transitions are stationary, and hence they preserve the rooted property.
\end{proof}

\section{Observational equivalence on lambda-terms} \label{sec:obs-equiv-lambda}

Let $\Oprex = \Oprex_\checkmark \uplus \Oprex_\lightning$ be the set of operations given by $\Oprex_\checkmark = \N \cup \{\lambda,\keyw{tt},\keyw{ff},()\}$ and $\Oprex_\lightning = \{ +,-,-_1,\lrapp,\keyw{ref},\mathit{!},\mathit{:=},\mathit{=} \}$.
The copying UAM $\UAM(\Oprex,B_\Oprex)$ provides operational semantics of the extended lambda-calculus (\autoref{fig:Lamex}).
Using contextual equivalence on hypernets (\autoref{def:CtxtRefEquiv}), we can define a notion of observational equivalence on lambda-terms. The notion only concerns the coincidence of termination, which is standard given that the extended lambda-calculus is untyped.
\begin{defi}[Observational equivalence on lambda-terms] \label{def:obs-equiv-lambda}
 Let $\Gamma \mid \Delta \vdash t$ and
 $\Gamma \mid \Delta \vdash u$ be two derivable judgements.
 The lambda-terms $t$ and $u$ are said to be
 \emph{observationally equivalent}, written as
 $\Gamma \mid \Delta \vDash t \simeq^\ddag u$, if
 $(\Gamma \mid \Delta \vdash t)^\ddag
 \simeq^{\mathbb{C}_{\Oprex\cbf}}_{\N \times \N}
 (\Gamma \mid \Delta \vdash t)^\ddag$
 holds.
\end{defi}

This definition uses the specific contextual equivalence $\simeq^{\mathbb{C}_{\Oprex\cbf}}_{\N \times \N}$.
Firstly, the use of the universal relation $\N \times \N$ makes the number of transitions until termination irrelevant.
Secondly, the set $\mathbb{C}_{\Oprex\cbf}$ is the set of all \emph{binding-free} contexts.
\begin{defi}[Binding-free contexts] \label{def:binding-free-ctxt}
A focus-free context $\mathcal{C}$ is said to be \emph{binding-free}
if there exists no path, at any depth, from a source of a contraction,
atom, box or hole edge, to a source of a hole edge.
\end{defi}
For example, \[\includegraphics[align=c,scale=.25]{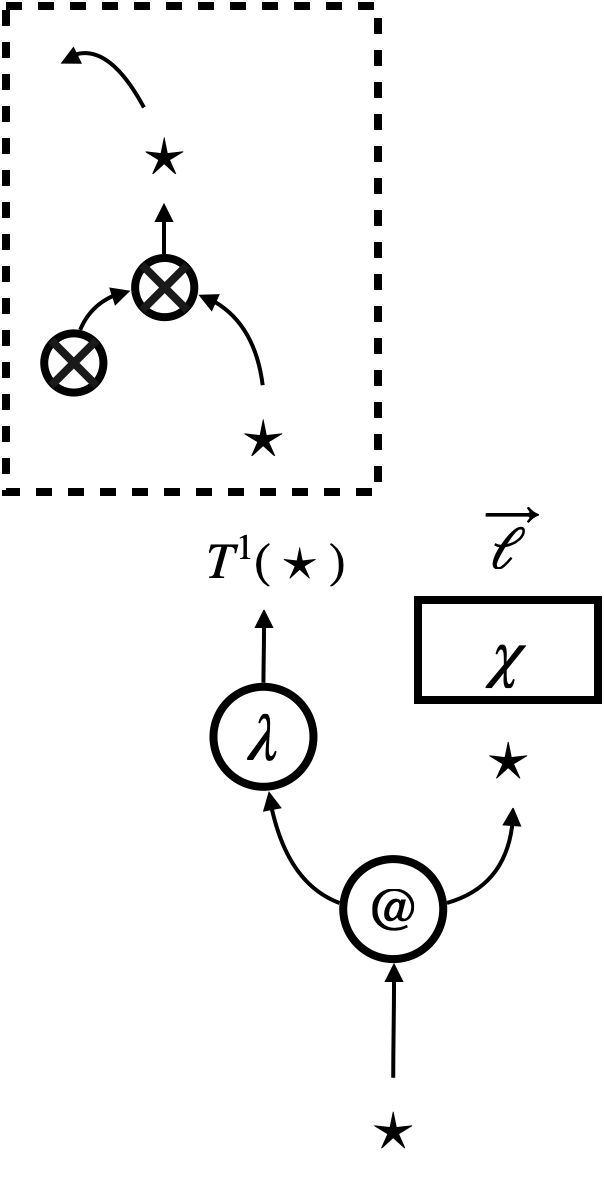}\] is a binding-free context where $\chi$ is a hole. Syntactically it would coincide with a context $(\lambda x. x)\ \square$.

Note that the contextual equivalence $\simeq^{\mathbb{C}_{\Oprex\cbf}}_{\N \times \N}$ is a larger relation than the contextual equivalence $\simeq^{\mathbb{C}_{\Oprex}}_{\N \times \N}$ with respect to the set $\mathbb{C}_{\Oprex}$ of all contexts. 

The restriction to binding-free contexts can be justified by the fact that the observational equivalence $\simeq^\ddag$ is a congruence relation with respect to lambda-contexts defined by the following grammar:
\begin{align*}
 C \mathrel{::=} [\,] \mid \lambda x.\,C \mid C\ t \mid t\ C
 \mid \mathop{\$_1} C \mid C \mathbin{\$_2} t \mid t \mathbin{\$_2} C
\end{align*}
where $\mathord{\$_1} \in \{ \keyw{ref},\mathord{!},\mathord{-_1} \}$
and $\mathord{\$_2} \in
\{ \mathord{=},\mathord{:=},\mathord{+},\mathord{-} \}$.
This congruence property will be formalised as
\autoref{lem:LamexObsEquivCongruence} below.

We can extend the translation $(-)^\ddag$ to accommodate lambda-contexts.
To do so, we first adapt the formation rules of lambda-terms in \autoref{fig:Lamex} to lambda-contexts, by annotating the hole `$[\,]$' as
`$[\,]_{\Gamma \mid \Delta}$' and adding a formation rule
$\infer{\Gamma \mid \Delta \vdash [\, ]_{\Gamma \mid \Delta}}{}$.
We write $C[\, ]_{\Gamma \mid \Delta}$ when the hole of $C$ is annotated
with $\Gamma \mid \Delta$.
The translation $(-)^\ddag$ of lambda-terms into hypernets in
\autoref{fig:transl-lambda-ex} can then be extended accordingly,
by translating the additional formation rule
$\infer{\Gamma \mid \Delta \vdash [\, ]_{\Gamma \mid \Delta}}{}$
into a path hypernet
$\chi : \star \To
\star^{\otimes k} \otimes \diamond^{\otimes h}$,
where $k$ and $h$ are the length of $\Gamma$ and $\Delta$
respectively.

The translation of lambda-contexts yields hypernets that are
binding-free contexts, and consequently,
the observational equivalence $\simeq^\ddag$ on lambda-terms is
indeed a congruence relation with respect to lambda-contexts.
\begin{lem}
 \label{lem:LamexObsEquivCongruence}
 Let $\Gamma \mid \Delta \vdash t$ and
 $\Gamma \mid \Delta \vdash u$ be two derivable judgements.
 If $\Gamma \mid \Delta \vDash t \simeq^\ddag u$ holds,
 then 
 for any lambda-context $C$ such that
 $\Gamma' \mid \Delta' \vdash_\Lambda C[\,]_{\Gamma \mid \Delta}$
 is derivable,
 $\Gamma' \mid \Delta' \vDash C[t] \simeq^\ddag C[u]$ holds.
\end{lem}
\begin{proof}[Proof outline]
 The proof is a combination of
 the congruence property of the contextual equivalence
 $\simeq^{\mathbb{C}_{\Oprex\cbf}}_{\N \times \N}$ with respect to
 binding-free contexts, with two key properties of the translation of
 lambda-contexts. The two properties, stated below, can be proved
 by induction on lambda-contexts.
 
 The first property is that the translation
 $(\infer{\Gamma' \mid \Delta' \vdash
 C[\, ]_{\Gamma \mid \Delta}}{})^\ddag$
 is a binding-free context (as a hypernet). To check this property,
 one needs to examine paths to the sole source of the unique hole edge
 that appears in the translation. These paths are in fact always operation
 paths, noting that paths never go across the boundary of boxes
 (by \autoref{def:path-reachability}~(\ref{item:def:path})).

 The second property is that
 $\Gamma' \mid \Delta' \vdash C[t]$ and
 $\Gamma' \mid \Delta' \vdash C[u]$ are both derivable, and
 moreover, their translations can be decomposed as follows:
 \begin{align*}
  (\Gamma' \mid \Delta' \vdash C[t])^\ddag &=
  (\Gamma' \mid \Delta' \vdash
  C[\,]_{\Gamma \mid \Delta})^\ddag%
  [(\Gamma \mid \Delta \vdash t)^\ddag], \\
  (\Gamma' \mid \Delta' \vdash C[u])^\ddag &=
  (\Gamma' \mid \Delta' \vdash
  C[\,]_{\Gamma \mid \Delta})^\ddag%
  [(\Gamma \mid \Delta \vdash u)^\ddag].
  \qedhere
 \end{align*}
\end{proof}

\section{Example equivalences} \label{sec:ex-law}

In this section we use the sufficiency-of-robustness theorem
(\autoref{thm:MetaThm}) and prove some example equivalences.
The first kind of example is \emph{Weakening laws}.
\begin{prop}[Weakening laws]
 \label{prop:WeakeningLaw}
 \noindent
 \begin{itemize}
  \item Given a derivable judgement
	$\Gamma_1, x, \Gamma_2 \mid \Delta \vdash t$
	such that $x \notin \FV(t)$,
	\begin{equation*}
	 \includegraphics[align=c,scale=.25]{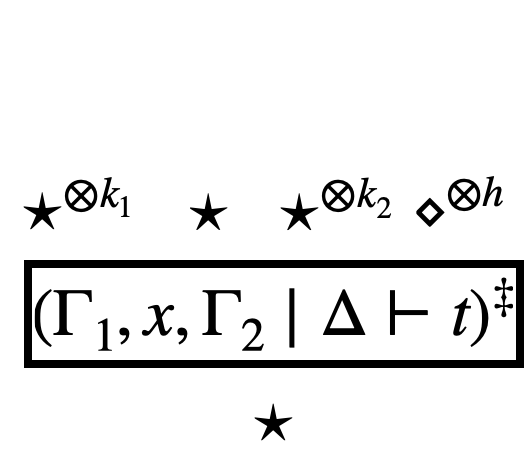}
	  \ \simeq^{\mathbb{C}_{\Oprex}}_{=_\N}\ 
	  \includegraphics[align=c,scale=.25]{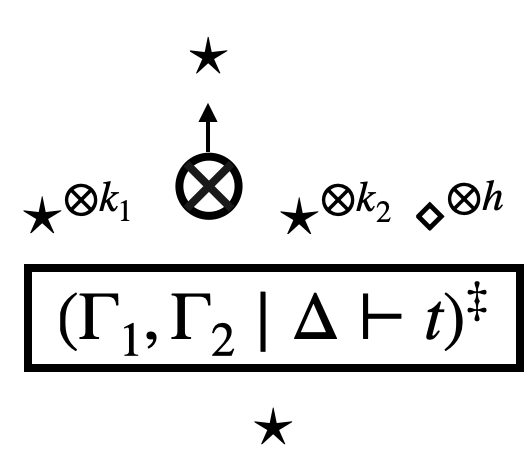}		 
	\end{equation*}
  \item Given a derivable judgement
	$\Gamma \mid \Delta_1, a, \Delta_2 \vdash t$
	such that $a \notin \FA(t)$,
	\begin{equation*}
	 \includegraphics[align=c,scale=.25]{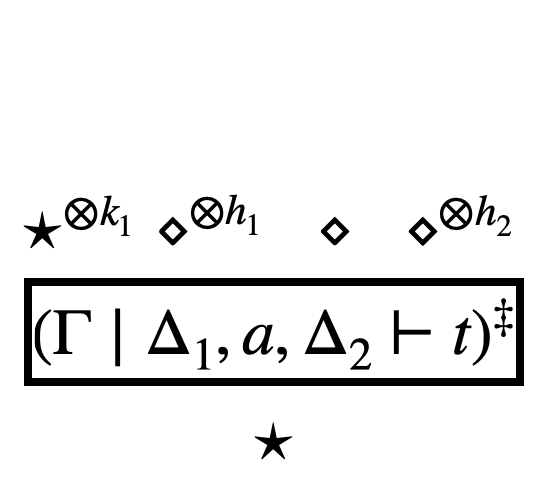}
	  \ \simeq^{\mathbb{C}_{\Oprex}}_{=_\N}\ 
	  \includegraphics[align=c,scale=.25]{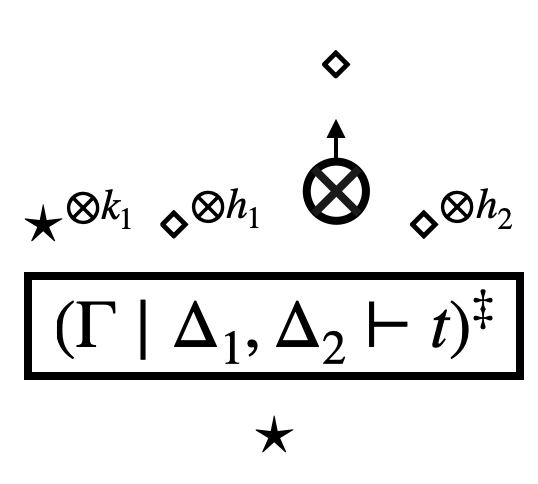}		 
	\end{equation*}
 \end{itemize}
 where $k,k_1,k_2,h,h_1,h_2$ are the lengths of
 $\Gamma,\Gamma_1,\Gamma_2,\Delta,\Delta_1,\Delta_2$, respectively.
\end{prop}

The next example equivalence is an instance of \emph{parametricity}.
This example originates in the Idealised Algol literature~\cite{OHearnT97algol}, a
call-by-name language with ground-type \emph{local} variables, although we state the example for the untyped call-by-value
lambda-calculus extended primarily with store (\autoref{fig:Lamex}).
Note that the example uses the standard call-by-value variable binding `$\keyw{let}$' and sequential composition `$;$', which are both defined by syntactic sugar (see \autoref{fig:Lamex}).
\begin{prop}
 \label{prop:ParametricityLaw}
 For any finite sequence of distinct variables $\Gamma$
 and any finite sequence of distinct variables $\Delta$,
 \begin{equation} \label{eq:param-law}
  \Gamma \mid \Delta \vDash
   \letin{x}{\keyw{ref}\ 1}{
   \lambda f.\, (f \  (); !x)}
   \ \simeq^\ddag\ 
   \lambda f.\, (f \  (); 1).
 \end{equation}
\end{prop}

In the left hand side (lhs) of (\ref{eq:param-law}), a store is created by `$\keyw{ref}\ 1$', and any
access to the store simply fetches the stored value, due to `$!x$',
without modifying it.
As a consequence, the fetched value is always expected to be the
original stored value `$1$', and hence the whole computation involving
the particular state is expected to have the same result as just
having the value `$1$' in the first place as in the right hand side (rhs) of (\ref{eq:param-law}).

The equivalence (\ref{eq:param-law}) is a typical challenging example in the literature~\cite{OHearnT97algol}, which has been proved using \emph{parametricity}. We take a different approach based on step-wise local reasoning.
The equivalence is not an example that is meant to show the full power of our approach; it is a simple yet motivating example that requires building of a whole proof infrastructure and a non-trivial proof methodology.

The rest of this section is organised as follows.
\autoref{sec:prerequisites} shows prerequisites on the copying UAM $\UAM(\Oprex, B_{\Oprex})$.
\autoref{sec:using-char-thm} defines necessary pre-templates, and proves that these imply contextual refinements, using the sufficiency-of-robustness theorem (\autoref{thm:MetaThm}).
\autoref{sec:combining-templates} then combines the resultant contextual refinements to prove the example equivalences (\autoref{prop:WeakeningLaw} \& \autoref{prop:ParametricityLaw}).
Finally, \autoref{sec:design-pre-templates} describes design process of some of the pre-templates.

\subsection{Prerequisites} \label{sec:prerequisites}

Here we establish that the particular copying UAM $\UAM(\Oprex, B_{\Oprex})$ is deterministic and refocusing, which enables us to apply the sufficiency-of-robustness theorem (\autoref{thm:MetaThm}).
\begin{defi}[the behaviour $B_{\Oprex}$] \label{def:ex-behaviour}
 The behaviour $B_{\Oprex}$ is defined locally via the following rewrite rules.
 \begin{itemize}
  \item Rewrite rules for function application are in \autoref{fig:rewrite-micro-beta}, where $G_S$ is additionally required to be stable.
  \item Rewrite rules for reference manipulation are in \autoref{fig:behaviour-ex-store}, where $G_S$ is additionally required to be stable.
  \item Rewrite rules for arithmetic are in \autoref{fig:rewrite-arith}.
 \end{itemize}
\end{defi}

The extra requirement of stable hypernets reflects the call-by-value nature of the extended lambda-calculus.

\begin{lem}
 \label{lem:determ-refocus} 
 The copying UAM $\UAM(\Oprex, B_{\Oprex})$ is deterministic and refocusing.
\end{lem}
\begin{proof}
 The proof boils down to show determinism, and preservation of the rooted property, of the compute transitions for $\Oprex_\lightning$, thanks to \autoref{lem:determ-refocusing-copying}.

 Compute transitions of operations
$\{ \lrapp,\keyw{ref},\mathord{+},\mathord{-},\mathord{-_1} \}$ are
deterministic, because at most one rewrite rule can be applied to
each state. In particular, the stable hypernet $G_S$ in the figures is
uniquely determined
(by \autoref{lem:StableShape}(\ref{item:StableMaximum})).

Compute transitions of name-accessing operations
$\{ \mathord{=},\mathord{:=},\mathord{!} \}$ are deterministic for the
same reason as copy transitions (see the proof of \autoref{lem:determ-refocusing-copying}).

Compute transitions of all the operations $\Oprex_\lightning$ are
stationary, and hence they preserve the rooted property.
The stationary property can be checked using local rewrite rules.
Namely, in each rewrite rule $\focussed{H} \mapsto \focussed{H'}$
of the operations, only one input of $|\focussed{H}|$ has type
$\star$, and $\focussed{H} = {\lightning;|\focussed{H}|}$ and
 $\focussed{H'} = {?;|\focussed{H'}|}$.
Moreover, any output of $|\focussed{H}|$ with type $\star$ is a target
of an atom edge or a box edge (by definition of stable hypernets),
which implies $|\focussed{H}|$ is one-way.
\end{proof}

\begin{rem}[Requirement of stable hypernets in \autoref{def:ex-behaviour}]
Because any initial state is rooted, given that all transitions
preserve the rooted property, we can safely assume that any state that
arises in an execution is rooted.
This means that the additional requirement of stable hypernets in \autoref{def:ex-behaviour} is in fact guaranteed to be satisfied in any execution
(by \autoref{lem:SearchSeqQueried},
\autoref{lem:AnswerAccessiblePaths} and
\autoref{lem:StablePathStableNet}).
\bqed
\end{rem}

\subsection{Pre-templates and robustness} \label{sec:using-char-thm}

\begin{figure}
 \centering
  \includegraphics[scale=.25]{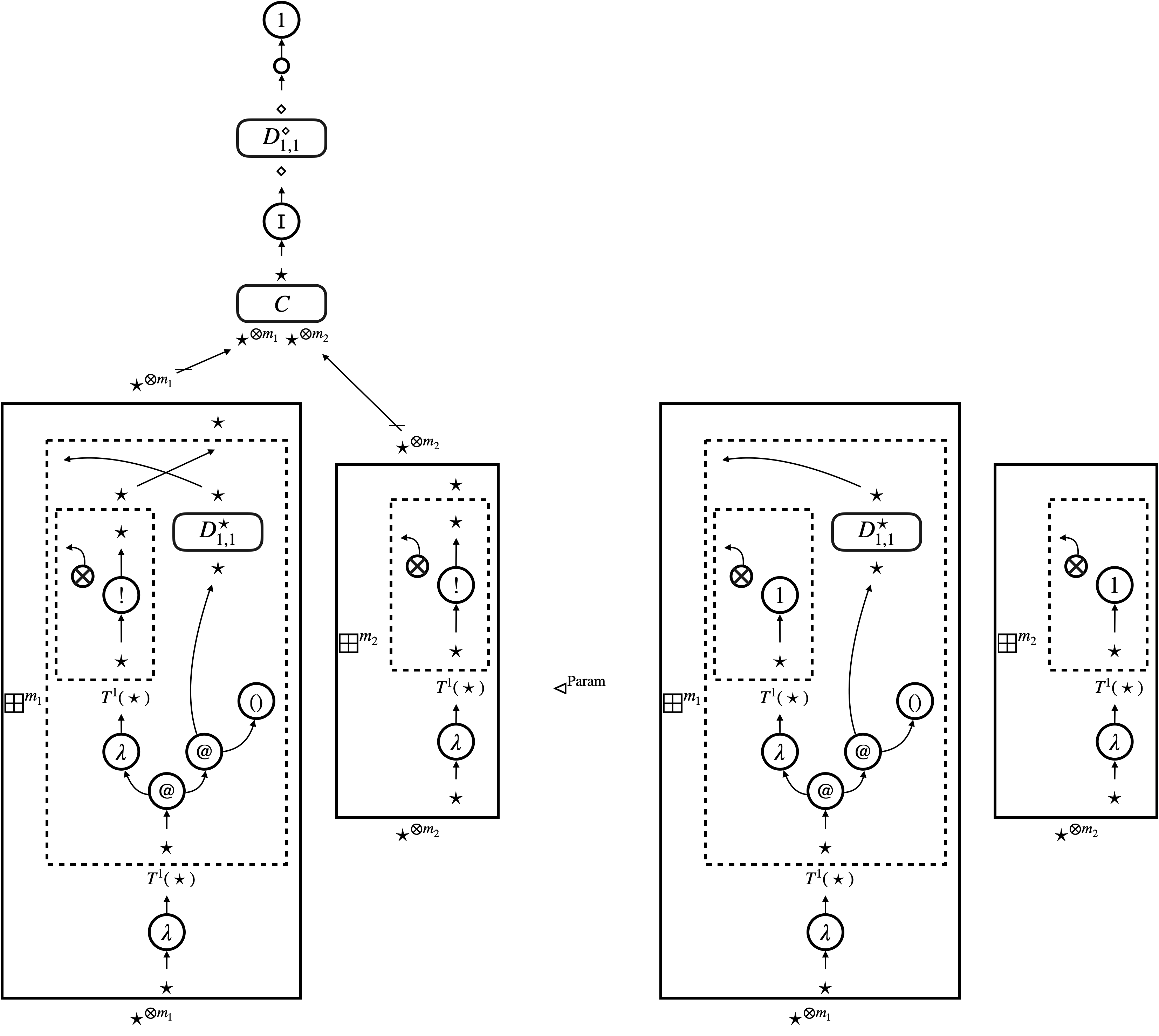}
  \caption{The parametricity pre-template where $C$ is a contraction tree}
 \label{fig:param}
\end{figure}

\begin{figure}[t]
 \includegraphics[scale=.2]{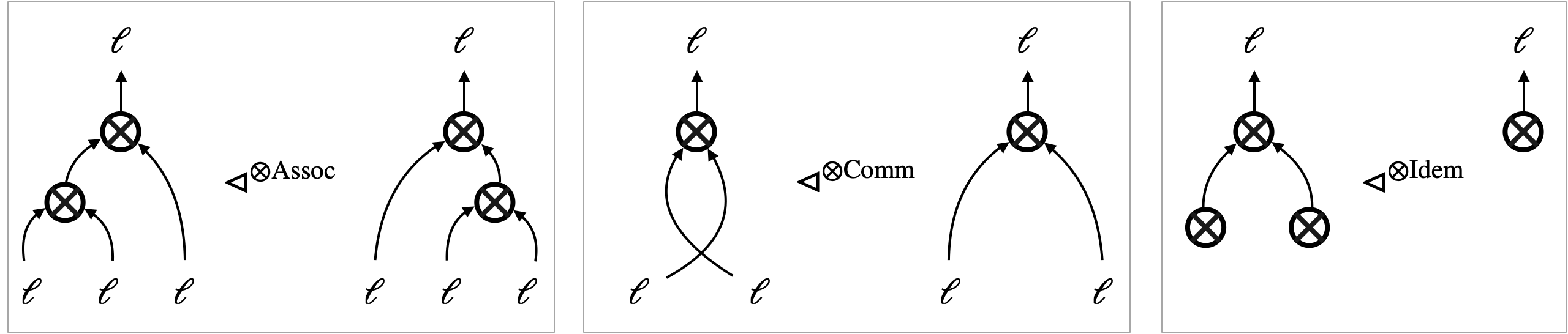} \\
 \includegraphics[scale=.2]{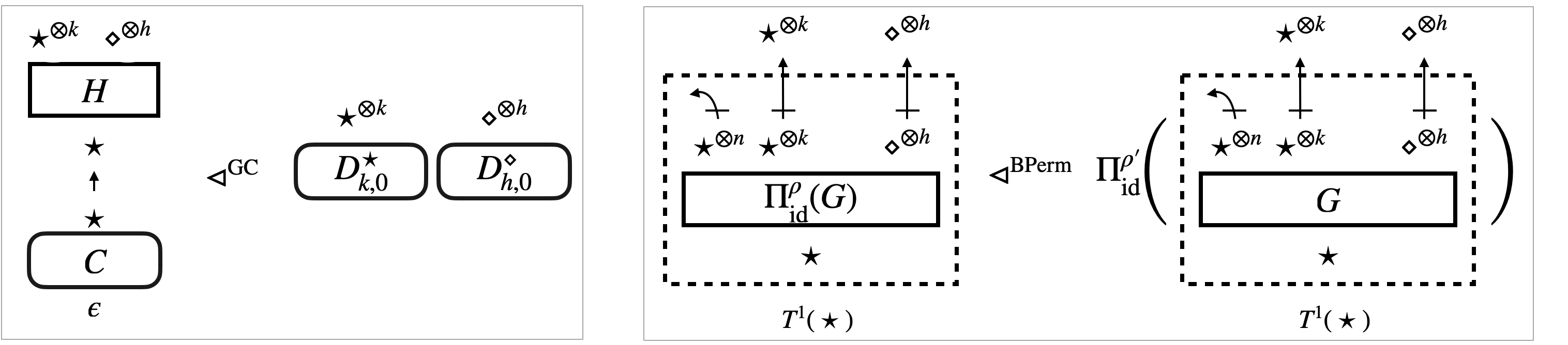} \\
 \includegraphics[scale=.2]{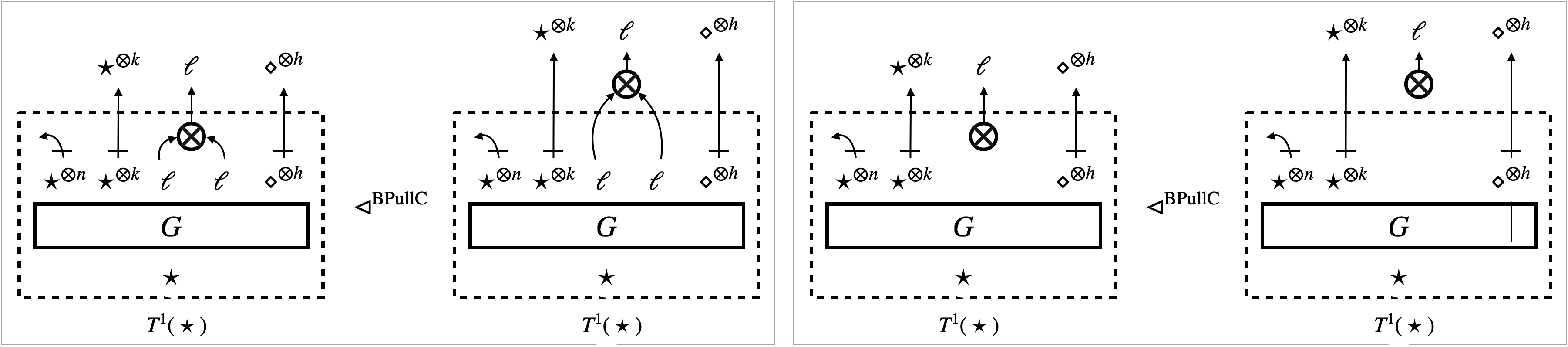}
 \caption{Structural pre-templates where $C : \epsilon \To \star$ is a contraction
 tree, $H$ is a copyable hypernet, $G$ is a hypernet, and $(\rho,\rho')$
 is a box-permutation pair}
 \label{fig:structural}
\end{figure}

We now define necessary pre-templates, which are the \emph{parametricity} pre-template $\vartriangleleft^{\mathrm{Param}}$, two \emph{operational} pre-templates $\vartriangleleft^{\keyw{ref}}$ and $\vartriangleleft^{\lrapp}$, and \emph{structural} pre-templates.
\pagebreak[5]
\begin{defi}[Pre-templates]
 \noindent
 \begin{itemize}
  \item The \emph{parametricity} pre-template $\vartriangleleft^{\mathrm{Param}}$ is as shown in \autoref{fig:param}.
  \item The \emph{micro-beta} pre-template $\vartriangleleft^{\lrapp}$ is derived from rewrite rules as follows:
	$|\focussed{G_1}| \vartriangleleft^{\lrapp} |\focussed{G_2}|$ if
	$\focussed{G_1} \mapsto \focussed{G_2}$ is a micro-beta rewrite rule
	(\autoref{fig:rewrite-micro-beta}) where $G_S$ (see the figure) is
	not an arbitrary hypernet but a stable hypernet.
  \item The \emph{reference-creation} pre-template $\vartriangleleft^{\keyw{ref}}$ is derived from rewrite rules as follows:
	$|\focussed{G_1}| \vartriangleleft^{\keyw{ref}} |\focussed{G_2}|$ if
	$\focussed{G_1} \mapsto \focussed{G_2}$ is a reference-creation
	rewrite rule (\autoref{fig:rewrite-ref}) where $G_S$ (see the figure) is
	not an arbitrary hypernet but a stable hypernet.
  \item \emph{Structural} pre-templates are as shown in \autoref{fig:structural}, using the following notion of \emph{box-permutation pair}.
 \end{itemize}
\end{defi}
\begin{defi}[Box-permutation pair]
 \label{def:box-permutation-pair}
 For any $n,k,h \in \N$,
 let $\rho$ and $\rho'$ be bijections on sets $\{ 1,\ldots,n+k+h \}$
 and $\{ 1,\ldots,k+h \}$, respectively.
 These bijections form a \emph{box-permutation} pair $(\rho,\rho')$ if,
 for each $i \in \{ 1,\ldots,n+k+h \}$, the following holds:
 \begin{enumerate}
  \item $\rho(i) = i$ if $1 \leq i \leq n$,
  \item $\rho(i) = \rho'(i-n)$ if $n < i \leq n+k+h$,
  \item $1 \leq \rho'(i-n) \leq k$ if $n < i \leq n+k$,
  \item $k < \rho'(i-n) \leq k+h$ if $n+k < i \leq n+k+h$.
 \end{enumerate}
\end{defi}

The rhs of $\vartriangleleft^{\mathrm{Param}}$ is straightforward. It simply consists of a bunch of
encodings of two function abstractions, namely
`$\lambda f.\, (\lambda w.\, 1) \ (f \ ())$' and `$\lambda w.\, 1$'. The empty
store becomes absent in the graphical representation.

The lhs of $\vartriangleleft^{\mathrm{Param}}$ contains a bunch of encodings of two function abstractions,
i.e.\ `$\lambda f.\, (\lambda w.\, !x) \ (f \ ())$' and `$\lambda w.\, !x$', and
also the graphical representation of the store `$\{ a \mapsto 1 \}$' that
consists of an atom edge and an edge labelled with the value `$1$'.
The variable `$x$' refers to the name `$a$', and therefore, the
encodings of function abstractions are all connected to the atom edge via a
contraction tree.

All structural pre-templates (\autoref{fig:structural})
but $\vartriangleleft^\mathrm{BPerm}$ concern contractions and weakenings. The pre-template $\vartriangleleft^\mathrm{BPerm}$ lets us permute outputs of a box.

\begin{table*}
 \centering\scriptsize
 \renewcommand{\arraystretch}{1.8}
 \newcommand{\oneH}{1.8}
 \newcommand{\twoH}{3.6}
 \newcommand{\robW}{85pt}
 \newcommand{\depW}{108pt}
 \newcommand{\upto}[1]{\cyan{#1}}
 \begin{tabular}{|c||c|c|c|c||c|}
  \hline
  & template
  & \multicolumn{2}{c|}{robustness} && \\ \cline{3-4}
  & (input-safety)
  & of $\vartriangleleft$ & of $\vartriangleleft^{-1}$
  & dependency
  & implication of $H_1 \vartriangleleft H_2$ \\ \hline \hline
  $\vartriangleleft^{\contr \mathrm{Assoc}}$
  & $\mathbb{C}_{\Oprex},=,=$
  & $\mathbb{C}_{\Oprex},=,=,=$
  & $\mathbb{C}_{\Oprex},=,=,=$
  & --- 
  & $H_1 \simeq^{\mathbb{C}_{\Oprex}}_{=_\N} H_2$ \\ \hline
  $\vartriangleleft^{\contr \mathrm{Comm}}$
  & $\mathbb{C}_{\Oprex},=,=$
  & $\mathbb{C}_{\Oprex},=,=,=$
  & $\mathbb{C}_{\Oprex},=,=,=$
  & --- 
  & $H_1 \simeq^{\mathbb{C}_{\Oprex}}_{=_\N} H_2$ \\ \hline
  $\vartriangleleft^{\contr \mathrm{Idem}}$
  & $\square,\square,\square$
  & $\mathbb{C}_{\Oprex},=,=,=$
  & $\mathbb{C}_{\Oprex},=,=,=$
  & --- 
  & $H_1 \simeq^{\mathbb{C}_{\Oprex}}_{=_\N} H_2$ \\ \hline
  $\vartriangleleft^{\contr}$
  & $\mathbb{C}_{\Oprex},\geq,\upto{=}$
  & $\mathbb{C}_{\Oprex},=,=,\upto{=}$
  & $\mathbb{C}_{\Oprex},=,\upto{=},=$
  & $\begin{array}{c}\vartriangleleft^{\contr \mathrm{Assoc}}\\
  \vartriangleleft^{\contr \mathrm{Comm}}\end{array}$
  & $H_1 \preceq^{\mathbb{C}_{\Oprex}}_{\geq_\N} H_2,$ \\
  &&&&& $H_2 \preceq^{\mathbb{C}_{\Oprex}}_{\leq_\N} H_1$ \\ \hline
  $\vartriangleleft^{\mathrm{GC}}$
  & $\square,\square,\square$
  & $\mathbb{C}_{\Oprex},=,=,=$
  & $\mathbb{C}_{\Oprex},=,=,=$
  & --- 
  & $H_1 \simeq^{\mathbb{C}_{\Oprex}}_{=_\N} H_2$ \\ \hline
  $\vartriangleleft^{\mathrm{BPerm}}$
  & $\square,\square,\square$
  & $\mathbb{C}_{\Oprex},=,=,=$
  & $\mathbb{C}_{\Oprex},=,=,=$
  & --- 
  & $H_1 \simeq^{\mathbb{C}_{\Oprex}}_{=_\N} H_2$ \\ \hline
  $\vartriangleleft^{\mathrm{BPullC}}$
  & $\square,\square,\square$
  & $\mathbb{C}_{\Oprex},=,=,\upto{=}$
  & $\mathbb{C}_{\Oprex},=,\upto{=},=$
  & $\begin{array}{c}\vartriangleleft^{\contr \mathrm{Assoc}}\\
  \vartriangleleft^{\contr \mathrm{Comm}}\\
  \vartriangleleft^{\contr \mathrm{Idem}}\end{array}$
  & $H_1 \simeq^{\mathbb{C}_{\Oprex}}_{=_\N} H_2$ \\ \hline
  $\vartriangleleft^{\mathrm{BPullW}}$
  & $\square,\square,\square$
  & $\mathbb{C}_{\Oprex},=,=,\upto{=}$
  & $\mathbb{C}_{\Oprex},=,\upto{=},=$
  & $\vartriangleleft^{\contr \mathrm{Idem}}$
  & $H_1 \simeq^{\mathbb{C}_{\Oprex}}_{=_\N} H_2$ \\ \hline \hline
  $\vartriangleleft^{\lrapp}$
  & $\mathbb{C}_{\Oprex},\geq,=$
  & $\mathbb{C}_{\Oprex\cbf},=,=,=$
  & $\mathbb{C}_{\Oprex\cbf},=,=,=$
  & \multirow{2}{*}{---} 
  & $H_1 \preceq^{\mathbb{C}_{\Oprex\cbf}}_{\geq_\N} H_2,$
  \\ \cline{2-2}
  & $\mathbb{C}_{\Oprex\cbf},\geq,=$ & & &
  & $H_2 \preceq^{\mathbb{C}_{\Oprex\cbf}}_{\leq_\N} H_1$
  \\ \hline
  $\vartriangleleft^{\keyw{ref}}$
  & $\mathbb{C}_{\Oprex},\geq,=$
  & $\mathbb{C}_{\Oprex\cbf},=,=,=$
  & $\mathbb{C}_{\Oprex\cbf},=,=,=$
  & \multirow{2}{*}{---} 
  & $H_1 \preceq^{\mathbb{C}_{\Oprex\cbf}}_{\geq_\N} H_2,$
  \\ \cline{2-2}
  & $\mathbb{C}_{\Oprex\cbf},\geq,=$ & & &
  & $H_2 \preceq^{\mathbb{C}_{\Oprex\cbf}}_{\leq_\N} H_1$
  \\ \hline \hline
  $\vartriangleleft^{\mathrm{Param}}$
  & $\mathbb{C}_{\Oprex},=,=$
  & $\mathbb{C}_{\Oprex},\geq,\upto{\geq},=$
  & $\mathbb{C}_{\Oprex},\leq,=,\upto{\leq}$
  & $\begin{array}{c}\vartriangleleft^{\contr \mathrm{Assoc}} \\
  \vartriangleleft^{\contr \mathrm{Idem}}\end{array}$
  & $H_1 \preceq^{\mathbb{C}_{\Oprex}}_{\geq_\N} H_2$ \\ \cline{3-3}
  && $\mathbb{C}_{\Oprex},\geq,\geq,\geq$
  && $\begin{array}{c}\vartriangleleft^{\contr}\\
  \vartriangleleft^{\mathrm{GC}}\end{array}$
  & $H_2 \preceq^{\mathbb{C}_{\Oprex}}_{\leq_\N} H_1$
  \\ \hline
 \end{tabular}
 \caption{Templates, with their robustness and implied contextual
 refinements/equivalences ($\square$ denotes anything)}
 \label{tab:TemplateAnalysis}
\end{table*}

Thanks to \autoref{lem:determ-refocus}, we can apply the sufficiency-of-robustness theorem (\autoref{thm:MetaThm}) and obtain contextual equivalences (on hypernets) as follows.
\begin{lem}
 \label{lem:pre-templ-ctxt-refi}
 \noindent
 \begin{enumerate}
  \item The micro-beta pre-template $\vartriangleleft^{\lrapp}$ implies contextual equivalence $\simeq^{\mathbb{C}_{\Oprex\cbf}}_{\N \times \N}$.
  \item The reference-creation pre-template $\vartriangleleft^{\keyw{ref}}$ implies contextual equivalence $\simeq^{\mathbb{C}_{\Oprex\cbf}}_{\N \times \N}$.
  \item Each structural pre-template implies contextual equivalence $\simeq^{\mathbb{C}_{\Oprex}}_{\N \times \N}$.
 \end{enumerate}
\end{lem}
\begin{proof}[Proof outline]
\autoref{tab:TemplateAnalysis} summarises how we use
\autoref{thm:MetaThm} on the pre-templates.
For example, $\vartriangleleft^{\contr}$ is a
$(\mathbb{C}_{\Oprex},\geq_\N,=_\N)$-template,
as shown in the ``template'' column, and both itself and
its converse are
$(\mathbb{C}_{\Oprex},=_\N,=_\N,=_\N)$-robust relative to all
rewrite transitions,
as shown in the ``robustness'' columns.
Thanks to monotonicity of robustness with respect to $Q$, we can then use
\autoref{thm:MetaThm}(\ref{item:forward})
with a reasonable triple $(\geq_\N,=_\N,=_\N)$,
and \autoref{thm:MetaThm}(\ref{item:backward})
with a reasonable triple $(\leq_\N,=_\N,=_\N)$.
Consequently,
$H_1 \vartriangleleft^{\contr} H_2$ implies
$H_1 \preceq^{\mathbb{C}_{\Oprex}}_{\geq_\N} H_2$ and
$H_2 \preceq^{\mathbb{C}_{\Oprex}}_{\leq_\N} H_1$,
which is shown in the ``implication of $H_1 \vartriangleleft H_2$''
column.

Output-closure of the pre-templates can be easily checked,
typically by spotting that an input or an output, of type $\star$, is
a source or a target of a contraction, atom or box edge.

Pre-templates that relate hypernets with no input of type
$\star$ are trivially a $(\mathbb{C},Q,Q')$-template for any
$\mathbb{C}$, $Q$ and $Q'$. The table uses
`$\square,\square,\square$' to represent this situation.

Typically,
a reasonable triple for a pre-template can be found by selecting
``bigger'' parameters from those of input-safety and robustness,
thanks to monotonicity of input-safety and robustness with respect to $(Q,Q',Q'')$.
However, the parametricity pre-template
$\vartriangleleft^{\mathrm{Param}}$ requires non-trivial use of the
monotonicity. This is because the parameter $(\geq,\geq,=)$ that makes
the pre-template robust, as the upper row in
the ``robustness'' column shows, is not itself a reasonable triple.
The lower row shows the alternative, bigger, parameter
$(\geq,\geq,\geq)$ to which \autoref{thm:MetaThm} can be applied.

In the table, cyan symbols indicate
where a proof of input-safety or robustness relies on contextual
refinement.
The ``dependency'' column indicates which pre-templates can be used to
prove the necessary contextual refinement, given that these
pre-templates imply contextual refinement as shown elsewhere in the
table.
This reliance specifically happens in finding a quasi-specimen,
using contextual refinements/equivalences
via \autoref{lem:TriggerToRefinement}.
In the case of $\vartriangleleft^{\contr}$, its input-safety and
robustness are proved under the assumption that
$\vartriangleleft^{\contr \mathrm{Assoc}}$ and
$\vartriangleleft^{\contr \mathrm{Comm}}$ imply contextual equivalence
$\simeq^{\mathbb{C}_{\Oprex}}_{=_\N}$.

Detailed proofs of input-safety and robustness are in \autoref{sec:transfer-param} and \autoref{appsec:ExampleProofSketch}.
\end{proof}

\begin{rem}[Necessity of binding-free contexts] \label{def:binding-free-ctxt}
The restriction to binding-free contexts plays a crucial role only in
robustness regarding the operational pre-templates
$\vartriangleleft^{\lrapp}$ and $\vartriangleleft^{\keyw{ref}}$.
In fact, these pre-templates are input-safe with respect to both
$\mathbb{C}_{\Oprex}$ and $\mathbb{C}_{\Oprex\cbf}$.
This gap reflects duplication behaviour on atom edges, which is
only encountered in a proof of robustness.

In fact, robustness of the micro-beta pre-template $\vartriangleleft^{\lrapp}$ is \emph{not} guaranteed in the presence of copy transitions, which apply contraction rules.
Starting from a pair of states given by a specimen
$(\focussed{\mathcal{C}};\vec{H^1};\vec{H^2})$,
it may be the case that some copy transitions are possible
without reaching another (quasi-)specimen.
An example scenario is when the specimen yields the following two
states, where the context $\focussed{\mathcal{C}}$ is indicated by
magenta:
\begin{equation*}
 \textcolor{magenta}{\focussed{\mathcal{C}}}[\vec{H^1}] =
  \includegraphics[align=c,scale=.2]{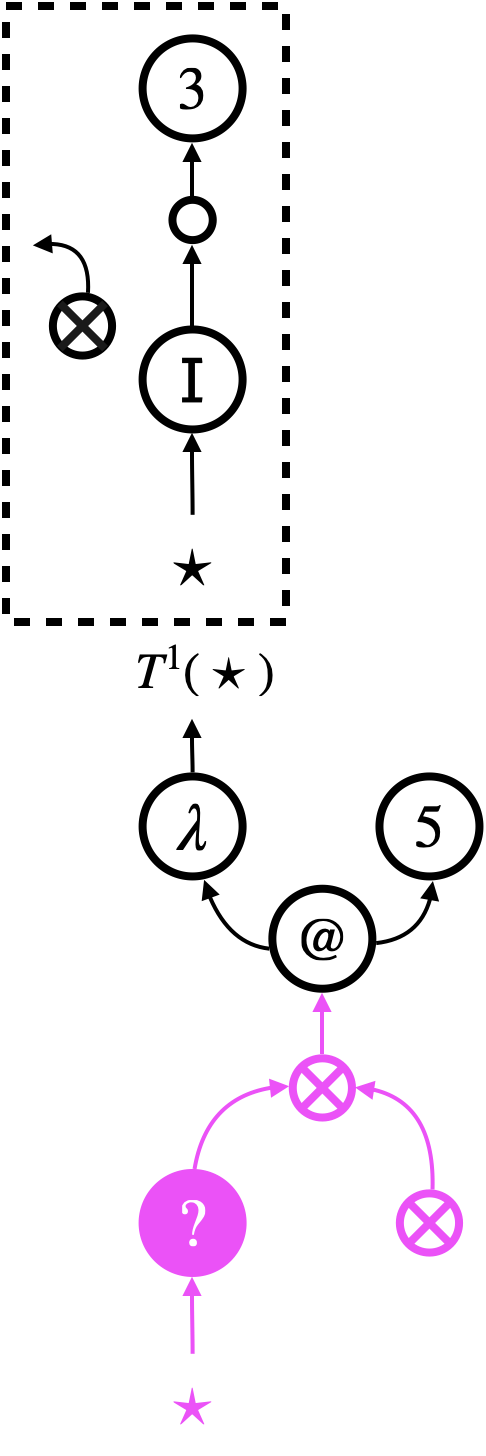}
  \quad, \qquad
  \textcolor{magenta}{\focussed{\mathcal{C}}}[\vec{H^2}] =
  \includegraphics[align=c,scale=.2]{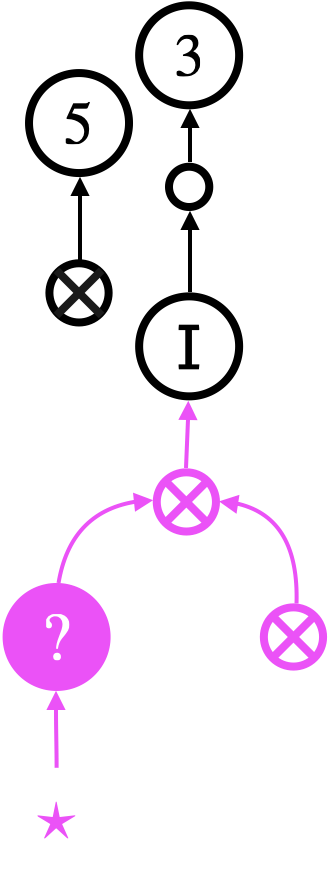}
  \quad.
\end{equation*}
Transitions from the state $\focussed{\mathcal{C}}[\vec{H^1}]$
eventually duplicate the application edge ($@$), the abstraction edge
($\lambda$) and also the entire box connected to the abstraction
edge. In particular, these transitions duplicate the atom edge
contained in the box.
However, transitions from the state
$\focussed{\mathcal{C}}[\vec{H^2}]$ can never duplicate the atom
edge, because the edge is shallow in this state.
This mismatch of duplication prevents the micro-beta pre-template from
being robust relative to a copy transition.

This is why we restrict contexts to be binding-free, when it comes to robustness of operational pre-templates.
If the context $|\focussed{\mathcal{C}}|$ is binding-free, the
situation explained above would never happen.
Application of a contraction rule can involve the hypernets
$\vec{H^1}$ and $\vec{H^2}$ only as a part of box contents that are
duplicated as a whole.
Any specimen of $\vartriangleleft^{\lrapp}$ is therefore turned into another specimen whose context
possibly has more holes, by a single copy transition.
Note that this also explains why we allow the context of a specimen to
have multiple holes.
\bqed
\end{rem}

\subsection{Combining templates} \label{sec:combining-templates}

We now combine the contextual equivalences and prove \autoref{prop:WeakeningLaw} and \autoref{prop:ParametricityLaw}.
We start with combining the structural templates
$\vartriangleleft^{\contr \mathrm{Idem}}$ and
$\vartriangleleft^{\mathrm{BPullW}}$, and prove the Weakening laws.

\begin{proof}[Proof outline of {\autoref{prop:WeakeningLaw}}]
 The proof is by induction on derivations.
 Base cases are for variables, atoms and constants. The proof for these cases
 are trivial, because any distributor $D^\ell_{k,0}$ with no inputs is simply a
 bunch of weakening edges (see \autoref{def:distributors}).

 In inductive cases, we need to identify a single weakening edge with a certain
 (sub-)hypernet, namely:
 (i) a distributor $D^\ell_{1,1}$ whose sole input is connected to a weakening
 edge,
 (ii) a distributor $D^\ell_{1,2}$ whose two inputs are connected to weakening
 edges, and
 (iii) a distributor $D^\ell_{1,1}$ whose sole input is connected to a box edge,
 in which a weakening edge is connected to the corresponding output.
 The first two situations are for unary/binary operations and function
 application. These can be handled with
 the contextual equivalence $\simeq^{\mathbb{C}_{\Oprex}}_{=_\N}$ implied by
 $\vartriangleleft^{\contr \mathrm{Idem}}$.
 The third situation is for function abstraction, and it boils down to the first
 situation, thanks to
 the contextual equivalence $\simeq^{\mathbb{C}_{\Oprex}}_{=_\N}$ implied by
 $\vartriangleleft^{\mathrm{BPullW}}$.
\end{proof}

We can then prove the equivalence (\ref{eq:param-law}) as follows.
\begin{figure}[pt]
 \newcommand{\msc}{.15}
 \begin{gather*}
  \bigl( - \mid - \vdash
  \bigl( \lambda x.\,\lambda f.\, (\lambda w.\, !x) \ (f \ ()) \bigr)
  \ (\keyw{ref}\ 1)
  \bigr)^\ddag = \\
  \includegraphics[align=c,scale=\msc]{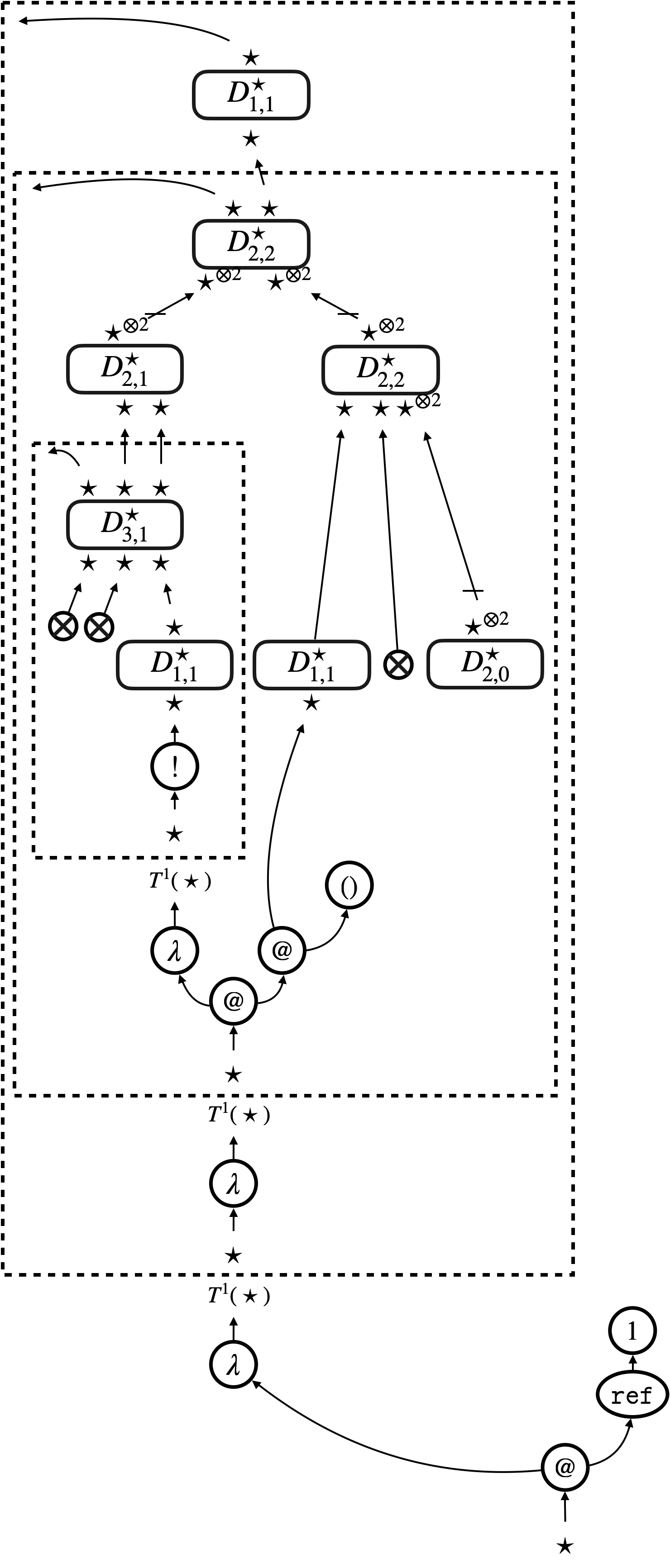}
  \begin{array}{c}
   \simeq^{\mathbb{C}_{\Oprex}} \\
   \left(\begin{array}{c}
    \vartriangleleft^{\otimes\mathrm{Assoc}} \\
	  \vartriangleleft^{\otimes\mathrm{Comm}} \\
	  \vartriangleleft^{\otimes\mathrm{Idem}} \\
	  \vartriangleleft^{\mathrm{BPerm}} \\
	  \vartriangleleft^{\mathrm{BPullC}} \\
	  \vartriangleleft^{\mathrm{BPullW}}
	   \end{array}\right)
  \end{array}
  \includegraphics[align=c,scale=\msc]{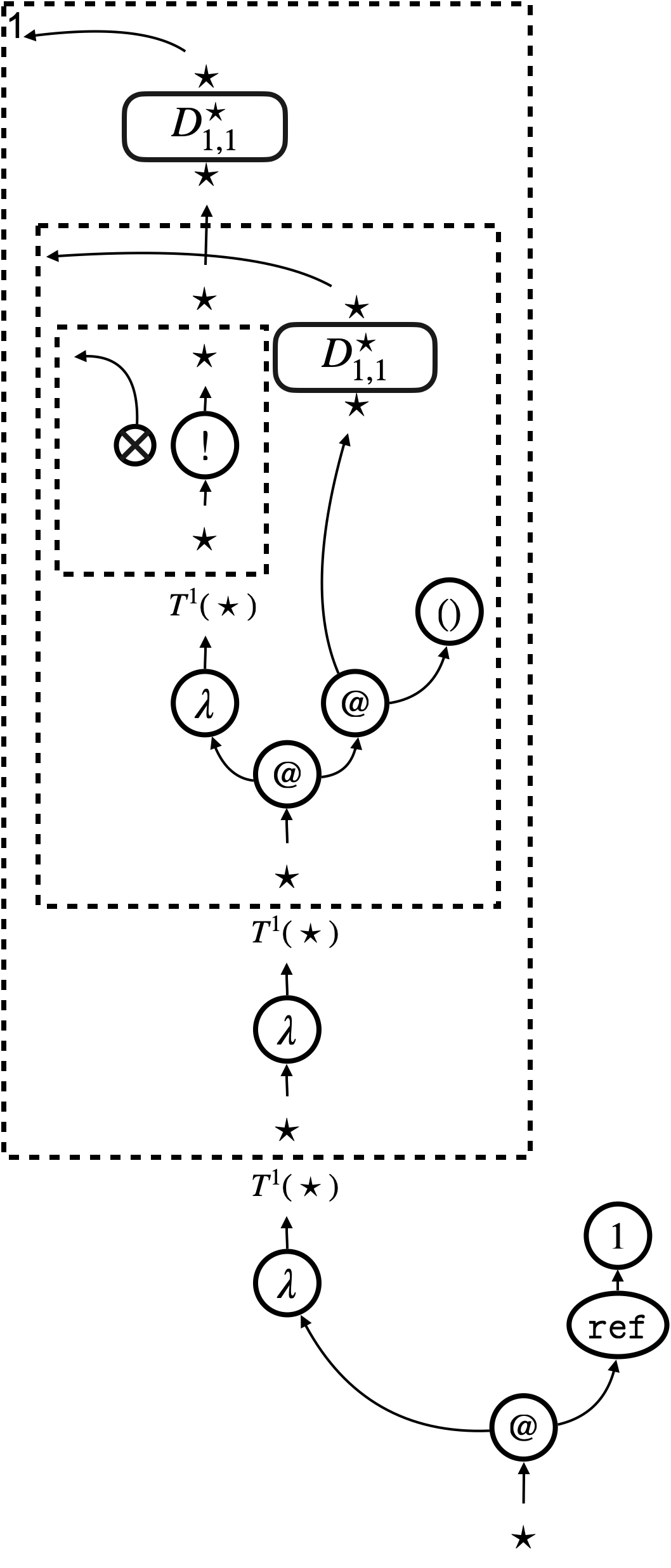}
  \begin{array}{c}
   \simeq^{\mathbb{C}_{\Oprex\cbf}} \\
   \left(\begin{array}{c}
	  \vartriangleleft^{\keyw{ref}}
	   \end{array}\right)
  \end{array}
  \includegraphics[align=c,scale=\msc]{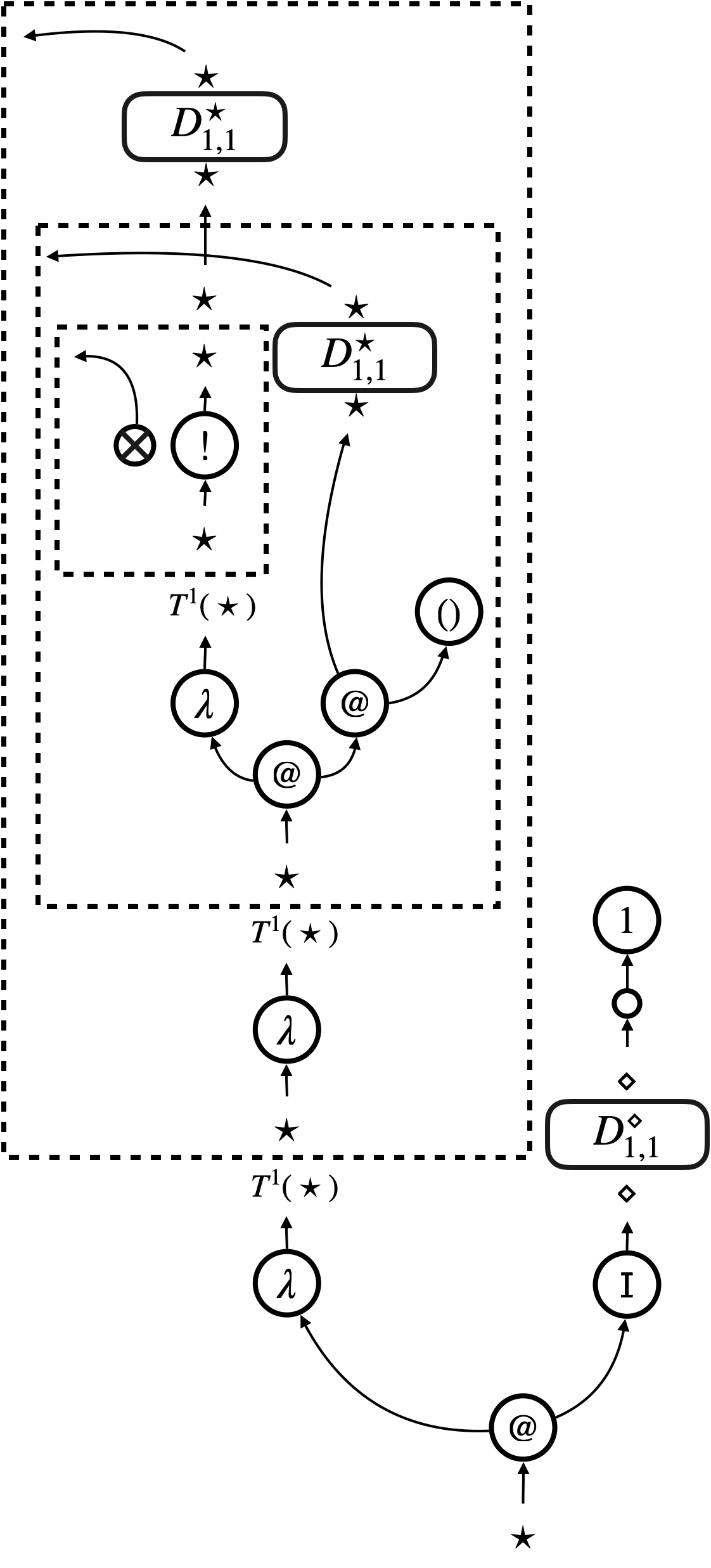} \\
  \begin{array}{c}
   \simeq^{\mathbb{C}_{\Oprex\cbf}} \\
   \left(\begin{array}{c}
	  \vartriangleleft^{\lrapp}
	   \end{array}\right)
  \end{array}
  \includegraphics[align=c,scale=\msc]{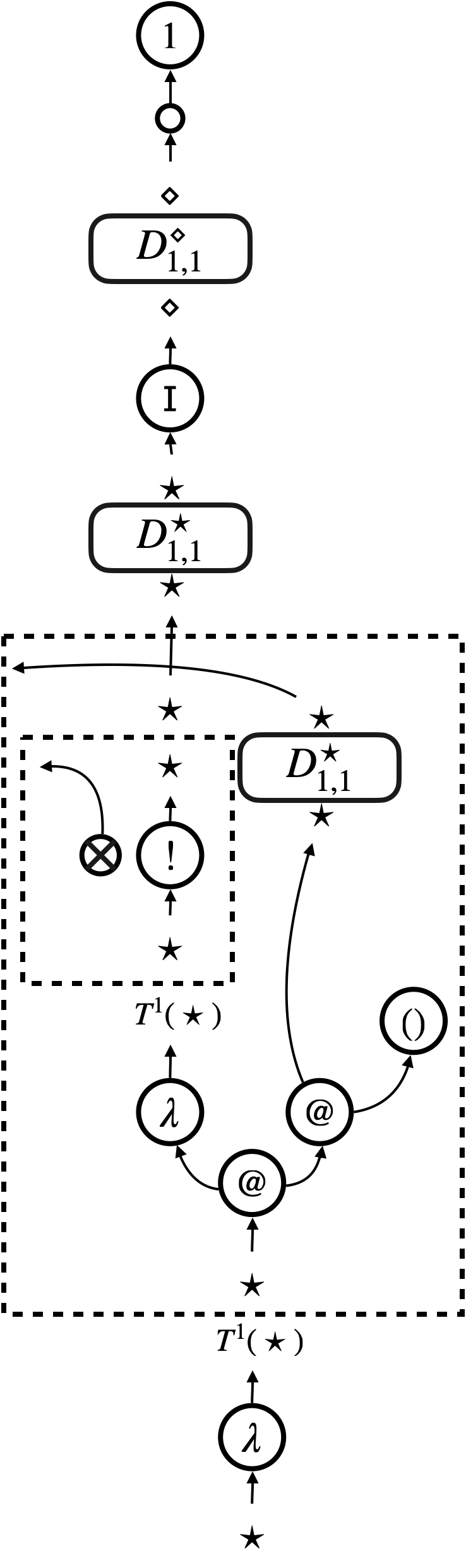}
  \begin{array}{c}
   \simeq^{\mathbb{C}_{\Oprex}} \\
   \left(\begin{array}{c}
	  \vartriangleleft^{\mathrm{Param}}
	   \end{array}\right)
  \end{array}
   \includegraphics[align=c,scale=\msc]{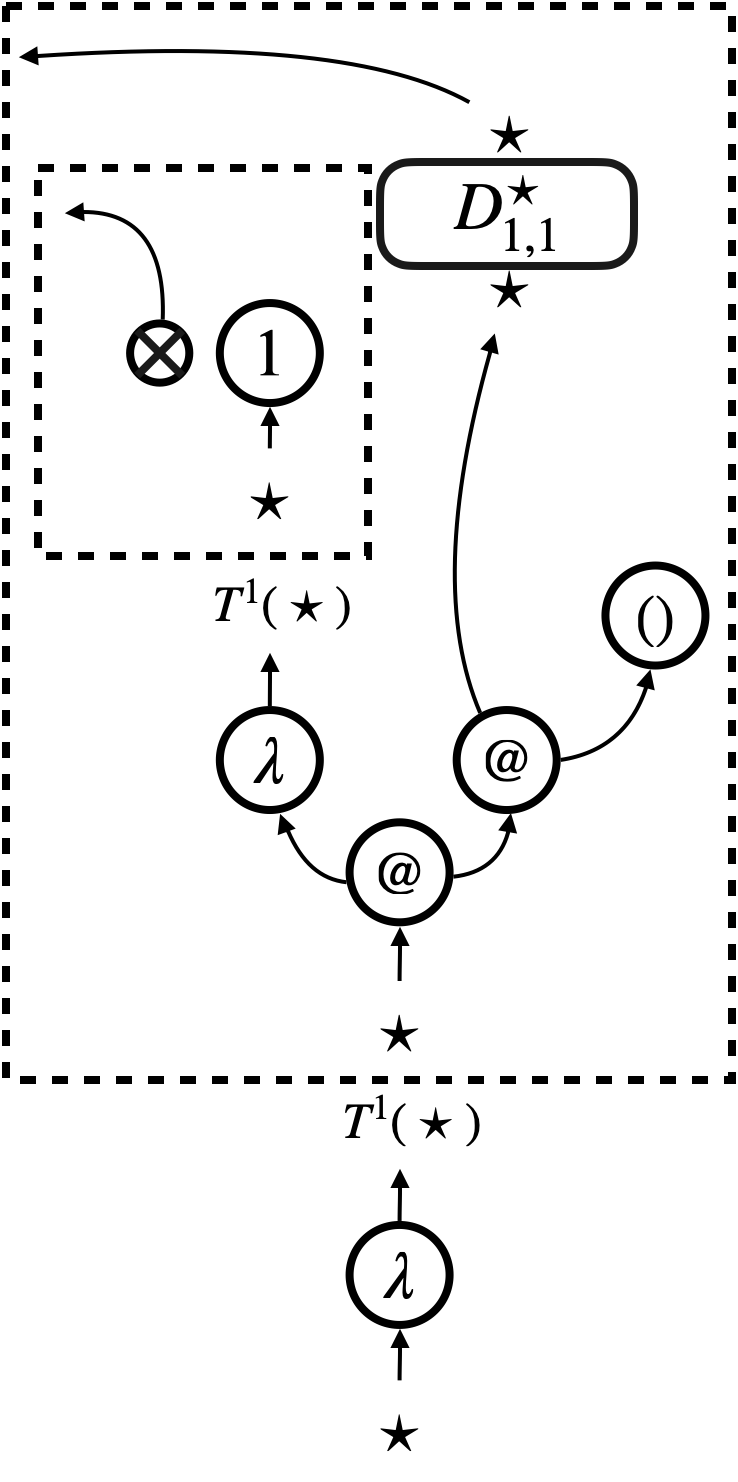}
  \begin{array}{c}
   \simeq^{\mathbb{C}_{\Oprex}} \\
   \left(\begin{array}{c}
    \vartriangleleft^{\otimes\mathrm{Assoc}} \\
	  \vartriangleleft^{\otimes\mathrm{Comm}} \\
	  \vartriangleleft^{\otimes\mathrm{Idem}} \\
	  \vartriangleleft^{\mathrm{BPullW}}
	   \end{array}\right)
  \end{array}
  \includegraphics[align=c,scale=\msc]{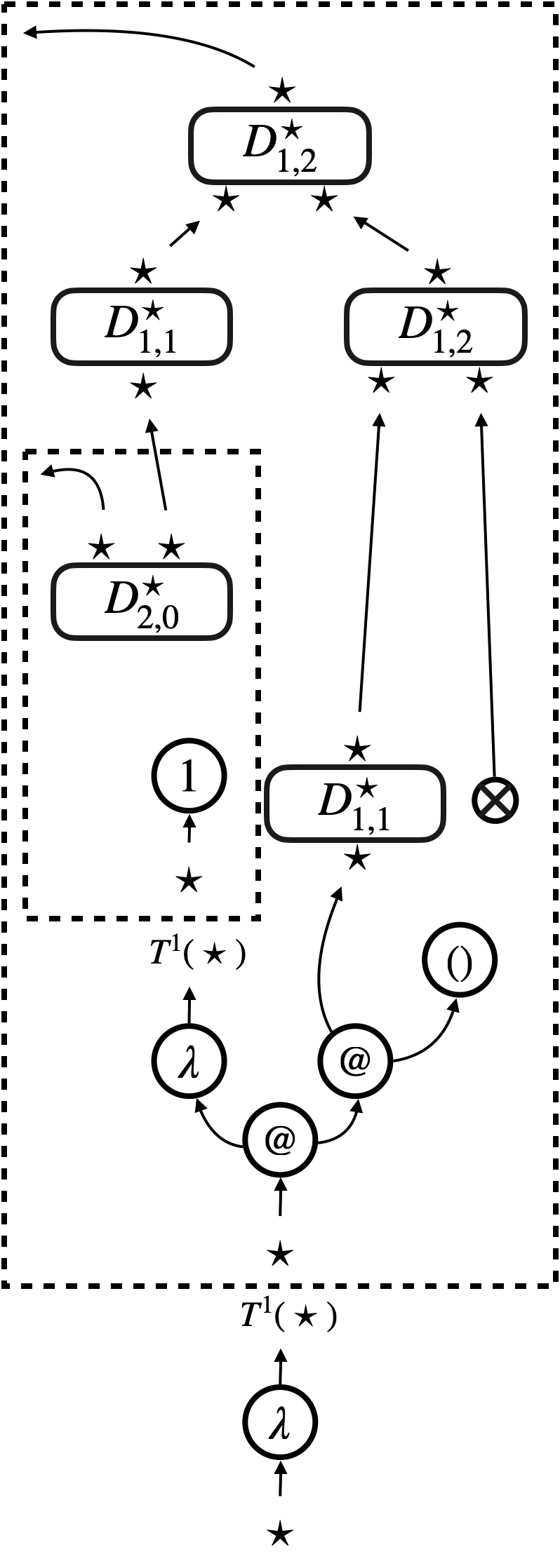} \\
  = ( - \mid - \vdash \lambda f.\, (\lambda w.\, 1) \ (f \ ()))^\ddag
 \end{gather*}
 \caption{A proof illustration of \autoref{prop:ParametricityLaw}, for the empty environment}
 \label{fig:NE4inUse}
\end{figure}

\begin{proof}[Proof of \autoref{prop:ParametricityLaw}]
 Let $P_\mathsf{L}$ and $P_\mathsf{R}$ be the de-sugared version of the left-hand
 side and the right-hand side of (\ref{eq:param-law}), i.e.:
  \begin{align*}
  P_\mathsf{L} &\equiv
  \bigl( \lambda x.\,\lambda f.\, (\lambda w.\, !x) \ (f \ ()) \bigr)
  \ (\keyw{ref}\ 1) \\
  P_\mathsf{R} &\equiv
  \lambda f.\, (\lambda w.\, 1) \ (f \ ()).
 \end{align*}
 The equivalence (\ref{eq:param-law}) can be obtained as a chain of contextual equivalences
 whose outline is as follows.
 \begin{equation}
  \label{eq:ParamLawGraphicalChain}
   \includegraphics[align=c,scale=.25]{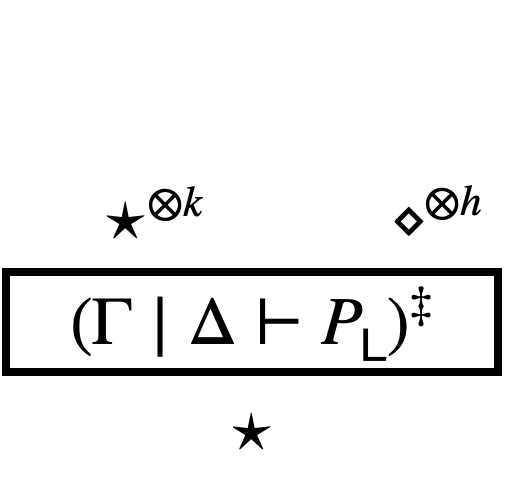}
   \ \simeq^{\mathbb{C}_{\Oprex}}_{=_\N}\ 
   \includegraphics[align=c,scale=.25]{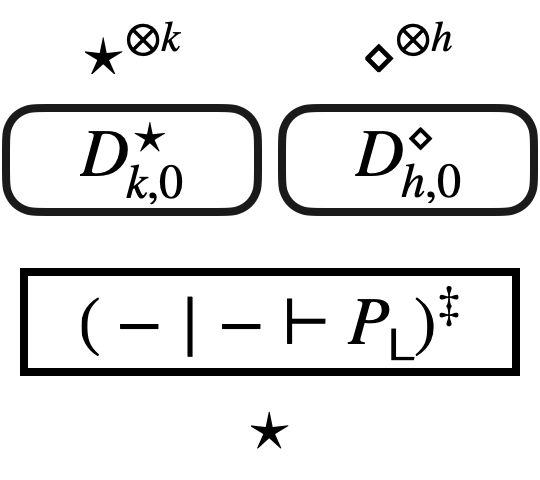}
   \ \simeq^{\mathbb{C}_{\Oprex\cbf}}_{\N \times \N}\ 
   \includegraphics[align=c,scale=.25]{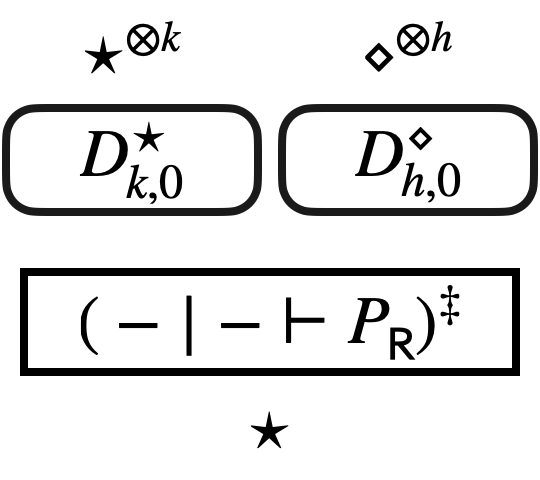}		 
   \ \simeq^{\mathbb{C}_{\Oprex}}_{=_\N}\ 
   \includegraphics[align=c,scale=.25]{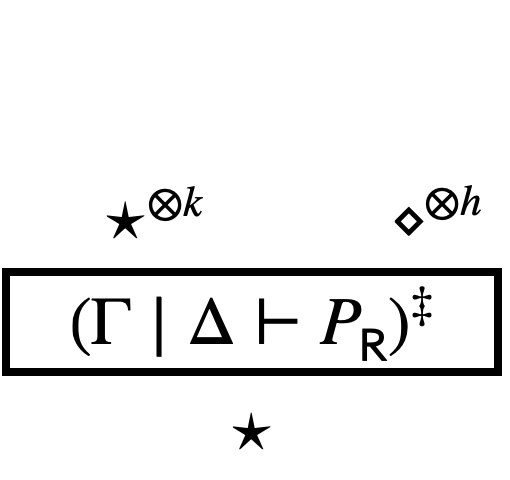}
 \end{equation}
 The leftmost and rightmost contextual equivalences are consequences of the
 Weakening
 laws (\autoref{prop:WeakeningLaw}), because the terms $P_\mathsf{L}$ and
 $P_\mathsf{R}$ have no free variables nor free atoms.
 The middle contextual equivalence follows from the special case of (\ref{eq:param-law})
 where the environment is empty, i.e.\ 
 $- \mid - \vDash P_\mathsf{L} \simeq^\ddag P_\mathsf{R}$.
 This contextual equivalence is namely derived from another chain of contextual
 equivalences that is shown in \autoref{fig:NE4inUse}, via the binding-free
 context that consists of a hole of type $\star \To \epsilon$ and weakening
 edges (i.e.\ $D^\star_{k,0}$ and $D^\diamond_{h,0}$).
 In \autoref{fig:NE4inUse}, each
 contextual equivalence is accompanied by relevant templates, and
 preorders on natural numbers are omitted.
 %
\end{proof}

 \newcommand{\dep}{$\circ$}
 \newcommand{\hid}{$\bullet$}
 \newcommand{\hidd}{}
\begin{table*}
 \centering\scriptsize
 \begin{tabular}{|c||c|c|c|c|c|c|c|c||c|c||c|}
  \hline
  & ${\contr \mathrm{Assoc}}$ & ${\contr \mathrm{Comm}}$
  & ${\contr \mathrm{Idem}}$
  & ${\contr}$ & ${\mathrm{GC}}$
  & ${\mathrm{BPerm}}$
  & ${\mathrm{BPullC}}$ & ${\mathrm{BPullW}}$
  & ${\lrapp}$ & ${\keyw{ref}}$
  & ${\mathrm{Param}}$ \\ \hline\hline
  \autoref{prop:WeakeningLaw}
  &&&\dep \hidd&&&&&\dep&&& \\ \hline
  \autoref{prop:ParametricityLaw}
  &\dep \hidd&\dep \hidd&\dep \hidd, (W)&\hid&\hid
  &\dep&\dep&\dep, (W)&\dep&\dep&\dep \\ \hline
 \end{tabular}
 \caption{Dependency of example equivalences on templates}
 \label{tab:ObsEquivAnalysis}
\end{table*}

This concludes our development leading to the proof our example equivalences.

\autoref{tab:ObsEquivAnalysis} summarises dependency of the example equivalences on templates, which can be observed in the above proofs.
The symbol `\dep' indicates \emph{direct} dependency on templates,
in the sense that an equivalence can be proved by combining contextual
equivalences implied by these templates. For example, the Weakening laws are
obtained by combining two templates $\vartriangleleft^{\contr \mathrm{Idem}}$
and $\vartriangleleft^{\mathrm{BPullW}}$.
Because the proof of \autoref{prop:ParametricityLaw} uses the Weakening law, it also depends on these two templates, which is indicated by `(W)' in the table.

Note that two templates
$\vartriangleleft^{\contr}$ and $\vartriangleleft^{\mathrm{GC}}$ do not directly
appear in the proof of \autoref{prop:ParametricityLaw}, in particular in the chain shown
in \autoref{fig:NE4inUse}. They are however necessary for robustness of the
parametricity template (see \autoref{tab:TemplateAnalysis}). \autoref{prop:ParametricityLaw} depends on the two templates only \emph{indirectly}, which is indicated by
the symbol `\hid' in \autoref{tab:ObsEquivAnalysis}.

\subsection{Designing pre-templates}
\label{sec:design-pre-templates}

\begin{figure}[tp]
 Chosen terms and contexts: \hspace*{\fill}
 \begin{align*}
  P_\mathsf{L} &\equiv
  \bigl( \lambda x.\,\lambda f.\, (\lambda w.\, !x) \ (f \ ()) \bigr)
  \ (\keyw{ref}\ 1)
  \tag{de-sugared version of the lhs of the law} \\
  P_\mathsf{R} &\equiv
  \lambda f.\, (\lambda w.\, 1) \ (f \ ())
  \tag{de-sugared version of the rhs of the law} \\
  C_\mathsf{1} &\equiv
  \bigl(
  \lambda g.\,(g \ (\lambda y.\, y)) + (g \ (\lambda z.\,0 )) \bigr)
  \ [\,]
 \end{align*}

 Informal reduction of a program $C_\mathsf{1}[P_\mathsf{L}]$ with empty
 store $\emptyset$:
 \hspace*{\fill}
 \begin{equation*}
  \begin{array}{l|lcc} \hline
    \texttt{L0} & \bigl(
    \lambda g.\,(g \ (\lambda y.\, y)) + (g \ (\lambda z.\,0)) \bigr)
    \ \Bigl(
    \bigl( \lambda x.\,\lambda f.\, (\lambda w.\, !x) \ (f \ ()) \bigr)
    \ (\keyw{ref}\ 1) \Bigr) \hfill , & \emptyset & \dashrightarrow \\
   \texttt{L0'} & \bigl(
    \lambda g.\,(g \ (\lambda y.\, y)) + (g \ (\lambda z.\,0)) \bigr)
    \ \Bigl(
    \bigl( \lambda x.\,\lambda f.\, (\lambda w.\, !x) \ (f \ ()) \bigr)
    \ a \Bigr) \hfill , & \{a \mapsto 1\} & \dashrightarrow \\
   \texttt{L1} & \bigl(
    \lambda g.\,(g \ (\lambda y.\, y)) + (g \ (\lambda z.\,0)) \bigr)
    \ \Mag{\bigl(\lambda f.\, (\lambda w.\, !a) \ (f \ ()) \bigr)}
    \hfill , & \Mag{\{a \mapsto 1\}} & \dashrightarrow \\ \hline
   \texttt{L2} &
    \bigl( \Mag{\bigl(\lambda f.\, (\lambda w.\, !a) \ (f \ ()) \bigr)}
    \ (\lambda y.\, y) \bigr) +
    \bigl( \Mag{\bigl(\lambda f.\, (\lambda w.\, !a) \ (f \ ()) \bigr)}
    \ (\lambda z.\,0) \bigr)
    \hfill , & \Mag{\{a \mapsto 1\}} & \dashrightarrow \\ \hline
    \texttt{L3} &
    \bigl( \Mag{(\lambda w.\, !a)} \ ((\lambda y.\, y) \ ()) \bigr) +
    \bigl( \Mag{\bigl(\lambda f.\, (\lambda w.\, !a) \ (f \ ()) \bigr)}
    \ (\lambda z.\,0) \bigr)
    \hfill , & \Mag{\{a \mapsto 1\}} & \dashrightarrow \\ \hline
   \texttt{L4} &
    \bigl( \Mag{(\lambda w.\, !a)} \ () \bigr) +
    \bigl( \Mag{\bigl(\lambda f.\, (\lambda w.\, !a) \ (f \ ()) \bigr)}
    \ (\lambda z.\,0) \bigr)
    \hfill , & \Mag{\{a \mapsto 1\}} & \dashrightarrow \\ \hline
   \texttt{L4'} &
    {!a} +
    \bigl( \bigl(\lambda f.\, (\lambda w.\, !a) \ (f \ ()) \bigr)
    \ (\lambda z.\,0) \bigr)
    \hfill , & \{a \mapsto 1\} & \dashrightarrow \\
   \texttt{L5} &
    1 +
    \bigl( \Mag{\bigl(\lambda f.\, (\lambda w.\, !a) \ (f \ ()) \bigr)}
    \ (\lambda z.\,0) \bigr)
    \hfill , & \Mag{\{a \mapsto 1\}} & \dashrightarrow \\ \hline
    \texttt{L6} &
    1 + \bigl( \Mag{(\lambda w.\, !a)} \ ((\lambda z.\, 0) \ ()) \bigr)
    \hfill , & \Mag{\{a \mapsto 1\}} & \dashrightarrow \\ \hline
   \texttt{L7} &
    1 + \bigl( \Mag{(\lambda w.\, !a)} \ 0 \bigr)
    \hfill , & \Mag{\{a \mapsto 1\}} & \dashrightarrow \\ \hline
   \texttt{L7'} &
    1 + {!a}
    \hfill , & \{a \mapsto 1\} & \dashrightarrow \\
   \texttt{L8} &
    1 + 1
    \hfill , & \Mag{\{a \mapsto 1\}} & \dashrightarrow \\ \hline
    \texttt{L9} &
    2
    \hfill , & \Mag{\{a \mapsto 1\}} & \\ \hline
  \end{array}
 \end{equation*}

 Informal reduction of a program $C_\mathsf{1}[P_\mathsf{R}]$ with empty
 store $\emptyset$:
 \hspace*{\fill}
 \begin{equation*}
  \begin{array}{l|lcc} \hline
   \texttt{R1} & \bigl(
    \lambda g.\,(g \ (\lambda y.\, y)) + (g \ (\lambda z.\,0)) \bigr)
    \ \Mag{\bigl( \lambda f.\, (\lambda w.\, 1) \ (f \ ()) \bigr)}
    \hfill , & \Mag{\emptyset} & \dashrightarrow \\ \hline
   \texttt{R2} &
    \bigl( \Mag{\bigl(\lambda f.\, (\lambda w.\, 1) \ (f \ ()) \bigr)}
    \ (\lambda y.\, y) \bigr) +
    \bigl( \Mag{\bigl(\lambda f.\, (\lambda w.\, 1) \ (f \ ()) \bigr)}
    \ (\lambda z.\,0) \bigr)
    \hfill , & \Mag{\emptyset} & \dashrightarrow \\ \hline
    \texttt{R3} &
    \bigl( \Mag{(\lambda w.\, 1)} \ ((\lambda y.\, y) \ ()) \bigr) +
    \bigl( \Mag{\bigl(\lambda f.\, (\lambda w.\, 1) \ (f \ ()) \bigr)}
    \ (\lambda z.\,0) \bigr)
    \hfill , & \Mag{\emptyset} & \dashrightarrow \\ \hline
   \texttt{R4} &
    \bigl( \Mag{(\lambda w.\, 1)} \ () \bigr) +
    \bigl( \Mag{\bigl(\lambda f.\, (\lambda w.\, 1) \ (f \ ()) \bigr)}
    \ (\lambda z.\,0) \bigr)
    \hfill , & \Mag{\emptyset} & \dashrightarrow \\ \hline
   \texttt{R5} &
    1 +
    \bigl( \Mag{\bigl(\lambda f.\, (\lambda w.\, 1) \ (f \ ()) \bigr)}
    \ (\lambda z.\,0) \bigr)
    \hfill , & \Mag{\emptyset} & \dashrightarrow \\ \hline
    \texttt{R6} &
    1 + \bigl( \Mag{(\lambda w.\, 1)} \ ((\lambda z.\, 0) \ ()) \bigr)
    \hfill , & \Mag{\emptyset} & \dashrightarrow \\ \hline
   \texttt{R7} &
    1 + \bigl( \Mag{(\lambda w.\, 1)} \ 0 \bigr)
    \hfill , & \Mag{\emptyset} & \dashrightarrow \\ \hline
   \texttt{R8} &
    1 + 1
    \hfill , & \Mag{\emptyset} & \dashrightarrow \\ \hline
    \texttt{R9} &
    2
    \hfill , & \Mag{\emptyset} \\ \hline
  \end{array}
 \end{equation*}

 \caption{The equivalence (\ref{eq:param-law}): an example scenario}
 \label{fig:ParamLawExampleScenario}
\end{figure}

We conclude this section with informal description of how some of the pre-templates are designed, using conventional reduction semantics.
This description will reveal how graphical representation is better suited for local reasoning compared to textual representation (see e.g.\ (\ref{eq:TextualParamTemplate})).
 
Instead of turning (\ref{eq:param-law}) directly into a single pre-template, we
decompose the equivalence into several, more primitive, pre-templates. Once we apply the
sufficiency-of-robustness theorem to the pre-templates, the obtained contextual
equivalences can be composed to yield (\ref{eq:param-law}), as we saw in \autoref{sec:combining-templates}.
This approach increases the possibility of reusing parts of a proof of one law
in a proof of another law. For example, most of the pre-templates that are
used to prove (\ref{eq:param-law}) can be reused for proving the call-by-value
Beta law (see~\cite[Section~4.5]{Muroya20PhD} for details).

The idea of the pre-templates is that they can describe all the possible
differences that may arise during execution of any two programs whose
differences is given precisely by (\ref{eq:param-law}).
As an illustration of the design process, we compare informal reduction sequences of two
example programs, as summarised in \autoref{fig:ParamLawExampleScenario}.

We choose the context $C_\mathsf{1}$, which expects a function
in the hole and uses it twice. It generates two programs, by receiving the terms
$P_\mathsf{L}$ and $P_\mathsf{R}$ that are a de-sugared version of the two sides
of (\ref{eq:param-law}).

Each informal reduction step $\dashrightarrow$ updates a term and its associated
store. The step is either the standard call-by-value beta-reduction, addition of
numbers (for `$+$'), reference creation (for `$\keyw{ref}$'), or dereferencing
(for `$!$'). Reference creation is the only step that modifies store. It
replaces a (sub-)term of
the form `$\keyw{ref}\ n$' with a fresh name, say `$a$', and extends the store
with `$a \mapsto n$'. The empty store is denoted by `$\emptyset$'.
Each reduction step in \autoref{fig:ParamLawExampleScenario} is given a tag,
such as \texttt{L0}, \texttt{L0'}, \texttt{R1}. We use the tags for referring to
a corresponding term and store, and also to the reduction step, if any, from the
term and store.

Before explaining the colouring scheme of
\autoref{fig:ParamLawExampleScenario}, let us observe the (in-)corre-spondence
between the reduction sequences of $C_\mathsf{1}[P_\mathsf{L}]$ and
$C_\mathsf{1}[P_\mathsf{R}]$.
The reduction sequence of $C_\mathsf{1}[P_\mathsf{R}]$
consists of seven beta-reduction steps
(\texttt{R1}--\texttt{R7}) and one addition step (\texttt{R8}).
The other reduction sequence, of $C_\mathsf{1}[P_\mathsf{L}]$, in fact contains
steps that correspond to these seven beta-reduction steps and the addition step,
as suggested by the tags (\texttt{L1}, \texttt{L2}, \texttt{L3}, \texttt{L4},
\texttt{L5}, \texttt{L6}, \texttt{L7}, \texttt{L8}).
The sequence has four
additional steps, namely: one reference-creation step (\texttt{L0}),
one application step (\texttt{L0'}) of a function to the created name `$a$',
and two dereferencing steps (\texttt{L4'}, \texttt{L7'}).
The two sequences result in the same term, but in different store (\texttt{L9},
\texttt{R9}).

The colouring scheme is as follows.
In the eight matching steps and the final result, differences between two sides
(e.g.\ \texttt{L1} and \texttt{R1}) are highlighted in magenta. Note that it is
not the minimum difference that are highlighted.
Highlighted parts are chosen in such a way that they capture the smallest
difference ``on the surface''. Sub-terms \emph{on the surface} are
those that are outside of any lambda-abstraction, which can be graphically
represented by sub-hypernets that are outside of any boxes.
For example, after the name creation (i.e.\ in \texttt{L1}, \texttt{R1}), the
function abstractions (`$\lambda f.\, (\lambda w.\, !a) \ (f \ ())$',
`$\lambda f.\, (\lambda w.\, 1) \ (f \ ())$') are highlighted, instead of the
minimum difference (`$!a$', `$1$').

In summary, the reduction sequences of $C_\mathsf{1}[P_\mathsf{L}]$ and
$C_\mathsf{1}[P_\mathsf{R}]$ have eight steps corresponding with each other
(\texttt{L1}--\texttt{L8}, \texttt{R1}--\texttt{R8}) where the differences of
their results can be described by means of store and sub-terms on the surface.
The extracted differences are namely function abstractions
(`$\lambda f.\, (\lambda w.\, !a) \ (f \ ())$' and `$\lambda w.\, !a$',
against
`$\lambda f.\, (\lambda w.\, 1) \ (f \ ())$' and `$\lambda w.\, 1$') and store
(`$\{ a \mapsto 1 \}$' against `$\emptyset$').
By simply collecting these differences, we can obtain our first pre-template:
\emph{parametricity} pre-template $\vartriangleleft^{\mathrm{Param}}$. It is the
essential, key, pre-template for (\ref{eq:param-law}). Textually, and
intuitively, it looks like the following.
\begin{multline}
 \label{eq:TextualParamTemplate}
 {\Mag{\lambda f.\, (\lambda w.\, !a) \ (f \ ())}, \enspace
 \ldots, \enspace \Mag{\lambda f.\, (\lambda w.\, !a) \ (f \ ())}, \enspace
 \Mag{\lambda w.\, !a}, \enspace
 \ldots, \enspace \Mag{\lambda w.\, !a}, \enspace
 \Mag{\{a \mapsto 1\}}
 } \\
 \quad \vartriangleleft^{\mathrm{Param}} \quad
 {\Mag{\lambda f.\, (\lambda w.\, 1) \ (f \ ())}, \enspace
  \ldots, \enspace \Mag{\lambda f.\, (\lambda w.\, 1) \ (f \ ())}, \enspace
  \Mag{\lambda w.\, 1}, \enspace \ldots, \enspace \Mag{\lambda w.\, 1}, \enspace
  \Mag{\emptyset}
  }
\end{multline}

Graphically, the parametricity pre-template is simply a relation between
hypernets (\autoref{fig:param}). In
particular, the lhs of the pre-template can be represented as a
single hypernet, thanks to the graphical representation where mentions of a name
`$a$' become connection. Hypernets of the function
abstractions are all connected to the hypernet of the store
`$\{ a \mapsto 1 \}$'.
This graphical representation naturally entails the crucial piece of
information, which is
invisible in the informal textual
representation~(\ref{eq:TextualParamTemplate}), that the function abstractions
are the all and only parts of a program that have access to the name `$a$'.

The choice of smallest differences on the surface, instead of absolute minimum
difference, also plays a key role here. Should we choose minimum differences
that may be inside lambda-abstraction, their hypernet representations cannot be
directly combined to yield a single valid hypernet. This is due to the box
structure of hypernets, which are used to represent function abstraction. In
other words, to connect
the hypernet of the store `$\{ a \mapsto 1 \}$' to the hypernet of the sub-term
`$!a$' that appears inside the function abstraction `$\lambda w.\, !a$', we must
first make a connection to the hypernet of the whole function abstraction, which
contains the hypernet of `$!a$' inside a box.

Recall that the parametricity pre-template $\vartriangleleft^{\mathrm{Param}}$
collects
differences in matching steps (eight steps each, i.e.\ \texttt{L1}--\texttt{L8},
\texttt{R1}--\texttt{R8}). From the first two unmatched steps
(\texttt{L0}, \texttt{L0'}), we extract the two operational pre-templates
$\vartriangleleft^{\keyw{ref}}$ and $\vartriangleleft^{\lrapp}$. These are
both
induced by the reductions, namely: $\vartriangleleft^{\keyw{ref}}$ by reference
creation (\texttt{L0}) and $\vartriangleleft^{\lrapp}$ by
beta-reduction (\texttt{L0'}). For instance, these pre-templates relate the
sub-terms that are yielded by the two reductions (\texttt{L0}, \texttt{L0'}),
informally as follows.
\begin{align*}
 {\keyw{ref}\ 1,\enspace \emptyset}
 & \quad \vartriangleleft^{\keyw{ref}} \quad {a,\enspace \{ a \mapsto 1 \}} \\
 {\bigl( \lambda x.\,\lambda f.\, (\lambda w.\, !x) \ (f \ ()) \bigr) \ a}
 & \quad \vartriangleleft^{\lrapp} \quad
 \lambda f.\, (\lambda w.\, !a) \ (f \ ())
\end{align*}

The three pre-templates $\vartriangleleft^{\mathrm{Param}}$,
$\vartriangleleft^{\keyw{ref}}$, $\vartriangleleft^{\lrapp}$ are all the key
pre-templates we needed for (\ref{eq:param-law}).
Once the sufficiency-of-robustness theorem is applied to these pre-templates,
the equivalence can be obtained as a chain of the induced contextual equivalences.
The chain roughly looks as follows for the particular programs
$C_\mathsf{1}[P_\mathsf{L}]$ and $C_\mathsf{1}[P_\mathsf{R}]$,
where we use $\simeq$ to denote the informal textual counterpart of contextual
equivalence that is between terms accompanied by store.
\begin{equation}
 \label{eq:ParamLawTextualChain}
  \begin{aligned}
   C_\mathsf{1}[P_\mathsf{L}], \enspace \emptyset
   &\quad\equiv\quad
   C_\mathsf{1}[
   \bigl( \lambda x.\,\lambda f.\, (\lambda w.\, !x) \ (f \ ()) \bigr)
   \ (\keyw{ref}\ 1)
   ], \enspace \emptyset \\
   &\quad\simeq\quad
   C_\mathsf{1}[%
   \bigl( \lambda x.\,\lambda f.\, (\lambda w.\, !x) \ (f \ ()) \bigr)
   \ a
   ], \enspace \{a \mapsto 1\}
   &\text{(induced by $\vartriangleleft^{\keyw{ref}}$)} \\
   &\quad\simeq\quad
   C_\mathsf{1}[%
   \lambda f.\, (\lambda w.\, !a) \ (f \ ())
   ], \enspace \{a \mapsto 1\}
   &\text{(induced by $\vartriangleleft^{\lrapp}$)} \\
   &\quad\simeq\quad
   C_\mathsf{1}[%
   \lambda f.\, (\lambda w.\, 1) \ (f \ ())
   ], \enspace
   \emptyset
   &\text{(induced by $\vartriangleleft^{\mathrm{Param}}$)} \\
   &\quad\equiv\quad
   C_\mathsf{1}[P_\mathsf{R}],\enspace \emptyset
  \end{aligned}
\end{equation}

Finally, in the formal proof that uses hypernets and focussed graph rewriting
rather than terms and reductions, we additionally needed auxiliary, structural pre-templates (\autoref{fig:structural}).
These enable us to simplify or identify certain contractions and
weakenings, which are
not present in the textual representation but important in the graphical
representation as hypernets.
Contextual equivalences induced by the structural pre-templates enable
simplification of certain hypernets that involve contractions and weakenings.
The simplification appeared in the formal counterpart of the
chain~(\ref{eq:ParamLawTextualChain}), which is shown in \autoref{fig:NE4inUse}. It primarily applies to the hypernets
produced by the encoding $(-)^\ddag$.
Additionally, the pre-templates helped us prove
input-safety and robustness of other pre-templates.

\section{Related and future work} \label{sec:related-future-work}

\subsection{Proof methodologies for observational equivalence}

This work deals with fragility of observational equivalence and of its proof methodologies. Dreyer et al.\ address this issue by carefully distinguishing between various kinds of operations (state vs.\ control)~\cite{DBLP:journals/jfp/DreyerNB12}. Dreyer et al.~\cite{Dreyer10} use Kripke relations to go beyond an enumerative classification of effects; they use \emph{characterisation} of effects in the aid of reasoning. Their notion of \emph{island} has similar intuitions to our robustness property.
More radical approaches are down to replacing the concept of syntactic context with an \emph{epistemic} context, akin to a Dolev-Yao-style attacker~\cite{DBLP:journals/entcs/GhicaT12}, and characterising combinatorially the interaction between a term and its context as is the case with the game semantics~\cite{AbramskyJM00,DBLP:journals/iandc/HylandO00} or the trace semantics~\cite{jeffrey2005java}.
We propose a new approach to the problem of fragility, namely directly reasoning about \emph{robustness} of observational equivalence, using a uniform graph representation of a program and its context.

We are not the first one to take a coinductive approach to observational equivalence. \emph{Applicative bisimilarity} and its successor, \emph{environmental bisimilarity}, have been successfully used to prove observational equivalence in various effectful settings~\cite{Abramsky90AppBisim,KoutavasLS11,DalLagoGL17,SimpsonV18}. 
Typically, one first constructs an applicative (or environmental) bisimulation, and then proves it is a congruence using Howe's method~\cite{Howe96}. In contrast, in our approach, one first constructs a relation that is closed, by definition, under term (graph) construction, and then proves it is a counting simulation up-to. Our approach does not need Howe's method.

We argue that our approach is both flexible and elementary. A specific version~\cite{
DBLP:conf/lics/MuroyaCG18} of this formalism has been used to prove, for example, the soundness of exotic operations involved in (a functional version of) Google's TensorFlow machine learning library. Even though the proofs can seem complicated, this is in part due to the graph-based formalism being new, and in part due to the fact that proofs of equivalences are lengthy case-based analyses. However, herein lies the simplicity and the robustness of the approach, avoiding the discovery of clever-but-fragile language invariants which can be used to streamline proofs.

Our tedious-but-elementary proofs on the other hand seem highly suitable for automation. The idea of elementary case analysis can be adopted in term rewriting, instead of hypernet rewriting, for a call-by-value lambda-calculus equipped with \emph{effect handlers}~\cite{Pretnar15}. In this setting, the case analysis can be formalised as \emph{critical pair analysis}, which is a fundamental and automatable technique in term rewriting, and indeed automated~\cite{MuroyaH24}.

\subsection{Focussed hypernet rewriting}

Focussed hypernet rewriting is a radically new approach to defining effectful programming languages and proving observational equivalence.
%
We are not so much interested in \emph{simulated} effects, which are essentially the encoding of effectful behaviour into pure languages, and which can be achieved via monads~\cite{wadler1998marriage}, but we are interested in genuine \emph{native} effects which happen outside of the language. Semantically this has been introduced by Plotkin and Power~\cite{DBLP:journals/entcs/PlotkinP08} and more recently developed by M{\o}gelberg and Staton~\cite{mogelberg2011linearly}. The (copying) UAM takes the idea to the extreme by situating all operations (pure or effectful) outside of the primitives, and by keeping as intrinsic to the language only the structural aspects of copying vs.\ sharing, and scheduling of computation via thunking.

The (copying) UAM presented in this paper is an extension of the \emph{Dynamic Geometry of Interaction Machine (DGoIM)}~\cite{DBLP:conf/csl/MuroyaG17} to effects. The DGoIM builds on operational machinery~\cite{DanosR96,Mackie95,DBLP:conf/csl/HoshinoMH14,DBLP:conf/popl/MuroyaHH16} inspired by Girard's \emph{Geometry of Interaction (GoI)}~\cite{Girard89GoI1}.
Unlike these ``conventional'' GoI-inspired operational semantics, the DGoIM and hence the UAM are modified so that the underlying hypernet can be rewritten during execution.

The original motivation of the DGoIM was to produce an abstract machine that expresses the computational intuitions of the GoI while correctly modelling the cost of evaluation, particularly for call-by-value and call-by-need.
The ability to rewrite its own hypernet makes the DGoIM \emph{efficient}, in the sense of Accattoli et al.~\cite{DBLP:conf/icfp/AccattoliBM14}, for common reduction strategies (namely, call-by-value and call-by-need).
It also gives the DGoIM the ability to model exotic effects, e.g.\ transforming stateful into pure computation by abstracting the state~\cite{DBLP:conf/lics/MuroyaCG18}.
As an extension of the DGoIM, the UAM is designed for new reasoning principles and methods that arise out of GoI-inspired operational semantics.

Although the UAM does not aim at efficiency at the moment, one can think of a cost model of the UAM in a similar way as the DGoIM.
Moreover, the indexing of observational equivalence with a preorder representing the number of steps gives a direct avenue for modelling and comparing computation costs. For example, the micro-beta law induced by the pre-template $\vartriangleleft^{\lrapp}$ is indexed by the normal order $\geq$ on $\mathbb N$ (cf.\ \autoref{tab:TemplateAnalysis}), which indicates that one side always requires fewer steps than the other in the evaluation process. The only details to be resolved are associating costs (time and space) with steps, in particular different costs for different operations. 

The UAM is motivated by a need for a flexible and expressive framework in which a wide variety of effects can be given a cost-accurate model. As discussed, the UAM opens the door to a uniform study of operations and their interactions. Defining new styles of abstract machines is a rich and attractive vein of research. The \emph{monoidal computer} of Pavlovic~\cite{pavlovic2013monoidal} or the \emph{evolving algebras} of Gurevich~\cite{gurevich2018evolving} are such examples. What sets the UAM apart is the fact that it can be used, rather conveniently, for reasoning robustly about observational equivalence. 

\subsection{Hypernets}

The hierarchy of hypernets is inspired by the \emph{exponential boxes} of
proof nets, a graphical representation of linear logic
proofs~\cite{Girard87LL} and have an informal connection to Milner's \emph{bigraphs}~\cite{milner2001bigraphical}. Exponential boxes can be formalised by
parameterising an agent (which corresponds to an edge in our setting)
by a net, as indicated by Lafont~\cite{Lafont95}.
In the framework of interaction nets~\cite{Lafont90} that subsume proof
nets, agents can be coordinated to represent a boundary of a box, as
suggested by Mackie~\cite{Mackie98}. An alternative representation of boxes
that use extra links between agents
is proposed by Accattoli and Guerrini~\cite{AccattoliG09}.

Our graphical formulation of
boxes shares the idea with the first parameterising approach, but we
have flexibility regarding types of a box edge itself and its content
(i.e.\ the hypernet that labels it).  We use box edges to
represent thunks, and a box edge can have less targets than outputs of
its contents, reflecting the number of bound variables a thunk has.
This generalised box structure is also studied by Drewes et al.~\cite{DrewesHP02} as
\emph{hierarchical graphs}, in the context of double-pushout graph
transformation (DPO)~\cite{GraphTransformationHandbook1}, an
well-established algebraic approach to graph rewriting. 
More recently, Alvarez-Picallo et al. have formulated DPO rewriting for a class of hypernets similar to those used here~\cite{Alvarez-PicalloGSZ22};
their work further relates hypernets with string diagrams with \emph{functorial boxes} in the style of Melli{\`{e}}s~\cite{Mellies06}.

Interaction nets are another established framework of graph rewriting,
in which various evaluations of pure lambda-calculus can be
implemented~\cite{Sinot05,Sinot06}. The idea of having the token to
represent an evaluation strategy can be found in \textit{loc.\ cit.},
which suggests that our focussed rewriting on hypernets could be
implemented using interaction nets.  However, the local reasoning we
are aiming at with focussed rewriting does not seem easy in the
setting of interaction nets, because of technical subtleties observed
in \textit{loc.\ cit.}; namely, a status of evaluation is remembered
by not only the token but also some other agents around an interaction
net.

\subsection{Future work}

One direction is the introduction of a more meaningful type system for hypernets. The current type system of hypernets is very weak, just ensuring well-formedness. We consider it a strength of the approach that equivalences can be proved without the aid of a powerful type infrastructure. On the other hand, in order to avoid stuck configurations and ensure safety of evaluation, more expressive types are required. The usage of more expressive types is perfectly compatible with focussed hypernet rewriting, and is something we intend to explore. In particular we would like to study notions of typing which are germane to focussed hypernet rewriting, capturing its concepts of locality and robustness.

Beyond types, if we look at logics there are some appealing similarities between hypernet rewriting and \emph{separation logic}~\cite{reynolds2002separation}. The division of nodes into copying nodes via variables and sharing nodes via atoms is not accidental, and their different contraction properties match those from \emph{bunched implications}~\cite{o1999logic}. On a deeper level, the concepts of locality and in particular robustness developed here are related to the \emph{frame} rule of separation logic. 

Finally, our formulation of equivalence has some self-imposed limitations needed to limit the complexity of the technical presentation. We are hereby concerned with \emph{sequential} and \emph{deterministic} computation. Future work will show how these restrictions can be relaxed.
Parallelism and concurrency can be naturally simulated using multi-token reductions, 
as inspired by the  multi-token GoI machine of Dal Lago et al.~\cite{DBLP:conf/lics/LagoTY17}, whereas nondeterminism (or probabilistic evaluation) requires no changes to the machinery but rather a new definition of observational equivalence. This is work we are aiming to undertake in the future. A first step towards nondeterminism has been made~\cite{MuroyaSU24} in which the notion of counting simulation is extended from branching-free transition systems to nondeterministic automata.

\bibliographystyle{alphaurl}
\bibliography{spartan,ref}

\clearpage
\appendix

\section{An alternative definition of hypernets} \label{sec:hypernets-alt}

Informally, hypernets are nested hypergraphs, and one hypernet can
contain nested hypergraphs up to different depths. This intuition is
reflected by \autoref{def:Hypernets} of hypernets, in particular the
big union in
$\mathcal{H}_{k+1}(L,M) = \mathcal{H}\Bigl(L,
M \cup \bigcup_{i \leq k}\mathcal{H}_i(L,M)\Bigr)$.
In fact, the definition can be replaced by a simpler, but possibly
less intuitive, definition below that does not explicitly deal with
the different depths of nesting.
\begin{defi}
 Given sets $L$ and $M$,
 a set $\mathcal{H}'_k(L,M)$ is defined by induction on $k \in \N$:
 \begin{align*}
  \mathcal{H}'_0(L, M) &:= \mathcal{H}(L, M) \\
  \mathcal{H}'_{k+1}(L, M) &:=
  \mathcal{H}\Bigl(L, M \cup \mathcal{H}'_k(L, M)\Bigr)
 \end{align*}
 and hence a set
 $\mathcal{H}'_\omega(L,M) := \bigcup_{i \in \N}\mathcal{H}'_i(L,M)$.
\end{defi}
\begin{lem}
 \label{lem:AltHypernetLevels}
 Given arbitrary sets $L$ and $M$,
 any two numbers $k,k' \in \N$ satisfy
 $\mathcal{H}'_k(L,M) \subseteq \mathcal{H}'_{k+k'}(L,M)$.
\end{lem}
\begin{proof}
 If $k' = 0$, the inclusion trivially holds. If not, i.e.\ $k' > 0$,
 it can be proved by induction on $k \in \N$. The key reasoning
 principle we use is that $M \subseteq M'$ implies
 $\mathcal{H}(L,M) \subseteq \mathcal{H}(L,M')$.

 In the base case, when $k = 0$ (and $k' > 0$), we have
 \begin{align*}
  \mathcal{H}'_0(L,M) &= \mathcal{H}(L,M) \\
  &\subseteq
  \mathcal{H}\Bigl(L, M \cup \mathcal{H}'_{k'-1}(L, M)\Bigr)
  = \mathcal{H}'_{k'}(L,M).
 \end{align*}
 In the inductive case, when $k > 0$ (and $k' > 0$), we have
 \begin{align*}
  \mathcal{H}'_k(L,M)
  &= \mathcal{H}\Bigl(L, M \cup \mathcal{H}'_{k-1}(L, M)\Bigr) \\
  &\subseteq
  \mathcal{H}\Bigl(L, M \cup \mathcal{H}'_{k-1+k'}(L, M)\Bigr)
  = \mathcal{H}'_{k+k'}(L,M)
 \end{align*}
 where the inclusion is by induction hypothesis on $k-1$.
\end{proof}

\begin{prop}
 \label{prop:AltDefHypernets}
 Any sets $L$ and $M$ satisfy
 $\mathcal{H}_k(L,M) = \mathcal{H}'_k(L,M)$ for any $k \in \N$,
 and hence $\mathcal{H}_\omega(L,M) = \mathcal{H}'_\omega(L,M)$.
\end{prop}
\begin{proof}
 We first prove
 $\mathcal{H}_k(L,M) \subseteq \mathcal{H}'_k(L,M)$ by induction on
 $k \in \N$.
 The base case, when $k = 0$, is trivial.
 In the inductive case, when $k > 0$, we have
 \begin{align*}
  \mathcal{H}_k(L,M)
  &= \mathcal{H}\Bigl(L,
  M \cup \bigcup_{i \leq k-1}\mathcal{H}_i(L,M)\Bigr) \\
  &\subseteq \mathcal{H}\Bigl(L,
  M \cup \bigcup_{i \leq k-1}\mathcal{H}'_i(L,M)\Bigr)
  \tag{by I.H.} \\
  &= \mathcal{H}\Bigl(L, M \cup \mathcal{H}'_{k-1}(L,M)\Bigr)
  \tag{by \autoref{lem:AltHypernetLevels}} \\
  &= \mathcal{H}'_k(L,M).
 \end{align*}

 The other direction, i.e.\
 $\mathcal{H}'_k(L,M) \subseteq \mathcal{H}_k(L,M)$, can be also
 proved by induction on $k \in \N$.
 The base case, when $k = 0$, is again trivial.
 In the inductive case, we have
 \begin{align*}
  \mathcal{H}'_k(L,M)
  &= \mathcal{H}\Bigl(L, M \cup \mathcal{H}'_{k-1}(L,M)\Bigr) \\
  &\subseteq \mathcal{H}\Bigl(L, M \cup \mathcal{H}_{k-1}(L,M)\Bigr)
  \tag{by I.H.} \\
  &\subseteq \mathcal{H}\Bigl(L,
  M \cup \bigcup_{i \leq k-1}\mathcal{H}_i(L,M)\Bigr) \\
  &= \mathcal{H}_k(L,M).
  \qedhere
 \end{align*}
\end{proof}

Given a hypernet $G$, by \autoref{lem:AltHypernetLevels} and
\autoref{prop:AltDefHypernets}, there exists a minimum number $k$
such that $G \in \mathcal{H}'_k(L,M)$, which we call the ``minimum
level'' of $G$.

\begin{lem}
 \label{lem:FiniteEdges}
 Any hypernet has a finite number of shallow edges, and a finite
 number of deep edges.
\end{lem}
\begin{proof}
 Any hypernet has a finite number of shallow edges by definition.
 We prove that any hypernet $G$ has a finite number of deep edges,
 by induction on minimum level $k$ of the hypernet.

 When $k = 0$, the hypernet has ho deep edges.

 When $k > 0$, each hypernet $H$ that labels a shallow edge of $G$
 belongs to $\mathcal{H}'_{k-1}(L,M)$, and therefore its minimum level
 is less than $k$. By induction hypothesis, the labelling hypernet $H$
 has a finite number of deep edges, and also a finite number of
 shallow edges.
 Deep edges of $G$ are given by edges, at any depth, of any hypernet
 that labels a shallow edge of $G$.
 Because there is a finite number of the hypernets that label the
 shallow edges of $G$, the number of deep edges of $G$ is finite.
\end{proof}

\section{Plugging} \label{app:plugging}

An interfaced labelled monoidal hypergraph can be given by data of the
following form:
$((V \uplus I \uplus O, E), (S, T), (f_V, f_E))$
where $I$ is the input list, $O$ is the output list, $V$ is the set of
all the other vertices, $E$ is the set of edges, $(S,T)$ defines
source and target lists, and $(f_V, f_E)$ is
labelling functions.
\begin{defi}[Plugging]
 \label{def:Plugging}
 Let
 $\mathcal{C}[\vec{\chi^1},\chi,\vec{\chi^2}] =
 ((V \uplus I \uplus O, E), (S, T), (f_V, f_E))$
 and
 $\mathcal{C}'[\vec{\chi^3}] =
 ((V' \uplus I' \uplus O', E'), (S', T'),
 (f'_V, f'_E))$
 be contexts, such that
 the hole $\chi$ and the latter context $\mathcal{C}'$ have the same
 type and
 $\vec{\chi^1} \cap \vec{\chi^2} \cap \vec{\chi^3} = \emptyset$.
 The \emph{plugging}
 $\mathcal{C}[\vec{\chi^1},\mathcal{C}',\vec{\chi^2}]$
 is a hypernet given by data
 $((\hat{V}, \hat{E}), (\hat{S}, \hat{T}),
 (\hat{f}_V, \hat{f}_E))$ such that:
 \begin{align*}
  \hat{V} &= V \uplus V' \uplus I \uplus O \\
  \hat{E} &= (E \backslash \{e_\chi\}) \uplus E' \\
  \hat{S}(e) &=
  \begin{cases}
   S(e) & \text{(if $e \in E \backslash \{e_\chi\}$)} \\
   g^*(S'(e)) & \text{(if $e \in E'$)}
  \end{cases} \\
  \hat{T}(e) &=
  \begin{cases}
   T(e) & \text{(if $e \in E \backslash \{e_\chi\}$)} \\
   g^*(T'(e)) & \text{(if $e \in E'$)}
  \end{cases} \\
  g(v) &=
  \begin{cases}
   v & \text{(if $v \in V'$)} \\
   (S(e_\chi))_i & \text{(if $v = (I')_i$)} \\
   (T(e_\chi))_i & \text{(if $v = (O')_i$)}
  \end{cases} \\
  \hat{f}_V(v) &=
  \begin{cases}
   f_V(v) & \text{(if $v \in V$)} \\
   f'_V(v) & \text{(if $v \in V'$)}
  \end{cases} \\
  \hat{f}_E(e) &=
  \begin{cases}
   f_E(e) & \text{(if $e \in E \backslash \{e_\chi\}$)} \\
   f'_E(e) & \text{(if $e \in E'$)}
  \end{cases}
 \end{align*}
 where $e_\chi \in E$ is the hole edge labelled with $\chi$,
 and $(-)_i$ denotes the $i$-th element of a list.
\end{defi}

In the resulting context
$\mathcal{C}[\vec{\chi'},\mathcal{C}',\vec{\chi''}]$,
each edge comes from either $\mathcal{C}$ or $\mathcal{C}'$.
If a path in $\mathcal{C}$ does not contain the hole edge $e_\chi$,
the path gives a path in 
$\mathcal{C}[\vec{\chi'},\mathcal{C}',\vec{\chi''}]$.
Conversely, if a path in
$\mathcal{C}[\vec{\chi'},\mathcal{C}',\vec{\chi''}]$ consists of edges
from $\mathcal{C}$ only, the path gives a path in $\mathcal{C}$.

Any path in $\mathcal{C}'$ gives a path in 
$\mathcal{C}[\vec{\chi'},\mathcal{C}',\vec{\chi''}]$.
However, if a path in
$\mathcal{C}[\vec{\chi'},\mathcal{C}',\vec{\chi''}]$ consists of edges
from $\mathcal{C}'$ only, the path does not necessarily give a path in
$\mathcal{C}'$.
The path indeed gives a path in $\mathcal{C}'$, if sources and targets
of the hole edge $e_\chi$ are distinct in $\mathcal{C}$ (i.e.\ the
hole edge $e_\chi$ is not a self-loop).

\section{Rooted states} \label{app:rooted-states}

\begin{lem}
 \label{lem:DeterministicAbstRWSystem}
 Let $(X, \rightarrowtriangle)$ is an abstract rewriting system that
 is deterministic.
 \begin{enumerate}
  \item \label{item:UniqueNormalisation}
	For any $x,y,y' \in X$ such
	that $y$ and $y'$ are normal forms, and for any $k,h \in \N$,
	if there exist two sequences $x \rightarrowtriangle^k y$ and
	$x \rightarrowtriangle^h y'$, then these sequences are exactly
	the same.
  \item \label{item:DistinctNormalisation}
	For any $x,y \in X$ such that $y$ is a normal form,
	and for any $i,j,k \in \N$ such that $i \neq j$ and
	$i,j \in \{ 1,\ldots,k \}$,
	if there exists a sequence $x \rightarrowtriangle^k y$, then
	its $i$-th rewrite $z \rightarrowtriangle z'$
	and $j$-th rewrite $w \rightarrowtriangle w'$ satisfy
	$z \neq w$.
 \end{enumerate}
\end{lem}
\begin{proof}
 The point (\ref{item:UniqueNormalisation}) is proved
 by induction on $k + h \in \N$.
 In the base case, when $k + h = 0$ (i.e.\ $k = h = 0$), the two
 sequences are both the empty sequence, and $x = y = y'$.
 The inductive case, when $k + h > 0$, falls into one of the following
 two situations.
 The first situation, where $k = 0$ or $h = 0$, boils down to the base
 case, because $x$ must be a normal form itself, which means
 $k = h = 0$.
 In the second situation, where $k > 0$ and $h > 0$, there exist
 elements $z,z' \in X$ such that
 $x \rightarrowtriangle z \rightarrowtriangle^{k-1} y$ and
 $x \rightarrowtriangle z' \rightarrowtriangle^{h-1} y'$.
 Because $\rightarrowtriangle$ is deterministic, $z = z'$ follows, and
 hence by induction hypothesis on $(k-1)+(h-1)$, these two sequences
 are the same.

 The point (\ref{item:DistinctNormalisation}) is proved by
 contradiction.
 The sequence $x \rightarrowtriangle^k y$ from $x$ to the normal form
 $y$ is unique, by the point (\ref{item:UniqueNormalisation}).
 If its $i$-th rewrite $z \rightarrowtriangle z'$ and
 $j$-th rewrite $w \rightarrowtriangle w'$ satisfy
 $z = w$, determinism of the system implies that these two rewrites
 are the same. This means that the sequence
 $x \rightarrowtriangle^k y$ has a cyclic sub-sequence, and by
 repeating the cycle different times, one can yield different
 sequences of rewrites $x \rightarrowtriangle^* y$ from $x$ to $y$.
 This contradicts the uniqueness of the original sequence
 $x \rightarrowtriangle^k y$. 
\end{proof}

\begin{lem}
 \label{lem:RootedUniqueDistinctSearch}
 If a state $\focussed{G}$ is rooted, a search sequence
 $?;|\focussed{G}| \tobul^* \focussed{G}$ from the initial state
 $?;|\focussed{G}|$ to the state $\focussed{G}$ is unique.
 Moreover, for any $i$-th search transition and $j$-th search
 transition in the sequence such that $i \neq j$, these transitions do
 not result in the same state.
\end{lem}
\begin{proof}
 Let $X$ be the set of states with the $?$-focus or the $\checkmark$-focus.
 We can define an abstract rewriting system $(X,\rightarrowtriangle)$
 of ``reverse search'' by:
 $\focussed{H} \rightarrowtriangle \focussed{H'}$ if
 $\focussed{H'} \tobul \focussed{H}$.
 Any search sequence corresponds to a sequence of rewrites in this
 rewriting system.

 The rewriting system is deterministic, i.e.\ if
 $\focussed{H'} \tobul \focussed{H}$ and
 $\focussed{H''} \tobul \focussed{H}$ then
 $\focussed{H'} = \focussed{H''}$, because the inverse $\mapsto^{-1}$
 of the interaction rules (\autoref{fig:interaction-subst-rules}) is
 deterministic.
 
 If a search transition changes a focus to the $?$-focus, the
 resulting $?$-focus always has an incoming operation edge. This
 means that, in the rewriting system $(X,\rightarrowtriangle)$,
 initial states are normal forms.
 Therefore, by
 \autoref{lem:DeterministicAbstRWSystem}(\ref{item:UniqueNormalisation}),
 if there exist two search sequences from the initial state
 $?;|\focussed{G}|$ to the state $\focussed{G}$, these search
 sequences are exactly the same.
 The rest is a consequence of
 \autoref{lem:DeterministicAbstRWSystem}(\ref{item:DistinctNormalisation}).
\end{proof}

\begin{lem}
 \label{lem:UniqueDistinctOprPathFromInput}
 For any hypernet $N$, if there exists an operation path from an
 input to a vertex, the path is unique. Moreover, no edge appears
 twice in the operation path.
\end{lem}
\begin{proof}
 Given the hypernet $N$ whose set of (shallow) vertices is $X$, we can
 define an abstract rewriting system
 $(X,\rightarrowtriangle)$ of ``reverse connection'' by:
 $v \rightarrowtriangle v'$ if there exists an operation edge whose
 unique source is $v'$ and targets include $v$.
 Any operation path from an input to a vertex in $N$ corresponds to a
 sequence of rewrites in this rewriting system.

 This rewriting system is deterministic, because each vertex can have
 at most one incoming edge in a hypergraph
 (\autoref{def:labelled-monoidal-hypergraphs}) and each operation edge has exactly one
 source. Because inputs of the hypernet $N$ have no incoming edges,
 they are normal forms in this rewriting system.
 Therefore, by
 \autoref{lem:DeterministicAbstRWSystem}(\ref{item:UniqueNormalisation}),
 an operation path from any input to any vertex is unique.

 The rest is proved by contradiction.
 We assume that, in an operation path $P$ from an input to a
 vertex,
 the same operation edge $e$ appears twice. The edge $e$ has one
 source, which either is an input of the hypernet $N$ or has an
 incoming edge.
 In the former case, the edge $e$ can only appear as the first edge of
 the operation path $P$, which is a contradiction.
 In the latter case, the operation edge $e$ has exactly one incoming
 edge $e'$ in the hypernet $N$. In the operation path $P$,
 each appearance of the operation edge $e$ must be preceded by this
 edge $e'$ via the same vertex. This contradicts
 \autoref{lem:DeterministicAbstRWSystem}(\ref{item:DistinctNormalisation}).
\end{proof}

\begin{lem}
 \label{lem:RootedThenOprPath}
 For any rooted state $\focussed{G}$, if its focus source (i.e.\ the
 source of the focus) does not coincide with the unique input, then
 there exists an operation path from the input to the focus source.
\end{lem}
\begin{proof}
 By \autoref{lem:RootedUniqueDistinctSearch},
 the rooted state $\focussed{G}$ has a unique search sequence
 $?;|\focussed{G}| \tobul^* \focussed{G}$.
 The proof is by the length $k$ of this sequence.

 In the base case, where $k = 0$, the state $\focussed{G}$ itself is
 an initial state, which means the input and focus source coincide in
 $\focussed{G}$.

 In the inductive case, where $k > 0$, there exists a state
 $\focussed{G'}$ such that
 $?;|\focussed{G}| \tobul^{k-1} \focussed{G'} \tobul \focussed{G}$.
 The proof here is by case analysis on the interaction rule used in
 $\focussed{G'} \tobul \focussed{G}$.
 \begin{itemize}
  \item When the interaction rules in \autoref{fig:interaction-contr-L},~\ref{fig:interaction-contr-R},~\ref{fig:interaction-I},~\ref{fig:interaction-opr-none} are used, the transition
	$\focussed{G'} \tobul \focussed{G}$ only changes a focus
	label.
  \item When the interaction rule in \autoref{fig:interaction-opr-first} is used, the transition
	$\focussed{G'} \tobul \focussed{G}$ turns the focus and its
	outgoing operation edge $e_{G'}$ into an operation edge $e_G$
	and its outgoing focus.
	By induction hypothesis	on $\focussed{G'}$, the focus source
	coincides with its input, or there exists an operation path
	from the input to the focus source, in $\focussed{G'}$.

	In the former case, in $\focussed{G}$, the source of the
	operation edge $e_G$ coincides with the input.
	The edge $e_G$ itself gives the desired operation path in
	$\focussed{G}$.

	In the latter case, the operation path $P_{G'}$ from the input
	to the focus source in $\focussed{G'}$ does not contain the
	outgoing operation edge $e_{G'}$ of the focus; otherwise, the
	edge $e_{G'}$ must be preceded by the focus edge in the
	operation path $P_{G'}$, which is
	a contradiction. Therefore, the operation path $P_{G'}$
	in $\focussed{G'}$ is inherited in
	$\focussed{G}$, becoming a path $P_G$ from the input to the
	source of the incoming operation edge $e_G$ of the focus.
	In the state $\focussed{G}$, the path $P_G$ followed by the
	edge $e_G$ yields the desired operation path.
  \item When the interaction rule in \autoref{fig:interaction-opr-next} is used, the transition
	$\focussed{G'} \tobul \focussed{G}$ changes the focus from a
	$(k+1)$-th outgoing edge of an operation edge $e$ to a
	$(k+2)$-th outgoing edge of the same operation edge $e$, for
	some $k \in \N$.
	In $\focussed{G'}$, the focus source is not an input, and
	therefore, there exists an operation path $P_{G'}$ from the
	input to the focus source, by induction hypothesis.
	
	The operation path $P_{G'}$ ends with the operation edge $e$,
	and no outgoing edge of the edge $e$ is involved in the path
	$P_{G'}$; otherwise, the edge $e$ must appear more than once
	in the path $P_{G'}$, which is a contradiction by
	\autoref{lem:UniqueDistinctOprPathFromInput}.
	Therefore, the path $P_{G'}$ is inherited exactly as it is in
	$\focussed{G}$, and it gives the desired operation path.
  \item When the interaction rule in \autoref{fig:interaction-opr-last} is used, by the same reasoning
	as in the case of \autoref{fig:interaction-opr-next}, $\focussed{G'}$ has an operation
	path $P_{G'}$ from the input to the focus source, where
	the incoming operation edge $e_{G'}$ of the focus appears
	exactly once, at the end.
	Removing the edge $e_{G'}$ from the path $P_{G'}$ yields
	another operation path $P$ from the input in $\focussed{G'}$,
	and it also gives an operation path from the input to the
	focus source in $\focussed{G}$.
	\qedhere
 \end{itemize}
\end{proof}

\begin{lem}
 \label{lem:SearchSeqQueried}
 For any state $\focussed{G}$ with a $\mathsf{t}$-focus such
 that $\mathsf{t} \neq \mathord{?}$, if $\focussed{G}$ is rooted, then
 there exists a search sequence
 $?;|\focussed{G}| \tobul^*
 \langle \focussed{G} \rangle_{?/\mathsf{t}} \tobul^+
 \focussed{G}$.
\end{lem}
\begin{proof}
 By \autoref{lem:RootedUniqueDistinctSearch},
 the rooted state $\focussed{G}$ has a unique search sequence
 $?;|\focussed{G}| \tobul^* \focussed{G}$.
 The proof is to show that a transition from the state
 $\langle \focussed{G} \rangle_{?/\mathsf{t}}$ appears in this search
 sequence, and it is by the length $k$ of the search sequence.

 Because $\focussed{G}$ does not have the $?$-focus, $k = 0$ is
 impossible, and therefore the base case is when $k = 1$.
 The search transition $?;|\focussed{G}| \tobul \focussed{G}$ must use
 one of the interaction rules in \autoref{fig:interaction-contr-L},~\ref{fig:interaction-contr-R},~\ref{fig:interaction-I},~\ref{fig:interaction-opr-none}. This means
 $?;|\focussed{G}| = \langle \focussed{G} \rangle_{?/\mathsf{t}}$.

 In the inductive case, where $k > 0$, there exists a state
 $\focussed{G'}$ such that
 $?;|\focussed{G}| \tobul^{k-1} \focussed{G'} \tobul \focussed{G}$.
 The proof here is by case analysis on the interaction rule used in
 $\focussed{G'} \tobul \focussed{G}$.
 \begin{itemize}
  \item When the interaction rule in \autoref{fig:interaction-contr-L},~\ref{fig:interaction-contr-R},~\ref{fig:interaction-I},~\ref{fig:interaction-opr-none} is used,
	$?;|\focussed{G}| =
	\langle \focussed{G} \rangle_{?/\mathsf{t}}$.
  \item Because $\focussed{G}$ does not have the $?$-focus, the
	interaction rules in \autoref{fig:interaction-opr-first},~\ref{fig:interaction-opr-next} can be never used in
	$\focussed{G'} \tobul \focussed{G}$.
  \item When the interaction rule in \autoref{fig:interaction-opr-last} is used, $\focussed{G'}$ has the $\checkmark$-focus, which is a $(k+1)$-th outgoing edge of an
	operation edge $e$, for some $k \in \N$.
	The operation edge $e$ becomes the outgoing edge of the focus
	in $\focussed{G}$.
	By induction hypothesis on $\focussed{G'}$, we have
	\begin{equation*}
	 ?;|\focussed{G}| \tobul^*
	  \langle \focussed{G'} \rangle_{?/\checkmark} \tobul^+
	  \focussed{G'} \tobul \focussed{G}.
	  \tag{A}
	\end{equation*}

	If $k = 0$, in $\focussed{G'}$, the focus is the only outgoing
	edge of the operation edge $e$.
	Because $\langle \focussed{G'} \rangle_{?/\checkmark}$ is not
	an initial state, it must be a result of the interaction rule
	in \autoref{fig:interaction-opr-first}, which means the search sequence (A) is factored through
	as:
	\begin{equation*}
	 ?;|\focussed{G}| \tobul^*
	  \langle \focussed{G} \rangle_{?/\mathsf{t}} \tobul
	  \langle \focussed{G'} \rangle_{?/\checkmark} \tobul^+
	  \focussed{G'} \tobul \focussed{G}.
	\end{equation*}

	If $k > 0$,
	for each $m \in \{ 0,\ldots,k \}$, let $\focussed{N_m}$ be a
	state with the $?$-focus, such that
	$|\focussed{N_m}| = |\focussed{G'}|$ and the
	focus is an $(m+1)$-th outgoing edge of the operation edge
	$e$.
	This means
	$\focussed{N_k}
	= \langle \focussed{G'} \rangle_{?/\checkmark}$.
	The proof concludes by combining the following internal lemma
	with (A), taking $k$ as $m$.
	\qedhere
	\begin{lem}
	 For any $m \in \{ 0,\ldots,k \}$,
	 if there exists $h < k$ such that
	 $?;|\focussed{G}| \tobul^h \focussed{N_m}$,
	 then it is factored through as
	 $?;|\focussed{G}| \tobul^*
	 \langle \focussed{G} \rangle_{?/\mathsf{t}} \tobul^+
	 \focussed{N_m}$.
	\end{lem}
	\begin{proof}
	 By induction on $m$.
	 In the base case, when $m = 0$, the focus of
	 $\focussed{N_m}$ is the first outgoing edge of the operation
	 edge $e$. This state is not initial, and therefore must be a
	 result of the interaction rule in \autoref{fig:interaction-opr-first}, which means
	 \begin{equation*}
	  ?;|\focussed{G}| \tobul^*
	   \langle \focussed{G} \rangle_{?/\mathsf{t}} \tobul
	   \focussed{N_m}.
	 \end{equation*}

	 In the inductive case, when $m > 0$, the state
	 $\focussed{N_m}$ is not an initial state and must be a result
	 of the interaction rule in \autoref{fig:interaction-opr-next}, which means
	 \begin{equation*}
	  ?;|\focussed{G}| \tobul^*
	   \langle \focussed{N_{m-1}} \rangle_{\checkmark/?} \tobul
	   \focussed{N_m}.
	 \end{equation*}
	 The first half of this search sequence, namely
	 $?;|\focussed{G}| \tobul^*
	 \langle \focussed{N_{m-1}} \rangle_{\checkmark/?}$,
	  consists of $h-1 < k$ transitions. Therefore, by (outer)
	 induction hypothesis on $h-1$, we have
	 \begin{equation*}
	  ?;|\focussed{G}| \tobul^*
	   \focussed{N_{m-1}} \tobul^+
	   \langle \focussed{N_{m-1}} \rangle_{\checkmark/?} \tobul
	   \focussed{N_m}.
	 \end{equation*}
	 The first part, namely
	 $?;|\focussed{G}| \tobul^* \focussed{N_{m-1}}$,
	 consists of less than $k$ transitions. Therefore, by (inner)
	 induction hypothesis on $m-1$, we have
	 \begin{equation*}
	  ?;|\focussed{G}| \tobul^*
	   \langle \focussed{G} \rangle_{?/\mathsf{t}} \tobul^+
	   \focussed{N_{m-1}} \tobul^+
	   \langle \focussed{N_{m-1}} \rangle_{\checkmark/?} \tobul
	   \focussed{N_m}.
	   \qedhere
	 \end{equation*}
	\end{proof}
 \end{itemize}
\end{proof}

\begin{lem}
 \label{lem:NotOprPathNotRooted}
 \noindent
 \begin{enumerate}
  \item \label{item:State}
	For any state $\focussed{N}$, if it has a path to the focus
	source that is not an operation path, then it is not rooted.
  \item \label{item:OneWay}
	For any focus-free hypernet $H$ and any focussed context
	$\focussed{\mathcal{C}}[\chi]$ with one hole edge, such that
	$\focussed{\mathcal{C}}[H]$ is a
	state, if the hypernet $H$ is one-way and the context
	$\focussed{\mathcal{C}}$ has a path to the focus source that
	is not an operation path, then the state
	$\focussed{\mathcal{C}}[H]$ is not rooted.
  \item \label{item:ExitingSafe}
	For any $\mathbb{C}$-specimen
	$(\focussed{\mathcal{C}}[\vec{\chi}];\vec{G};\vec{H})$ of an
	output-closed pre-template $\vartriangleleft$,
	if the context $\focussed{\mathcal{C}}[\vec{\chi}]$ has a path
	to the focus source that is not an operation path, then at
	least one of the states $\focussed{\mathcal{C}}[\vec{G}]$ and
	$\focussed{\mathcal{C}}[\vec{H}]$ is not rooted.
 \end{enumerate}
\end{lem}
\begin{proof}[Proof of the point~(\ref{item:State})]
 Let $P$ be the path in $\focussed{N}$ to the focus source
 that is not an operation path.
 The proof is by contradiction; we assume that $\focussed{N}$ is a
 rooted state.

 Because of $P$, the focus source is not an input.
 Therefore by \autoref{lem:RootedThenOprPath}, the state
 $\focussed{N}$ has an operation path from its unique input to the
 focus source.
 This operation path contradicts the path $P$, which is not an
 operation path, because each
 operation edge has only one source and each vertex has at most one
 incoming edge.
\end{proof}
\begin{proof}[Proof of the point~(\ref{item:OneWay})]
 Let $P$ be the path in $\focussed{\mathcal{C}}$ to the focus source
 that is not an operation path.

 If the path $P$ contains no hole edge, it gives a path
 in the state $\focussed{\mathcal{C}}[H]$ to the focus source
 that is not an operation path. By the point~(\ref{item:State}), the
 state is not rooted.

 Otherwise, i.e.\ if the path $P$ contains a hole edge,
 we give a proof by contradiction; we assume that the state
 $\focussed{\mathcal{C}}[H]$ is rooted.
 We can take a suffix of the path $P$, so that it gives
 a path from a target of a hole edge to the focus
 source in $\focussed{\mathcal{C}}$, and moreover, gives
 a path $P'$ from a source of an edge from $H$ to the focus source in
 $\focussed{\mathcal{C}}[H]$.
 This implies the focus source is not an input, and therefore by
 \autoref{lem:RootedThenOprPath},
 the state $\focussed{\mathcal{C}}[H]$ has an operation path from its
 unique input to the focus source.
 This operation path must have $P'$ has a suffix, meaning $P'$ is also
 an operation path, because each
 operation edge has only one source and each vertex has at most one
 incoming edge.
 Moreover, $H$ must have an operation path from an input to an output,
 such that the input and the output have type $\star$ and the path
 ends with the first edge of the path $P'$.
 This contradicts $H$ being one-way.
\end{proof}
\begin{proof}[Proof of the point~(\ref{item:ExitingSafe})]
 Let $P$ be the path in $\focussed{\mathcal{C}}$ to the focus source
 that is not an operation path.
 
 If the path $P$ contains no hole edge, it gives a path
 in the states $\focussed{\mathcal{C}}[\vec{G}]$ and
 $\focussed{\mathcal{C}}[\vec{H}]$ to the focus source
 that is not an operation path. By the point~(\ref{item:State}), the
 states are not rooted.

 Otherwise, i.e.\ if the path $P$ contains a hole edge, we can take a
 suffix of $P$ that gives a path $P'$ from a source of a hole edge $e$
 to the focus source in $\focussed{\mathcal{C}}$, so that the path
 $P'$ does not contain any hole edge.
 We can assume that the hole edge $e$ is labelled with $\chi_1$,
 without loss of generality.
 The path $P'$ gives paths $P'_G$ and $P'_H$ to the focus source,
 in contexts
 $\focussed{\mathcal{C}}[\chi_1, \vec{G} \backslash \{ G_1 \}]$ and
 $\focussed{\mathcal{C}}[\chi_1, \vec{H} \backslash \{ H_1 \}]$,
 respectively.
 The paths $P'_G$ and $P'_H$ are not an
 operation path, because they start with the hole edge $e$ labelled
 with $\chi_1$.

 Because $\vartriangleleft$ is output-closed, $G_1$ or $H_1$ is
 one-way. 
 By the point~(\ref{item:OneWay}), at least one of the states
 $\focussed{\mathcal{C}}[\vec{G}]$ and
 $\focussed{\mathcal{C}}[\vec{H}]$ is not rooted.
\end{proof}

\begin{lem}
 \label{lem:StationaryRWPreserveRooted}
 If a rewrite transition $\focussed{G} \to \focussed{G'}$ is
 stationary, it preserves the rooted property, i.e.\
 $\focussed{G}$ being rooted implies $\focussed{G'}$ is also rooted.
\end{lem}
\begin{proof}
 The stationary rewrite transition
 $\focussed{G} \to \focussed{G'}$ is in the form of
 $\mathcal{C}[\lightning ;_i H] \to \mathcal{C}[? ;_i H']$, where
 $\mathcal{C}$ is a focus-free simple context, $H$ is a focus-free
 one-way hypernet, $H'$ is a focus-free hypernet and $i \in \N$.
 We assume $\mathcal{C}[\lightning ;_i H]$ is rooted, and
 prove that $\mathcal{C}[? ;_i H']$ is rooted, i.e.\
 ${? ; \mathcal{C}[H']} \tobul^* \mathcal{C}[? ;_i H']$.
 By \autoref{lem:SearchSeqQueried}, there exists a number $k \in \N$
 such that:
 \begin{equation*}
  {? ; \mathcal{C}[H]} \tobul^k \mathcal{C}[? ;_i H]
   \tobul^+ \mathcal{C}[\lightning ;_i H].
 \end{equation*}
 The rest of the proof is by case analysis on the number $k$.
 \begin{itemize}
  \item When $k = 0$, i.e.\
	${? ; \mathcal{C}[H]} = \mathcal{C}[? ;_i H]$, the unique
	input and the $i$-th source of the hole	coincide in the
	simple context $\mathcal{C}$.
	Therefore, ${? ; \mathcal{C}[H']} = \mathcal{C}[? ;_i H']$,
	which means $\mathcal{C}[? ;_i H']$ is rooted.
  \item When $k > 0$, there exists a state $\focussed{N}$ such that
	${? ; \mathcal{C}[H]} \tobul^{k-1} \focussed{N} \tobul
	\mathcal{C}[? ;_i H]$.
	By the following internal lemma
	(\autoref{intlem:StationaryRWPreserveRooted}),
	there exists a focussed simple context
	$\focussed{\mathcal{C}_N}$, whose focus is not entering nor
	exiting, and we have two search sequences:
	\begin{align*}
	 {? ; \mathcal{C}[H]} &\tobul^{k-1}
	 \focussed{\mathcal{C}_N}[H] \tobul
	 \mathcal{C}[? ;_i H], \\
	 {? ; \mathcal{C}[H']} &\tobul^{k-1}
	 \focussed{\mathcal{C}_N}[H'].
	\end{align*}
	The last search transition
	$\focussed{\mathcal{C}_N}[H] \tobul \mathcal{C}[? ;_i H]$,
	which yields the $?$-focus, must use the interaction rule in \autoref{fig:interaction-opr-first},~\ref{fig:interaction-opr-next}.
	Because the focus is not entering nor exiting in the
	simple context $\focussed{\mathcal{C}_N}$, either of the two
	interaction rules acts on the focus and an edge of the
	context. This means that the same interaction
	is possible in the state $\focussed{\mathcal{C}_N}[H']$,
	yielding:
	\begin{equation*}
	 {? ; \mathcal{C}[H']} \tobul^{k-1}
	 \focussed{\mathcal{C}_N}[H'] \tobul
	 \mathcal{C}[? ;_i H'],
	\end{equation*}
	which means $\mathcal{C}[? ;_i H']$ is rooted.
	\qedhere
	\begin{lem}
	 \label{intlem:StationaryRWPreserveRooted}
	 For any $m \in \{0,\ldots,k-1\}$ and any state $\focussed{N}$
	 such that
	 ${? ; \mathcal{C}[H]} \tobul^m \focussed{N} \tobul^{k-m}
	 \mathcal{C}[? ;_i H]$,
	 the following holds.

	 (A)
	 If there exists a focussed simple context
	 $\focussed{\mathcal{C}_N}$
	 such that $\focussed{N} = \focussed{\mathcal{C}_N}[H]$,
	 the focus of the context $\focussed{\mathcal{C}_N}$ is
	 not entering.

	 (B)
	 If there exists a focussed simple context
	 $\focussed{\mathcal{C}_N}$
	 such that $\focussed{N} = \focussed{\mathcal{C}_N}[H]$,
	 the focus of the context $\focussed{\mathcal{C}_N}$ is
	 not exiting.

	 (C)
	 There exists a focussed simple context
	 $\focussed{\mathcal{C}_N}$
	 such that $\focussed{N} = \focussed{\mathcal{C}_N}[H]$, and
	 ${? ; \mathcal{C}[H']} \tobul^m
	 \focussed{\mathcal{C}_N}[H']$ holds.
	\end{lem}
	\begin{proof}
	 Firstly, because search transitions do not change an
	 underlying hypernet,
	 if there exists a focussed simple context
	 $\focussed{\mathcal{C}_N}$
	 such that $\focussed{N} = \focussed{\mathcal{C}_N}[H]$,
	 $|\focussed{\mathcal{C}_N}| = \mathcal{C}$ necessarily holds.

	 The point (A) is proved by contradiction; we assume that
	 the context $\focussed{\mathcal{C}_N}$ has an entering focus.
	 This means that there exist a number $p \in \N$ and a focus
	 label
	 $\mathsf{t} \in
	 \{ \mathord{?},\mathord{\checkmark},\mathord{\lightning} \}$
	 such that
	 $\focussed{\mathcal{C}_N} = \mathcal{C}[\mathsf{t} ;_p H]$.
	 By \autoref{lem:SearchSeqQueried}, there
	 exists a number $h$ such that $h \leq m$ and:
	 \begin{equation*}
	  {? ; \mathcal{C}[H]} \tobul^h
	   \mathcal{C}[? ;_p H] \tobul^{k-h}
	 \mathcal{C}[? ;_i H] \tag{$\$$}.
	 \end{equation*}
	 We derive a contradiction by case analysis on the numbers $p$
	 and $h$.
	 \begin{itemize}
	  \item If $p = i$ and $h = 0$, the state
		$\mathcal{C}[? ;_i H]$ must be initial, but it is a
		result of a search transition because $k-h > 0$. This
		is a contradiction.
	  \item If $p = i$ and $h > 0$, two different transitions in
		the search sequence ($\$$) result in the same state,
		because of $h > 0$ and $k-h > 0$,
		which contradicts
		\autoref{lem:RootedUniqueDistinctSearch}.
	  \item If $p \neq i$, by \autoref{def:StationaryRW}, there
		exists a state $\focussed{N'}$ with the $\lightning$-focus
		such that
		$\mathcal{C}[? ;_p H] \tobul \focussed{N'}$.
		This contradicts the search sequence ($\$$), because
		$k-h > 0$ and search transitions are deterministic.
	 \end{itemize}

	 The point (B) follows from the contraposition of
	 \autoref{lem:NotOprPathNotRooted}(\ref{item:OneWay}),
	 because $H$ is one-way and $\focussed{N}$ is rooted.
	 The rooted property of $\focussed{N}$ follows from the fact
	 that search transitions do not change underlying hypernets.

	 The point (C) is proved by induction on
	 $m \in \{ 0,\ldots,k-1 \}$.
	 In the base case, when $m = 0$,
	 we have
	 $?;\mathcal{C}[H] = \focussed{N}$, and therefore
	 the context $?;\mathcal{C}$ can be taken as
	 $\focussed{\mathcal{C}_N}$.
	 This means
	 $?;\mathcal{C}[H'] = \focussed{\mathcal{C}_N}[H']$.

	 In the inductive case, when $m > 0$, there exists a state
	 $\focussed{N'}$ such that
	 \begin{equation*}
	  {? ; \mathcal{C}[H]} \tobul^{m-1}
	   \focussed{N'} \tobul
	   \focussed{N} \tobul^{k-m}
	  \mathcal{C}[? ;_i H].
	 \end{equation*}
	 By the induction hypothesis,
	 there exists a focussed simple context
	 $\focussed{\mathcal{C}_{N'}}$ such that
	 $\focussed{N'} = \focussed{\mathcal{C}_{N'}}[H]$ and
	 \begin{align*}
	  {? ; \mathcal{C}[H]} &\tobul^{m-1}
	  \focussed{\mathcal{C}_{N'}}[H] \tobul
	  \focussed{N} \tobul^{k-m}
	  \mathcal{C}[? ;_i H], \\
	  {? ; \mathcal{C}[H']} &\tobul^{m-1}
	  \focussed{\mathcal{C}_{N'}}[H'].
	 \end{align*}
	 Our goal here is to find a focussed simple context
	 $\focussed{\mathcal{C}_{N}}$, such that
	 $\focussed{N} = \focussed{\mathcal{C}_{N}}[H]$ and
	 $\focussed{\mathcal{C}_{N'}}[H'] \tobul
	 \focussed{\mathcal{C}_N}[H']$.

	 In the search transition
	 $\focussed{\mathcal{C}_{N'}}[H] \tobul \focussed{N}$,
	 the only change happens to the focus and its incoming or
	 outgoing edge $e$ in the state
	 $\focussed{\mathcal{C}_{N'}}[H]$.
	 By the points (A) and (B), the focus is not entering nor
	 exiting in the context $\focussed{\mathcal{C}_{N'}}$, which
	 means the edge $e$ must be from the context, not from $H$.

	 Now that no edge from $H$ is changed in
	 $\focussed{\mathcal{C}_{N'}}[H] \tobul \focussed{N}$,
	 there exists
	 a focussed simple context
	 $\focussed{\mathcal{C}_N}$ such that
	 $\focussed{N} = \focussed{\mathcal{C}_{N}}[H]$, and moreover,
	 $\focussed{\mathcal{C}_{N'}}[H'] \tobul
	 \focussed{\mathcal{C}_N}[H']$.
	\end{proof}
 \end{itemize}
\end{proof}

\section{Stable hypernets} \label{app:stable-hypernets}

\begin{defi}[Accessible paths]
 \noindent
 \begin{itemize}
  \item A path of a hypernet is said to be \emph{accessible} if it
	consists of edges whose all sources have type~$\star$.
  \item An accessible path is called \emph{stable} if the labels
	of its edges are included in
	$\{ \mathsf{I} \} \cup \Opr_\checkmark$.
  \item An accessible path is called \emph{active} if it starts with
	one active operation edge and possibly followed by a stable
	path.
 \end{itemize}
\end{defi}

A stable hypernet always has at least one edge, and any non-output
vertex is labelled with $\star$. It has a tree-like shape.
\begin{lem}[Shape of stable hypernets]
 \label{lem:StableShape}
 \begin{enumerate}
  \item \label{item:StableUniquePath}
	In any stable hypernet, if a vertex $v'$ is reachable from
	another
	vertex $v$ such that $v \neq v'$, there exists a unique path
	from the vertex $v$ to the vertex $v'$.
  \item \label{item:StableNoCycle}
	Any stable hypernet has no cyclic path, i.e.\ a path from a
	vertex to itself.
  \item \label{item:StableMaximum}
	Let $\mathcal{C} : \star \To \otimes_{i=1}^m \ell_i$ be a
	simple context such that: its hole has
	one source and at least one outgoing edge; and its
	unique input is the hole's source.
	There are no two stable hypernets $G$ and $G'$ that satisfy
	$G = \mathcal{C}[G']$.
 \end{enumerate}
\end{lem}
\begin{proof}
 To prove the point~(\ref{item:StableUniquePath}), assume there are
 two different paths from the vertex $v$ to the vertex $v'$. These
 paths, i.e.\ non-empty sequences of edges, have to involve an edge
 with more than one source, or two different edges that share the same
 target. However, neither of these is possible in a stable hypernet,
 because both a passive operation edge and an instance edge have only
 one source and vertices can have at most one incoming edge. The
 point~(\ref{item:StableUniquePath}) follows from this by
 contradiction.

 If a stable hypernet has a cyclic path from a vertex $v$ to
 itself, there must be infinitely many paths from the input to the
 vertex $v$, depending on how many times the cycle is included. This
 contradicts the point~(\ref{item:StableUniquePath}).

 The point~(\ref{item:StableMaximum}) is also proved by contradiction.
 Assume that there exist two stable hypernets $G$ and $G'$ that
 satisfy
 $G = \mathcal{C}[G']$ for the simple context $\mathcal{C}$.
 In the stable hypernet $G$, a vertex is always labelled with $\star$
 if it is not an output. However, in the simple context
 $\mathcal{C}$, there exists at least one target of the hole that is
 not an output of the context but not labelled
 with $\star$ either. This contradicts $\mathcal{C}[G']$ being a
 stable hypernet.
\end{proof}

\begin{lem}
 \label{lem:StableActivePathAcyclic}
 For any state $\focussed{N}$, and its vertex $v$, such that
 the vertex $v$ is not a target of an instance edge or a passive
 operation edge, if an accessible path from the vertex $v$ is stable
 or active, then the path has no multiple occurrences of a single
 edge.
\end{lem}
\begin{proof}
 Any stable or active path consists of edges that has only one
 source.
 As a consequence, except for the first edge, no edge appears twice in
 the stable path.
 If the stable path is from the vertex $v$, its first edge also
 does not appear twice, because $v$ is not a target of an instance
 edge or a passive operation edge.
\end{proof}

\begin{lem}
 \label{lem:StablePathStableNet}
 For any state $\focussed{N}$, and its vertex $v$, such that
 the vertex $v$ is not a target of an instance edge or a passive
 operation edge, the following are equivalent.

 (A) There exist a focussed simple context
 $\focussed{\mathcal{C}}[\chi]$ and a stable hypernet $G$, such that
 $\focussed{N} = \focussed{\mathcal{C}}[G]$, where the vertex $v$ of
 $\focussed{N}$ corresponds to a unique source of the hole edge in
 $\focussed{\mathcal{C}}$.

 (B) Any accessible path from the vertex $v$ in $\focussed{N}$ is a
 stable path.
\end{lem}
\begin{proof}[Proof of (A) $\To$ (B)]
 Because no output of a stable hypernet has
 type $\star$, any path from the vertex $v$ in
 $\focussed{\mathcal{C}}[G]$ gives a path from the unique input
 in $G$.
 In the stable hypernet $G$, any path from the unique input is
 a stable path.
\end{proof}
\begin{proof}[Proof of (B) $\To$ (A)]
 In the state $\focussed{N}$, the focus target has to be a source of
 an edge, which forms an accessible path itself.
 By \autoref{lem:StableActivePathAcyclic},
 in the state $\focussed{N}$, we can take maximal stable paths
 from the vertex $v$, in the sense that appending any edge to these
 paths, if possible, does not give a stable path.

 If any of these maximal stable paths is to some
 vertex, the vertex does not have type $\star$; this can be confirmed
 as follows.
 If the vertex has type $\star$, it is not an output, so it is a
 source of an instance, focus, operation or contraction edge.
 The case of an instance or passive operation edge contradicts the
 maximality.
 The other case yields a non-stable accessible path that
 contradicts the assumption (B).

 Collecting all edges contained by the maximal stable paths,
 therefore, gives the desired hypernet $G$.
 These edges are necessarily all shallow, because of the vertex $v$ of
 $\focussed{N}$.
 The focussed context $\focussed{\mathcal{C}}[\chi]$, whose hole is
 shallow, can be made of
 all the other edges (at any depth) of the state $\focussed{N}$.
\end{proof}

\begin{lem}
 \label{lem:SwapTokenOprStableActivePath}
 Let $\focussed{N}$ be a state, where the focus is an incoming edge of
 an operation edge $e$, whose label $\phi$ takes at least one eager
 arguments.
 Let $k$ denote the number of eager arguments of $\phi$.

 For each $i \in \{ 1,\ldots,k \}$, let
 $\mathit{sw}_i(\focussed{N})$ be a state such that:
 both states
 $\mathit{sw}_i(\focussed{N})$ and $\focussed{N}$ have the same focus
 label and the same underlying hypernet, and
 the focus in $\mathit{sw}_i(\focussed{N})$ is the $i$-th outgoing
 edge of the operation edge $e$.

 For each $i \in \{ 1,\ldots,k \}$, the following are equivalent.

 (A) In $\focussed{N}$,
 any accessible path from an $i$-th target of the operation edge
 $e$ is a stable (resp.\ active) path.

 (B) In $\mathit{sw}_i(\focussed{N})$,
 any accessible path from the focus target
 is a stable (resp.\ active) path.
\end{lem}
\begin{proof}
 The only difference between $\focussed{N}$ and
 $\mathit{sw}_i(\focussed{N})$ is the swap of the focus with the
 operation edge $e$,
 and these two edges form an accessible
 path in the states $\focussed{N}$ and
 $\mathit{sw}_i(\focussed{N})$, individually or together
 (in an appropriate order).
 Therefore, there is one-to-one correspondence between
 accessible paths from an $i$-th target of the edge $e$
 in $\focussed{N}$, and
 accessible paths from the focus target in
 $\mathit{sw}_i(\focussed{N})$.

 When (A) is the case, in $\focussed{N}$,
 any accessible paths from an $i$-th target of the edge $e$
 does not contain the focus nor the edge $e$; otherwise there would be
 an accessible path that contains the focus and hence not stable nor
 active, which is a contradiction.
 This means that, in $\mathit{sw}_i(\focussed{N})$,
 any accessible path from the focus target also
 does not contain the focus nor the edge $e$, and the path must
 be a stable (resp.\ active) path.

 When (B) is the case, the proof takes the same reasoning in the
 reverse way.
\end{proof}

\begin{lem}
 \label{lem:AnswerAccessiblePaths}
 Let $\focussed{N}$ be a rooted state with the $?$-focus, such that
 the focus is not an incoming edge of a contraction edge.
 \begin{enumerate}
  \item \label{item:ValueToken}
	$\focussed{N} \tobul^+
	\langle \focussed{N} \rangle_{\checkmark/?}$, if and only if
	any accessible path from the focus target in $\focussed{N}$
	is a stable path.
  \item \label{item:RewriteToken}
	$\focussed{N} \tobul^+
	\langle \focussed{N} \rangle_{\lightning/?}$, if and only if
	any accessible path from the focus target in $\focussed{N}$
	is an active path.
 \end{enumerate}
\end{lem}
\begin{proof}[Proof of the forward direction]
 Let $\mathsf{t}$ be either `$\checkmark$' or `$\lightning$'.
 The assumption is
 $\focussed{N} \tobul^*
 \langle \focussed{N} \rangle_{\mathsf{t}/?}$.
 We prove the following, by induction on the length $n$ of this search
 sequence:
 \begin{itemize}
  \item any accessible path from the focus target in $\focussed{N}$
	is a stable path,
	when $\mathsf{t} = \mathord{\checkmark}$, and
  \item any accessible path from the focus target in $\focussed{N}$
	is an active path,
	when $\mathsf{t} = \mathord{\lightning}$.
 \end{itemize}

 In the base case, where $n = 1$, because the focus is not an incoming
 edge of a contraction edge,
 the focus target is a source of an
 instance edge, or an operation edge labelled with
 $\phi \in \Opr_\mathsf{t}$ that takes no eager argument.
 In either situation, the outgoing edge of the focus
 gives the only possible accessible path from the
 focus target. The path is stable when
 $\mathsf{t} = \mathord{\checkmark}$, and active when
 $\mathsf{t} = \mathord{\lightning}$.

 In the inductive case, where $n > 1$, the focus target is a source of
 an operation edge $e_\phi$ labelled with an operation
 $\phi \in \Opr_\mathsf{t}$ that takes at least one eager argument.

 Let $k$ denote the number of eager arguments of $\phi_\mathsf{t}$,
 and $i$ be an arbitrary number in $\{ 1,\ldots,k \}$.
 Let $\mathit{sw}_i(\focussed{N})$ be the state as defined in
 \autoref{lem:SwapTokenOprStableActivePath}.
 Because $\focussed{N}$ is rooted, by \autoref{lem:SearchSeqQueried},
 the given search sequence gives the following search sequence
 (proof by induction on $k-i$):
 \begin{equation*}
  ?;|\focussed{N}| \tobul^*
  \focussed{N} \tobul^+
   \mathit{sw}_i(\focussed{N}) \tobul^+
   \langle \mathit{sw}_i(\focussed{N}) \rangle_{\checkmark/?} \tobul^+
   \langle \focussed{N} \rangle_{\mathsf{t}/?}.
 \end{equation*}
 By induction hypothesis on the intermediate sequence
 $\mathit{sw}_i(\focussed{N}) \tobul^+
 \langle \mathit{sw}_i(\focussed{N}) \rangle_{\checkmark/?}$,
 any accessible path from the focus target in
 $\mathit{sw}_i(\focussed{N})$ is a stable path.
 By \autoref{lem:SwapTokenOprStableActivePath},
 any accessible path from an $i$-th target of the operation edge
 $e_\phi$ in $\focussed{N}$ is a stable path.

 In $\focussed{N}$, any accessible path from the focus
 target is given by the operation edge $e_\phi$ followed by an
 accessible path, which is proved to be stable above, from a target of
 $e_\phi$.
 Any accessible path from the focus target is therefore stable when
 $\mathsf{t} = \mathord{\checkmark}$, and active when
 $\mathsf{t} = \mathord{\lightning}$. 
\end{proof}
\begin{proof}[Proof of the backward direction]
 Let $\mathsf{t}$ be either `$\checkmark$' or `$\lightning$'.
 The assumption is the following:
 \begin{itemize}
  \item any accessible path from the focus target in $\focussed{N}$
	is a stable path,
	when $\mathsf{t} = \mathord{\checkmark}$, and
  \item any accessible path from the focus target in $\focussed{N}$
	is an active path,
	when $\mathsf{t} = \mathord{\lightning}$.
 \end{itemize}
 Our goal is to show
 $\focussed{N} \tobul^*
 \langle \focussed{N} \rangle_{\mathsf{t}/?}$.

 In the state $\focussed{N}$, the focus target has to be a source of
 an edge, which forms an accessible path itself.
 By \autoref{lem:StableActivePathAcyclic}, we can define
 $r(\focussed{N})$ by the maximum length of stable paths from the
 focus target.
 This number $r(\focussed{N})$ is well-defined and positive.
 We prove
 $\focussed{N} \tobul^*
 \langle \focussed{N} \rangle_{\mathsf{t}/?}$
 by induction on $r(\focussed{N})$.

 In the base case, where $r(\focussed{N}) = 1$,
 the outgoing edge of the focus is the only possible accessible path
 from the focus target.
 The outgoing edge is not a contraction edge by the assumption, and
 hence it is an instance edge, or an operation edge labelled with
 $\phi \in \Opr_\mathsf{t}$ that takes no eager argument.
 We have
 $\focussed{N} \tobul
 \langle \focussed{N} \rangle_{\mathsf{t}/?}$.
 
 In the inductive case, where $r(\focussed{N}) > 1$, the outgoing edge
 of the focus is an operation edge $e_\phi$ labelled with
 $\phi \in \Opr_\mathsf{t}$ that takes at least one eager argument.
 Any accessible path from the focus target in $\focussed{N}$ is
 given by the edge $e_\phi$ followed by a stable path from a
 target of \nolinebreak $e_\phi$.

 Let $k$ denote the number of eager arguments of $\phi_\mathsf{t}$,
 and $i$ be an arbitrary number in $\{ 1,\ldots,k \}$.
 Let $\mathit{sw}_i(\focussed{N})$ be the state as defined in
 \autoref{lem:SwapTokenOprStableActivePath}.

 By the assumption, any accessible path from an $i$-th
 target of the operation edge $e_\phi$ in $\focussed{N}$ is a
 stable path.
 Therefore by \autoref{lem:SwapTokenOprStableActivePath},
 in $\mathit{sw}_i(\focussed{N})$,
 any accessible path from the focus target is a stable
 path.
 Moreover, these paths in $\focussed{N}$ and
 $\mathit{sw}_i(\focussed{N})$ correspond to each other.
 By \autoref{lem:StableActivePathAcyclic}, we can define
 $r(\mathit{sw}_i(\focussed{N}))$
 by the maximum length of stable paths from the
 focus target.
 This number
 $r(\mathit{sw}_i(\focussed{N}))$ is well-defined, and
 satisfies
 $r(\mathit{sw}_i(\focussed{N})) < r(\focussed{N})$.
 By induction hypothesis on this number, we have:
 \begin{equation*}
  \mathit{sw}_i(\focussed{N}) \tobul^*
   \langle \mathit{sw}_i(\focussed{N}) \rangle_{\checkmark/?}.
 \end{equation*}
 Combining this search sequence with the following possible search
 transitions concludes the proof:
 \begin{gather*}
  \focussed{N} \tobul \mathit{sw}_1(\focussed{N}), \\
  \langle \mathit{sw}_i(\focussed{N}) \rangle_{\checkmark/?} \tobul
  \mathit{sw}_{i+1}(\focussed{N}), \\
  \tag{when $k \neq 1$ and $i < k$} \\
  \langle \mathit{sw}_k(\focussed{N}) \rangle_{\checkmark/?} \tobul
  \langle \focussed{N} \rangle_{\mathsf{t}/?}.
  \qedhere
 \end{gather*}
\end{proof}

\section{Parameterised (contextual) refinement and equivalence}

\begin{lem}
 \label{lem:BFCtxtClosedUnderPlugging}
 For any focus-free contexts
 $\mathcal{C}_1[\vec{\chi'},\chi,\vec{\chi''}]$ and $\mathcal{C}_2$
 such that
 $\mathcal{C}_1[\vec{\chi'},\mathcal{C}_2,\vec{\chi''}]$ is defined,
 if both $\mathcal{C}_1$ and $\mathcal{C}_2$ are binding-free, then
 $\mathcal{C}_1[\vec{\chi'},\mathcal{C}_2,\vec{\chi''}]$ is also
 binding-free.
\end{lem}
\begin{proof}
 Let $\mathcal{C}$ denote
 $\mathcal{C}_1[\vec{\chi'},\mathcal{C}_2,\vec{\chi''}]$,
 and $e_\chi$ denote the hole edge of $\mathcal{C}_1$ labelled with
 $\chi$.

 The proof is by contradiction. We assume that there exists a path $P$
 in $\mathcal{C}$, from a source of a contraction, atom, box or hole
 edge $e$, to a source of a hole edge $e'$.
 We derive a contradiction by case analysis on the path $P$.
 \begin{itemize}
  \item When $e'$ comes from $\mathcal{C}_1$, and the path
	$P$ consists of edges from $\mathcal{C}_1$ only, the path $P$
	gives a path in $\mathcal{C}_1$ that contradicts
	$\mathcal{C}_1$ being binding-free.
  \item When $e'$ comes from $\mathcal{C}_1$, and the path
	$P$ contains an edge from $\mathcal{C}_2$, by finding the last
	edge from $\mathcal{C}_2$ in $P$, we can take a	suffix of $P$
	that gives a path from a target of the hole edge
	$e_\chi$ to a source of a hole edge, in $\mathcal{C}_1$.
	Adding the hole edge $e_\chi$ at the beginning yields a path
	in $\mathcal{C}_1$ that contradicts $\mathcal{C}_1$ being
	binding-free.
  \item When both $e$ and $e'$ come from $\mathcal{C}_2$, and the path
	$P$ gives a path in $\mathcal{C}_2$, this contradicts
	$\mathcal{C}_2$ being binding-free.
  \item When both $e$ and $e'$ come from $\mathcal{C}_2$, and the path
	$P$ does not give a single path in $\mathcal{C}_2$,
	there exists a path from a source of the hole edge $e_\chi$ to
	a source of the hole edge $e_\chi$, in $\mathcal{C}_1$.
	This path contradicts $\mathcal{C}_1$ being binding-free.
  \item When $e$ comes from $\mathcal{C}_1$ and $e'$ comes from
	$\mathcal{C}_2$, by finding the first edge from
	$\mathcal{C}_2$ in $P$, we can take a prefix of $P$ that
	gives a path from a source of a contraction, atom,
	box or hole edge to a source of the hole edge $e_\chi$, in
	$\mathcal{C}_1$.
	This path contradicts $\mathcal{C}_1$ being binding-free.
	\qedhere
 \end{itemize}
\end{proof}

\begin{lem}
 \label{lem:QReflTransSym}
 For any set $\mathbb{C}$ of contexts that is closed under plugging,
 and any preorder $Q$ on natural numbers, the following holds.
 \begin{itemize}
  \item $\dotrel{\preceq}_Q$ and $\preceq^\mathbb{C}_Q$ are reflexive.
  \item $\dotrel{\preceq}_Q$ and $\preceq^\mathbb{C}_Q$ are
	transitive.
  \item $\dotrel{\simeq}_Q$ and $\simeq^\mathbb{C}_Q$ are
	equivalences.
 \end{itemize}
\end{lem}
\begin{proof}
 Because $\dotrel{\simeq}_Q$ and $\simeq^\mathbb{C}_Q$ are defined as
 a symmetric subset of
 $\dotrel{\preceq}_Q$ and $\preceq^\mathbb{C}_Q$, respectively,
 $\dotrel{\simeq}_Q$ and $\simeq^\mathbb{C}_Q$ are equivalences if
 $\dotrel{\preceq}_Q$ and $\preceq^\mathbb{C}_Q$ are preorders.

 Reflexivity and transitivity of $\dotrel{\preceq}_Q$ is a direct
 consequence of those of the preorder $Q$.

 For any focus-free hypernet $H$, and any focus-free context
 $\mathcal{C}[\chi] \in \mathbb{C}$ such that $?; \mathcal{C}[H]$ is a
 state,
 ${? ; \mathcal{C}[H]} \dotrel{\preceq}_Q {? ; \mathcal{C}[H]}$
 because of reflexivity of $\dotrel{\preceq}_Q$.

 For any focus-free hypernets $H_1$, $H_2$ and $H_3$, and any
 focus-free context $\mathcal{C}[\chi] \in \mathbb{C}$, such that
 $H_1 \preceq^\mathbb{C}_Q H_2$, $H_2 \preceq^\mathbb{C}_Q H_3$, and
 both
 $? ; \mathcal{C}[H_1]$ and $? ; \mathcal{C}[H_3]$ are states,
 our goal is to show
 ${? ; \mathcal{C}[H_1]} \dotrel{\preceq}_Q {? ; \mathcal{C}[H_3]}$.
 Because $H_1 \preceq^\mathbb{C}_Q H_2$ and
 $H_2 \preceq^\mathbb{C}_Q H_3$, all three
 hypernets $H_1$, $H_2$ and $H_3$ have the same type, and hence
 $? ; \mathcal{C}[H_2]$ is also a state.
 Therefore, we have
 ${? ; \mathcal{C}[H_1]} \dotrel{\preceq}_Q {? ; \mathcal{C}[H_2]}$
 and
 ${? ; \mathcal{C}[H_2]} \dotrel{\preceq}_Q {? ; \mathcal{C}[H_3]}$,
 and the transitivity of $\dotrel{\preceq}_Q$
 implies
 ${? ; \mathcal{C}[H_1]} \dotrel{\preceq}_Q {? ; \mathcal{C}[H_3]}$.
\end{proof}

\begin{lem}
 \label{lem:BoundedIdentityCtxtEquiv}
 For any set $\mathbb{C}$ of contexts that is closed under plugging,
 and any preorder $Q$ on natural numbers, the following holds.
 \begin{enumerate}
  \item For any hypernets $H_1$ and $H_2$,
	$H_1 \simeq^\mathbb{C}_{Q \cap Q^{-1}} H_2$ implies
	$H_1 \simeq^\mathbb{C}_{Q} H_2$.
  \item If all compute transitions are deterministic,
	for any hypernets $H_1$ and $H_2$,
	$H_1 \simeq^\mathbb{C}_{Q} H_2$ implies
	$H_1 \simeq^\mathbb{C}_{Q \cap Q^{-1}} H_2$.
 \end{enumerate}
\end{lem}
\begin{proof}
 Because $(Q \cap Q^{-1}) \subseteq Q$, the point (1) follows from the
 monotonicity of contextual equivalence.

 For the point (2),
 $H_1 \simeq^\mathbb{C}_{Q} H_2$ means that any
 focus-free context $\mathcal{C}[\chi] \in \mathbb{C}$, such that
 $? ; \mathcal{C}[H_1]$ and $? ; \mathcal{C}[H_2]$ are states,
 yields
 ${? ; \mathcal{C}[H_1]} \dotrel{\preceq}_{Q}
 {? ; \mathcal{C}[H_2]}$ and
 ${? ; \mathcal{C}[H_2]} \dotrel{\preceq}_{Q}
 {? ; \mathcal{C}[H_1]}$.
 If the state ${? ; \mathcal{C}[H_1]}$ terminates at a final state
 after $k_1$ transitions, there exists $k_2$ such that
 $k_1 \mathrel{Q} k_2$
 and the state ${? ; \mathcal{C}[H_2]}$ terminates at a final state
 after $k_2$ transitions.
 Moreover, there exists $k_3$ such that $k_2 \mathrel{Q} k_3$
 and the state ${? ; \mathcal{C}[H_1]}$ terminates at a final state
 after $k_3$ transitions.

 Because search transitions and copy transitions are deterministic, if
 all compute transitions are deterministic, states and transitions
 comprise a deterministic abstract rewriting system, in which final
 states are normal forms. By \autoref{lem:DeterministicAbstRWSystem},
 $k_1 = k_3$ must hold.
 This means $k_1 \mathrel{Q \cap Q^{-1}} k_2$, and
 ${? ; \mathcal{C}[H_1]} \dotrel{\preceq}_{Q \cap Q^{-1}}
 {? ; \mathcal{C}[H_2]}$.
 Similarly, we can infer
 ${? ; \mathcal{C}[H_2]} \dotrel{\preceq}_{Q \cap Q^{-1}}
 {? ; \mathcal{C}[H_1]}$,
 and hence
 $H_1 \simeq^\mathbb{C}_{Q \cap Q^{-1}} H_2$.
\end{proof}

\section{Proof of \autoref{thm:MetaThm}}
\label{sec:ProofMetaThm}

This section details the coinductive proof of \autoref{thm:MetaThm},
with respect to the UAM $\UAM(\Opr,B_\Opr)$ parameterised by $\Opr$
and $B_\Opr$.

At the core of the proof is step-wise reasoning,
or transition-wise reasoning, using a lax variation of simulation dubbed \emph{counting simulation}.
Providing a simulation boils down to case analysis on transitions,
namely on possible interactions between the focus and parts of states
contributed by a pre-template.
While output-closure helps us disprove some cases,
input-safety and robustness give the cases that are specific to a
pre-template and an operation set.

We also employ the so-called \emph{up-to} technique in the use of
quasi-specimens. We use counting simulations up to state
refinements, with a quantitative restriction implemented by the notion
of reasonable triple. This restriction is essential to make
this particular up-to technique work, in combination with counting simulations.
A similar form of up-to technique is studied categorically
by Bonchi et al.~\cite{DBLP:journals/acta/BonchiPPR17}, but for the ordinary notion
of (weak) simulation, without this quantitative restriction.


The counting simulation up-to we use is namely
\emph{$(Q,Q',Q'')$-simulation}, parameterised by a triple $(Q,Q',Q'')$.
This provides a sound approach to prove state
refinement $\dotrel{\preceq}_Q$, using $\dotrel{\preceq}_{Q'}$ and
$\dotrel{\preceq}_{Q''}$,
given that all transitions are deterministic and $(Q,Q',Q'')$ forms a
reasonable triple.
\begin{defi}[$(Q,Q',Q'')$-simulations]
 \label{def:QSimulation}
 Let $R$ be a binary relation on states, and $(Q,Q',Q'')$ be a triple
 of preorders on $\N$.
       The binary relation $R$ is a
 \emph{$Q$-counting simulation up-to $(Q',Q'')$} (\emph{$(Q,Q',Q'')$-simulation} in short)
	if, for any two related states
	$\focussed{G_1} \mathrel{R} \focussed{G_2}$, the following
	(A) and (B) hold:
\\[1.5ex]
	(A) If $\focussed{G_1}$ is final, $\focussed{G_2}$ is also
	final.
\\[1.5ex]
	(B) If there exists a state $\focussed{G'_1}$ such that
	$\focussed{G_1} \to \focussed{G'_1}$, one of the following
	(I) and (II) holds:
\\[1.5ex]
	\mbox{}\quad (I)
	There exists a stuck state $\focussed{G''_1}$ such that
	$\focussed{G'_1} \to^* \focussed{G''_1}$.
\\[1.5ex]
	\mbox{}\quad (II)
	There exist two states $\focussed{H_1}$ and $\focussed{H_2}$,
	and numbers $k_1,k_2 \in \N$, such that \\
	$\focussed{H_1} \mathrel{(\mathord{\dotrel{\preceq}_{Q'}}
	\circ R \circ \mathord{\dotrel{\preceq}_{Q''}})}
	\focussed{H_2}$,
	$(1 + k_1) \mathrel{Q} k_2$,
	$\focussed{G'_1} \to^{k_1} \focussed{H_1}$, and
	$\focussed{G_2} \to^{k_2} \focussed{H_2}$.
\end{defi}

\begin{prop}
 \label{prop:SimulationToRefinement}
 When the universal abstract machine $\UAM(\Opr,B_\Opr)$ is
 deterministic, it satisfies the following.

 For any binary relation $R$ on states, and any reasonable triple
 $(Q,Q',Q'')$,
 if $R$ is a $(Q,Q',Q'')$-simulation,
 then $R$ implies refinement up to $Q$, i.e.\ any
 $\focussed{G_1} \mathrel{R} \focussed{G_2}$ implies
 $\focussed{G_1} \dotrel{\preceq}_Q \focussed{G_2}$.

\end{prop}
\begin{proof}
 Our goal is to show the following:
 for any states $\focussed{G_1} \mathrel{R} \focussed{G_2}$,
 any number $k_1 \in \N$ and any final state $\focussed{N_1}$,
 such that $\focussed{G_1} \to^{k_1} \focussed{N_1}$,
 there exist a number $k_2 \in \N$ and a final state
 $\focussed{N_2}$ such that $k_1 \mathrel{Q} k_2$ and
 $\focussed{G_2} \to^{k_2} \focussed{N_2}$.
 The proof is by induction on $k_1 \in \N$.

 In the base case, when $k_1 = 0$, the state $\focussed{G_1}$ is
 itself final because $\focussed{G_1} = \focussed{N_1}$.
 Because $R$ is a $(Q,Q',Q'')$-simulation, $\focussed{G_2}$ is also a
 final state, which means we can take $0$ as $k_2$ and
 $\focussed{G_2}$ itself as $\focussed{N_2}$.
 Because $(Q,Q',Q'')$ is a reasonable triple, $Q$ is a preorder and
 $0 \mathrel{Q} 0$ holds.

 In the inductive case, when $k_1 > 0$, we assume the induction
 hypothesis on any $h \in \N$ such that $h < k_1$.
 Now that $k_1 > 0$, there exists a state $\focussed{G'_1}$ such that
 $\focussed{G_1} \to \focussed{G'_1} \to^{k-1} \focussed{N_1}$.
 Because all intrinsic transitions are deterministic, the assumption
 that compute transitions are all deterministic implies that states
 and transitions
 comprise a deterministic abstract rewriting system, in which final
 states and stuck states are normal forms. By
 \autoref{lem:DeterministicAbstRWSystem}, we can conclude that
 there exists no stuck state $\focussed{G''_1}$ such that
 $\focussed{G'_1} \to^* \focussed{G''_1}$.

 Therefore, by $R$ being a $(Q,Q',Q'')$-simulation, there exist
 two states $\focussed{H_1}$ and $\focussed{H_2}$, and numbers
 $l_1,l_2 \in \N$, such that
 $\focussed{H_1}
 \mathrel{(\mathord{\dotrel{\preceq}_{Q'}} \circ R \circ
 \mathord{\dotrel{\preceq}_{Q''}})} \focussed{H_2}$,
 $(1 + l_1) \mathrel{Q} l_2$,
 $\focussed{G'_1} \to^{l_1} \focussed{H_1}$, and
 $\focussed{G_2} \to^{l_2} \focussed{H_2}$.
 By the determinism, $1 + l_1 \leq k_1$ must hold; if $\focussed{H_1}$
 is a final state, $\focussed{G'_1} \to^{l_1} \focussed{H_1}$ must
 coincide with $\focussed{G'_1} \to^{k-1} \focussed{N_1}$; otherwise,
 $\focussed{G'_1} \to^{l_1} \focussed{H_1}$ must be a suffix of
 $\focussed{G'_1} \to^{k-1} \focussed{N_1}$.
 There exist two states $\focussed{H_3}$ and $\focussed{H_4}$, and we
 have the following situation,
 where the relations $R$, $\dotrel{\preceq}_{Q'}$ and
 $\dotrel{\preceq}_{Q''}$ are represented
 by vertical dotted lines from top to bottom.
 \begin{equation*}
  \xymatrix@R=2ex{
   {\focussed{G_1}} \ar[r] \ar@{.}[ddd]_{R}
   & {\focussed{G'_1}} \ar[r]^>{l_1}
   \ar@{}[ddd]|*+[F]{(1 + l_1)\,Q\,l_2}
   & *+<1em>{\focussed{H_1}} \ar[r]^>{k_1 - 1 - l_1}
   \ar@{.}[d]^{\dotrel{\preceq}_{Q'}}
   & *+<3em>{\focussed{N_1}} \\
  && {\focussed{H_3}} \ar@{.}[d]^{R} \\
  && {\focussed{H_4}} \ar@{.}[d]^{\dotrel{\preceq}_{Q''}} \\
  {\focussed{G_2}} \ar[rr]^>{l_2}
   && *+<1em>{\focussed{H_2}}
   }
 \end{equation*}

 We expand the above diagram as below (indicated by magenta), in three
 steps.
 \begin{equation*}
  \xymatrix@R=2ex{
   {\focussed{G_1}} \ar[r] \ar@{.}[ddd]_{R}
   & {\focussed{G'_1}} \ar[r]^>{l_1}
   \ar@{}[ddd]|*+[F]{(1 + l_1)\,Q\,l_2}
   & *+<1em>{\focussed{H_1}} \ar[r]^>{k_1 - 1 - l_1}
   \ar@{.}[d]^{\dotrel{\preceq}_{Q'}}
   & *+<3em>{\focussed{N_1}}
   \ar@{}[d]|*+[magenta][F:magenta]{(k_1 - 1 - l_1)\,Q'\,m_3} \\
  && {\focussed{H_3}} \ar@{.}[d]^{R} \ar@[magenta][r]^>*[magenta]{m_3}
   & *+<3em>[magenta]{\focussed{N_3}}
   \ar@{}[d]|*+[magenta][F:magenta]{m_3\,Q\,m_4} \\
  && {\focussed{H_4}} \ar@{.}[d]^{\dotrel{\preceq}_{Q''}}
   \ar@[magenta][r]^>*[magenta]{m_4}
   & *+<3em>[magenta]{\focussed{N_4}}
   \ar@{}[d]|*+[magenta][F:magenta]{m_4\,Q''\,m_2} \\
  {\focussed{G_2}} \ar[rr]^>{l_2}
   && *+<1em>{\focussed{H_2}} \ar@[magenta][r]^>*[magenta]{m_2}
   & *+<3em>[magenta]{\focussed{N_2}}
   }
 \end{equation*}
 Firstly, by definition of state refinement,
 there exist a number $m_3 \in \N$ and a final state $\focussed{N_3}$
 such that $(k_1 - 1 - l_1) \mathrel{Q'} m_3$ and
 $\focussed{H_3} \to^{m_3} \focussed{N_3}$.
 Because $(Q,Q',Q'')$ is a reasonable triple, $Q' \subseteq \geq_\N$,
 and hence $k_1 > {k_1 - 1 - l_1} \geq m_3$.
 Therefore, secondly, by induction hypothesis on $m_3$,
 there exist a number $m_4 \in \N$ and a final state $\focussed{N_4}$
 such that $m_3 \mathrel{Q} m_4$ and
 $\focussed{H_4} \to^{m_4} \focussed{N_4}$.
 Thirdly, by definition of state refinement,
 there exist a number $m_2 \in \N$ and a final state $\focussed{N_2}$
 such that $m_4 \mathrel{Q''} m_2$ and
 $\focussed{H_2} \to^{m_2} \focussed{N_2}$.

 Now we have $(k_1 - 1 - l_1) \mathrel{Q'} m_3$,
 $m_3 \mathrel{Q} m_4$ and
 $m_4 \mathrel{Q''} m_2$, which means
 $(k_1 - 1 - l_1) \mathrel{(Q' \circ Q \circ Q'')} m_2$.
 Because $(Q,Q',Q'')$ is a reasonable triple, this implies
 $(k_1 - 1 - l_1) \mathrel{Q} m_2$, and moreover,
 $k_1 \mathrel{Q} (l_2 + m_2)$. We can take $l_2 + m_2$ as $k_2$.
\end{proof}

The focus in a focussed context $\focussed{\mathcal{C}}$ is said to be
\emph{remote}, if it is not the entering $\lightning$-focus.
The procedure of \emph{contextual lifting} reduces a proof of
contextual refinement down to that of state refinement.
\begin{defi}[Contextual lifting]
 Let
 $\mathbb{C} \subseteq \mathcal{H}_\omega(L, M_\Opr \cup \MVar)$
 be a set of contexts.
 Given a pre-template $\vartriangleleft$ on focus-free hypernets
 $\mathcal{H}_\omega(L, M_\Opr \backslash
 \{?,\checkmark,\lightning\})$,
 its \emph{$\mathbb{C}$-contextual lifting}
 $\overline{\vartriangleleft}^\mathbb{C}$ is a binary relation on
 states defined by:
 $\focussed{G_1} \mathrel{\overline{\vartriangleleft}^\mathbb{C}}
 \focussed{G_2}$
 if there exists a $\mathbb{C}$-specimen
 $(\focussed{\mathcal{C}};\vec{H^1};\vec{H^2})$ of $\vartriangleleft$,
 such that
 the focus of $\focussed{\mathcal{C}}$ is remote,
 $\focussed{G_p} = \focussed{\mathcal{C}}[\vec{H^p}]$,
 and $\focussed{\mathcal{C}}[\vec{H^p}]$ is rooted,
 for each $p \in \{ 1,2 \}$.
\end{defi}

The contextual lifting $\overline{\vartriangleleft}^\mathbb{C}$ is by
definition a binary relation on rooted states.

\begin{prop}
 \label{prop:StateToCtxt}
 For any set
 $\mathbb{C} \subseteq \mathcal{H}_\omega(L, M_\Opr \cup \MVar)$
 of contexts that is closed under plugging,
 any preorder $Q$ on $\N$,
 and any pre-template $\vartriangleleft$ on focus-free hypernets
 $\mathcal{H}_\omega(L, M_\Opr \backslash
 \{?,\checkmark,\lightning\})$,
 if the $\mathbb{C}$-contextual lifting
 $\overline{\vartriangleleft}^\mathbb{C}$ implies
 refinement $\dotrel{\preceq}_Q$
 (resp.\ equivalence $\dotrel{\simeq}_Q$), then
 $\vartriangleleft$ implies contextual refinement
 $\preceq^\mathbb{C}_Q$
 (resp.\ contextual equivalence $\simeq^\mathbb{C}_Q$).
\end{prop}
\begin{proof}[Proof of refinement case]
 Our goal is to show that, for any $H_1 \vartriangleleft H_2$ and any
 focus-free context $\mathcal{C}[\chi] \in \mathbb{C}$ such that
 ${?;\mathcal{C}[H_1]}$ and ${?;\mathcal{C}[H_2]}$ are states,
 we have refinement
 ${?;\mathcal{C}[H_1]} \dotrel{\preceq}_Q {?;\mathcal{C}[H_2]}$.

 Because
 $?;\mathcal{C}[H_p] = ?;(\mathcal{C}[H_p]) = (?;\mathcal{C})[H_p]$
 for $p \in \{1,2\}$,
 and $|?;\mathcal{C}| = \mathcal{C} \in \mathbb{C}$,
 the triple $((?;\mathcal{C});H_1;H_2)$ is a $\mathbb{C}$-specimen of
 $\vartriangleleft$ with the $?$-focus.
 Moreover the states $?;\mathcal{C}[H_1]$ and $?;\mathcal{C}[H_2]$ are
 trivially rooted.
 Therefore,
 ${?;\mathcal{C}[H_1]}
 \mathrel{\overline{\vartriangleleft}^\mathbb{C}}
 {?;\mathcal{C}[H_2]}$, and by the assumption,
 ${?;\mathcal{C}[H_1]} \dotrel{\preceq}_Q {?;\mathcal{C}[H_2]}$.
\end{proof}
\begin{proof}[Proof of equivalence case]
 It suffices to show that, for any $H_1 \vartriangleleft H_2$ and any
 focus-free context $\mathcal{C}[\chi] \in \mathbb{C}$ such that
 ${?;\mathcal{C}[H_1]}$ and ${?;\mathcal{C}[H_2]}$ are states,
 we have refinements
 ${?;\mathcal{C}[H_1]} \dotrel{\preceq}_Q {?;\mathcal{C}[H_2]}$ and
 ${?;\mathcal{C}[H_2]} \dotrel{\preceq}_Q {?;\mathcal{C}[H_1]}$,
 i.e.\ equivalence
 ${?;\mathcal{C}[H_1]} \dotrel{\simeq}_Q {?;\mathcal{C}[H_2]}$.

 Because
 $?;\mathcal{C}[H_p] = ?;(\mathcal{C}[H_p]) = (?;\mathcal{C})[H_p]$
 for $p \in \{1,2\}$,
 and $|?;\mathcal{C}| = \mathcal{C} \in \mathbb{C}$,
 the triple $((?;\mathcal{C});H_1;H_2)$ is a $\mathbb{C}$-specimen of
 $\vartriangleleft$ with the $?$-focus.
 Moreover the states $?;\mathcal{C}[H_1]$ and $?;\mathcal{C}[H_2]$ are
 trivially rooted.
 Therefore,
 ${?;\mathcal{C}[H_1]}
 \mathrel{\overline{\vartriangleleft}^\mathbb{C}}
 {?;\mathcal{C}[H_2]}$, and by the assumption,
 ${?;\mathcal{C}[H_1]} \dotrel{\simeq}_Q {?;\mathcal{C}[H_2]}$.
\end{proof}

\begin{lem}
 \label{lem:TemplateToSimulationSharedCases}
 For any set
 $\mathbb{C} \subseteq \mathcal{H}_\omega(L, M_\Opr \cup \MVar)$
 of contexts that is closed under plugging,
 any pre-template $\vartriangleleft$ on focus-free hypernets
 $\mathcal{H}_\omega(L, M_\Opr \backslash
 \{?,\checkmark,\lightning\})$,
 and any $\mathbb{C}$-specimen
 $(\focussed{\mathcal{C}}[\vec{\chi}];\vec{H^1};\vec{H^2})$ of
 $\vartriangleleft$, the following holds.
 \begin{enumerate}
  \item \label{item:Final}
	The state $\focussed{\mathcal{C}}[\vec{H^1}]$ is final
	(resp.\ initial)
	if and only if
	the state $\focussed{\mathcal{C}}[\vec{H^2}]$ is final
	(resp.\ initial).
  \item \label{item:RootedNotExitToken}
	If $\vartriangleleft$ is output-closed, and
	$\focussed{\mathcal{C}}[\vec{H^1}]$ and
	$\focussed{\mathcal{C}}[\vec{H^2}]$ are both rooted states,
	then the focus of $\focussed{\mathcal{C}}$ is not exiting.
  \item \label{item:ValueOrNonEnteringSearchToken}
	If $\vartriangleleft$ is output-closed,
	$\focussed{\mathcal{C}}[\vec{H^1}]$ and
	$\focussed{\mathcal{C}}[\vec{H^2}]$ are both rooted states,
	the focus of $\focussed{\mathcal{C}}$ is the $\checkmark$-focus or
	the non-entering $?$-focus, and
	a transition is possible from
	$\focussed{\mathcal{C}}[\vec{H^1}]$ or
	$\focussed{\mathcal{C}}[\vec{H^2}]$,
	then there exists a focussed context $\focussed{\mathcal{C}'}$
	with a remote focus such that
	$|\focussed{\mathcal{C}'}| = |\focussed{\mathcal{C}}|$ and
	$\focussed{\mathcal{C}}[\vec{H^p}] \to
	\focussed{\mathcal{C}'}[\vec{H^p}]$ for each
	$p \in \{ 1,2 \}$.
 \end{enumerate}
\end{lem}
\begin{proof}[Proof of point~(\ref{item:Final})]
 Let $(p,q)$ be an arbitrary element of a set $\{ (1,2), (2,1) \}$.
 If $\focussed{\mathcal{C}}[\vec{H^p}]$ is final (resp.\ initial), the
 focus source is
 an input in $\focussed{\mathcal{C}}[\vec{H^p}]$.
 Because input lists of
 $\focussed{\mathcal{C}}[\vec{H^p}]$, $\focussed{\mathcal{C}}$ and
 $\focussed{\mathcal{C}}[\vec{H^q}]$ all coincide,
 the focus source must be an input in $\focussed{\mathcal{C}}$, and
 in $\focussed{\mathcal{C}}[\vec{H^q}]$ too.
 This means $\focussed{\mathcal{C}}[\vec{H^q}]$ is also a final
 (resp.\ initial) state.
\end{proof}
\begin{proof}[Proof of point~(\ref{item:RootedNotExitToken})]
 This is a consequence of the contraposition of
 \autoref{lem:NotOprPathNotRooted}(\ref{item:ExitingSafe}).
\end{proof}
\begin{proof}[Proof of the
 point~(\ref{item:ValueOrNonEnteringSearchToken})]
 The transition possible from $\focussed{\mathcal{C}}[\vec{H^1}]$ or
 $\focussed{\mathcal{C}}[\vec{H^2}]$ is necessarily a search
 transition.
 By case analysis on the focus of $\focussed{\mathcal{C}}$, we can
 confirm that the search transition applies an interaction rule to the
 focus and an edge from $\focussed{\mathcal{C}}$.
 \begin{itemize}
  \item When the focus of $\focussed{\mathcal{C}}$ is the $\checkmark$-focus,
	the transition can only change the focus and its incoming
	operation edge.
	Because $\vartriangleleft$ is output-closed,
	by the point~(\ref{item:RootedNotExitToken}), the focus of
	$\focussed{\mathcal{C}}$ is not exiting.
	This implies that the incoming operation edge of the focus is
	from $\focussed{\mathcal{C}}$ in both states
	$\focussed{\mathcal{C}}[\vec{H^1}]$ and
	$\focussed{\mathcal{C}}[\vec{H^2}]$.
  \item When the focus of $\focussed{\mathcal{C}}$ is the non-entering
	$?$-focus, the transition can only change the focus and its
	outgoing edge.
	Because the focus is not entering in $\focussed{\mathcal{C}}$,
	the outgoing edge is from $\focussed{\mathcal{C}}$ in both
	states $\focussed{\mathcal{C}}[\vec{H^1}]$ and
	$\focussed{\mathcal{C}}[\vec{H^2}]$.
 \end{itemize}
 Therefore, there exist a focus-free simple context
 $\mathcal{C}_0[\chi,\vec{\chi}]$ and an interaction rule
 $\focussed{N_0} \mapsto \focussed{N'_0}$, such that
 $\focussed{\mathcal{C}} =
 \mathcal{C}_0[\focussed{N_0},\vec{\chi}]$,
 and $\mathcal{C}_0[\focussed{N'_0},\vec{\chi}]$ is a focussed
 context.

 Examining interaction rules confirms
 $|\focussed{N_0}| = |\focussed{N'_0}|$, and hence
 $|\focussed{\mathcal{C}}| =
 |\mathcal{C}_0[\focussed{N_0},\vec{\chi}]| =
 |\mathcal{C}_0[\focussed{N'_0},\vec{\chi}]|$.
 By definition of search transitions, we have:
 \begin{equation*}
  \focussed{\mathcal{C}}[\vec{H^p}] =
   \mathcal{C}_0[\focussed{N_0},\vec{H^p}] \to
   \mathcal{C}_0[\focussed{N'_0},\vec{H^p}]
 \end{equation*}
 for each $p \in \{ 1,2 \}$.

 The rest of the proof is to check that
 $\mathcal{C}_0[\focussed{N'_0},\vec{\chi}]$ has a remote focus,
 namely that, if its focus is the $\lightning$-focus, the focus is not
 entering. This is done by inspecting interaction rules.
 \begin{itemize}
  \item When the interaction rule
	$\focussed{N_0} \mapsto \focussed{N'_0}$ changes the $\checkmark$-focus
	to the $\lightning$-focus, this must be the interaction rule in \autoref{fig:interaction-opr-last},
	which means
	$\focussed{N'_0}$ consists of the $\lightning$-focus and its
	outgoing operation edge.
	The operation edge remains to be a (unique) outgoing edge of
	the focus in
	$\mathcal{C}_0[\focussed{N'_0},\vec{\chi}]$, and hence the
	focus is not entering in
	$\mathcal{C}_0[\focussed{N'_0},\vec{\chi}]$.
  \item When the interaction rule
	$\focussed{N_0} \mapsto \focussed{N'_0}$ changes the $?$-focus
	to the $\lightning$-focus, this must be the interaction rule
	in \autoref{fig:interaction-contr-L},~\ref{fig:interaction-contr-R},~\ref{fig:interaction-opr-none}, which means
	$\focussed{N'_0} =
	\langle \focussed{N_0} \rangle_{\lightning/?}$.
	Because the focus is not entering in
	$\mathcal{C}_0[\focussed{N_0},\vec{\chi}] =
	\focussed{\mathcal{C}}$,
	the focus is also not entering in
	$\mathcal{C}_0[\focussed{N'_0},\vec{\chi}] =
	\langle \mathcal{C}_0[\focussed{N_0},\vec{\chi}]
	\rangle_{\lightning/?}$.
	\qedhere
 \end{itemize} 
\end{proof}

\begin{prop}
 \label{prop:TemplateToSimulation}
 When the universal abstract machine $\UAM(\Opr,B_\Opr)$ is
 deterministic and refocusing, it satisfies the following,
 for any set
 $\mathbb{C} \subseteq \mathcal{H}_\omega(L, M_\Opr \cup \MVar)$
 of contexts that is closed under plugging,
 any reasonable triple $(Q,Q',Q'')$,
 and any pre-template $\vartriangleleft$ on focus-free hypernets
 $\mathcal{H}_\omega(L, M_\Opr \backslash
 \{?,\checkmark,\lightning\})$.
 \begin{enumerate}
  \item \label{item:forward2}
	If $\vartriangleleft$ is a $(\mathbb{C},Q,Q')$-template and
	$(\mathbb{C},Q,Q',Q'')$-robust relative to all rewrite
	transitions, then the $\mathbb{C}$-contextual lifting
	$\overline{\vartriangleleft}^\mathbb{C}$ is a
	$(Q,Q',Q'')$-simulation.
  \item \label{item:backward2}
	If $\vartriangleleft$ is a $(\mathbb{C},Q^{-1},Q')$-template
	and the converse $\vartriangleleft^{-1}$ is
	$(\mathbb{C},Q,Q',Q'')$-robust relative to all rewrite
	transitions, then the $\mathbb{C}$-contextual lifting
	$\overline{\vartriangleleft^{-1}}^\mathbb{C}$ of the converse
	is a $(Q,Q',Q'')$-simulation.
 \end{enumerate}
\end{prop}
\begin{proof}[Proof prelude]
 Let $(\focussed{\mathcal{C}};\vec{H^1};\vec{H^2})$ be an arbitrary
 $\mathbb{C}$-specimen of $\vartriangleleft$,
 such that the focus of $\focussed{\mathcal{C}}$ is remote, and
 $\focussed{G_p} := \focussed{\mathcal{C}}[\vec{H^p}]$ is
 a rooted state for each $p \in \{ 1,2 \}$.
 By definition of contextual lifting,
 $\focussed{G_1} \mathrel{\overline{\vartriangleleft}^\mathbb{C}}
 \focussed{G_2}$, and equivalently,
 $\focussed{G_2}
 \mathrel{(\overline{\vartriangleleft}^\mathbb{C})^{-1}}
 \focussed{G_1}$.
 Note that
 $\overline{\vartriangleleft^{-1}}^\mathbb{C} =
 (\overline{\vartriangleleft}^\mathbb{C})^{-1}$.

 Because $\vartriangleleft$ is output-closed,
 by
 \autoref{lem:TemplateToSimulationSharedCases}(\ref{item:RootedNotExitToken}),
 the focus is not exiting in $\focussed{\mathcal{C}}$.
 This implies that, if the focus has an incoming edge in
 $\focussed{G_1}$ or $\focussed{G_2}$, the incoming edge must be from
 $\focussed{\mathcal{C}}$.

 Because the machine is deterministic and refocusing, rooted states
 and transitions
 comprise a deterministic abstract rewriting system, in which final
 states and stuck states are normal forms. By
 \autoref{lem:DeterministicAbstRWSystem}, from any state, a sequence
 of transitions that result in a final state or a stuck state is
 unique, if any.

 Because $(Q,Q',Q'')$ is a reasonable triple, $Q'$ and $Q''$ are
 reflexive.
 By \autoref{lem:QReflTransSym}, this implies that
 $\dotrel{\preceq}_{Q'}$ and $\dotrel{\preceq}_{Q''}$ are reflexive,
 and hence
 $\overline{\vartriangleleft}^\mathbb{C} \subseteq
 {\mathord{\dotrel{\preceq}_{Q'}}
 \circ \overline{\vartriangleleft}^\mathbb{C}
 \circ \mathord{\dotrel{\preceq}_{Q''}}}$, and
 $(\overline{\vartriangleleft}^\mathbb{C})^{-1} \subseteq
 {\mathord{\dotrel{\preceq}_{Q'}}
 \circ (\overline{\vartriangleleft}^\mathbb{C})^{-1}
 \circ \mathord{\dotrel{\preceq}_{Q''}}}$.
\end{proof}
\begin{proof}[Proof of the point~(\ref{item:forward})]
 Our goal is to check conditions (A) and (B) of
 \autoref{def:QSimulation} for the states
 $\focussed{G_1} \mathrel{\overline{\vartriangleleft}^\mathbb{C}}
 \focussed{G_2}$.

 If $\focussed{G_1}$ is final, by
 \autoref{lem:TemplateToSimulationSharedCases}(\ref{item:Final}),
 $\focussed{G_2}$ is also final.
 The condition (A) of \autoref{def:QSimulation} is fulfilled.

 If there exists a state $\focussed{G'_1}$ such that
 $\focussed{G_1} \to \focussed{G'_1}$, we show that one of the
 conditions (I) and (II) of \autoref{def:QSimulation} is fulfilled,
 by case analysis of the focus in $\focussed{\mathcal{C}}$.
 \begin{itemize}
  \item When the focus is the $\checkmark$-focus,
	or the $?$-focus that is not entering,
	by
	\autoref{lem:TemplateToSimulationSharedCases}(\ref{item:ValueOrNonEnteringSearchToken}),
	there exists a focussed context
	$\focussed{\mathcal{C}'}$ with a remote focus, such that
	$|\focussed{\mathcal{C}'}| = |\focussed{\mathcal{C}}|$ and
	$\focussed{G_p} = \focussed{\mathcal{C}}[\vec{H^p}] \to
	\focussed{\mathcal{C}'}[\vec{H^p}]$ for each
	$p \in \{ 1,2 \}$.
	We have the following situation, namely the black part of the
	diagram below. Showing the magenta part confirms that
	the condition (II) of \autoref{def:QSimulation} is fulfilled.
	\begin{equation*}
	 \begin{gathered}
	  \xymatrix@R=2ex{
	  {\focussed{G_1}
	  = \focussed{\mathcal{C}}[\vec{H^1}]} \ar[r]
	  \ar@{.}[d]_{\overline{\vartriangleleft}^\mathbb{C}}
	  \ar@{}[dr]|*+[magenta][F:magenta]{1\,Q\,1}
	  & {\focussed{\mathcal{C}'}[\vec{H^1}]
	  \textcolor{magenta}{\text{$\,= \focussed{G'_1}$}}}
	  \ar@{.}@[magenta][d]^*[magenta]{
	  \overline{\vartriangleleft}^\mathbb{C}} \\
	  {\focussed{G_2}
	  = \focussed{\mathcal{C}}[\vec{H^2}]} \ar[r]
	  & {\focussed{\mathcal{C}'}[\vec{H^2}]}
	  }
	 \end{gathered}
	\end{equation*}
	By the determinism,
	$\focussed{\mathcal{C}'}[\vec{H^1}] = \focussed{G'_1}$.
	Because $(Q,Q',Q'')$ is a reasonable triple, $Q$ is a preorder
	and $1 \mathrel{Q} 1$.
	The context $\focussed{\mathcal{C}'}$ satisfies
	$|\focussed{\mathcal{C}'}| = |\focussed{\mathcal{C}}| \in
	\mathbb{C}$, so
	$(\focussed{\mathcal{C}'};\vec{H^1};\vec{H^2})$ is a
	$\mathbb{C}$-specimen of $\vartriangleleft$.
	The context $\focussed{\mathcal{C}'}$ has a remote focus, and
	the states
	$\focussed{\mathcal{C}'}[\vec{H^1}]$ and
	$\focussed{\mathcal{C}'}[\vec{H^2}]$ are both rooted.
	Therefore, we have
	$\focussed{\mathcal{C}'}[\vec{H^1}]
	\mathrel{\overline{\vartriangleleft}^\mathbb{C}}
	\focussed{\mathcal{C}'}[\vec{H^2}]$.
  \item When the focus is the $?$-focus that is entering in
	$\focussed{\mathcal{C}}$,
	because $\vartriangleleft$ is
	$(\mathbb{C},Q,Q')$-input-safe, we have one of the
	following three situations corresponding to (I), (II) and
	(III) of \autoref{def:InputSafe}.
	\begin{itemize}
	 \item There exist two stuck states $\focussed{N_1}$ and
	       $\focussed{N_2}$ such that
	       $\focussed{G_p} \to^* \focussed{N_p}$ for each
	       $p \in \{ 1,2 \}$.
	       By the determinism of transitions, we have
	       $\focussed{G_1} \to \focussed{G'_1}
	       \to^* \focussed{N_1}$, which means the condition (I) of
	       \autoref{def:QSimulation} is satisfied.
	 \item There exist a $\mathbb{C}$-specimen
	       $(\focussed{\mathcal{C}'};\vec{H'^1};\vec{H'^2})$
	       of $\vartriangleleft$
	       and two numbers $k_1,k_2 \in \N$, such that
	       the focus of
	       $\focussed{\mathcal{C}'}$ is the $\checkmark$-focus or the non-entering $?$-focus, 
	       $(1 + k_1) \mathrel{Q} k_2$,
	       $\focussed{\mathcal{C}}[\vec{H^1}] \to^{1 + k_1}
	       \focussed{\mathcal{C}'}[\vec{H'^1}]$, and
	       $\focussed{\mathcal{C}}[\vec{H^2}] \to^{k_2}
	       \focussed{\mathcal{C}'}[\vec{H'^2}]$.
	       By the determinism of transitions, we have the
	       following situation, namely the black part of the
	       diagram below.
	       Showing the magenta part confirms that
	       the condition (II) of \autoref{def:QSimulation} is
	       fulfilled.
	       \begin{equation*}
		\begin{gathered}
		 \xymatrix@R=2ex{
		 {\focussed{G_1}
		 = \focussed{\mathcal{C}}[\vec{H^1}]} \ar[r]
		 \ar@{.}[d]_{\overline{\vartriangleleft}^\mathbb{C}}
		 \ar@{}[drr]|*+[black][F:black]{(1 + k_1)\,Q\,k_2}
		 & {\focussed{G'_1}} \ar[r]^>{k_1}
		 & {\focussed{\mathcal{C}'}[\vec{H'^1}]}
		 \ar@{.}@[magenta][d]^*[magenta]{
		 \overline{\vartriangleleft}^\mathbb{C}} \\
		 {\focussed{G_2}
		 = \focussed{\mathcal{C}}[\vec{H^2}]}
		 \ar[rr]^>{k_2}
		 && {\focussed{\mathcal{C}'}[\vec{H'^2}]}
		 }
		\end{gathered}
	       \end{equation*}
	       The context $\focussed{\mathcal{C}'}$ has a remote
	       focus, and states
	       $\focussed{\mathcal{C}'}[\vec{H'^1}]$ and
	       $\focussed{\mathcal{C}'}[\vec{H'^2}]$ are rooted.
	       Therefore,
	       $\focussed{\mathcal{C}'}[\vec{H'^1}]
	       \mathrel{\overline{\vartriangleleft}^\mathbb{C}}
	       \focussed{\mathcal{C}'}[\vec{H'^2}]$.
	 \item There exist a quasi-$\mathbb{C}$-specimen
	       $(\focussed{N_1},\focussed{N_2})$ of $\vartriangleleft$
	       up to $(\dot{\simeq}_{Q'},\dot{\simeq}_{Q'})$, whose
	       focus is not the $\lightning$-focus,
	       and two numbers $k_1,k_2 \in \N$, such that
	       $(1 + k_1) \mathrel{Q} (1 + k_2)$,
	       $\focussed{\mathcal{C}}[\vec{H^1}] \to^{1 + k_1}
	       \focussed{N_1}$, and
	       $\focussed{\mathcal{C}}[\vec{H^2}] \to^{1 + k_2}
	       \focussed{N_2}$.
	       By the determinism of transitions,
	       we have the following situation, namely the black part
	       of the diagram below.
	       Showing the magenta part confirms that
	       the condition (II) of \autoref{def:QSimulation} is
	       fulfilled.
	       \begin{equation*}
		\begin{gathered}
		 \xymatrix@R=2ex@C=40pt{
		 {\focussed{G_1}
		 = \focussed{\mathcal{C}}[\vec{H^1}]} \ar[r]
		 \ar@{.}[d]_{\overline{\vartriangleleft}^\mathbb{C}}
		 \ar@{}[drr]|*+[black][F:black]{
		 (1 + k_1)\,Q\,(1 + k_2)}
		 & {\focussed{G'_1}} \ar[r]^>{k_1}
		 & {\focussed{N_1}}
		 \ar@{.}@[magenta][d]^*[magenta]{
		 \mathord{\dotrel{\preceq}_{Q'}}
		 \circ \overline{\vartriangleleft}^\mathbb{C}
		 \circ \mathord{\dotrel{\preceq}_{Q''}}} \\
		 {\focussed{G_2}
		 = \focussed{\mathcal{C}}[\vec{H^2}]}
		 \ar[rr]^>>>{1 + k_2}
		 && {\focussed{N_2}}
		 }
		\end{gathered}
	       \end{equation*}
	       Because
	       $(\focussed{N_1},\focussed{N_2})$ is a
	       quasi-$\mathbb{C}$-specimen of $\vartriangleleft$
	       up to $(\dot{\simeq}_{Q'},\dot{\simeq}_{Q'})$, and
	       states
	       $\focussed{N_1}$ and $\focussed{N_2}$ are rooted,
	       there exists a $\mathbb{C}$-specimen
	       $(\focussed{\mathcal{C}'};\vec{H'^1};\vec{H'^2})$
	       of $\vartriangleleft$
	       with a non-$\lightning$ focus, such that
	       $\focussed{\mathcal{C}'}[\vec{H'^1}]$ and
	       $\focussed{\mathcal{C}'}[\vec{H'^2}]$ are also rooted,
	       $\focussed{N_1}
	       \dotrel{\simeq}_{Q'}
	       \focussed{\mathcal{C}'}[\vec{H'^1}]$, and
	       $\focussed{\mathcal{C}'}[\vec{H'^2}]
	       \dotrel{\simeq}_{Q'}
	       \focussed{N_2}$.
	       Because $(Q,Q',Q'')$ is a reasonable triple,
	       $Q' \subseteq Q''$, and hence
	       $\mathord{\dot{\simeq}_{Q'}} \subseteq
	       \mathord{\dot{\simeq}_{Q''}}$.
	       Therefore, we have:
	       \begin{equation*}
		\focussed{N_1} \dotrel{\preceq}_{Q'}
		 \focussed{\mathcal{C}'}[\vec{H'^1}]
		 \mathrel{\overline{\vartriangleleft}^\mathbb{C}}
		 \focussed{\mathcal{C}'}[\vec{H'^2}]
		 \dotrel{\preceq}_{Q''} \focussed{N_2}.
	       \end{equation*}
	\end{itemize}
  \item When the focus is the $\lightning$-focus,
	$\focussed{G_1} \to \focussed{G'_1}$ is a rewrite transition,
	and by definition of contextual
	lifting, the focus is not entering in
	$\focussed{\mathcal{C}}$.
	Because $\vartriangleleft$ is $(\mathbb{C},Q,Q',Q'')$-robust
	relative to all rewrite transitions, and
	$\focussed{G_1}$ and $\focussed{G_2}$ are
	rooted, we have one of the following two situations
	corresponding to (II) and (III) of \autoref{def:Robustness}.
	\begin{itemize}
	 \item There exists a stuck state $\focussed{N}$ such that
	       $\focussed{G'_1} \to^* \focussed{N}$.
	       The condition (I) of \autoref{def:QSimulation} is
	       satisfied.
	 \item There exist a quasi-$\mathbb{C}$-specimen
	       $(\focussed{N_1},\focussed{N_2})$ of $\vartriangleleft$
	       up to $(\dot{\preceq}_{Q'},\dot{\preceq}_{Q''})$, whose
	       focus is not the $\lightning$-focus,
	       and two numbers $k_1,k_2 \in \N$, such that
	       $(1 + k_1) \mathrel{Q} k_2$,
	       $\focussed{G'_1} \to^{k_1} \focussed{N_1}$, and
	       $\focussed{G_2} \to^{k_2} \focussed{N_2}$.
	       We have the following situation, namely the black part
	       of the diagram below.
	       Showing the magenta part confirms that
	       the condition (II) of \autoref{def:QSimulation} is
	       fulfilled.
	       \begin{equation*}
		\begin{gathered}
		 \xymatrix@R=2ex{
		 {\focussed{G_1}
		 = \focussed{\mathcal{C}}[\vec{H^1}]} \ar[r]
		 \ar@{.}[d]_{\overline{\vartriangleleft}^\mathbb{C}}
		 \ar@{}[drr]|*+[black][F:black]{(1 + k_1)\,Q\,k_2}
		 & {\focussed{G'_1}} \ar[r]^>{k_1}
		 & {\focussed{N_1}}
		 \ar@{.}@[magenta][d]^*[magenta]{
		 \mathord{\dotrel{\preceq}_{Q'}}
		 \circ \overline{\vartriangleleft}^\mathbb{C}
		 \circ \mathord{\dotrel{\preceq}_{Q''}}} \\
		 {\focussed{G_2}
		 = \focussed{\mathcal{C}}[\vec{H^2}]}
		 \ar[rr]^>{k_2}
		 && {\focussed{N_2}}
		 }
		\end{gathered}
	       \end{equation*}
	       Because
	       $(\focussed{N_1},\focussed{N_2})$ is a
	       quasi-$\mathbb{C}$-specimen of $\vartriangleleft$
	       up to $(\dot{\preceq}_{Q'},\dot{\preceq}_{Q''})$, and
	       states
	       $\focussed{N_1}$ and $\focussed{N_2}$ are rooted,
	       there exists a $\mathbb{C}$-specimen
	       $(\focussed{\mathcal{C}'};\vec{H'^1};\vec{H'^2})$
	       of $\vartriangleleft$
	       with a non-$\lightning$ focus, such that
	       $\focussed{\mathcal{C}'}[\vec{H'^1}]$ and
	       $\focussed{\mathcal{C}'}[\vec{H'^2}]$ are also rooted,
	       $\focussed{N_1}
	       \dotrel{\preceq}_{Q'}
	       \focussed{\mathcal{C}'}[\vec{H'^1}]$, and
	       $\focussed{\mathcal{C}'}[\vec{H'^2}]
	       \dotrel{\preceq}_{Q''}
	       \focussed{N_2}$.
	       This means
	       $\focussed{\mathcal{C}'}[\vec{H'^1}]
	       \mathrel{\overline{\vartriangleleft}^\mathbb{C}}
	       \focussed{\mathcal{C}'}[\vec{H'^2}]$,
	       and hence:
	       \begin{equation*}
		\focussed{N_1} \dotrel{\preceq}_{Q'}
		 \focussed{\mathcal{C}'}[\vec{H'^1}]
		 \mathrel{\overline{\vartriangleleft}^\mathbb{C}}
		 \focussed{\mathcal{C}'}[\vec{H'^2}]
		 \dotrel{\preceq}_{Q''} \focussed{N_2}.
		 \qedhere
	       \end{equation*}
	\end{itemize}
 \end{itemize}
\end{proof}
\begin{proof}[Proof of the point~(\ref{item:backward2})]
 It suffices to check the ``reverse'' of conditions (A) and (B) of
 \autoref{def:QSimulation} for the states
 $\focussed{G_2}
 \mathrel{(\overline{\vartriangleleft}^\mathbb{C})^{-1}}
 \focussed{G_1}$,
 namely the following conditions (A') and (B').

 (A') If $\focussed{G_2}$ is final, $\focussed{G_1}$ is also
 final.

 (B') If there exists a state $\focussed{G'_2}$ such that
 $\focussed{G_2} \to \focussed{G'_2}$, one of the following
 (I') and (II') holds.

 \quad (I')
 There exists a stuck state $\focussed{G''_2}$ such that
 $\focussed{G'_2} \to^* \focussed{G''_2}$.

 \quad (II')
 There exist two states $\focussed{N_2}$ and $\focussed{N_1}$,
 and numbers $k_2,k_1 \in \N$, such that
 $\focussed{N_2} \mathrel{(\mathord{\dotrel{\preceq}_{Q'}}
 \circ (\overline{\vartriangleleft}^\mathbb{C})^{-1}
 \circ \mathord{\dotrel{\preceq}_{Q'}})}
 \focussed{N_1}$,
 $(1 + k_2) \mathrel{Q} k_1$,
 $\focussed{G'_2} \to^{k_2} \focussed{N_2}$, and
 $\focussed{G_1} \to^{k_1} \focussed{N_1}$.

 The proof is mostly symmetric to the point~(\ref{item:forward}).
 Note that there is a one-to-one correspondence between
 $\mathbb{C}$-specimens of $\vartriangleleft$ and
 $\mathbb{C}$-specimens of $\vartriangleleft^{-1}$;
 any $\mathbb{C}$-specimen
 $(\focussed{\mathcal{C}_0};\vec{H^{01}};\vec{H^{02}})$
 of $\vartriangleleft$ gives a $\mathbb{C}$-specimen
 $(\focussed{\mathcal{C}_0};\vec{H^{02}};\vec{H^{01}})$
 of $\vartriangleleft^{-1}$.
 Because $\vartriangleleft$ is output-closed,
 its converse $\vartriangleleft^{-1}$ is also output-closed.

 If $\focussed{G_2}$ is final, by
 \autoref{lem:TemplateToSimulationSharedCases}(\ref{item:Final}),
 $\focussed{G_1}$ is also final.
 The condition (A') is fulfilled.

 If there exists a state $\focussed{G'_2}$ such that
 $\focussed{G_2} \to \focussed{G'_2}$, we show that one of the
 conditions (I') and (II') above is fulfilled,
 by case analysis of the focus in $\focussed{\mathcal{C}}$.
 \begin{itemize}
  \item When the focus is the $\checkmark$-focus,
	or the $?$-focus that is not entering,
	by
	\autoref{lem:TemplateToSimulationSharedCases}(\ref{item:ValueOrNonEnteringSearchToken}),
	there exists a focussed context
	$\focussed{\mathcal{C}'}$ with a remote focus, such that
	$|\focussed{\mathcal{C}'}| = |\focussed{\mathcal{C}}|$ and
	$\focussed{G_p} = \focussed{\mathcal{C}}[\vec{H^p}] \to
	\focussed{\mathcal{C}'}[\vec{H^p}]$ for each
	$p \in \{ 1,2 \}$.
	We have the following situation, namely the black part of the
	diagram below. Showing the magenta part confirms that
	the condition (II') is fulfilled.
	\begin{equation*}
	 \begin{gathered}
	  \xymatrix@R=2ex{
	  {\focussed{G_2}
	  = \focussed{\mathcal{C}}[\vec{H^1}]} \ar[r]
	  \ar@{.}[d]_{(\overline{\vartriangleleft}^\mathbb{C})^{-1}}
	  \ar@{}[dr]|*+[magenta][F:magenta]{1\,Q\,1}
	  & {\focussed{\mathcal{C}'}[\vec{H^2}]
	  \textcolor{magenta}{\text{$\,= \focussed{G'_2}$}}}
	  \ar@{.}@[magenta][d]^*[magenta]{
	  (\overline{\vartriangleleft}^\mathbb{C})^{-1}} \\
	  {\focussed{G_1}
	  = \focussed{\mathcal{C}}[\vec{H^1}]} \ar[r]
	  & {\focussed{\mathcal{C}'}[\vec{H^1}]}
	  }
	 \end{gathered}
	\end{equation*}
	By the determinism,
	$\focussed{\mathcal{C}'}[\vec{H^2}] = \focussed{G'_2}$.
	Because $(Q,Q',Q'')$ is a reasonable triple, $Q$ is a preorder
	and $1 \mathrel{Q} 1$.
	The context $\focussed{\mathcal{C}'}$ satisfies
	$|\focussed{\mathcal{C}'}| = |\focussed{\mathcal{C}}| \in
	\mathbb{C}$, so
	$(\focussed{\mathcal{C}'};\vec{H^2};\vec{H^1})$ is a
	$\mathbb{C}$-specimen of $\vartriangleleft^{-1}$.
	The context $\focussed{\mathcal{C}'}$ has a remote focus, and
	the states
	$\focussed{\mathcal{C}'}[\vec{H^1}]$ and
	$\focussed{\mathcal{C}'}[\vec{H^2}]$ are both rooted.
	Therefore, we have
	$\focussed{\mathcal{C}'}[\vec{H^2}]
	\mathrel{(\overline{\vartriangleleft}^\mathbb{C})^{-1}}
	\focussed{\mathcal{C}'}[\vec{H^1}]$.
  \item When the focus is the $?$-focus that is entering in
	$\focussed{\mathcal{C}}$,
	because $\vartriangleleft$ is
	$(\mathbb{C},Q^{-1},Q')$-input-safe, we have one of the
	following three situations corresponding to (I), (II) and
	(III) of \autoref{def:InputSafe}.
	\begin{itemize}
	 \item There exist two stuck states $\focussed{N_1}$ and
	       $\focussed{N_2}$ such that
	       $\focussed{G_p} \to^* \focussed{N_p}$ for each
	       $p \in \{ 1,2 \}$.
	       By the determinism of transitions, we have
	       $\focussed{G_2} \to \focussed{G'_2}
	       \to^* \focussed{N_2}$, which means the condition (I')
	       is satisfied.
	 \item There exist a $\mathbb{C}$-specimen
	       $(\focussed{\mathcal{C}'};\vec{H'^1};\vec{H'^2})$
	       of $\vartriangleleft$
	       and two numbers $k_1,k_2 \in \N$, such that
	       the focus of
	       $\focussed{\mathcal{C}'}$ is the $\checkmark$-focus or the non-entering $?$-focus, 
	       $(1 + k_1) \mathrel{Q^{-1}} k_2$,
	       $\focussed{\mathcal{C}}[\vec{H^1}] \to^{1 + k_1}
	       \focussed{\mathcal{C}'}[\vec{H'^1}]$, and
	       $\focussed{\mathcal{C}}[\vec{H^2}] \to^{k_2}
	       \focussed{\mathcal{C}'}[\vec{H'^2}]$.
	       We have the following situation, namely the black part
	       of the diagram below.
	       \begin{equation*}
		\begin{gathered}
		 \xymatrix@R=4ex{
		 {\focussed{G_2}
		 = \focussed{\mathcal{C}}[\vec{H^2}]}
		 \ar[rr]^>{k_2}
		 \ar@{.}[d]_{
		 (\overline{\vartriangleleft}^\mathbb{C})^{-1}}
		 \ar@{}[drr]|*+[black][F:black]{k_2\,Q\,(1 + k_1)}
		 && {\quad\focussed{\mathcal{C}'}[\vec{H'^2}]}
		 \ar@{.}@[magenta][d]^*[magenta]{
		 (\overline{\vartriangleleft}^\mathbb{C})^{-1}} \\
		 {\focussed{G_1}
		 = \focussed{\mathcal{C}}[\vec{H^1}]}
		 \ar[rr]^>{1 + k_1}
		 && {\quad\focussed{\mathcal{C}'}[\vec{H'^1}]}
		 }
		\end{gathered}
	       \end{equation*}
	       The magenta part holds, because
	       the focus of $\focussed{\mathcal{C}'}$ is not the $\lightning$-focus
	       and not entering, and because states
	       $\focussed{\mathcal{C}'}[\vec{H'^1}]$ and
	       $\focussed{\mathcal{C}'}[\vec{H'^2}]$ are rooted.
	       We check the condition (II') by case analysis on the
	       number $k_2$.
	       \begin{itemize}
		\item When $k_2 > 0$, by the determinism of
		      transitions, we have the
		      following diagram, which means the condition
		      (II') is fulfilled.
		      \begin{equation*}
		       \begin{gathered}
			\xymatrix@R=4ex{
			{\focussed{G_2}
			= \focussed{\mathcal{C}}[\vec{H^2}]}
			\ar[r]
			\ar@{.}[d]_{
			(\overline{\vartriangleleft}^\mathbb{C})^{-1}}
			\ar@{}[drr]|*+[black][F:black]{
			k_2\,Q\,(1 + k_1)}
			& {\focussed{G'_2}} \ar[r]^>{k_2 - 1}
			& {\quad\focussed{\mathcal{C}'}[\vec{H'^2}]}
			\ar@{.}@[black][d]^*[black]{
			(\overline{\vartriangleleft}^\mathbb{C})^{-1}}
			\\
			{\focussed{G_1}
			= \focussed{\mathcal{C}}[\vec{H^1}]}
			\ar[rr]^>{1 + k_1}
			&& {\quad\focussed{\mathcal{C}'}[\vec{H'^1}]}
			}
		       \end{gathered}
		      \end{equation*}
		\item When $k_2 = 0$,
		      $\focussed{G_2} =
		      \focussed{\mathcal{C}}[\vec{H^2}] =
		      \focussed{\mathcal{C}'}[\vec{H'^2}]$,
		      and we have the following situation,
		      namely the black part of the diagram below.
		      \begin{equation*}
		       \begin{gathered}
			\xymatrix@R=4ex{
			{\focussed{G_2}
			= \focussed{\mathcal{C}}[\vec{H^2}]}
			\ar[rr]^>{0}
			\ar@{.}[d]_{
			(\overline{\vartriangleleft}^\mathbb{C})^{-1}}
			\ar@{}[drr]|*+[black][F:black]{0\,Q\,(1 + k_1)}
			&& {\enspace\focussed{G_2}
			= \focussed{\mathcal{C}'}[\vec{H'^2}]}
			\ar@[magenta][r]
			\ar@{.}@[black][d]^*[black]{
			(\overline{\vartriangleleft}^\mathbb{C})^{-1}}
			\ar@{}[dr]|-*+[magenta][F:magenta]{1\,Q\,1}
			& *[magenta]{\enspace\focussed{G'_2} =
			\focussed{\mathcal{C}''}[\vec{H'^1}]}
			\ar@{.}@[magenta][d]^*[magenta]{
			(\overline{\vartriangleleft}^\mathbb{C})^{-1}}
			\\
			{\focussed{G_1}
			= \focussed{\mathcal{C}}[\vec{H^1}]}
			\ar[rr]^>{1 + k_1}
			&& {\quad\focussed{\mathcal{C}'}[\vec{H'^1}]}
			\ar@[magenta][r]
			& *[magenta]{\enspace
			\focussed{\mathcal{C}''}[\vec{H'^1}]}
			}
		       \end{gathered}
		      \end{equation*}
		      Because $\focussed{G_2} \to \focussed{G'_2}$,
		      and the focus of $\focussed{\mathcal{C}'}$
		      is the $\checkmark$-focus, or the non-entering $?$-focus,
		      by
		      \autoref{lem:TemplateToSimulationSharedCases}(\ref{item:ValueOrNonEnteringSearchToken}),
		      there exists a focussed context
		      $\focussed{\mathcal{C}''}$ with a remote focus,
		      such that
		      $|\focussed{\mathcal{C}''}| =
		      |\focussed{\mathcal{C}'}|$
		      and
		      $\focussed{\mathcal{C}'}[\vec{H'^p}] \to
		      \focussed{\mathcal{C}''}[\vec{H'^p}]$
		      for each $p \in \{ 1,2 \}$.
		      By the determinism of transitions, 
		      $\focussed{G'_2} =
		      \focussed{\mathcal{C}''}[\vec{H'^1}]$.
		      Because $(Q,Q',Q'')$ is a reasonable triple,
		      $Q$ is a preorder and $1 \mathrel{Q} 1$.
		      The context $\focussed{\mathcal{C}''}$ satisfies
		      $|\focussed{\mathcal{C}''}| =
		      |\focussed{\mathcal{C}'}| \in
		      \mathbb{C}$, so
		      $(\focussed{\mathcal{C}''};
		      \vec{H'^2};\vec{H'^1})$
		      is a $\mathbb{C}$-specimen of
		      $\vartriangleleft^{-1}$.
		      The context $\focussed{\mathcal{C}''}$ has a
		      remote focus, and the states
		      $\focussed{\mathcal{C}''}[\vec{H'^1}]$ and
		      $\focussed{\mathcal{C}''}[\vec{H'^2}]$ are both
		      rooted.
		      Therefore, we have
		      $\focussed{\mathcal{C}''}[\vec{H'^2}]
		      \mathrel{
		      (\overline{\vartriangleleft}^\mathbb{C})^{-1}}
		      \focussed{\mathcal{C}''}[\vec{H'^1}]$.
		      Finally, because $(Q,Q',Q'')$ is a reasonable
		      triple, $Q$ is closed under addition, and hence
		      $1 \mathrel{Q} (2 + k_1)$.
		      The condition (II') is fulfilled.
	       \end{itemize}
	 \item There exist a quasi-$\mathbb{C}$-specimen
	       $(\focussed{N_1},\focussed{N_2})$ of $\vartriangleleft$
	       up to $(\dot{\simeq}_{Q'},\dot{\simeq}_{Q'})$, whose
	       focus is not the $\lightning$-focus,
	       and two numbers $k_1,k_2 \in \N$, such that
	       $(1 + k_1) \mathrel{Q^{-1}} (1 + k_2)$,
	       $\focussed{\mathcal{C}}[\vec{H^1}] \to^{1 + k_1}
	       \focussed{N_1}$, and
	       $\focussed{\mathcal{C}}[\vec{H^2}] \to^{1 + k_2}
	       \focussed{N_2}$.
	       By the determinism of transitions,
	       we have the following situation, namely the black part
	       of the diagram below.
	       Showing the magenta part confirms that
	       the condition (II') is fulfilled.
	       \begin{equation*}
		\begin{gathered}
		 \xymatrix@R=2ex@C=40pt{
		 {\focussed{G_2}
		 = \focussed{\mathcal{C}}[\vec{H^2}]} \ar[r]
		 \ar@{.}[d]_{
		 (\overline{\vartriangleleft}^\mathbb{C})^{-1}}
		 \ar@{}[drr]|*+[black][F:black]{
		 (1 + k_2)\,Q\,(1 + k_1)}
		 & {\focussed{G'_2}} \ar[r]^>{k_2}
		 & {\focussed{N_2}}
		 \ar@{.}@[magenta][d]^*[magenta]{
		 \mathord{\dotrel{\preceq}_{Q'}}
		 \circ (\overline{\vartriangleleft}^\mathbb{C})^{-1}
		 \circ \mathord{\dotrel{\preceq}_{Q''}}} \\
		 {\focussed{G_1}
		 = \focussed{\mathcal{C}}[\vec{H^1}]}
		 \ar[rr]^>>>{1 + k_1}
		 && {\focussed{N_1}}
		 }
		\end{gathered}
	       \end{equation*}
	       Because
	       $(\focussed{N_1},\focussed{N_2})$ is a
	       quasi-$\mathbb{C}$-specimen of $\vartriangleleft$
	       up to $(\dot{\simeq}_{Q'},\dot{\simeq}_{Q'})$, and
	       states
	       $\focussed{N_1}$ and $\focussed{N_2}$ are rooted,
	       there exists a $\mathbb{C}$-specimen
	       $(\focussed{\mathcal{C}'};\vec{H'^1};\vec{H'^2})$
	       of $\vartriangleleft$
	       with a non-$\lightning$ focus, such that
	       $\focussed{\mathcal{C}'}[\vec{H'^1}]$ and
	       $\focussed{\mathcal{C}'}[\vec{H'^2}]$ are also rooted,
	       $\focussed{N_1}
	       \dotrel{\simeq}_{Q'}
	       \focussed{\mathcal{C}'}[\vec{H'^1}]$, and
	       $\focussed{\mathcal{C}'}[\vec{H'^2}]
	       \dotrel{\simeq}_{Q'}
	       \focussed{N_2}$.
	       Because $(Q,Q',Q'')$ is a reasonable triple,
	       $Q' \subseteq Q''$, and hence
	       $\mathord{\dot{\simeq}_{Q'}} \subseteq
	       \mathord{\dot{\simeq}_{Q''}}$.
	       Therefore, we have:
	       \begin{equation*}
		\focussed{N_2} \dotrel{\preceq}_{Q'}
		 \focussed{\mathcal{C}'}[\vec{H'^2}]
		 \mathrel{
		 (\overline{\vartriangleleft}^\mathbb{C})^{-1}}
		 \focussed{\mathcal{C}'}[\vec{H'^1}]
		 \dotrel{\preceq}_{Q''} \focussed{N_1}.
	       \end{equation*}
	\end{itemize}
  \item When the focus is the $\lightning$-focus,
	$\focussed{G_2} \to \focussed{G'_2}$ is a rewrite transition,
	and by definition of contextual
	lifting, the focus is not entering in
	$\focussed{\mathcal{C}}$.
	Because $\vartriangleleft^{-1}$ is
	$(\mathbb{C},Q,Q',Q'')$-robust
	relative to all rewrite transitions, and
	$\focussed{G_1}$ and $\focussed{G_2}$ are
	rooted, we have one of the following two situations
	corresponding to (II) and (III) of \autoref{def:Robustness}.
	\begin{itemize}
	 \item There exists a stuck state $\focussed{N}$ such that
	       $\focussed{G'_2} \to^* \focussed{N}$.
	       The condition (I') is satisfied.
	 \item There exist a quasi-$\mathbb{C}$-specimen
	       $(\focussed{N_2},\focussed{N_1})$ of
	       $\vartriangleleft^{-1}$
	       up to $(\dot{\preceq}_{Q'},\dot{\preceq}_{Q''})$, whose
	       focus is not the $\lightning$-focus,
	       and two numbers $k_2,k_1 \in \N$, such that
	       $(1 + k_2) \mathrel{Q} k_1$,
	       $\focussed{G'_2} \to^{k_2} \focussed{N_2}$, and
	       $\focussed{G_1} \to^{k_1} \focussed{N_1}$.
	       We have the following situation, namely the black part
	       of the diagram below.
	       Showing the magenta part confirms that
	       the condition (II') is fulfilled.
	       \begin{equation*}
		\begin{gathered}
		 \xymatrix@R=2ex{
		 {\focussed{G_2}
		 = \focussed{\mathcal{C}}[\vec{H^2}]}
		 \ar[r]
		 \ar@{.}[d]_{
		 (\overline{\vartriangleleft}^\mathbb{C})^{-1}}
		 \ar@{}[drr]|*+[black][F:black]{(1 + k_2)\,Q\,k_1}
		 & {\focussed{G'_2}} \ar[r]^>{k_2}
		 & {\focussed{N_2}}
		 \ar@{.}@[magenta][d]^*[magenta]{
		 \mathord{\dotrel{\preceq}_{Q'}}
		 \circ (\overline{\vartriangleleft}^\mathbb{C})^{-1}
		 \circ \mathord{\dotrel{\preceq}_{Q''}}} \\
		 {\focussed{G_1}
		 = \focussed{\mathcal{C}}[\vec{H^1}]} \ar[rr]^>{k_1}
		 && {\focussed{N_1}}
		 }
		\end{gathered}
	       \end{equation*}
	       Because
	       $(\focussed{N_2},\focussed{N_1})$ is a
	       quasi-$\mathbb{C}$-specimen of $\vartriangleleft^{-1}$
	       up to $(\dot{\preceq}_{Q'},\dot{\preceq}_{Q''})$, and
	       states
	       $\focussed{N_2}$ and $\focussed{N_1}$ are rooted,
	       there exists a $\mathbb{C}$-specimen
	       $(\focussed{\mathcal{C}'};\vec{H'^2};\vec{H'^1})$
	       of $\vartriangleleft^{-1}$
	       with a non-$\lightning$ focus, such that
	       $\focussed{\mathcal{C}'}[\vec{H'^2}]$ and
	       $\focussed{\mathcal{C}'}[\vec{H'^1}]$ are also rooted,
	       $\focussed{N_2}
	       \dotrel{\preceq}_{Q'}
	       \focussed{\mathcal{C}'}[\vec{H'^2}]$, and
	       $\focussed{\mathcal{C}'}[\vec{H'^1}]
	       \dotrel{\preceq}_{Q''}
	       \focussed{N_1}$.
	       This means
	       $\focussed{\mathcal{C}'}[\vec{H'^2}]
	       \mathrel{\overline{\vartriangleleft^{-1}}^\mathbb{C}}
	       \focussed{\mathcal{C}'}[\vec{H'^1}]$,
	       and hence:
	       \begin{equation*}
		\focussed{N_2} \dotrel{\preceq}_{Q'}
		 \focussed{\mathcal{C}'}[\vec{H'^2}]
		 \mathrel{
		 (\overline{\vartriangleleft}^\mathbb{C})^{-1}}
		 \focussed{\mathcal{C}'}[\vec{H'^1}]
		 \dotrel{\preceq}_{Q''} \focussed{N_1}.
		 \qedhere
	       \end{equation*}
	\end{itemize}
 \end{itemize}
\end{proof}

\section{Sufficient conditions for robustness}
\label{app:SufficientCheck}

A proof of robustness becomes trivial for a specimen with a rewrite
token that gives a non-rooted state.
Thanks to the lemma below, we can show that a state is not rooted, by
checking paths from the token target.
\begin{defi}[Accessible paths]
 \noindent
 \begin{itemize}
  \item A path of a hypernet is said to be \emph{accessible} if it
	consists of edges whose all sources have type~$\star$.
  \item An accessible path is called \emph{stable} if the labels
	of its edges are included in
	$\{ \mathsf{I} \} \cup \Opr_\checkmark$.
  \item An accessible path is called \emph{active} if it starts with
	one active operation edge and possibly followed by a stable
	path.
 \end{itemize}
\end{defi}
Note that box edges and atom edges never appear in an accessible path.
\begin{lem}
 \label{lem:NotStablePathNotRooted}
 If a state has a rewrite token that is not an incoming edge of
 a contraction edge, then the state satisfies the following property:
 If there exists an accessible, but not active, path from the token
 target, then the state is not rooted.
\end{lem}
\begin{proof}
 This is a contraposition of a consequence of
 \autoref{lem:SearchSeqQueried} and
 \autoref{lem:AnswerAccessiblePaths}(\ref{item:RewriteToken}).
\end{proof}

Checking the condition (III) of robustness (see
\autoref{def:Robustness}) involves finding a
quasi-$\mathbb{C}$-specimen of $\vartriangleleft$ up to
$(\dot{\preceq}_{Q'},\dot{\preceq}_{Q''})$,
namely checking the condition (B) of
\autoref{def:SpecimenQuasiSpecimen}(\ref{item:QuasiSpecimen}).
The following lemma enables us to use contextual refinement
$\preceq^\mathbb{C}_{Q'}$ to yield state refinement
$\dotrel{\preceq}_{Q'}$, via single $\mathbb{C}$-specimens of
a certain pre-template $\vartriangleleft$.
\begin{defi}
 A pre-template $\vartriangleleft$ is a \emph{trigger} if it satisfies
 the following:
\\[1.5ex]
 (A) For any single $\mathbb{C}$-specimen
 $(\focussed{\mathcal{C}}[\chi];H^1;H^2)$ of
 $\vartriangleleft$, such that $\focussed{\mathcal{C}}$ has an
 entering search token,
 $\focussed{\mathcal{C}}[H^p] \to
 \langle \focussed{\mathcal{C}}[H^p] \rangle_{\lightning/?}$
 for each $p \in \{ 1,2 \}$.
\\[1.5ex]
 (B) For any hypernets $H^1 \vartriangleleft H^2$, both $H_1$ and
 $H_2$ are one-way.
\end{defi}
\begin{lem}
 \label{lem:TriggerToRefinement}
 Let $\mathbb{C}$ be a set of contexts, and
 $Q'$ be a binary relation on $\N$ such that,
 for any $k_0,k_1,k_2 \in \N$,
 $(k_0 + k_1) \mathrel{Q'} (k_0 + k_2)$ implies
 $k_1 \mathrel{Q'} k_2$.
 Let $\vartriangleleft$ be a pre-template that is a trigger and
 implies contextual refinement $\preceq^\mathbb{C}_{Q'}$.
 For any single $\mathbb{C}$-specimen
 $(\focussed{\mathcal{C}}[\chi];H^1;H^2)$ of
 $\vartriangleleft$,
 if compute transitions are all deterministic, and
 one of states $\focussed{\mathcal{C}}[H^1]$ and
 $\focussed{\mathcal{C}}[H^2]$ is rooted, then
 the other state is also rooted, and moreover,
 $\focussed{\mathcal{C}}[H^1] \dotrel{\preceq}_{Q'}
 \focussed{\mathcal{C}}[H^2]$.
\end{lem}
\begin{proof}
 This is a corollary of \autoref{lem:TriggerToStateRefinement}.
\end{proof}

\begin{lem}
 \label{lem:TriggerToStateRefinement}
 Let $\mathbb{C}$ be a set of contexts, and
 $Q'$ be a binary relation on $\N$ such that,
 for any $k_0,k_1,k_2 \in \N$,
 $(k_0 + k_1) \mathrel{Q'} (k_0 + k_2)$ implies
 $k_1 \mathrel{Q'} k_2$.
 Let $\vartriangleleft$ be a pre-template that is a trigger and
 implies contextual refinement $\preceq^\mathbb{C}_{Q'}$.
 For any single $\mathbb{C}$-specimen
 $(\focussed{\mathcal{C}}[\chi];H^1;H^2)$ of
 $\vartriangleleft$, the following holds.

 \begin{enumerate}
  \item \label{item:SameStepsRooted}
	For any $k \in \N$,
	$?;|\focussed{\mathcal{C}}|[H^1] \tobul^k
	\focussed{\mathcal{C}}[H^1]$
	if and only if
	$?;|\focussed{\mathcal{C}}|[H^2] \tobul^k
	\focussed{\mathcal{C}}[H^2]$.
  \item \label{item:ToStateRefinement}
	If compute transitions are all deterministic, and
	one of states $\focussed{\mathcal{C}}[H^1]$ and
	$\focussed{\mathcal{C}}[H^2]$ is rooted, then the other state
	is also rooted, and moreover,
	$\focussed{\mathcal{C}}[H^1] \dotrel{\preceq}_{Q'}
	\focussed{\mathcal{C}}[H^2]$.
 \end{enumerate}
\end{lem}
\begin{proof}[Proof of the point~(\ref{item:SameStepsRooted})]
 Let $(p,q)$ be an arbitrary element of a set $\{ (1,2), (2,1) \}$.
 We prove that, for any $k \in \N$,
 $?;|\focussed{\mathcal{C}}|[H^p] \tobul^k
 \focussed{\mathcal{C}}[H^p]$
 implies
 $?;|\focussed{\mathcal{C}}|[H^q] \tobul^k
 \focussed{\mathcal{C}}[H^q]$.
 The proof is by case analysis on the number $k$.
 \begin{itemize}
  \item When $k = 0$, $\focussed{\mathcal{C}}[H^p]$ is initial, and by
	\autoref{lem:TemplateToSimulationSharedCases}(\ref{item:Final}),
	$\focussed{\mathcal{C}}[H^q]$ is also initial.
	Note that $\vartriangleleft$ is a trigger and hence
	output-closed.
  \item When $k > 0$, by the following internal lemma,
	$?;|\focussed{\mathcal{C}}|[H^q] \tobul^k
	\focussed{\mathcal{C}}[H^q]$
	follows from
	$?;|\focussed{\mathcal{C}}|[H^p] \tobul^k
	\focussed{\mathcal{C}}[H^p]$.
	\qedhere
	\begin{lem}
	 For any $m \in \{ 0,\ldots,k \}$,
	 there exists a focussed context
	 $\focussed{\mathcal{C}'}[\chi]$ such that
	 $|\focussed{\mathcal{C}'}| = |\focussed{\mathcal{C}}|$
	 and the following holds:
	 \begin{align*}
	  ?;|\focussed{\mathcal{C}}|[H^p] &\tobul^m
	  \focussed{\mathcal{C}'}[H^p] \tobul^{k-m}
	  \focussed{\mathcal{C}}[H^p], \\
	  ?;|\focussed{\mathcal{C}}|[H^q] &\tobul^m
	  \focussed{\mathcal{C}'}[H^q].
	 \end{align*}
	\end{lem}
	\begin{proof}
	 By induction on $m$.
	 In the base case, when $m = 0$, we can take
	 $?;|\focussed{\mathcal{C}}|$ as $\focussed{\mathcal{C}'}$.

	 In the inductive case, when $m > 0$, by induction hypothesis,
	 there exists a focussed context
	 $\focussed{\mathcal{C}'}[\chi]$ such that
	 $|\focussed{\mathcal{C}'}| = |\focussed{\mathcal{C}}|$
	 and the following holds:
	 \begin{align*}
	  ?;|\focussed{\mathcal{C}}|[H^p] &\tobul^{m-1}
	  \focussed{\mathcal{C}'}[H^p] \tobul^{k-m+1}
	  \focussed{\mathcal{C}}[H^p], \\
	  ?;|\focussed{\mathcal{C}}|[H^q] &\tobul^{m-1}
	  \focussed{\mathcal{C}'}[H^q].
	 \end{align*}
	 Because
	 $|\focussed{\mathcal{C}'}| = |\focussed{\mathcal{C}}|
	 \in \mathbb{C}$,
	 $(\focussed{\mathcal{C}'};H^1;H^2)$ is a single
	 $\mathbb{C}$-specimen of $\vartriangleleft$, which yields
	 rooted states.
	 Because $k-m+1 > 0$, $\focussed{\mathcal{C}'}$ cannot have a
	 rewrite token.
	 The rest of the proof is by case analysis on the token of
	 $\focussed{\mathcal{C}'}$.
	 \begin{itemize}
	  \item When $\focussed{\mathcal{C}'}$ has an entering search
		token, because $\vartriangleleft$ is a trigger,
		$\focussed{\mathcal{C}'}[H^r] \to
		\langle \focussed{\mathcal{C}'}[H^r]
		\rangle_{\lightning/?}$
		for each $r \in \{ p,q \}$.
		Because
		$\langle \focussed{\mathcal{C}'}[H^r]
		\rangle_{\lightning/?} =
		\langle \focussed{\mathcal{C}'}
		\rangle_{\lightning/?}[H^r]$,
		and search transitions are deterministic, we have the
		following:
		\begin{align*}
		 ?;|\focussed{\mathcal{C}}|[H^p] &\tobul^{m-1}
		 \focussed{\mathcal{C}'}[H^p] \tobul
		 \langle \focussed{\mathcal{C}'}
		 \rangle_{\lightning/?}[H^p] \tobul^{k-m}
		 \focussed{\mathcal{C}}[H^p], \\
		 ?;|\focussed{\mathcal{C}}|[H^q] &\tobul^{m-1}
		 \focussed{\mathcal{C}'}[H^q] \tobul
		 \langle \focussed{\mathcal{C}'}
		 \rangle_{\lightning/?}[H^q].
		\end{align*}
		We also have
		$|\langle \focussed{\mathcal{C}'}
		\rangle_{\lightning/?}| =
		|\focussed{\mathcal{C}'}| =
		|\focussed{\mathcal{C}}|$.
	  \item When $\focussed{\mathcal{C}'}$ has a value token, or a
		non-entering search token, because $\vartriangleleft$
		is output-closed, by
		\autoref{lem:TemplateToSimulationSharedCases}(\ref{item:ValueOrNonEnteringSearchToken}),
		there exists a focussed context
		$\focussed{\mathcal{C}''}$ such that
		$|\focussed{\mathcal{C}''}| =
		|\focussed{\mathcal{C}'}|$
		and
		$\focussed{\mathcal{C}'}[H^r] \to
		\focussed{\mathcal{C}''}[H^r]$
		for each $r \in \{ p,q \}$.
		The transition
		$\focussed{\mathcal{C}'}[H^r] \to
		\focussed{\mathcal{C}''}[H^r]$,
		for each $r \in \{ p,q \}$, is a search transition,
		and by the determinism of search transitions, we have
		the following:
		\begin{align*}
		 ?;|\focussed{\mathcal{C}}|[H^p] &\tobul^{m-1}
		 \focussed{\mathcal{C}'}[H^p] \tobul
		 \focussed{\mathcal{C}''}[H^p] \tobul^{k-m}
		 \focussed{\mathcal{C}}[H^p], \\
		 ?;|\focussed{\mathcal{C}}|[H^q] &\tobul^{m-1}
		 \focussed{\mathcal{C}'}[H^q] \tobul
		 \focussed{\mathcal{C}''}[H^q].
		 \qedhere
		\end{align*}
	 \end{itemize}
	\end{proof}
 \end{itemize}
\end{proof}
\begin{proof}[Proof of the point~(\ref{item:ToStateRefinement})]
 If one of states $\focussed{\mathcal{C}}[H^1]$ and
 $\focussed{\mathcal{C}}[H^2]$ is rooted, by the
 point~(\ref{item:SameStepsRooted}),
 the other state is also rooted, and moreover, there exists $k \in \N$
 such that
 $?;|\focussed{\mathcal{C}}|[H^r] \tobul^k
 \focussed{\mathcal{C}}[H^r]$
 for each $r \in \{ 1,2 \}$.

 Our goal is to prove that, for any $k_1 \in \N$ and any final state
 $\focussed{N_1}$ such that
 $\focussed{\mathcal{C}}[H^1] \to^{k_1} \focussed{N_1}$,
 there exist $k_2 \in \N$ and a final state $\focussed{N_2}$ such that
 $k_1 \mathrel{Q'} k_2$ and
 $\focussed{\mathcal{C}}[H^2] \to^{k_2} \focussed{N_2}$.
 Assuming $\focussed{\mathcal{C}}[H^1] \to^{k_1} \focussed{N_1}$,
 we have the following:
 \begin{align*}
  ?;|\focussed{\mathcal{C}}|[H^1] &\tobul^k
  \focussed{\mathcal{C}}[H^1] \to^{k_1} \focussed{N_1}, \\
  ?;|\focussed{\mathcal{C}}|[H^2] &\tobul^k
  \focussed{\mathcal{C}}[H^2].
 \end{align*}

 Because $\vartriangleleft$ implies contextual refinement
 $\preceq^\mathbb{C}_{Q'}$, and
 $|\focussed{\mathcal{C}}| \in \mathbb{C}$,
 we have state refinement
 $?;|\focussed{\mathcal{C}}|[H^1] \dotrel{\preceq}_{Q'}
 ?;|\focussed{\mathcal{C}}|[H^2]$.
 Therefore, there exist $l_2 \in \N$ and a final state
 $\focussed{N_2}$ such that
 $(k + k_1) \mathrel{Q'} l_2$ and
 $?;|\focussed{\mathcal{C}}|[H^2] \to^{l_2} \focussed{N_2}$.

 The assumption that compute transitions are all deterministic
 implies that all transitions, including intrinsic ones, are
 deterministic. Following from this are $l_2 \geq k$ and the
 following:
 \begin{align*}
  ?;|\focussed{\mathcal{C}}|[H^1] &\tobul^k
  \focussed{\mathcal{C}}[H^1] \to^{k_1} \focussed{N_1}, \\
  ?;|\focussed{\mathcal{C}}|[H^2] &\tobul^k
  \focussed{\mathcal{C}}[H^2] \to^{l_2 - k} \focussed{N_2}.
 \end{align*}
 By the assumption on $Q'$, $(k + k_1) \mathrel{Q'} l_2$ implies
 $k_1 \mathrel{Q'} (l_2 - k)$.
\end{proof}

\section{Local rewrite rules and transfer properties}
\label{app:Local-Transfer}

The sufficiency-of-robustness theorem reduces a proof of an observational
equivalence down to establishing robust templates. As illustrated in
\autoref{sec:ex-law}, this typically boils down to checking
input-safety of pre-templates, and checking robustness of pre-templates relative
to rewrite transitions.

The key part of checking input-safety or robustness of a pre-template is to
analyse how a rewrite transition involves edges (at any depth) of a state that
are contributed by the pre-template.
In this section, we focus on rewrite transitions that are locally specified by
means of the contraction rules or rewrite rules (e.g.\ the micro-beta rewrite rules),
and identify some situation where these transitions involve
the edges contributed by a pre-template in a \emph{safe} manner.
These situations can be formalised for arbitrary instances of the universal
abstract machine, including the particular instance
$\UAM(\Oprex, B_{\Oprex})$ that is used in \autoref{sec:ex-law} to prove the
Parametricity law.

This section proceeds as follows.
Firstly, Appendix~\ref{sec:transfer-properties} formalises the safe involvement in
terms of \emph{transfer} properties.
Appendix~\ref{sec:transfer-param} establishes transfer properties for the
particular pre-templates and local rewrite rules used in
\autoref{sec:ex-law} to prove the Parametricity law.
Finally, Appendix~\ref{appsec:ExampleProofSketch} demonstrates the use of
the transfer properties, by providing details of
checking input-safety and robustness to prove the Parametricity law.

\subsection{Transfer properties}
\label{sec:transfer-properties}

We refer to the contraction rules which locally specify copy transitions, and
rewrite rules that locally specify rewrite transitions for active operations,
altogether as \emph{$\lightning$-rules}.
To analyse how a $\lightning$-rule involves edges contributed by a pre-template,
one would first need to check all possible overlaps between the local rule and
the edges, and then observe how these overlaps are affected by application of
the local rule.
We identify \emph{safe} involvement of the pre-template in the
$\lightning$-rule, as the situation where the overlaps get only eliminated or
duplicated without any internal modification.

We will first formalise safe involvement for a single application
of $\lightning$-rules, and then for a pair of applications of
$\lightning$-rules. The latter can capture safe involvement of edges contributed
by a pre-template, which can be exploited to check input-safety and
robustness of pre-templates.

\begin{notation}
 Let $m \in \N$ and $m' \in \N$.
 Given a sequence $\vec{x} = x_1,\ldots,x_m$ of length $m$ and a function
 $f \colon \{ 1,\ldots,m' \} \to \{ 1,\ldots,m \}$, a sequence
 $f(\vec{x}) = x'_1,\ldots,x'_{m'}$ of
 length $m'$ is given by $x'_j = x_{f(j)}$ for each $j \in \{ 1,\ldots,m' \}$.
\end{notation}

\begin{defi}[Transfer of hypernets]
 \label{def:transfer-hypernets}
 Let $\mathbb{C}$ and $\mathbb{C'}$ be two sets of focus-free contexts, and
 $\mathbb{H}$ be a set of focus-free hypernets.
 A $\lightning$-rule $\focussed{N} \mapsto \focussed{N'}$ of a universal
 abstract machine $\UAM(\Opr, B_\Opr)$
 \emph{transfers $\mathbb{H}$ from $\mathbb{C}$ to $\mathbb{C'}$} if,
 for any $m \in \N$,
 any focussed context $\focussed{\mathcal{C}}[\chi_1,\ldots,\chi_m]$ such that
 $|\focussed{\mathcal{C}}| \in \mathbb{C}$,
 and any $m$ focus-free hypernets $G_i \in \mathbb{H}$
 ($i \in \{ 1,\ldots,m \}$) such that
 $\focussed{N} = \focussed{\mathcal{C}}[G_1,\ldots,G_m]$,
 there exist some $m' \in \N$,
 some focussed context $\focussed{\mathcal{C'}}[\chi'_1,\ldots,\chi'_{m'}]$,
 and some function $f \colon \{ 1,\ldots,m' \} \to \{ 1,\ldots,m \}$,
 and the following holds.
 \begin{itemize}
  \item $|\focussed{\mathcal{C'}}| \in \mathbb{C'}$.
  \item $\focussed{N'} = \focussed{\mathcal{C'}}[f(G_1,\ldots,G_m)]$.
  \item $\focussed{\mathcal{C}}[H_1,\ldots,H_m] \mapsto
	\focussed{\mathcal{C'}}[f(H_1,\ldots,H_m)]$ is a $\lightning$-rule,
	for any $m$ focus-free hypernets $H_i \in \mathbb{H}$
	($i \in \{ 1,\ldots,m \}$).	
 \end{itemize}
\end{defi}

This transfer property enjoys monotonicity in the following sense:
if a $\lightning$-rule transfers $\mathbb{H}$ from $\mathbb{C}$ to
$\mathbb{C'}$, and $\mathbb{C'} \subseteq \mathbb{C''}$, then
the $\lightning$-rule transfers $\mathbb{H}$ from $\mathbb{C}$ to
$\mathbb{C''}$ as well.
If a $\lightning$-rule transfers $\mathbb{H}$ from $\mathbb{C}$ to the same
$\mathbb{C}$, we say the $\lightning$-rule
\emph{preserves $\mathbb{H}$ in $\mathbb{C}$}.

Given an operation set $\Opr$, we will be particularly interested in the
following sets of hypernets and contexts for $\Opr$, some of which have already
been introduced elsewhere:
the set $\mathbb{H}_\Opr$ of all focus-free hypernets,
the set $\mathbb{H}_\otimes$ of contraction trees,
the set $\mathbb{C}_\Opr$ of all focus-free contexts,
the set $\mathbb{C}_{\Opr\cbf}$ of all binding-free contexts,
the set $\mathbb{C}_{\Opr\cdp}$ of all \emph{deep} contexts, i.e.\ focus-free
contexts whose holes are all deep.

\begin{exa}[Transfer/preservation of hypernets in contexts]
 \noindent
 \begin{itemize}
  \item When a $\lightning$-rule $\focussed{N} \mapsto \focussed{N'}$ preserves
	$\mathbb{H}_\Opr$ in $\mathbb{C}_{\Opr\cdp}$, any deep edge of
	$\focussed{N}$ also appears as a deep edge in $\focussed{N'}$, and it
	also retains its neighbours.
	This is trivially the case if the $\lightning$-rule involves no box
	edges (and hence deep edges) at all. It is also the case if the
	$\lightning$-rule only eliminates or duplicates box edges without
	modifying deep edges. The contraction rules are an example of
	duplicating boxes.
  \item Preservation of deep edges can be restricted to binding-free positions,
	which are specified by binding-free contexts.
	When a $\lightning$-rule $\focussed{N} \mapsto \focussed{N'}$ preserves
	$\mathbb{H}_\Opr$ in $\mathbb{C}_{\Opr\cdp} \cap \mathbb{C}_{\Opr\cbf}$,
	any deep edge of $\focussed{N}$ in a binding-free position also appears
	as a deep edge in a binding-free position in $\focussed{N'}$.
  \item When a $\lightning$-rule $\focussed{N} \mapsto \focussed{N'}$ transfers
	$\mathbb{H}_\Opr$ from $\mathbb{C}_{\Opr\cdp}$ to $\mathbb{C}_\Opr$, any
	deep edge of $\focussed{N}$ also appears as an edge in $\focussed{N'}$,
	retaining its neighbours, but not necessarily as a deep edge.
	This is preservation of deep edges in a weak sense.
	It is the case when a $\lightning$-rule replaces a box edge with its
	contents, turning some deep edges into shallow edges without modifying
	their connection. The micro-beta rewrite rules are an example of this
	situation.
  \item When a $\lightning$-rule $\focussed{N} \mapsto \focussed{N'}$ preserves
	$\mathbb{H}_\otimes$ in $\mathbb{C}_\Opr$, any contraction tree in
	$\focussed{N}$ also appears in $\focussed{N'}$. The contraction rules
	are designed to satisfy this preservation property.
	\bqed
 \end{itemize}
\end{exa}

\begin{defi}[Transfer of (rooted) specimens]
 \label{sec:transfer-specimens}
 Let $\mathbb{C}$ and $\mathbb{C'}$ be two sets of focus-free contexts, and
 $\vartriangleleft$ be a pre-template.
 \begin{itemize}
  \item A $\lightning$-rule $\focussed{N} \mapsto \focussed{N'}$ of a universal
	abstract machine $\UAM(\Opr, B_\Opr)$
	\emph{transfers specimens of $\vartriangleleft$ from $\mathbb{C}$ to
	$\mathbb{C'}$} if,
	for any $\mathbb{C}$-specimen of the form
	$(\mathcal{C}_1[\vec{\chi'}, \focussed{\mathcal{C}_2}[\vec{\chi''}]];
	\vec{G'},\vec{G''}; \vec{H'},\vec{H''})$
	such that $\focussed{N} = \focussed{\mathcal{C}_2}[\vec{G''}]$,
	there exist some focussed context $\focussed{\mathcal{C}'_2}$ and
	two sequences $\vec{G'''}$ and $\vec{H'''}$ of focus-free hypernets, and
	the following holds.
	\begin{itemize}
	 \item $\focussed{N'} = \focussed{\mathcal{C}'_2}[\vec{G'''}]$.
	 \item $\focussed{\mathcal{C}_2}[\vec{H''}] \mapsto
	       \focussed{\mathcal{C}'_2}[\vec{H'''}]$
	       is a $\lightning$-rule.
	 \item $(\mathcal{C}_1[\vec{\chi'},\focussed{\mathcal{C}'_2}];
	       \vec{G'},\vec{G'''}; \vec{H'},\vec{H'''})$
	       is a $\mathbb{C'}$-specimen of $\vartriangleleft$.
	\end{itemize}
  \item The $\lightning$-rule $\focussed{N} \mapsto \focussed{N'}$ is said to
	transfer \emph{rooted} specimens of $\vartriangleleft$ from $\mathbb{C}$
	to $\mathbb{C'}$ if, in the above definition, the $\mathbb{C}$-specimen
	$(\mathcal{C}_1[\vec{\chi'}, \focussed{\mathcal{C}_2}[\vec{\chi''}]];
	\vec{G'},\vec{G''}; \vec{H'},\vec{H''})$
	is restricted to yield two rooted states
	$\mathcal{C}_1[\vec{G'},\focussed{\mathcal{C}_2}[\vec{G''}]]$ and
	$\mathcal{C}_1[\vec{H'},\focussed{\mathcal{C}_2}[\vec{H''}]]$.
 \end{itemize}
\end{defi}

If a $\lightning$-rule transfers specimens of $\vartriangleleft$ from
$\mathbb{C}$ to the same $\mathbb{C}$, we say the $\lightning$-rule
\emph{preserves specimens of $\vartriangleleft$ in $\mathbb{C}$}.

We can prove that certain transfer properties of hypernets imply the
corresponding transfer properties of specimens, as stated in
\autoref{prop:Transfer} below.
These are primarily transfer of deep edges, and preservation of contraction
trees.
\autoref{prop:Transfer} below will simplify some part of establishing
input-safety and robustness of a pre-template, because it
enables us to analyse a single application of a
$\lightning$-rule on a state, instead of a pair of applications of a
$\lightning$-rule on two states induced by a specimen of the pre-template.

\begin{defi}[Root-focussed $\lightning$-rules]
 A $\lightning$-rule $\focussed{N} \mapsto \focussed{N'}$ is said to be
 \emph{root-focussed} if it satisfies the following.
 \begin{itemize}
  \item $\focussed{N}$ has only one input.
  \item $\focussed{N} = \lightning; |\focussed{N}|$ holds, i.e.\ the sole input
	of $\focussed{N}$ coincides with the source of the token.
  \item Every output of $\focussed{N}$ is reachable from the sole input of
	$\focussed{N}$.
 \end{itemize}
\end{defi}

\begin{prop}
 \label{prop:Transfer}
 For any $\lightning$-rule $\focussed{N} \mapsto \focussed{N'}$ of a universal
 abstract machine $\UAM(\Opr, B_\Opr)$, the following holds.
 \begin{enumerate}
  \item \label{item:DeepTransfer}
	If it transfers $\mathbb{H}_\Opr$
	from $\mathbb{C}_{\Opr\cdp}$ to $\mathbb{C}_{\Opr}$,
	it transfers specimens of any pre-template $\vartriangleleft$
	from $\mathbb{C}_{\Opr\cdp}$ to $\mathbb{C}_{\Opr}$.
  \item \label{item:DeepBFPreserve}
	If it preserves $\mathbb{H}_\Opr$
	in $\mathbb{C}_{\Opr\cdp} \cap \mathbb{C}_{\Opr\cbf}$,
	it preserves specimens of any pre-template
	$\vartriangleleft$
	in $\mathbb{C}_{\Opr\cdp} \cap \mathbb{C}_{\Opr\cbf}$.
  \item \label{item:DeepBFTransfer}
	If it is root-focussed and transfers $\mathbb{H}_\Opr$
	from $\mathbb{C}_{\Opr\cdp} \cap \mathbb{C}_{\Opr\cbf}$
	to $\mathbb{C}_{\Opr\cbf}$,
	it transfers rooted specimens of any output-closed pre-template
	$\vartriangleleft$
	from $\mathbb{C}_{\Opr\cdp} \cap \mathbb{C}_{\Opr\cbf}$
	to $\mathbb{C}_{\Opr\cbf}$.
  \item \label{item:ContrPreserve}
	If it preserves $\mathbb{H}_\otimes$
	in $\mathbb{C}_{\Opr}$,
	it preserves specimens of any pre-template
	$\mathord{\vartriangleleft} \subseteq
	\mathbb{H}_\otimes \times \mathbb{H}_\otimes$,
	which is on contraction trees,
	in $\mathbb{C}_{\Opr}$.
 \end{enumerate}
\end{prop}
\begin{proof}[Proof of the point~(\ref{item:DeepTransfer})]
 We take an arbitrary $\mathbb{C}_{\Opr\cdp}$-specimen of the form
 \[
 (\mathcal{C}_1[\vec{\chi'},\focussed{\mathcal{C}_2}[\vec{\chi''}]];
 \vec{G'},\vec{G''};\vec{H'},\vec{H''})
 \]
 of the pre-template $\vartriangleleft$, such that
 $\focussed{N} = \focussed{\mathcal{C}_2}[\vec{G''}]$.
 Because the context
 $\mathcal{C}_1[\vec{\chi'},\focussed{\mathcal{C}_2}[\vec{\chi''}]]$ of the
 specimen must be focussed, the token in the context is shallow. This means that
 the hole labelled with $\chi$ in the context
 $\mathcal{C}_1[\vec{\chi'}, \chi]$ must be shallow.
 On the other hand, the specimen satisfies
 $|\mathcal{C}_1[\vec{\chi'},\focussed{\mathcal{C}_2}]|
 \in \mathbb{C}_{\Opr\cdp}$,
 and hence
 $\mathcal{C}_1[\vec{\chi'},|\focussed{\mathcal{C}_2}|]
 \in \mathbb{C}_{\Opr\cdp}$.
 As a consequence, we have
 $|\focussed{\mathcal{C}_2}| \in \mathbb{C}_{\Opr\cdp}$.
 Now, by the assumption, there exist some focussed context
 $\focussed{\mathcal{C}'_2}$ and some function $f$, such that
 $|\focussed{\mathcal{C}'_2}| \in \mathbb{C}_{\Opr}$ and
 $\focussed{N'} = \focussed{\mathcal{C}'_2}[f(\vec{G''})]$ hold, and moreover,
 $\focussed{\mathcal{C}_2}[\vec{H''}]
 \mapsto \focussed{\mathcal{C}'_2}[f(\vec{H''})]$
 is also a $\lightning$-rule.
 We obtain a triple
 $(\mathcal{C}_1[\vec{\chi'},\focussed{\mathcal{C}'_2}];
 \vec{G'},f(\vec{G''});\vec{H'},f(\vec{H''}))$.
 It satisfies
 $|\mathcal{C}_1[\vec{\chi'},\focussed{\mathcal{C}'_2}]| \in \mathbb{C}_{\Opr}$,
 and is a $\mathbb{C}_{\Opr}$-specimen of the pre-template $\vartriangleleft$.
\end{proof}
\begin{proof}[Proof of the point~(\ref{item:DeepBFPreserve})]
 We take an arbitrary
 $(\mathbb{C}_{\Opr\cdp} \cap \mathbb{C}_{\Opr\cbf})$-specimen of the form
 \[
 (\mathcal{C}_1[\vec{\chi'},\focussed{\mathcal{C}_2}[\vec{\chi''}]];
 \vec{G'},\vec{G''};\vec{H'},\vec{H''})
 \]
 of the pre-template $\vartriangleleft$, such that
 $\focussed{N} = \focussed{\mathcal{C}_2}[\vec{G''}]$.

 We first check that
 $|\focussed{\mathcal{C}_2}|
 \in \mathbb{C}_{\Opr\cdp} \cap \mathbb{C}_{\Opr\cbf}$
 follows from
 $|\mathcal{C}_1[\vec{\chi'},\focussed{\mathcal{C}_2}]|
 \in \mathbb{C}_{\Opr\cdp} \cap \mathbb{C}_{\Opr\cbf}$, as follows.
 \begin{itemize}
  \item Because the context $\mathcal{C}_1[\vec{\chi'},\focussed{\mathcal{C}_2}]$
	must be focussed, the token in the context is shallow. This means that
	the hole labelled with $\chi$ in the context
	$\mathcal{C}_1[\vec{\chi'}, \chi]$ must be shallow.
	This, combined with
	$|\mathcal{C}_1[\vec{\chi'},\focussed{\mathcal{C}_2}]|
	= \mathcal{C}_1[\vec{\chi'},|\focussed{\mathcal{C}_2}|]
	\in \mathbb{C}_{\Opr\cdp}$,
	implies $|\focussed{\mathcal{C}_2}| \in \mathbb{C}_{\Opr\cdp}$.
  \item If the context $|\focussed{\mathcal{C}_2}|$ contains a path that makes
	it not binding-free, the path is also a path in the context
	$\mathcal{C}_1[\vec{\chi'},|\focussed{\mathcal{C}_2}|]$ and makes the
	context not binding-free.
	Therefore, because
	$|\mathcal{C}_1[\vec{\chi'},\focussed{\mathcal{C}_2}]|
	= \mathcal{C}_1[\vec{\chi'},|\focussed{\mathcal{C}_2}|]
	\in \mathbb{C}_{\Opr\cbf}$ holds,
	the context $|\focussed{\mathcal{C}_2}|$ is without any path that
	makes the context not binding-free.
	This means
	$|\focussed{\mathcal{C}_2}| \in \mathbb{C}_{\Opr\cbf}$.
 \end{itemize}
 
 By the assumption, there exist some focussed context
 $\focussed{\mathcal{C}'_2}$ and some function $f$, such that
 $|\focussed{\mathcal{C}'_2}|
 \in \mathbb{C}_{\Opr\cdp} \cap \mathbb{C}_{\Opr\cbf}$ and
 $\focussed{N'} = \focussed{\mathcal{C}'_2}[f(\vec{G''})]$ hold, and moreover,
 $\focussed{\mathcal{C}_2}[\vec{H''}]
 \mapsto \focussed{\mathcal{C}'_2}[f(\vec{H''})]$
 is also a $\lightning$-rule.
 We obtain a triple
 $(\mathcal{C}_1[\vec{\chi'},\focussed{\mathcal{C}'_2}];
 \vec{G'},f(\vec{G''});\vec{H'},f(\vec{H''}))$.

 To conclude the proof, it suffices to prove that this triple is a
 $(\mathbb{C}_{\Opr\cdp} \cap \mathbb{C}_{\Opr\cbf})$-specimen of the
 pre-template $\vartriangleleft$,
 which boils down to showing
 $|\mathcal{C}_1[\vec{\chi'},\focussed{\mathcal{C}'_2}]|
 = \mathcal{C}_1[\vec{\chi'},|\focussed{\mathcal{C}'_2}|]
 \in \mathbb{C}_{\Opr\cdp} \cap \mathbb{C}_{\Opr\cbf}$.

 We firstly prove $\mathcal{C}_1[\vec{\chi'},|\focussed{\mathcal{C}'_2}|]
 \in \mathbb{C}_{\Opr\cdp}$.
 Because $\mathcal{C}_1[\vec{\chi'},|\focussed{\mathcal{C}_2}|]
 \in \mathbb{C}_{\Opr\cdp}$ holds,
 the holes labelled with $\vec{\chi'}$ of the context
 $\mathcal{C}_1$ must be all deep. This, together with
 $|\focussed{\mathcal{C}'_2}| \in \mathbb{C}_{\Opr\cdp}$,
 implies $\mathcal{C}_1[\vec{\chi'},|\focussed{\mathcal{C}'_2}|]
 \in \mathbb{C}_{\Opr\cdp}$.
 
 We then prove $\mathcal{C}_1[\vec{\chi'},|\focussed{\mathcal{C}'_2}|]
 \in \mathbb{C}_{\Opr\cbf}$ by contradiction,
 which will conclude the whole proof.
 Assume that the context is not binding-free. It has a path $P$ from a source of
 an edge $e$ that is either a contraction edge, an atom edge, a box edge or a
 hole edge, to a source of an edge $e'$ that is a hole edge.
 Thanks to $\mathcal{C}_1[\vec{\chi'},|\focussed{\mathcal{C}'_2}|]
 \in \mathbb{C}_{\Opr\cdp}$ and
 $|\focussed{\mathcal{C}'_2}| \in \mathbb{C}_{\Opr\cdp}$,
 the hole edge $e'$ of the context
 $\mathcal{C}_1[\vec{\chi'},|\focussed{\mathcal{C}'_2}|]$
 must be deep.
 We will infer a contradiction by case analysis on the hole edge $e'$.
 There are two cases.
 \begin{itemize}
  \item One case is when the edge $e'$ of the context
	$\mathcal{C}_1[\vec{\chi'},|\focussed{\mathcal{C}'_2}|]$
	comes from the context $\mathcal{C}_1$.
	In this case, the edge $e'$ is one of the deep hole edges
	labelled with $\vec{\chi'}$.
	This means that the path $P$ in the context
	$\mathcal{C}_1[\vec{\chi'},|\focussed{\mathcal{C}'_2}|]$
	must consist of deep edges only, and these deep edges, together
	with the edge $e'$, must be contained in a box of the context
	$\mathcal{C}_1$.
	Therefore the path $P$ is also a path in the context
	$\mathcal{C}_1$, and it makes the context
	$\mathcal{C}_1[\vec{\chi'},|\focussed{\mathcal{C}_2}|]$ not
	binding-free. This contradicts
	$\mathcal{C}_1[\vec{\chi'},|\focussed{\mathcal{C}_2}|]
	\in \mathbb{C}_{\Opr\cbf}$.
  \item The other case is when the edge $e'$ of the context
	$\mathcal{C}_1[\vec{\chi'},|\focussed{\mathcal{C}'_2}|]$
	comes from the context $|\focussed{\mathcal{C}'_2}|$.
	In this case, the edge $e'$ is a hole edge of the context
	$|\focussed{\mathcal{C}'_2}| \in \mathbb{C}_{\Opr\cdp}$, and
	hence a deep edge.
	This means that the path $P$ is also a path in the context
	$|\focussed{\mathcal{C}'_2}|$, consisting of deep edges only.
	The path $P$ therefore makes the context
	$|\focussed{\mathcal{C}'_2}|$ not binding-free, which contradicts
	$|\focussed{\mathcal{C}'_2}| \in \mathbb{C}_{\Opr\cbf}$.
	\qedhere
 \end{itemize}	
\end{proof}
\begin{proof}[Proof of the point~(\ref{item:DeepBFTransfer})]
 We take an arbitrary
 $(\mathbb{C}_{\Opr\cdp} \cap \mathbb{C}_{\Opr\cbf})$-specimen of the form
 \[
 (\mathcal{C}_1[\vec{\chi'},\focussed{\mathcal{C}_2}[\vec{\chi''}]];
 \vec{G'},\vec{G''};\vec{H'},\vec{H''})
 \]
 of the pre-template $\vartriangleleft$, such that
 $\focussed{N} = \focussed{\mathcal{C}_2}[\vec{G''}]$ holds,
 and two states
 $\mathcal{C}_1[\vec{G'},\focussed{\mathcal{C}_2}[\vec{G''}]]$ and
 $\mathcal{C}_1[\vec{H'},\focussed{\mathcal{C}_2}[\vec{H''}]]$ are both rooted.

 We can first check
 $|\focussed{\mathcal{C}_2}|
 \in \mathbb{C}_{\Opr\cdp} \cap \mathbb{C}_{\Opr\cbf}$,
 in the same way as the proof of the point~(\ref{item:DeepBFPreserve}).
 By the assumption, there exist some focussed context
 $\focussed{\mathcal{C}'_2}$ and some function $f$, such that
 $|\focussed{\mathcal{C}'_2}| \in \mathbb{C}_{\Opr\cbf}$ and
 $\focussed{N'} = \focussed{\mathcal{C}'_2}[f(\vec{G''})]$ hold, and moreover,
 $\focussed{\mathcal{C}_2}[\vec{H''}]
 \mapsto \focussed{\mathcal{C}'_2}[f(\vec{H''})]$
 is also a $\lightning$-rule.
 We obtain a triple
 $(\mathcal{C}_1[\vec{\chi'},\focussed{\mathcal{C}'_2}];
 \vec{G'},f(\vec{G''});\vec{H'},f(\vec{H''}))$.

 To conclude the proof, it suffices to prove that this triple is a
 $\mathbb{C}_{\Opr\cbf}$-specimen of the pre-template $\vartriangleleft$,
 which boils down to showing
 $|\mathcal{C}_1[\vec{\chi'},\focussed{\mathcal{C}'_2}]|
 = \mathcal{C}_1[\vec{\chi'},|\focussed{\mathcal{C}'_2}|]
 \in \mathbb{C}_{\Opr\cbf}$.
 We prove this by contradiction, as follows.
 
 Assume that the context
 $\mathcal{C}_1[\vec{\chi'},|\focussed{\mathcal{C}'_2}|]$
 is not binding-free. It has a path $P$ from a source of
 an edge $e$ that is either a contraction edge, an atom edge, a box edge or a
 hole edge, to a source of an edge $e'$ that is a hole edge.
 We will infer a contradiction by case analysis on the edge $e'$.
 There are two cases.
 \begin{itemize}
  \item One case is when the edge $e'$ of the context
	$\mathcal{C}_1[\vec{\chi'},|\focussed{\mathcal{C}'_2}|]$
	comes from the context $\mathcal{C}_1$.
	In this case, the edge $e'$ is one of the hole edges labelled with
	$\vec{\chi'}$.
	Because of $\mathcal{C}_1[\vec{\chi'},|\focussed{\mathcal{C}_2}|]
	\in \mathbb{C}_{\Opr\cdp}$,
	the hole edge $e'$ must be deep.
	This means that the path $P$ must consist of deep edges contained in a
	box of the context $\mathcal{C}_1$. The path is therefore a path in
	the context $\mathcal{C}_1$, and also in the context
	$\mathcal{C}_1[\vec{\chi'},|\focussed{\mathcal{C}_2}|]$.
	This means
	$\mathcal{C}_1[\vec{\chi'},|\focussed{\mathcal{C}_2}|]
	\notin \mathbb{C}_{\Opr\cbf}$, which is a contradiction.
  \item The other case is when the edge $e'$ of the context
	$\mathcal{C}_1[\vec{\chi'},|\focussed{\mathcal{C}'_2}|]$
	comes from the context $|\focussed{\mathcal{C}'_2}|$.
	In this case, we will infer a contradiction by further case analysis on
	the edge $e$ and the path $P$. There are three (sub-)cases.
	\begin{itemize}
	 \item The first case is when the edge $e$ comes from the context
	       $|\focussed{\mathcal{C}'_2}|$ and $P$ is a path in the context.
	       In this case, the path $P$ makes the context
	       $|\focussed{\mathcal{C}'_2}|$ not binding-free, which contradicts
	       $|\focussed{\mathcal{C}'_2}| \in \mathbb{C}_{\Opr\cbf}$.
	 \item The second case is when the edge $e$ comes from the context
	       $|\focussed{\mathcal{C}'_2}|$ and $P$ does not give a single path
	       in the context.
	       In this case, the edges $e$ and $e'$ both come from the context
	       $|\focussed{\mathcal{C}'_2}|$, but $P$ is a valid path only in
	       the whole context
	       $\mathcal{C}_1[\vec{\chi'},|\focussed{\mathcal{C}'_2}|]$.
	       This means that, in the context
	       $\mathcal{C}_1[\vec{\chi'},\chi]$, a source of the hole edge
	       labelled with $\chi$ is reachable from a target of the same hole
	       edge.

	       Because the $\lightning$-rule
	       $\focussed{\mathcal{C}_2}[\vec{G''}]
	       \mapsto \focussed{\mathcal{C}'_2}[f(\vec{G''})]$
	       is root-focussed, the focussed hypernet
	       $\focussed{\mathcal{C}_2}[\vec{G''}]$ has only one input, the
	       input coincides with the source of the token, and
	       every output of the hypernet is reachable from the sole input.
	       Moreover, because of
	       $|\focussed{\mathcal{C}_2}| \in \mathbb{C}_{\Opr\cdp}$,
	       the same holds for the focussed context
	       $\focussed{\mathcal{C}_2}$ too, namely:
	       the context has only one input, the input coincides with the
	       source of the token, and every output of the
	       context is reachable from the sole input.

	       As a consequence, in the focussed context
	       $\mathcal{C}_1[\vec{\chi'},\focussed{\mathcal{C}_2}]$,
	       the token source is reachable from itself, via a cyclic path
	       that contains some edges coming from the context
	       $\focussed{\mathcal{C}_2}$ including the token edge.
	       This path is not an operation path. Therefore, by
	       \autoref{lem:NotOprPathNotRooted}(\ref{item:ExitingSafe}),
	       at least one of the states
	       $\mathcal{C}_1[\vec{G'},\focussed{\mathcal{C}_2}[\vec{G''}]]$ and
	       $\mathcal{C}_1[\vec{H'},\focussed{\mathcal{C}_2}[\vec{H''}]]$
	       is not rooted. This is a contradiction.
	 \item The last case is when the edge $e$ comes from the context
	       $\mathcal{C}_1$.
	       Recall that the edge $e'$ comes from the context
	       $|\focussed{\mathcal{C}'_2}|$.
	       In this case, the path $P$ in the context
	       $\mathcal{C}_1[\vec{\chi'},|\focussed{\mathcal{C}'_2}|]$
	       has a prefix that gives a path $P'$ in the context
	       $\mathcal{C}_1[\vec{\chi'},\chi]$, from the same source of the
	       edge $e$ as the path $P$, to a source of the hole
	       edge labelled with $\chi$.
	       Because the path $P'$ is given as a part of the path $P$ in the
	       context $\mathcal{C}_1[\vec{\chi'},|\focussed{\mathcal{C}'_2}|]$,
	       the path $P'$ in the context
	       $\mathcal{C}_1[\vec{\chi'},\chi]$ does not itself contain the
	       hole edge labelled with $\chi$.

	       Because the $\lightning$-rule
	       $\focussed{\mathcal{C}_2}[\vec{G''}]
	       \mapsto \focussed{\mathcal{C}'_2}[f(\vec{G''})]$
	       is root-focussed, the focussed hypernet
	       $\focussed{\mathcal{C}_2}[\vec{G''}]$ has only one input, and the
	       input coincides with the source of the token.
	       Moreover, because of
	       $|\focussed{\mathcal{C}_2}| \in \mathbb{C}_{\Opr\cdp}$,
	       the same holds for the focussed context
	       $\focussed{\mathcal{C}_2}$ too, namely:
	       the context has only one input, and the input coincides with the
	       source of the token.

	       As a consequence, the path $P'$ in turn gives a path in the
	       focussed context
	       $\mathcal{C}_1[\vec{\chi'},\focussed{\mathcal{C}_2}]$,
	       from the same source of the edge $e$ as the path $P$,
	       to the source of the token. The first edge $e$ of this path is
	       not an operation edge, and therefore the path is not an operation
	       path. By
	       \autoref{lem:NotOprPathNotRooted}(\ref{item:ExitingSafe}),
	       at least one of the states
	       $\mathcal{C}_1[\vec{G'},\focussed{\mathcal{C}_2}[\vec{G''}]]$ and
	       $\mathcal{C}_1[\vec{H'},\focussed{\mathcal{C}_2}[\vec{H''}]]$
	       is not rooted. This is a contradiction.
	       \qedhere
	\end{itemize}
 \end{itemize}
 
\end{proof}
\begin{proof}[Proof of the point~(\ref{item:ContrPreserve})]
 We take an arbitrary $\mathbb{C}_{\Opr}$-specimen of the form
 \[
 (\mathcal{C}_1[\vec{\chi'},\focussed{\mathcal{C}_2}[\vec{\chi''}]];
 \vec{G'},\vec{G''};\vec{H'},\vec{H''})
 \]
 of the pre-template
 $\mathord{\vartriangleleft}
 \subseteq \mathbb{H}_\otimes \times \mathbb{H}_\otimes$,
 such that $\focussed{N} = \focussed{\mathcal{C}_2}[\vec{G''}]$.
 All the hypernets in $\vec{G'},\vec{G''},\vec{H'},\vec{H''}$ are elements of
 $\mathbb{H}_\otimes$, i.e.\ contraction trees. It trivially holds that
 $|\focussed{\mathcal{C}_2}| \in \mathbb{C}_{\Opr}$.
 Therefore, by the assumption, there exist some focussed context
 $\focussed{\mathcal{C}'_2}$ and some function $f$, such that
 $|\focussed{\mathcal{C}'_2}| \in \mathbb{C}_{\Opr}$ and
 $\focussed{N'} = \focussed{\mathcal{C}'_2}[f(\vec{G''})]$ hold, and moreover,
 $\focussed{\mathcal{C}_2}[\vec{H''}]
 \mapsto \focussed{\mathcal{C}'_2}[f(\vec{H''})]$
 is also a $\lightning$-rule.
 We obtain a triple
 $(\mathcal{C}_1[\vec{\chi'},\focussed{\mathcal{C}'_2}];
 \vec{G'},f(\vec{G''});\vec{H'},f(\vec{H''}))$,
 and this is a $\mathbb{C}_{\Opr}$-specimen of the pre-template
 $\vartriangleleft$.
\end{proof}

\subsection{Transfer properties for the Parametricity law}
\label{sec:transfer-param}

We can now establish transfer properties of deep edges and contraction trees for
the particular machine $\UAM(\Oprex, B_{\Oprex})$ which is used to prove the
Parametricity law.

\begin{prop}
 \label{prop:OprexTransferHypernets}
 The universal abstract machine $\UAM(\Oprex, B_{\Oprex})$ satisfies the
 following.
 \begin{enumerate}
  \item \label{item:OprexDeepTransfer}
	The contraction rules and all rewrite rules
	transfer $\mathbb{H}_{\Oprex}$
	from $\mathbb{C}_{\Oprex\cdp}$ to $\mathbb{C}_{\Oprex}$.
  \item \label{item:OprexDeepBFPreserve}
	The contraction rules, and all rewrite rules except for the micro-beta rewrite
	rules,
	preserve $\mathbb{H}_\Oprex$
	in $\mathbb{C}_{\Oprex\cdp} \cap \mathbb{C}_{\Oprex\cbf}$.
  \item \label{item:OprexDeepBFTransfer}
	The micro-beta rewrite rules
	transfer $\mathbb{H}_\Oprex$
	from $\mathbb{C}_{\Oprex\cdp} \cap \mathbb{C}_{\Oprex\cbf}$
	to $\mathbb{C}_{\Oprex\cbf}$.
  \item \label{item:OprexContrPreserve}
	The contraction rules and all rewrite rules
	preserve $\mathbb{H}_\otimes$
	in $\mathbb{C}_{\Oprex}$.	
 \end{enumerate}
\end{prop}
\begin{proof}[Sketch of the proof]
 We can prove the four points by analysing each $\lightning$-rule, i.e.\ a
 contraction rule or a local rewrite rule for an active operation, of the
 universal abstract machine $\UAM(\Oprex, B_{\Oprex})$.

 Firstly, the only way in which a contraction rule involves deep edges is to
 have them inside the hypernet to be duplicated ($H$ in
 \autoref{fig:copy-rule}).
 The deep edges and their connection are all preserved, and replacing these
 edges with arbitrary deep edges still enables the contraction rule. The
 point~(\ref{item:OprexDeepTransfer}) therefore holds.
 Additionally, any path to a source of a
 deep edge must consist of deep edges only, and if such a path appears in the
 result $\focussed{N'}$ of a contraction rule
 $\focussed{N} \mapsto \focussed{N'}$, the path necessarily appears in the
 original hypernet $\focussed{N}$ too. Therefore, if the contraction rule
 moves deep edges out of binding-free positions, these edges must not be at
 binding-free positions beforehand. This is a contradiction, and
 the point~(\ref{item:OprexDeepBFPreserve}) holds.
 As for contraction trees, whenever a contraction rule involves a contraction
 tree, the tree is either deep and gets duplicated, or shallow and left
 unmodified. Replacing the contraction tree with another contraction tree still
 enables the contraction rule that duplicates the same hypernet.
 The point~(\ref{item:OprexContrPreserve}) therefore holds.
 
 Secondly, we analyse the micro-beta rewrite rules.
 Whenever deep edges are involved in a micro-beta rewrite rule, they must be inside
 the box edge that gets opened (i.e.\ $G$ in \autoref{fig:rewrite-micro-beta}).
 These deep edges may be turned into shallow edges, but their connection is
 unchanged. The difference of deep edges does not affect application of the
 rule, and hence the point~(\ref{item:OprexDeepTransfer}) holds.
 If these deep edges are at binding-free positions, they remain at binding-free
 positions after applying the micro-beta rewrite rule, for a similar reason as the
 contraction rules. The point~(\ref{item:OprexDeepBFTransfer}) therefore holds.
 As for contraction trees,
 the only way in which contraction trees get involved in a micro-beta rewrite rule is
 for them to be deep. The point~(\ref{item:OprexContrPreserve}) reduces to the
 point~(\ref{item:OprexDeepTransfer}) for micro-beta rewrite rules.

 The rest of the local rewrite rules involve no deep edges at all, and therefore
 points~(\ref{item:OprexDeepTransfer}) and~(\ref{item:OprexDeepBFPreserve})
 trivially hold. These rules either involve no contraction trees, or involve
 shallow contraction trees without any modification. The difference of
 contraction trees does not affect application of the rules. The
 point~(\ref{item:OprexContrPreserve}) therefore holds.
\end{proof}

\begin{cor}
 \label{cor:OprexTransferSpecimens}
 In the universal abstract machine $\UAM(\Oprex, B_{\Oprex})$, the contraction
 rules and all rewrite rules satisfies the following.
 \begin{enumerate}
  \item \label{item:Deep}
	The rules
	transfer specimens of any pre-template $\vartriangleleft$
	from $\mathbb{C}_{\Oprex\cdp}$ to $\mathbb{C}_{\Oprex}$.
  \item \label{item:DeepBF}
	The rules
	transfer rooted specimens of any output-closed pre-template
	$\vartriangleleft$
	from $\mathbb{C}_{\Oprex\cdp} \cap \mathbb{C}_{\Oprex\cbf}$
	to $\mathbb{C}_{\Oprex\cbf}$.
  \item \label{item:Contr}
	The rules
	preserve specimens of any pre-template
	$\mathord{\vartriangleleft} \subseteq
	\mathbb{H}_\otimes \times \mathbb{H}_\otimes$,
	which is on contraction trees,
	in $\mathbb{C}_{\Oprex}$.	
 \end{enumerate}
\end{cor}
\begin{proof}
 This is a consequence of \autoref{prop:OprexTransferHypernets} and
 \autoref{prop:Transfer}, noting that the micro-beta rewrite rules are root-focussed
 and preserving in $\mathbb{C}_{\Oprex\cdp} \cap \mathbb{C}_{\Oprex\cbf}$ implies
 transferring from $\mathbb{C}_{\Oprex\cdp} \cap \mathbb{C}_{\Oprex\cbf}$
 to $\mathbb{C}_{\Oprex\cbf}$.
\end{proof}

\subsection{Input-safety and robustness for the Parametricity law}
\label{appsec:ExampleProofSketch}

\begin{figure}
 \centering
 \begin{gather*}
  \includegraphics[scale=.2]{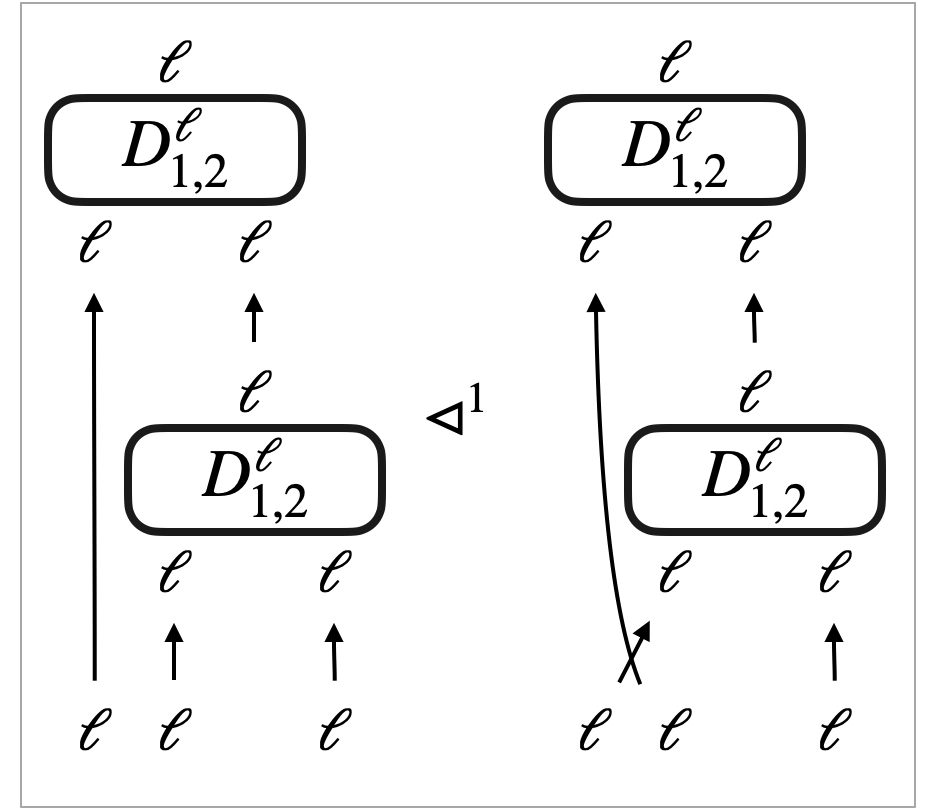} \qquad
  \includegraphics[scale=.2]{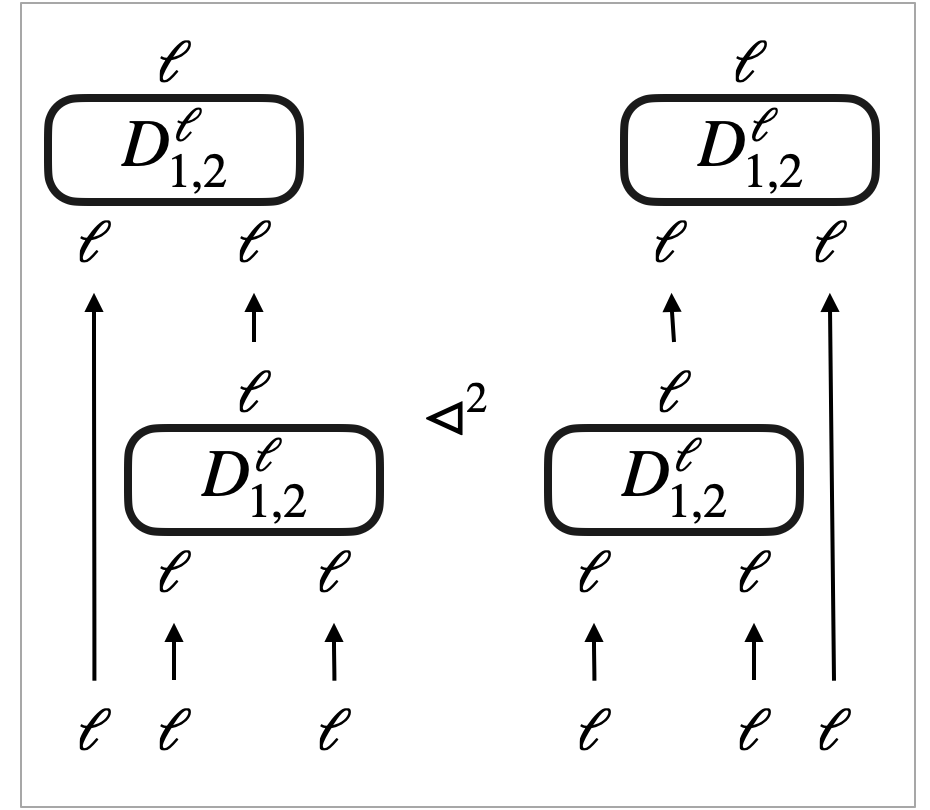} \qquad
  \includegraphics[scale=.2]{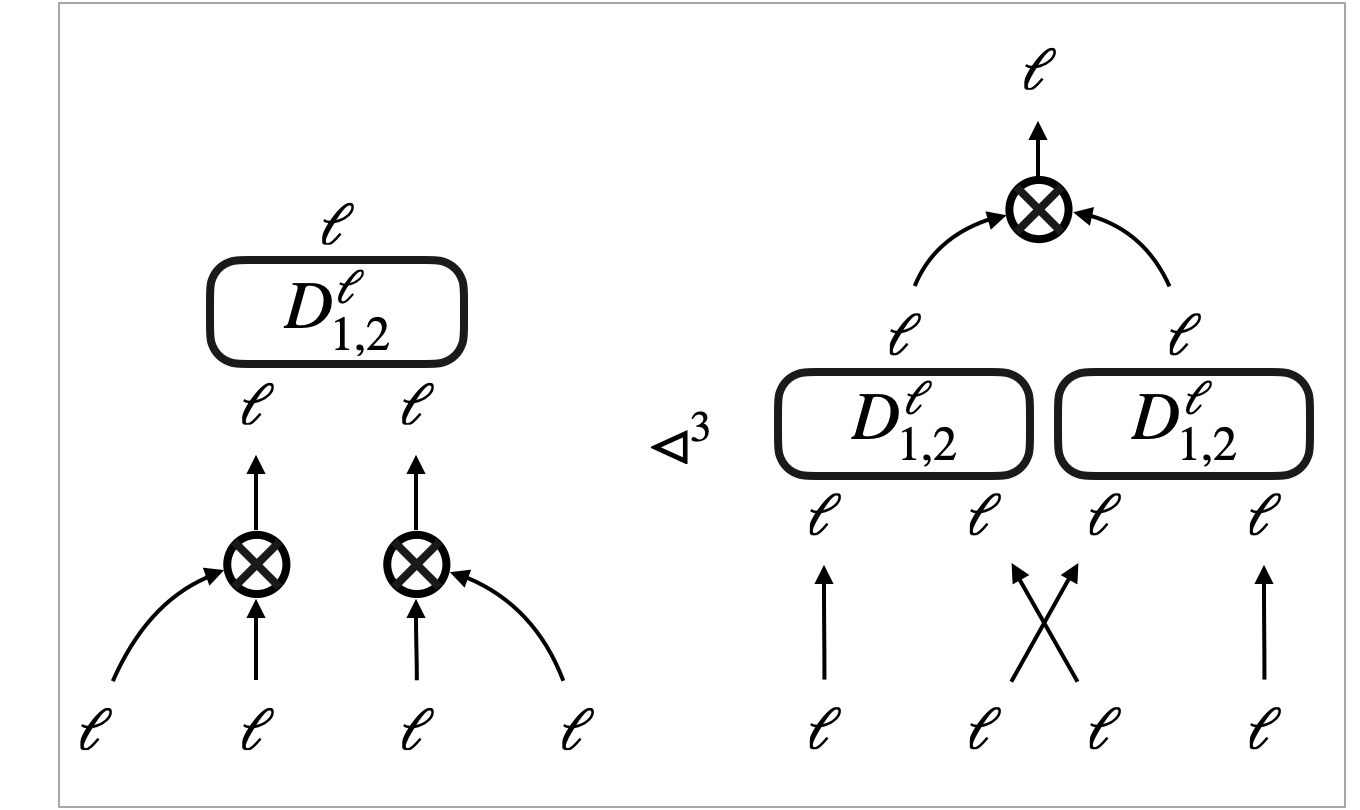} \\
  \includegraphics[scale=.2]{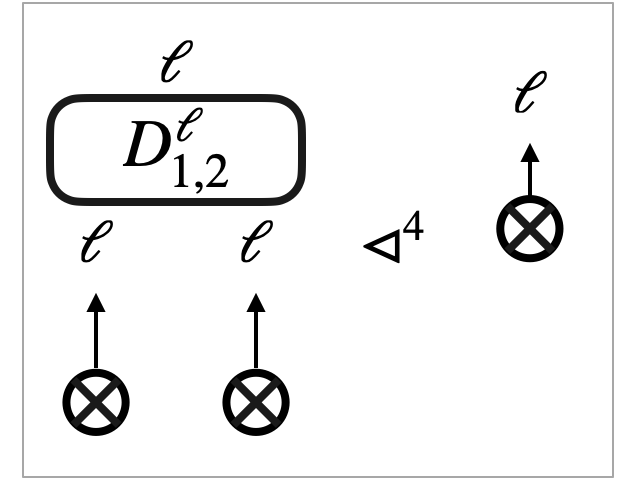} \qquad
  \includegraphics[scale=.2]{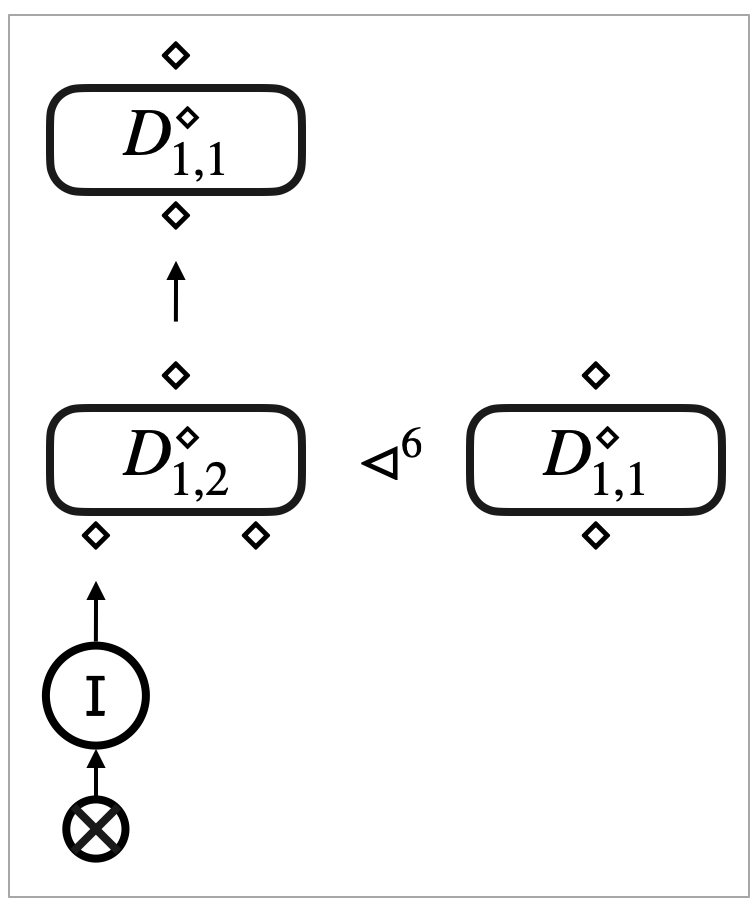} \qquad
  \includegraphics[scale=.2]{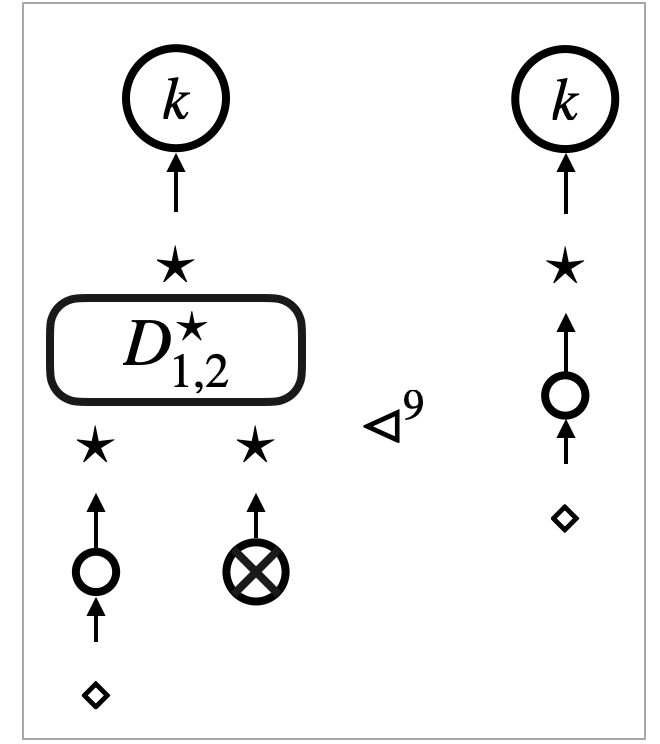}
 \end{gather*}
 \caption{Triggers where $k \in \N$}
 \label{fig:Trigger}
\end{figure}

\begin{table*}
 \centering
 \renewcommand{\arraystretch}{1.8}
 \begin{tabular}{|c||c|c||c|}
  \hline
  & dependency
  & implication of $H_1 \vartriangleleft H_2$
  & used for \\ \hline \hline
  $\vartriangleleft^1$
  & $\vartriangleleft^{\contr \mathrm{Assoc}},
  \vartriangleleft^{\contr \mathrm{Comm}}$
  & $H_1 \simeq^{\mathbb{C}_{\Oprex}}_{=_\N} H_2$
  & $\vartriangleleft^{\contr}$ \\ \hline
  $\vartriangleleft^2$
  & $\vartriangleleft^{\contr \mathrm{Assoc}},
  \vartriangleleft^{\contr \mathrm{Comm}}$
  & $H_1 \simeq^{\mathbb{C}_{\Oprex}}_{=_\N} H_2$
  & $\vartriangleleft^{\contr}$ \\ \hline
  $\vartriangleleft^3$
  & $\vartriangleleft^{\contr \mathrm{Assoc}},
  \vartriangleleft^{\contr \mathrm{Comm}},
  \vartriangleleft^{\contr \mathrm{Idem}}$
  & $H_1 \simeq^{\mathbb{C}_{\Oprex}}_{=_\N} H_2$
  & $\vartriangleleft^{\mathrm{BPullC}}$ \\ \hline
  $\vartriangleleft^4$
  & $\vartriangleleft^{\contr \mathrm{Idem}}$
  & $H_1 \simeq^{\mathbb{C}_{\Oprex}}_{=_\N} H_2$
  & $\vartriangleleft^{\mathrm{BPullW}}$ \\ \hline
  $\vartriangleleft^6$
  & $\vartriangleleft^{\contr \mathrm{Assoc}},
  \vartriangleleft^{\contr \mathrm{Idem}},
  \vartriangleleft^{\mathrm{GC}}$
  & $H_1 \simeq^{\mathbb{C}_{\Oprex}}_{=_\N} H_2$
  & $\vartriangleleft^{\mathrm{Param}}$ \\ \hline
  $\vartriangleleft^9$
  & $\vartriangleleft^{\contr},
  \vartriangleleft^{\mathrm{GC}}$
  & $H_1 \preceq^{\mathbb{C}_{\Oprex}}_{\geq_\N} H_2,\enspace
  H_2 \preceq^{\mathbb{C}_{\Oprex}}_{\leq_\N} H_1$
  & $\vartriangleleft^{\mathrm{Param}}$ \\ \hline
 \end{tabular}
 \caption{Triggers and their implied contextual
 refinements/equivalences}
 \label{tab:TriggerAnalysis}
\end{table*}

In this section we give some details of proving
input-safety and robustness of the pre-templates for the Parametricity law,
as indicated in \autoref{tab:TemplateAnalysis}.
The proofs exploit the transfer properties established in
\autoref{cor:OprexTransferSpecimens}.

\autoref{fig:Trigger} lists triggers that we use to prove
input-safety and robustness of some of the pre-templates\footnote{%
The numbering of triggers is according to the one used
in~\cite[Section~4.5.5]{Muroya20PhD}. Some of triggers in \textit{loc.\ cit.} are
for observational equivalences that we do not consider in this paper, and hence
not presented here.}.
\autoref{tab:TriggerAnalysis} shows contextual
refinements/equivalences implied by these triggers
(in the ``implication'' column), given that some
pre-templates (shown in the ``dependency'' column) imply contextual
refinement as shown in \autoref{tab:TemplateAnalysis}.
All the implications can be proved simply using the congruence
property and transitivity of contextual refinement.
\autoref{tab:TriggerAnalysis} shows which pre-template requires
each trigger in its proof of input-safety or robustness
(in the ``used for'' column).
Note that the converse of any trigger is again a trigger.

Recall that there is a choice of contraction trees upon applying a
contraction rule and some of the local rewrite rules.
The minimum choice is to collect only contraction edges whose target
is reachable from the token target.
The maximum choice is to take the contraction tree(s) so that no
contraction or weakening edge is incoming to the unique hole edge in a
context.

\subsubsection{Pre-templates on contraction trees}

First we check input-safety and robustness of
$\vartriangleleft^{\contr \mathrm{Assoc}}$,
$\vartriangleleft^{\contr \mathrm{Comm}}$ and
$\vartriangleleft^{\contr \mathrm{Idem}}$,
which are all on contraction trees.

Input-safety of
$\vartriangleleft^{\contr \mathrm{Assoc}}$ and
$\vartriangleleft^{\contr \mathrm{Comm}}$ can be checked as follows.
Given a $\mathbb{C}_{\Oprex}$-specimen
$(\focussed{\mathcal{C}};\vec{H^1};\vec{H^2})$ with an entering search
token,
because any input of a contraction tree is a source of a contraction
edge, we have:
\begin{equation*}
 \focussed{\mathcal{C}}[\vec{H^1}] \tobul
  \langle \focussed{\mathcal{C}}[\vec{H^1}] \rangle_{\lightning/?},
  \quad
  \focussed{\mathcal{C}}[\vec{H^2}] \tobul
  \langle \focussed{\mathcal{C}}[\vec{H^2}] \rangle_{\lightning/?}.
\end{equation*}
It can be observed that
a rewrite transition is possible in
$\langle \focussed{\mathcal{C}}[\vec{H^1}] \rangle_{\lightning/?}$
if and only if
a rewrite transition is possible in
$\langle \focussed{\mathcal{C}}[\vec{H^2}] \rangle_{\lightning/?}$.
When a rewrite transition is possible in both states, we can use
\autoref{cor:OprexTransferSpecimens}(\ref{item:Contr}), by considering
a maximal possible contraction rule.
The results of the rewrite transition can be given by a new
quasi-$\mathbb{C}_{\Oprex}$-specimen up to $(=,=)$ (here $=$ denotes
equality on states).
When no rewrite transition is possible, both of the states are not
final but stuck.

Robustness of the three pre-templates and their converse can also be
proved using \autoref{cor:OprexTransferSpecimens}(\ref{item:Contr}),
by considering a maximal possible local (contraction or rewrite) rule
in each case.

\subsubsection{Input-safety of pre-templates not on
contraction trees}

As mentioned in \autoref{sec:using-char-thm},
pre-templates that relate hypernets with no input of type $\star$ are
trivially input-safety for any parameter $(\mathbb{C},Q,Q')$.
This leaves us pre-templates $\vartriangleleft^{\contr}$,
$\vartriangleleft^{\lrapp}$, $\vartriangleleft^{\keyw{ref}}$
and $\vartriangleleft^{\mathrm{Param}}$
to check.

As for $\vartriangleleft^{\contr}$,
note that the pre-template $\vartriangleleft^{\contr}$ relates
hypernets with at least one input.
Any $\mathbb{C}_{\Oprex}$-specimen of $\vartriangleleft^{\contr}$
with an entering search token can be turned into the form
$(\mathcal{C}[(?;_j \chi),\vec{\chi}];H^1,\vec{H^1};H^2,\vec{H^2})$
where $j$ is a positive number.
The proof is by case analysis on the number $j$.
\begin{itemize}
 \item When $j = 1$, we have:
       \begin{align*}
	\mathcal{C}[(?;_j H^1),\vec{H^1}]
	&\tobul \mathcal{C}[(\lightning;_j H^1),\vec{H^1}] \\
	&\to \mathcal{C}[(?;_j H^2),\vec{H^1}].
       \end{align*}
       We can take
       $(\mathcal{C}[(?;_j H^2),\vec{\chi}];\vec{H^1};\vec{H^2})$
       as a $\mathbb{C}_{\Oprex}$-specimen, and
       the token in $\mathcal{C}[(?;_j H^2),\vec{\chi}]$ is not
       entering.
 \item When $j > 1$, the token target must be a source of a
       contraction edge. There exist
       a focus-free context $\mathcal{C}'[\chi']$,
       two focus-free hypernets
       $H'^1 \vartriangleleft^{\contr} H'^2$ and
       a focus-free hypernet $G$, such that
       \begin{align*}
	\mathcal{C}[(?;_j H^1),\vec{H^1}] 
	&\tobul \mathcal{C}[(\lightning;_j H^1),\vec{H^1}] \\
	&\to \mathcal{C}[(?;_j \mathcal{C}'[H'^1]),\vec{H^1}], \\
	\mathcal{C}[(?;_j H^2),\vec{H^2}]
	&\tobul \mathcal{C}[(\lightning;_j H^2),\vec{H^2}] \\
	&\to \mathcal{C}[(?;_j G),\vec{H^2}],
       \end{align*}
       and
       $\mathcal{C'}[H'^2] \dotrel{\simeq}_{=_\N} G$
       given by the trigger $\vartriangleleft^1$
       via \autoref{lem:TriggerToRefinement}.
       The results of these sequences give a
       quasi-$\mathbb{C}_{\Oprex}$-specimen
       up to $(\mathord{=},\mathord{\dot{\simeq}_{=_\N}})$.
\end{itemize}

A proof of input-safety of the operational pre-templates
$\vartriangleleft^{\lrapp}$ and $\vartriangleleft^{\keyw{ref}}$ is a
simpler version of that of $\vartriangleleft^{\contr}$,
because the operational pre-templates relate hypernets with only one
input.

Let $\mathbb{C}$ be either $\mathbb{C}_{\Oprex}$ or
$\mathbb{C}_{\Oprex\cbf}$.
Any $\mathbb{C}$-specimen of an operational pre-template
with an entering search token can be turned into the form
$(\mathcal{C}[(?; \chi),\vec{\chi}];H^1,\vec{H^1};H^2,\vec{H^2})$;
note that the parameter $j$ that we had for
$\vartriangleleft^{\contr}$ is redundant in $?; \chi$.
We have:
\begin{align*}
 \mathcal{C}[(?; H^1),\vec{H^1}]
 &\tobul \mathcal{C}[(\lightning; H^1),\vec{H^1}] \\
 &\to \mathcal{C}[(?; H^2),\vec{H^1}].
\end{align*}
We can take
$(\mathcal{C}[(?; H^2),\vec{\chi}];\vec{H^1};\vec{H^2})$
as a $\mathbb{C}_{\Oprex}$-specimen, and
the token in $\mathcal{C}[(?;_j H^2),\vec{\chi}]$ is not
entering.
This data gives a $\mathbb{C}_{\Oprex\cbf}$-specimen
when $\mathbb{C} = \mathbb{C}_{\Oprex\cbf}$,
which follows from the closedness of $\mathbb{C}_{\Oprex\cbf}$ with
respect to plugging (\autoref{lem:BFCtxtClosedUnderPlugging}).
Note that $?; H^1$ can be seen as a context with no holes, which is
trivially binding-free.

Finally, we look at the parametricity pre-template
$\vartriangleleft^{\mathrm{Param}}$.
Any $\mathbb{C}_{\Oprex}$-specimen of this pre-template,
with an entering search token, can be turned into the form
$(\mathcal{C}[(?;_j \chi),\vec{\chi}];$ $H^1,\vec{H^1};H^2,\vec{H^2})$
where $j$ is a positive number.
The token target is a source of an edge labelled with
$\lambda \in \Oprex_\checkmark$, so we have:
\begin{align*}
 \mathcal{C}[(?;_j H^1),\vec{H^1}] 
 &\tobul \mathcal{C}[(\checkmark;_j H^1),\vec{H^1}], \\
 \mathcal{C}[(?;_j H^2),\vec{H^2}]
 &\tobul \mathcal{C}[(\checkmark;_j H^2),\vec{H^2}].
\end{align*}
The results of these sequences give a
quasi-$\mathbb{C}_{\Oprex}$-specimen
up to $(\mathord{=},\mathord{=})$.

\subsubsection{Robustness of pre-templates not on
contraction trees: a principle}
\label{sec:RobustnessCheckPrinciple}


Robustness can be checked by inspecting rewrite transition
$\focussed{\mathcal{C}}[\vec{H^1}] \to \focussed{N'}$
from the state given by a specimen
$(\focussed{\mathcal{C}};\vec{H^1};\vec{H^2})$ of a pre-template,
where the token of $\focussed{\mathcal{C}}$ is not entering.
We in particular consider the minimum local (contraction or rewrite)
rule $\focussed{G} \mapsto \focussed{G'}$ applied in this transition.
This means that, in the hypernet $\focussed{G}$, every vertex is
reachable from the token target.

The inspection boils down to analyse how the minimum local rule
involves edges that come from the hypernets $\vec{H^1}$.
If all the involvement is deep, i.e.\ only deep edges from $\vec{H^1}$
are involved in the local rule, these deep edges must come via deep
holes in the context $\focussed{\mathcal{C}}$. We can use
\autoref{cor:OprexTransferSpecimens}(\ref{item:Deep}).

If the minimum local rule involves shallow edges that are from
$\vec{H^1}$, endpoints of these edges are reachable from the token
target. This means that, in the context $\focussed{\mathcal{C}}$, some
holes are shallow and their sources are reachable from the token
target.
Moreover, given that the token is not entering in
$\focussed{\mathcal{C}}$, the context has a path from the token target
to a source of a hole edge.

\begin{figure}[h]
 \centering
 \subfloat[A shallow overlap where $C$ is a contraction tree and $B_i$ are box edges
 \label{fig:ShallowOverlapEx}]{
 \includegraphics[scale=.2]{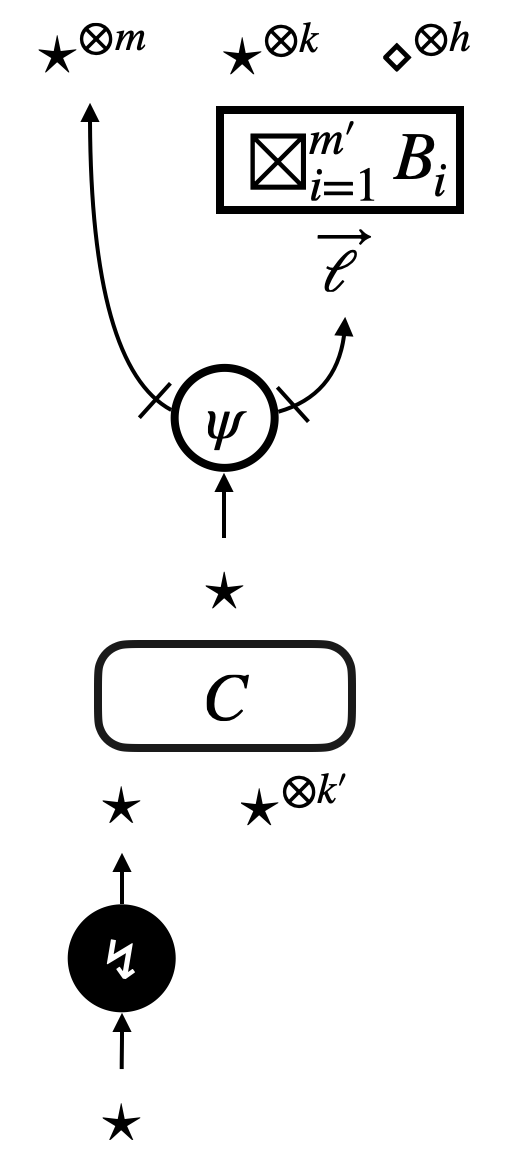}
 }
 \hfil
 \subfloat[A focussed context where $C$ is a contraction tree
 \label{tab:RobustContrCtxt}]{
 \includegraphics[scale=.2]{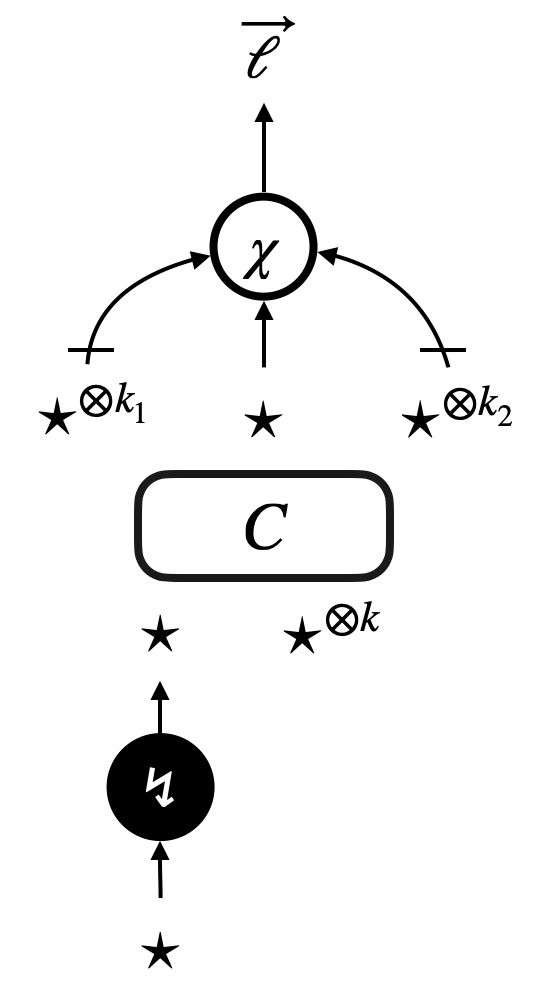}
 }
\end{figure}

For example, in checking robustness of
$\vartriangleleft^{\mathrm{BPerm}}$ with respect to copy transitions,
one situation of shallow overlaps is when $\focussed{G}$ is in the
form in \autoref{fig:ShallowOverlapEx}, and some of the box edges
$B_i$ are from $\vec{H^1}$.
Taking the minimum contraction rule means that $C$ in the graph is a
contraction tree that gives a path from the token target.
This path $C$ followed by the operation edge $\phi$ corresponds to
paths from the token target to hole sources in the context
$\focussed{\mathcal{C}}$.

So, if the minimum local rule involves shallow edges that
are from $\vec{H^1}$, the context $\focussed{\mathcal{C}}$ necessarily
has a path $P$ from the token target to a hole source.
The path becomes a path in the state
$\focussed{\mathcal{C}}[\vec{H^1}]$, from the token target to a source
of an edge $e$ that is from $\vec{H^1}$.
The edge $e$ is necessarily shallow, and also involved in the
application of the minimum local rule, because of the connectivity of
$\focussed{G}$.
Moreover, a source of the edge $e$ is an input, in the relevant
hypernet of $\vec{H^1}$.
By inspecting minimum local rules, we can enumerate possible
labelling of the path $P$ and the edge $e$, as summarised in
\autoref{tab:OverlapPaths}.
Explanation on the notation used in the table is to follow.
\begin{table}[h]
 \centering
 \renewcommand{\arraystretch}{1.2}
 \begin{tabular}{|c|c|c|}
  \hline
  local rule & labels of path $P$ & label of edge $e$ \\ \hline \hline
  contraction
  & $(\contr^\star_\mathsf{C})^+ \cdot \Oprex$ & box \\ \cline{2-3}
  & $(\contr^\star_\mathsf{C})^+$
  & $\contr^\star_\mathsf{C},\ \mathsf{I},\ \Oprex$ \\ \hline
  $\lrapp$
  & $\mathord{\lrapp} \cdot \lambda$ & box \\ \cline{2-3}
  & $\mathord{\lrapp} \cdot (\Oprex_\checkmark)^*$
  & $\Oprex_\checkmark,\ \mathsf{I}$ \\ \hline
  $\keyw{ref}$
  & $\mathord{\keyw{ref}} \cdot (\Oprex_\checkmark)^*$
  & $\Oprex_\checkmark,\ \mathsf{I}$ \\ \hline
  $=$
  & $\mathord{=}$ & $\mathsf{I}$ \\ \cline{2-3}
  & $\mathord{=} \cdot \mathsf{I} \cdot
  (\contr^\diamond_\mathsf{C})^*$
  & $\contr^\diamond_\mathsf{C},\ \circ$ \\ \hline
  $:=$
  & $\mathord{:=} \cdot (\Oprex_\checkmark)^* \cdot \mathsf{I} \cdot
  (\contr^\diamond_\mathsf{C})^*$
  & $\contr^\diamond_\mathsf{C},\ \circ$ \\ \cline{2-3}
  & $\mathord{:=} \cdot (\Oprex_\checkmark)^*$
  & $\Oprex_\checkmark,\ \mathsf{I}$ \\ \hline
  $!$
  & $\mathord{!}$ & $\mathsf{I}$ \\ \cline{2-3}
  & $\mathord{!} \cdot \mathsf{I} \cdot
  (\contr^\diamond_\mathsf{C})^*$
  & $\contr^\diamond_\mathsf{C},\ \circ$ \\ \hline
  $+$ & $+$ & $\N$ \\ \hline
  $-$ & $-$ & $\N$ \\ \hline
  $-_1$ & $-_1$ & $\N$ \\ \hline
 \end{tabular}
 \caption{Summary of paths that witness shallow overlaps}
 \label{tab:OverlapPaths}
\end{table}

We use the regular-expression like notation in
\autoref{tab:OverlapPaths}.
For example, $(\contr^\star_\mathsf{C})^+ \cdot \Oprex$ represents
finite sequences of edge labels, where more than one occurrences of
the label $\contr^\star_\mathsf{C}$ is followed by one operation
$\phi \in \Oprex$.
This characterises paths that inhabit the overlap shown in
\autoref{fig:ShallowOverlapEx}, i.e.\ the
contraction tree $C$ followed by the operation edge $\phi$.
Note that this regular-expression like notation is not a proper
regular expression, because it is over the infinite alphabet
$M_{\Oprex}$, the
edge label set, and it accordingly admits infinite alternation (aka.\
union) implicitly.

To wrap up, checking robustness of each pre-template that is not
on contraction trees boils down to using
\autoref{cor:OprexTransferSpecimens}(\ref{item:Deep}) and/or analysing
the cases enumerated in the table above.

\subsubsection{Robustness of $\vartriangleleft^{\otimes}$ and its converse}

Robustness check of the pre-template $\vartriangleleft^{\contr}$
with respect to copy transitions has two cases. The first case is when
one shallow overlap is caused by a path characterised by
$(\contr^\star_\mathsf{C})^+$, and the second case is when no shallow
overlaps are present and 
\autoref{cor:OprexTransferSpecimens}(\ref{item:Deep}) can be used.


In the first case, namely,
a $\mathbb{C}_{\Oprex}$-specimen with a non-entering rewrite token
can be turned into the form
$(\mathcal{C}[\focussed{\mathcal{C}'}[\chi'],\vec{\chi}];
H^1,\vec{H^1};H^2,\vec{H^2})$
where $j$ is a positive number, and $\focussed{\mathcal{C}'}$ is a
focussed context in the form
of \autoref{tab:RobustContrCtxt}.
A rewrite transition is possible on both states given by the specimen,
in which a contraction rule is applied to
$\focussed{\mathcal{C'}}[H^1]$ and
$\focussed{\mathcal{C'}}[H^2]$.
Results of the rewrite transition give a new
quasi-$\mathbb{C}_{\Oprex}$-specimen.
When $k_1 = 0$, this quasi-specimen is up to
$(\mathord{=},\mathord{\dot{\simeq}_{=_\N}})$,
using the trigger $\vartriangleleft^2$.
When $k_1 > 0$, the quasi-specimen is also up to
$(\mathord{=},\mathord{\dot{\simeq}_{=_\N}})$,
but using the trigger $\vartriangleleft^1$.

Robustness check of the pre-template $\vartriangleleft^{\contr}$
with respect to rewrite transitions always boils down to
\autoref{cor:OprexTransferSpecimens}(\ref{item:Deep}).
This is intuitively because no local rewrite rule of operations
involves any shallow contraction edge of type $\star$.

Robustness of $(\vartriangleleft^{\contr})^{-1}$ can be checked in a
similar manner.
Namely, using \autoref{tab:OverlapPaths}, shallow overlaps are
caused by paths:
\begin{align*}
 &(\contr^\star_\mathsf{C})^+,
 &&\mathord{=},
 && +, \\
 &\mathord{\lrapp} \cdot (\Oprex_\checkmark)^*,
 && \mathord{:=} \cdot (\Oprex_\checkmark)^*,
 && -, \\
 &\mathord{\keyw{ref}} \cdot (\Oprex_\checkmark)^*,
 && !,
 && -_1
\end{align*}
from the token target. All paths but
$(\contr^\star_\mathsf{C})^+$ gives rise to a state that is not
rooted, which can be checked using
\autoref{lem:NotStablePathNotRooted}.
This reduces the robustness check of
$(\vartriangleleft^{\contr})^{-1}$ to that of
$\vartriangleleft^{\contr}$.

\subsubsection[]{Robustness of
$\vartriangleleft^{\mathrm{GC}}$ and its converse}
\label{sec:GCNE2}

These two pre-templates both relate hypernets with no inputs.
Proofs of their robustness always boil down to
the use of \autoref{cor:OprexTransferSpecimens}(\ref{item:Deep}),
following the discussion in \autoref{sec:RobustnessCheckPrinciple}.
Namely, it is impossible to find the path $P$ in the context
$\focussed{\mathcal{C}}$ from the token target to a hole source.

\subsubsection[]{Robustness of
$\vartriangleleft^{\mathrm{BPerm}}$,
$\vartriangleleft^{\mathrm{BPullC}}$ and
$\vartriangleleft^{\mathrm{BPullW}}$, and their converse}

These six pre-templates all concern boxes.
Using \autoref{tab:OverlapPaths}, shallow overlaps are caused by
paths $(\contr^\star_\mathsf{C})^+ \cdot \Oprex$ and
$\mathord{\lrapp} \cdot \lambda$ from the token target.

Robustness check with respect to compute transitions of operations
$\Oprex_\lightning \backslash \{ \lrapp \}$ always boil down to
\autoref{cor:OprexTransferSpecimens}(\ref{item:Deep}).

As for compute transitions of the operation `$\lrapp$', either one
path $\mathord{\lrapp} \cdot \lambda$ causes one shallow overlap, or
all overlaps are deep.
The latter situation boils down to
\autoref{cor:OprexTransferSpecimens}(\ref{item:Deep}).
In the former situation, a micro-beta rule involves one box that is
contributed by a pre-template, and
states given by a $\mathbb{C}_{\Oprex}$-specimen
are turned into a quasi-$\mathbb{C}_{\Oprex}$-specimen up to
$(\mathord{=},\mathord{=})$, by one rewrite transition.

As for copy transitions, there are two possible situations.
\begin{itemize}
 \item Paths $(\contr^\star_\mathsf{C})^+ \cdot \Oprex$ cause some
       shallow overlaps and there are some deep overlaps too.
 \item All overlaps are deep, which boils down to
       \autoref{cor:OprexTransferSpecimens}(\ref{item:Deep}).
\end{itemize}
In the first situation,
some of the shallow boxes duplicated by a contraction rule are
contributed by a pre-template, and other duplicated boxes may have
deep edges contributed by the pre-template.
By tracking these shallow and deep contributions in a contraction
rule, it can be checked that one rewrite transition turns
states given by a $\mathbb{C}_{\Oprex}$-specimen
into a quasi-$\mathbb{C}_{\Oprex}$-specimen.
This quasi-specimen is up to the following, depending on pre-templates:
\begin{itemize}
 \item $(\mathord{=},\mathord{=})$ for
       $\vartriangleleft^{\mathrm{BPerm}}$ and its converse,
 \item $(\mathord{=},\mathord{\dot{\simeq}_{=_\N}})$
       for $\vartriangleleft^{\mathrm{BPullC}}$,
       and $(\mathord{\dot{\simeq}_{=_\N}},\mathord{=})$ for its
       converse,
       using the trigger $\vartriangleleft^3$, and
 \item $(\mathord{=},\mathord{\dot{\simeq}_{=_\N}})$
       for $\vartriangleleft^{\mathrm{BPullW}}$,
       and $(\mathord{\dot{\simeq}_{=_\N}},\mathord{=})$ for its
       converse,
       using the trigger $\vartriangleleft^4$.
\end{itemize}

\subsubsection[]{Robustness of operational pre-templates and their
converse}

For the operational pre-templates and their converse, we use the class
$\mathbb{C}_{\Oprex\cbf}$ of binding-free contexts.
This restriction is crucial to rule out some shallow overlaps.

Using \autoref{tab:OverlapPaths}, shallow overlaps
with the operational pre-templates $\vartriangleleft^{\lrapp}$ and
$\vartriangleleft^{\keyw{ref}}$ are caused by paths
$(\contr^\star_\mathsf{C})^+$ from the token context.
However, the restriction to binding-free contexts makes this situation
impossible, which means the robustness check always boils down to
\autoref{cor:OprexTransferSpecimens}(\ref{item:Deep}) and
\autoref{cor:OprexTransferSpecimens}(\ref{item:DeepBF}).

In checking robustness of the converse
$(\vartriangleleft^{\lrapp})^{-1}$ and
$(\vartriangleleft^{\keyw{ref}})^{-1}$,
shallow overlaps are caused by paths:
\begin{align*}
 &(\contr^\star_\mathsf{C})^+,
 &&\mathord{=},
 && +, \\
 &\mathord{\lrapp} \cdot (\Oprex_\checkmark)^*,
 && \mathord{:=} \cdot (\Oprex_\checkmark)^*,
 && -, \\
 &\mathord{\keyw{ref}} \cdot (\Oprex_\checkmark)^*,
 && !,
 && -_1
\end{align*}
from the token target.
Like the case of $(\vartriangleleft^{\contr})^{-1}$, all paths but
$(\contr^\star_\mathsf{C})^+$ give rise to a state that is not rooted,
which can be checked using \autoref{lem:NotStablePathNotRooted}.
The paths $(\contr^\star_\mathsf{C})^+$ are impossible because of the
binding-free restriction.
As a result, this robustness check also boils down to
\autoref{cor:OprexTransferSpecimens}(\ref{item:Deep}) and
\autoref{cor:OprexTransferSpecimens}(\ref{item:DeepBF}).

\subsubsection{Robustness of the parametricity pre-template
$\vartriangleleft^{\mathrm{Param}}$ and its converse}

These two pre-templates concern lambda-abstractions,
and they give rather rare examples of robustness check where we
compare different numbers of transitions.

Using \autoref{tab:OverlapPaths}, shallow overlaps
with these pre-templates are caused by paths:
\begin{align*}
 &(\contr^\star_\mathsf{C})^+,
 &&\mathord{\lrapp} \cdot (\Oprex_\checkmark)^*, \\
 &\mathord{\keyw{ref}} \cdot (\Oprex_\checkmark)^*,
 && \mathord{:=} \cdot (\Oprex_\checkmark)^*
\end{align*}
from the token target.

As for compute transitions of operations
$\Oprex_\lightning \backslash \{ \lrapp \}$, there are two possible
situations.
\begin{itemize}
 \item Shallow overlaps are caused by paths
       $\mathord{\keyw{ref}} \cdot (\Oprex_\checkmark)^*$ or
       $\mathord{:=} \cdot (\Oprex_\checkmark)^*$.
 \item There is no overlap at all, which boils down to
       \autoref{cor:OprexTransferSpecimens}(\ref{item:Deep}).
\end{itemize}
In the first situation,
a stable hypernet $G_S$ of a local rewrite rule (see e.g.\
\autoref{fig:rewrite-ref}) contains shallow edges,
labelled with $\lambda \in \Oprex_\checkmark$, that are
contributed by a pre-template.
The overlapped shallow contributions are
not modified at all by the rewrite rule, and consequently, one
rewrite transition results in a
quasi-$\mathbb{C}_{\Oprex}$-specimen up to
$(\mathord{=},\mathord{=})$.

As for copy transitions, either one path
$(\contr^\star_\mathsf{C})^+$ causes one shallow overlap, or all
overlaps are deep.
The latter situation boils down to
\autoref{cor:OprexTransferSpecimens}(\ref{item:Deep}).
In the former situation, one lambda-abstraction contributed by a
pre-template gets duplicated.
Namely,
a $\mathbb{C}_{\Oprex}$-specimen with a non-entering rewrite token
can be turned into the form
$(\mathcal{C}[\focussed{\mathcal{C}'}[\chi'],\vec{\chi}];
H^1,\vec{H^1};H^2,\vec{H^2})$
where $\focussed{\mathcal{C}'}$ is a
focussed context in the form of \autoref{tab:RobustContrCtxt}.
There exist a focussed context $\focussed{\mathcal{C}''}$ and
two hypernets $G^1 \vartriangleleft^{\mathrm{Param}} G^2$
such that:
\begin{align*}
 \mathcal{C}[\focussed{\mathcal{C}'}[H^1],\vec{H^1}]
 &\to \mathcal{C}[\focussed{\mathcal{C}''}[G^1],\vec{H^1}], \\
 \mathcal{C}[\focussed{\mathcal{C}'}[H^2],\vec{H^2}]
 &\to \mathcal{C}[\focussed{\mathcal{C}''}[G^2],\vec{H^2}].
\end{align*}
Results of these rewrite transitions give a new
quasi-$\mathbb{C}_{\Oprex}$-specimen up to
$(\mathord{=},\mathord{=})$.

As for compute transitions of the operation `$\lrapp$', there are two
possible situations.
\begin{itemize}
 \item One path $\lrapp$ causes a shallow overlap of the edge that has
       label $\lambda$ and gets eliminated by a micro-beta rewrite rule, and
       possibly some other paths
       $\mathord{\lrapp} \cdot (\Oprex_\checkmark)^*$
       cause shallow overlaps in the stable hypernet $G_S$ (see
       \autoref{fig:rewrite-micro-beta}).
 \item There are possibly deep overlaps, and paths
       $\mathord{\lrapp} \cdot (\Oprex_\checkmark)^*$
       may cause shallow overlaps in the stable hypernet $G_S$.
\end{itemize}

In the second situation, all overlaps are not modified at all
by the micro-beta rewrite rule, except for some deep overlaps turned
shallow. Consequently, one rewrite transition results in a
quasi-$\mathbb{C}_{\Oprex}$-specimen up to
$(\mathord{=},\mathord{=})$.

In the first situation, one lambda-abstraction contributed by the
pre-template is modified, while all the other shallow overlaps (if
any) are not.
We can focus on the lambda-abstraction.
The micro-beta rewrite acts on the lambda-abstraction, an edge labelled with
`$\lrapp$', and the stable hypernet $G_S$.

The involved lambda-abstraction can be in two forms (see
\autoref{fig:param}).
Firstly, it contains function application in its body. Application of the micro-beta
rule discloses the inner function application, whose function side is another
lambda-abstraction that can be related by the pre-template
$\vartriangleleft^{\mathrm{Param}}$ again. As a result, one rewrite transition
yields a quasi-$\mathbb{C}_{\Oprex}$-specimen up to
$(\mathord{=},\mathord{=})$.
Secondly, the involved lambda-abstraction consists of dereferencing `$!$' or
constant `$1$'. Application of the micro-beta rule discloses the dereferencing, or
constant, edge. When it is the dereferencing edge that is disclosed, the micro-beta
rule is followed by a few transitions to perform dereferencing and produce the
same constant `$1$'. As a result, we compare nine transitions with one
transition, and obtain a quasi-$\mathbb{C}_{\Oprex}$-specimen up to
$(\mathord{\dot{\preceq}_{\geq_\N}},\mathord{=})$, using triggers
$\vartriangleleft^6$ and $\vartriangleleft^9$.

The case of the converse of $\vartriangleleft^{\mathrm{Param}}$ is similar.
The only difference is that, in the last situation described above where we
compare nine transitions with one transition and obtain a
quasi-$\mathbb{C}_{\Oprex}$-specimen up to
$(\mathord{\dot{\preceq}_{\geq_\N}},\mathord{=})$, we obtain a
quasi-$\mathbb{C}_{\Oprex}$-specimen up to
$(\mathord{=},\mathord{\dot{\preceq}_{\leq_\N}})$ as a result of the symmetrical
comparison of transitions.

On a final note, let us recall \autoref{sec:design-pre-templates}, where we
observed some situations of robustness for $\vartriangleleft^{\mathrm{Param}}$
using informal reduction semantics on terms.
We namely observed situations where parts related by the pre-template is subject
to the standard call-by-value beta reduction, either as an argument or a
function.
The involvement as a function corresponds to one of the robustness
situations described in this section, namely: a shallow overlap with a micro-beta
rewrite rule that is caused by a path $\lrapp$ and modified by the rewrite rule.
The other involvement, which is as an argument, corresponds to a
combination of two situations described in this section, namely: any overlaps
with a contraction rule, and shallow overlaps with a micro-beta rule that are caused
by paths $\mathord{\lrapp} \cdot (\Oprex_\checkmark)^*$ and preserved. The
combination is due to the fact that the universal abstraction machine decomposes
the beta reduction into the micro-beta rewrite rule and contraction rules, making
substitution explicit and not eager.

\end{document}